%% file: sampling_filter_TR.tex
\documentclass[preprint,12pt,3p]{elsarticle}

\usepackage{amssymb}
\usepackage[colorlinks,bookmarksopen,bookmarksnumbered,citecolor=red,urlcolor=red]{hyperref}
\usepackage{moreverb,dblfloatfix}
\usepackage{algorithmic}
\usepackage{algorithm}
\usepackage{amssymb}
\usepackage{graphicx,color}
\usepackage{float}	
\usepackage{amsmath,amssymb}
\usepackage{subfigure}
\usepackage{rotating}
\usepackage{multirow}
\usepackage{placeins}
\usepackage[normalem]{ulem}
\usepackage{appendix}

\newcommand{\x}{\mathbf{x}}
\newcommand{\xa}{\mathbf{x}^\textnormal{a}}
\newcommand{\xb}{\mathbf{x}^\textnormal{b}}
\newcommand{\xf}{\mathbf{x}^\textnormal{f}}
\newcommand{\xtin}{\mathbf{x}_0}
\newcommand{\xk}{\mathbf{x}_k}
\newcommand{\xbarf}{\overline{\textbf{x}}^{\rm f}}

\newcommand{\y}{\mathbf{y}}
\newcommand{\yk}{\mathbf{y}_k}
\newcommand{\nobs}{\textsc{n}_\textnormal{obs}}
\newcommand{\nens}{\textsc{n}_\textnormal{ens}}
\newcommand{\nvar}{\textsc{n}_\textnormal{var}}
\newcommand{\p}{\mathbf{p}}

\newcommand{\bk}{\mathbf{b}_\textnormal{k}}
\newcommand{\z}{\mathbf{z}}

\journal{QJRMS}

\begin{document}
\include{logo}
\begin{frontmatter}

\title{A Hybrid Monte Carlo Sampling Filter \\ for Non-Gaussian Data Assimilation}

\author[labela]{Ahmed Attia}
\author[labela]{Adrian Sandu}
\address[labela]{Computational Science Laboratory \\
Department of Computer Science   \\
Virginia Polytechnic Institute and State University \\
2201 Knowledgeworks II, 2202 Kraft Drive, Blacksburg, VA 24060, USA \\
Phone: 540-231-2193, Fax: 540-231-9218    \\
E-mail: sandu@cs.vt.edu}

%

%
%

\begin{abstract}
Data assimilation combines information from models, measurements, and priors to estimate the state of a dynamical system such as the atmosphere. 
The Ensemble Kalman filter (EnKF) is a family of ensemble-based data assimilation approaches that has gained wide popularity due its simple formulation, ease of 
implementation, and good practical results.  Most EnKF algorithms assume that the underlying probability distributions are Gaussian. Although this assumption is well accepted, it is too restrictive when applied to large nonlinear models, nonlinear observation operators, and large levels of uncertainty. Several approaches have been proposed in order to avoid the Gaussianity assumption. One of the most successful strategies is the maximum likelihood ensemble filter (MLEF) which computes a maximum a posteriori estimate of the state assuming the posterior distribution is Gaussian. MLEF is designed to work with nonlinear and even non-differentiable observation operators, and shows good practical performance. However, there are limits to the degree of nonlinearity that MLEF can handle.
This paper proposes a new ensemble-based data assimilation method, named the ``\textit{sampling filter}", which obtains the analysis by sampling directly from the posterior distribution. The sampling strategy is based on a Hybrid Monte Carlo (HMC) approach that can handle non-Gaussian probability distributions. 
Numerical experiments are carried out using the Lorenz-96 model and observation operators with different levels of non-linearity and differentiability. 
The proposed filter is also tested with shallow water model on a sphere with linear observation operator.
The results show that the sampling filter can perform well even in highly nonlinear situations were EnKF and MLEF filters diverge. 
\end{abstract}

\begin{keyword}
     {Data assimilation, variational methods, ensemble filters, Markov chain, hybrid Monte-Carlo}
\end{keyword}     

\end{frontmatter}
%
%
\newpage

\tableofcontents
\newpage

\section{Introduction}
\label{sec0}

%
Data assimilation is the process of combining information from models, measurements, and priors - all with associated uncertainties - in order to obtain the best estimate of the state of a physical system.
Two families of methods, variational and ensemble based filters, have proved very successful in real applications. Variational methods, rooted in control theory, require costly developments of tangent linear
and adjoint models~\citep{Kalnay_2002_book}. Ensemble-based sequential data assimilation schemes are rooted in statistical estimation theory. The ensemble Kalman Filter was introduced by Evensen~\citep{Evensen_1994}
and has undergone considerable developments since then. EnKF formulations fall in one  of two classes, namely stochastic or deterministic  formulations~\citep{Tippett_2003_EnSRF}. In the stochastic approach,
each ensemble member is updated using a perturbed version of the observation vector~\citep{Burgers_1998,Houtekamer_1998a}. In the deterministic formulation (which leads to square root 
ensemble filters~\citep{anderson2001ensemble,Bishop_2001a,ott2004local,Tippett_2003_EnSRF,Whitaker_2002a} no observation noise is added, but transformations of the covariance matrix are applied such as to 
recover the correct analysis statistics. 

All variants of the EnKF work well in case of linear observations~\citep{Evensen_2007_book}, however in real applications the observation operators are in general nonlinear.
EnKF can accommodate nonlinear observation operators using linearization, in the spirit of the extended Kalman filter ~\citep{zupanski2005maximum}.
An alternative approach to handle the non-linearity of observation operators is to use the difference between nonlinear operators evaluated at two states instead of the linearized version;
this approach can result in mathematical inconsistencies ~\citep{zupanski2005maximum}. 
A different approach to deal with nonlinear observations is to pose a nonlinear estimation problem  in a subspace spanned by the ensemble members, and to compute the maximum a posteriori estimate in that subspace.
This leads to the maximum likelihood ensemble filter (MLEF) proposed by Zupanski~\citep{zupanski2005maximum}.  MLEF minimizes a cost function that depends on nonlinear observation operators. 
MLEF doesn't require the observation operator to be differentiable and uses a difference approximation of the Jacobian of the observation operator. However, this approach may diverge if the observation operator is highly nonlinear.
In addition it is inherently assumed that the posterior distribution is Gaussian; the MLEF maximum a posteriori probability estimate may face difficulties in case of multimodal distributions. 

The current advances in sampling algorithms make it feasible to directly sample from the posterior probability distribution of the system state. 
A promising step towards efficient sequential Monte Carlo sampling from the posterior density is the implicitly particle filter~\citep{Chorin2010implicit}. 
This algorithm directs the sampling towards the regions of high density areas in the posterior. This helps to control the number of particles in case of of very high dimensional state spaces. 
The implicit sampling filter, however, is expensive: it requires an optimization step for each particle and each ensemble member is generated by solving a set of algebraic equations.

This work seeks to develop an ensemble-based data assimilation filtering technique that can accommodate non-Gaussian posterior distributions and can be efficiently applied in operational situations.
Our approach is based on directly sampling the posterior probability density using a Markov Chain Monte Carlo (MCMC) strategy that generates a Markov chain whose invariant (stationary) distribution is
the target probability density. Specifically, we employ the hybrid Markov Chain Monte Carlo (HMCMC) algorithm, a variant of MCMC sampling that incorporates an auxiliary variable and takes advantage of
the properties of Hamiltonian system dynamics~\citep{duane1987hybrid}.
This sampling scheme turns out to be very useful in case of complex high dimensional distribution. The new fully nonlinear sampling filter can accommodate nonlinear
observation operators and it does not require the target probability distribution to be Gaussian.

The paper is organized as follows. An overview of data assimilation problem and widely-used solution strategies is given in Section \ref{sec:DA}. 
     Sampling MCMC and HMC algorithms are summarized in Section \ref{sec:MCMC}.
     The proposed sampling filter is presented in Section \ref{sec:Sampling_filter}.
     Numerical experiments, and a comparison of the sampling filter against traditional EnKF and MLEF methods, are given in Section \ref{sec:Results}.
     Conclusions are drawn in Section \ref{sec:conclusions}.

\section{Data Assimilation}\label{sec:DA}

This section provides a brief overview of the data assimilation (DA) problem and of several solution strategies, and highlights the motivation behind the present research.

     \subsection{Problem formulation}\label{subsec:DA_formulation}
     Data assimilation combines information from prior (background) knowledge, a numerical model, and observations, all with associated errors, to obtain a
     statistically best estimate of the state $\x^{\rm true}$ of a physical system.

The background represents the best estimate of the true state \textit{prior}
to any measurement being available. The background errors (uncertainties) are generally assumed to have a Gaussian distribution
${\xb - \x^{\rm true}}  \in \mathcal{N}(0,\mathbf{B})$, where $\mathbf{B}$ is the background error covariance matrix.
The Gaussian assumption is widely used and we will follow it as well.

The numerical model propagates the initial model state (initial condition) $\xtin\in \mathbb{R}^{\nvar}$ at time $t_0$ to future states 
$\xk \in \mathbb{R}^{\nvar}$ at times $t_k$:
\begin{equation}
\label{eqn:forward_model}
   \xk = \mathcal{M}_{t_0 \rightarrow t_k} (\xtin)\ ,\ \ t_0 \leq t_k \leq t_F \,,
\end{equation}
where $t_0$ and $t_F$ are the beginning and the end points of the simulation time interval.    
The model solution operator $\mathcal{M}$ represents, for example, a discrete approximation of the partial differential equations that govern the evolution of the 
dynamical system (e.g., the atmosphere). The state space is typically large, e.g., $\nvar\sim 10^6 - 10^9$ variables for atmospheric simulations.

Small perturbations $\delta \x$ of the state of the system evolve according to the tangent linear model: 
\begin{equation}
\label{eqn:tangent_forward_model}
   \delta \xk = \mathbf{M}_{t_0 \rightarrow t_k} (\xtin ) \cdot \delta \xtin\ ,\ \ t_0 \leq t_k \leq t_F,
\end{equation}
where $\mathbf{M} = \mathcal{M}'$ is the linearized model solution operator.

Observations of the true state are available at discrete time instants $t_k$,  $t_0 \leq t_k \leq t_F$, 
\[
\yk=\textbf{y}(t_k) = \mathcal{H}_k(\xk) + \varepsilon_k, \quad k = 0,1,\ldots,\nobs-1.
\] 
The observation operator $\mathcal{H}_k$ maps the state space to the observation space at time $t_k$. The observations are corrupted by measurement and representativeness errors~\citep{Cohn_1997}, 
which are also assumed to have a normal distribution, $\varepsilon_k \in \mathcal{N}(0,\mathbf{R}_k)$, where $\mathbf{R}_k$ is the observation error   covariance matrix at time $t_k$.

Data assimilation combines the background estimate, the measurements , and the model to obtain an improved estimate
$\xa$, called the ``\textit{analysis}'' (or \textit{posterior}), of the true state $\x^{\rm true}$. 
Two approaches for solving the data assimilation problem have gained widespread popularity, variational and ensemble-based methods. The sampling filter proposed in this  paper belongs to the latter family. 
We will compare the new methodology with two existing algorithms in this family, the ensemble Kalman filter and the maximum likelihood ensemble filter, which are reviewed next.

     \subsection{The ensemble Kalman filter}\label{subsec:EnKF}
 Kalman filters (KF)~\citep{Kalman_1960,Kalman_1961} are sequential data assimilation methodologies, where measurements are incorporated 
 at the time moment when they become available.   Sequential data assimilation algorithms proceed in two steps, namely, \textit{forecast} and \textit{analysis}. In the forecast step, the state of the system is propagated forward by the model equations \eqref{eqn:forward_model} to the next time point where observations are available, producing a forecast of the state of the system, and a forecast error covariance matrix is presented to quantify the uncertainty of the forecast.    
 
The ensemble Kalman filter (EnKF)~\citep{Burgers_1998,Evensen_1994,Evensen_2003,Houtekamer_1998a}  takes a Monte-Carlo approach to representing the uncertainty. An ensemble of $\nens$ states ($\xa_{k-1}(e)$, $e = 1,\ldots,\nens$) is used to sample the analysis probability distribution at time $t_{k-1}$. Each member of the ensemble is propagated to $t_k$ using the nonlinear model \eqref{eqn:forward_model}  to obtain the "forecast" ensemble 
  \begin{subequations}
     \label{eqn:local_EnKF}
     \begin{equation}
    \label{eq:EnKF_forecast}
 \xf_k(e) = \mathcal{M}_{t_{k-1}\rightarrow t_k}(\xa_{k-1}(e)) + \eta_k(e),\ \ e=1,\ldots, \nens .
\end{equation}
To simulate the fact that the model is an imperfect representation of reality model errors are added. They are typically considered Gaussian random variables, $\eta_k \in \mathcal{N}(0,\mathbf{Q}_k)$. The ensemble mean and covariance approximate the background estimate and the background error covariance of the state at the next time point $t_k$:
    \begin{eqnarray}
 \xbarf_k   &=&  \frac{1}{\nens} \sum_{e=1}^{\nens}{\xf_k(e) } \,, \\
 \mathbf{X}^{\rm f}_k   &=&  [\xf_k(1)- \xbarf_k, \ldots,  \xf_k(\nens)- \xbarf_k] \,, \\
   \mathbf{B}_k   &=&  \left( \frac{1}{\nens-1} \left( \mathbf{X}^{\rm f}_k \left( \mathbf{X}^{\rm f}_k \right)^T \right) \right) \circ \rho. \label{eq:EnKF_localization}
     \end{eqnarray}
 \end{subequations}
To reduce sampling error due to the small ensemble size, localization ~\citep{Hamill_2001,Houtekamer_2001,Whitaker_2002a} is performed by taking the point-wise product of the ensemble covariance and a  decorrelation  matrix $\rho$.

Each member of the forecast (ensemble of forecast states $\{ \xf_k(e) \}_{e=1,\ldots,\nens}$) is analyzed separately using the Kalman filter formulas~\citep{Burgers_1998,Evensen_1994}
 \begin{subequations}
\label{eqn:EnKF_Analysi_and_gain}
\begin{eqnarray}
    \xa_k(e)  &=&  \xf_k(e) + \mathbf{K}_k \left( \left[\yk + \zeta_k(e)\right] - \mathcal{H}_k(\xf_k(e)) \right),\ \\
    \mathbf{K}_k    &=&  \mathbf{B}_k \mathbf{H}^T_k { \left(\mathbf{H}_k \mathbf{B}_k \mathbf{H}^T_k + \mathbf{R}_k \right)}^{-1}.
\end{eqnarray}
 \end{subequations}
The stochastic (``perturbed observations'' ) version~\citep{Burgers_1998} of the ensemble Kalman filter adds a different realization of the observation noise $\zeta_k \in \mathcal{N}(0,\mathbf{R}_k)$ to each individual assimilation. The Kalman gain matrix $\mathbf{K}_k$ makes use of the linearized observation operator
$\mathbf{H}_k = \mathcal{H}_k'(\overline{\x}^{\rm f}_k) $. The same Kalman gain is used for all ensemble members.

 Square root versions (deterministic formulations) of EnKF~\citep{Tippett_2003_EnSRF} avoid adding random noise to observations, and thus avoid additional sampling errors. They also avoid the explicit construction of the full covariance matrices and work by
 updating only a matrix of state deviations from the mean. A detailed discussion of EnKF and variants can be found in~\citep{Evensen_2007_book}.
 
 The main shortcomings of the ensemble Kalman filter are the Gaussianity assumption on which the Kalman updates are based. The filter is optimal only when the observation operators are linear, and both the forecast and the observation errors   are Gaussian.    
 

     \subsection{The maximum likelihood ensemble filter}\label{subsec:MLEF}
The maximum likelihood ensemble filter (MLEF)~\citep{zupanski2005maximum} seeks to alleviate the limitations of the Gaussian assumptions by computing the maximum likelihood estimate of the state in the ensemble space. Specifically, it maximizes the posterior probability density, or equivalently, minimizes the following nonlinear objective  function over the ensemble subspace~\citep{lorenc1986analysis,zupanski2005maximum}: 
 \begin{subequations}
 \label{eqn:MLEF-problem}
  \begin{eqnarray}
 \label{eqn:MLEF-optimization}
 \x_k^{\rm opt} &=& \arg\min_\x\, \mathcal{J}(\x), \\
 \label{eqn:MLEF-costfun}
 \mathcal{J}(\x) &=& \frac{1}{2} {( \x - \xb_k )^T\, \mathbf{B}_k^{-1}\, ( \x - \xb_k )}  \\
    &+& \frac{1}{2} {( \y_k - \mathcal{H}_k(\x) )^T\, \mathbf{R}_k^{-1}\, ( \y_k - \mathcal{H}_k(\x)  ) }\,, \nonumber
 \end{eqnarray}
 \end{subequations}
 and then updates the analysis error covariance matrix based on the fact that it is approximately equal to the inverse of the Hessian matrix
 at the minimum~\citep{fisher1995estimating}.
 
 The MLEF algorithms operates sequentially by applying a forecast step and an analysis step. Let $\x^{\rm opt}_{\rm k-1}$ be the optimal solution
 at the previous time point $t_{k-1}$, and let 
 \begin{equation}
  \mathbf{A}_{k-1}^{1/2} = \left[ \mathbf{a}_{k-1}(1),\mathbf{a}_{k-1}(2),\ ,\ldots, \mathbf{a}_{k-1}({\nens})  \right] \,,
 \end{equation}
 be the matrix of scaled perturbations corresponding to the analysis ensemble at $t_{k-1}$, such that the analysis covariance matrix
 is $\mathbf{A}_{k-1} = \mathbf{A}_{k-1}^{1/2} \mathbf{A}_{k-1}^{T/2}$.
 
 %
 %
 %
 The forecast step provides the background state $\xb_k$ and a square root of the background covariance matrix at the current time point $t_k$ as follows:
 \begin{subequations}
\begin{eqnarray}
 \xb_k     &=& \mathcal{M}_{t_{k-1} \rightarrow t_k}(\x^{\rm opt}_{\rm k-1}) \,, \\
\bk(e)     &=& \mathcal{M}_{t_{k-1} \rightarrow t_k}(\x^{\rm opt}_{\rm k-1} + \mathbf{a}_{\rm k-1}(e))  \\
           &-& \mathcal{M}_{t_{k-1} \rightarrow t_k}(\x^{\rm opt}_{\rm k-1});~~e=1,\ldots,\nens \,, \nonumber \\
  \mathbf{B}_k^{1/2} &=& \left[ \bk(1),\ \bk(2),\ ,\ldots, \bk({\nens})  \right].
\end{eqnarray}
 \end{subequations}
 To speed up the optimization problem \eqref{eqn:MLEF-optimization}
 Hessian preconditioning is carried out through the change of variables
 \begin{equation}
 \label{eqn:change-of-vars}
  \x_k(\xi) = \xb_k + \mathbf{B}_k^{1/2} \bigl(\mathbf{I+C}(0)\bigr)^{-\frac{T}{2}} \xi,
 \end{equation}
 where $\xi$ is a vector of control variables in the ensemble space and
 \begin{subequations}
\begin{eqnarray}
 \mathbf{C}(\zeta)  &=& \mathbf{Z}(\zeta)^T \,  \mathbf{Z}(\zeta), \\
\mathbf{Z}(\zeta)   &=& \left[ \z(\zeta,1),\ \z(\zeta,2),\ \ldots,\ \z(\zeta,{\nens}) \right] \,, \\
     \z(\zeta,e)   
 &=& \mathbf{R}_k^{-\frac{1}{2}} \left( \yk - \mathcal{H}_k \bigl(\x_k(\zeta)\bigr) \right)   \\
     &-&  \mathbf{R}_k^{-\frac{1}{2}} \left(\yk - \mathcal{H}_k \bigl(\x_k(\zeta) + \bk\left(e\right) \bigr) \right) \,.    \label{eqn:MLEF_Z}  \nonumber
\end{eqnarray}
 \end{subequations}
 The matrix $\mathbf{C}(0)$ in \eqref{eqn:change-of-vars} is obtained by using $\x_k(0) \equiv \xb_k$ in formula \eqref{eqn:MLEF_Z}. 
  
 After replacing \eqref{eqn:change-of-vars} in \eqref{eqn:MLEF-costfun} 
 the optimal solution is found by solving the following minimization problem in the ensemble subspace:
 \begin{eqnarray}
 \xi^{\rm opt} &=& \arg\min\, \mathcal{J}(\xi),  \\
    \mathcal{J}(\xi) &=& \frac{1}{2} \xi^T \bigl(\mathbf{I+C}(0)\bigr)^{-1} \xi  \\ 
                     &+& \frac{1}{2}  \left(\yk-\mathcal{H}_k\left(\x_k(\xi) \right) \right)^T \mathbf{R}_k^{-1} \left(\yk-\mathcal{H}_k\left(\x_k(\xi) \right) \right). \nonumber
 \end{eqnarray}
 The gradient reads:
 \begin{eqnarray}
   \nabla_{\xi} \mathcal{J}(\xi) 
 &=& \bigl( \mathbf{I+C}(0) \bigr)^{-1} \xi  \\
 &-& \bigl( \mathbf{I+C}(0) \bigr)^{-1/2}\, \mathbf{Z}(\xi)^T\, \mathbf{R}_k^{-\frac{1}{2}} \left(\yk-\mathcal{H}_k\left(\x_k(\xi) \right) \right). \nonumber
 \end{eqnarray}
 %
 %
 The optimal solution in the model subspace is given by:
 \begin{equation}
    \x^{\rm opt}_k = \xb_k + \mathbf{B}_k^{1/2} \, \bigl( \mathbf{I+C}(0) \bigr)^{-\frac{T}{2}} \xi^{\rm opt}.
 \end{equation}
 The matrix of scaled perturbations representing the analysis is updated as:
 \begin{subequations}
 \begin{eqnarray}
    \mathbf{A}_k^{1/2} &=&  \mathbf{B}_k^{1/2} \bigl(  \mathbf{I}+\mathbf{C}(\xi^{\rm opt}) \bigr)^{-\frac{T}{2}}.
 \end{eqnarray}
 \end{subequations}
 %
  
 An important advantage of the algorithm is that the observation operator is not linearized. 
 Consequently MLEF can work efficiently with non-linear observation operators (without the requirement of differentiability and without using
 finite-difference approximations of the Jacobian of the observation operators)~\citep{zupanski2005maximum}.) 
 The cost function \eqref{eqn:MLEF-costfun} to minimize implicitly assumes that the posterior distribution is Gaussian. The method is unlikely to give good results when the posterior distributions are multimodal.

\section{Hybrid Markov Chain Monte Carlo}\label{sec:MCMC}
     Markov Chain Monte Carlo (MCMC) algorithms~\citep{neal1993probabilistic},  introduced by Metropolis \textit{et. al}~\citep{metropolis1953equation}, can sample from distributions with complex probability densities $\pi(\x)$. They generate a Markov chain $\{\x(i)\}_{i\geq 0}$ for which $\pi(\x)$
     is the invariant (stationary) distribution, given that $\pi(\x)$ is known up to a multiplicative constant~\citep{neal1993probabilistic}.
     MCMC methods work by generating a random walk using a proposal PDF and an ``acceptance/rejection" criterion to decide whether proposed samples
     should be accepted as part of the Markov chain or should just be rejected. These algorithms  are generally powerful, but may take a long time
     to explore the whole state space or even to converge~\citep{tierney1994Markov}. This section starts with a review of the Hybrid  MCMC sampling (HMCMC) then presents the sampling filter algorithm for data assimilation.

  Hybrid Monte Carlo (HMC) methods, also known an Hamiltonian Monte Carlo, originated in the physics literature~\citep{duane1987hybrid}. They attempt to handle the drawbacks of MCMC algorithms by incorporating an auxiliary variable such as to reduce the correlation
  between successive samples, to explore the entire space in very few steps, and to ensure high probability of acceptance for proposed samples
  in high dimensions~\citep{sanz2014Markov}.

%
     
\subsection{Hamiltonian dynamics}\label{subsec:Hamiltonian_Eqns}
 Hamiltonian dynamical systems operate in a phase space of points  $(\p,\x) \in \mathbb{R}^{2\nvar}$, where the individual variables are the position $\x \in \mathbb{R}^{\nvar}$ and the momentum $\p \in \mathbb{R}^{\nvar}$. The total energy of the system is described by the Hamiltonian  function $H(\p,\x)$.
 The dynamics of the system in time is described by the following ordinary differential equations:
\begin{eqnarray}
 \label{eqn:hamiltonian_equations}
\frac{d\x}{dt} = \nabla_\p\, H\,,\qquad
\frac{d\p}{dt} = - \nabla_\x\, H.
\end{eqnarray}
The time evolution of the system \eqref{eqn:hamiltonian_equations}  in state space is described by the flow ~\citep{neal2011mcmc,sanz1994numerical}
 \begin{equation}
 \label{eqn:hamiltonian_flow}
 \Phi_T:\mathbb{R}^{2\nvar} \rightarrow \mathbb{R}^{2\nvar}, \quad \Phi_T\bigl(\p(0),\x(0)\bigr)=\bigl(\p(T),\x(T)\bigr),
 \end{equation}
 which maps the initial state of the system $(\p(0),\x(0))$ to $(\p(T),\x(T))$, the state of the system at time $T$.
  
In practical computations the analytic flow $\Phi_T$ is replaced by a numerical solution using a time reversible and symplectic numerical integration method  ~\citep{sanz1994numerical,sanz2014Markov}. 
In this paper we use five different high order symplectic integrators based on Strang's splitting formula~\citep{sanz1994numerical}: Verlet (St{\"o}rmer, Leapfrog) algorithm \eqref{eqn:Verlet}~\citep{sanz1994numerical,sanz2014Markov}, 
higher order integrators namely, two-stage \eqref{eqn:two_stage}, three-stage \eqref{eqn:three_stage}, and four-stage \eqref{eqn:four_stage} position splitting integrators from~\citep{blanes2013numerical}, 
and the Hilbert space integrator \eqref{eqn:hilbert_integrator} from~\citep{beskos2011hybrid}. The methods are summarized in \ref{sec:numerical-integrators}.
To approximate $\Phi_T$ the integrator at hand takes $m$ steps of size $h=T/m$. With a slight abuse of notation we will also denote by $\Phi_T$ the flow of the numerical solution.

\subsection{HMCMC sampling algorithm}\label{subsec:HMCMC_Sampling}
In order to draw samples $\{\x(e)\}_{e\geq 0}$ from a given probability distribution $\pi(\x)$ HMC makes the following analogy with a Hamiltonian mechanical system \eqref{eqn:hamiltonian_equations}. The state $\x$ is viewed as a ``position variable'',  and an auxiliary ``momentum variable'' $\p$ is included.
The Hamiltonian function of the system is:
\begin{equation}     
\label{eqn:hamiltonian_function}
     H(\p,\x)  = \frac{1}{2} \, \p^T \mathbf{M}^{-1} \p -\log(\pi(\x)) = \frac{1}{2} \, \p^T \mathbf{M}^{-1} \p + \mathcal{J}(\x).
\end{equation}  
The negative logarithm of the target probability density $\mathcal{J}(\x)=-\log(\pi(\x))$ is viewed as the potential energy of the system. The kinetic energy of the system is given by the auxiliary momentum variable $\p$. The constant positive definite symmetric ``mass matrix'' $\mathbf{M}$ is yet to be defined~\citep{sanz1994numerical}.
Based on the Hamiltonian equations \eqref{eqn:hamiltonian_equations} the dynamics of the system is given by
\begin{eqnarray}
\label{eqn:hamiltonian_vector_dynamics}
\frac{d\x}{dt} =  \mathbf{M}^{-1} \p \,, \qquad
\frac{d\p}{dt} = -\nabla_\x \mathcal{J}(\x). 
\end{eqnarray}

The canonical  probability distribution of the state of the system $(\p,\x)$ in the phase space $\mathbb{R}^{2\nvar}$ is, up to a constant, equal to
 \begin{eqnarray}
 \label{eqn:Canonical_Pdf}
   \exp{( - H(\p,\x) )}  &=&  \exp{\left( -\frac{1}{2} \p^T \mathbf{M}^{-1} \p - \mathcal{J}(\x) \right)}  \\
    &=& \exp{\left( -\frac{1}{2} \p^T \mathbf{M}^{-1} \p \right)}\cdot \pi(\x). \nonumber
 \end{eqnarray}
 The product form of this joint probability distribution shows that the two variables $\p, \x$ are independent~\citep{sanz2014Markov}. 
 The distribution of the momentum variable is Gaussian, $\p \sim \mathcal{N}(0,\mathbf{M})$, 
 while the distribution of the position variable is the target probability density, $\x \sim \pi$~\citep{sanz2014Markov}.
 
The HMC sampling algorithm builds a Markov chain starting from an initial state $\xtin=\x(0)$.
Algorithm \ref{alg:HMC_sampling} summarizes the transition from the 
current Markov chain state $\x_k$ to a new state $\x_{k+1}$~\citep{sanz2014Markov}.
Practical issues are related to the choice of the numerical integrator, the time step, and the choice of the function $\mathcal{J}(\x)$
 that represents the PDF we wish to sample from. The construction of the mass matrix $\mathbf{M}$ does not impact the final distribution, but does affect the computational performance of the algorithm~\citep{girolami2011riemann}. The mass matrix $\mathbf{M}$ is symmetric and positive definite and is a parameter that is tuned by the user.
 It can be for example, a constant multiple of the identity~\citep{neal2011mcmc}, or a diagonal matrix whose entries are the background error variances~\citep{beskos2011hybrid,liu2008monte}. 
 We found that the latter approach is more efficient for the current application and used it in all numerical experiments reported here.

 \begin{algorithm}[H]
     \begin{algorithmic}[1]
    \STATE Draw a random vector $\p_k \sim \mathcal{N}(0,\mathbf{M})$.
    \STATE Use a symplectic numerical integrator (from \ref{sec:numerical-integrators}) to advance the current state $(\p_k,\x_k )$
     by a time increment $T$ to obtain a \textit{proposal} state $( \p^* , \x^* )$: 
     \begin{equation}
            ( \p^* , \x^* ) = \Phi_T\bigl(( \p_k , \x_k )\bigr).
     \end{equation}
    \STATE Evaluate the loss of energy based on the Hamiltonian function.
    For the standard Verlet \eqref{eqn:Verlet},  two-stage \eqref{eqn:two_stage}, three-stage \eqref{eqn:three_stage}, and four-stage \eqref{eqn:four_stage}  integrators~\citep{blanes2013numerical,sanz2014Markov} the loss of energy is computed as:
  \begin{equation}
  \label{eqn:loss_of_Energy_Verlet}
     \Delta H =  H( \p^* , \x^* ) - H( \p_k, \x_k ).
  \end{equation} 
  For the Hilbert space integrator  \eqref{eqn:hilbert_integrator} ~\citep{beskos2011hybrid}  the loss of energy is computed as:
  \begin{eqnarray}
  \label{eqn:loss_of_Energy_Hilbert}
     \Delta H &=& \phi(\x^*) - \phi(\x_k) \\
               &+& \frac{h^2}{8} \left( | \mathbf{M}^{-\frac{1}{2}} (-\nabla \phi(\x_k))|^2  - | \mathbf{M}^{-\frac{1}{2}} (-\nabla \phi(\x^{*}))|^2   \right)   \nonumber \\
 &+& h \sum_{i=1}^{m-1}{\left( \p_k^T  \left(-\nabla \phi(\x_k) \right)  \right) }     \nonumber \\
 &+& \frac{h}{2} \left( \p_k^T  \left(-\nabla \phi(\x_k) \right)  +  \left(\p^{*} \right)^T  \left(-\nabla \phi(\x^{*}) \right)   \right)   \,,  \nonumber
  \end{eqnarray}
  where $\phi(\x) = -\log{(\pi(\x))} $ and $h=T/m$ is the integration time step~\citep{sanz2014Markov}.
     \STATE Calculate the probability:
     \begin{equation}
     \label{eqn:acceptance_probability}
 a^{(k)} = 1 \wedge e^{-\Delta H}.
     \end{equation}
    \STATE Discard both $\p^*,\ \p_k$.
    \STATE \textbf{(Acceptance/Rejection)} Draw a uniform random variable $u^{(k)}\sim \mathcal{U}(0,1)$:
     \begin{enumerate}
  \item[i-  ] If $a^{(k)} > u^{(k)}$ accept the proposal as the next sample: $\x_{k+1} := \x^*$;
  \item[ii- ] If $a^{(k)} \leq u^{(k)}$ reject the proposal and continue with the current state: $\x_{k+1} := \x_k$.
     \end{enumerate}
    \STATE Repeat steps $1$ to $6$ until sufficiently many distinct samples are drawn.
     \end{algorithmic}
     \caption{HMCMC Sampling~\citep{sanz2014Markov}.}
 \label{alg:HMC_sampling}
 \end{algorithm}
 %
 %
%
\section{The Sampling Filter for Data Assimilation}
\label{sec:Sampling_filter}
The goal of this filter is to replace the analysis step in the traditional EnKF with a resampling procedure that draws representative ensemble members from
the posterior distribution $ \pi(\x) = \mathcal{P}^{\rm a}(\x)$. Even if the posterior may in general be non-Gaussian we 
assume, as most of the current ensemble-based data assimilation  algorithms, that the posterior has the form: 
\begin{eqnarray}
  \pi(\x) & = &      \mathcal{P}^{\rm a}(\x) \, \propto \, \exp{\Bigl( - \mathcal{J}(\x) \Bigr)} \,, \\
  \mathcal{J}(\x) &=& \frac{1}{2} {\left( \x - \xb \right)^T \mathbf{B}^{-1} \left( \x - \xb \right)}  \\
                  &+& \frac{1}{2} {\Bigl( \y - \mathcal{H}(\x)  \Bigr)^T \mathbf{R}^{-1} \Bigl( \y - \mathcal{H}(\x) \Bigr) } \, . \nonumber
\end{eqnarray}
where $\xb$ is the background state (forecast), $\y$ is the observation vector, and $\mathcal{H}$ is the observation operator that is generally non-linear.
     
For sampling at time $t_k$ the corresponding $\mathcal{J}(\x)$ is:
     \begin{eqnarray}
     \label{eqn:J-sampling-filter}
  \mathcal{J}(\x) &=& -\log{\Bigl(\mathcal{P}^{\rm a}(\x) \Bigr)}  \\
                  &=& \frac{1}{2} {\Bigl( \x - \xb_k \Bigr)^T\, \mathbf{B}^{-1}_k\, \Bigl( \x - \xb_k \Bigr)} \\
                  &+& \frac{1}{2} \, \Bigl( \y_k - \mathcal{H}_k(\x)\Bigr) ^T\, \mathbf{R}^{-1}_k\, \Bigl( \y_k - \mathcal{H}_k(\x)\Bigr) \, ,  \nonumber
     \end{eqnarray}     
    and its gradient has the form
     \begin{equation}
     \label{eqn:J-gradient-sampling-filter}
          \nabla_\x \mathcal{J}(\x) = \mathbf{B}^{-1}_k\, ( \x - \xb_k ) + \mathbf{H}_k^T \, \mathbf{R}^{-1}_k\, \bigl( \y_k - \mathcal{H}_k(\x)\bigr)\, ,
     \end{equation}
     where $\mathbf{H}_k = \mathcal{H}_k'(\x) $ is the linearized observation operator.

Algorithm \eqref{alg:HMC_sampling} is used to generate $\nens$ ensemble members drawn from the posterior distribution $\{ \xa_k(e)\sim \mathcal{P}^{\rm a}(\x)\}_{e=1,2,\ldots ,\nens}$.  The mean of this ensemble is an estimate of the \textit{analysis} state, and the ensemble covariance estimates the analysis error covariance matrix.
Note that the proposed sampling filter is not restricted to a specific form of the posterior PDF, and the Gaussian assumption \eqref{eqn:J-sampling-filter} can in principle be removed.  The remaining issue is to represent non-Gaussian probability density functions and their logarithm. In the next section we describe the proposed sampling filter as an alternative to the EnKF.
     diagonal

The sampling filter is described in Algorithm \ref{alg:sampling_filter}. Like most of the ensemble-based sequential data assimilation algorithms the sampling filter consists of two stages, namely, the \textit{forecast} step and the  \textit{analysis} step.
 
 Start with an ensemble $\{ \xa_{k-1}{(e)}\}_{e=1,\ldots ,\nens}$ describing the analysis PDF at time
 $t_{k-1}$. In the forecast step each ensemble member is propagated by the full model to the next time $t_{k-1}$ where observations are available, resulting in the forecast ensemble. In the analysis step the HMCMC algorithm is simply used to sample from the posterior PDF of the state, providing the new analysis ensemble $\{ \xa_{k}{(e)}\}_{e=1,\ldots ,\nens}$.
 \begin{algorithm}[H]
     \begin{algorithmic}[1]
    \STATE \textbf{Forecast step:} given an analysis ensemble $\{ \xa_{k-1}{(e)}\}_{e=1,2,\ldots ,\nens}$ at time $t_{k-1}$;
     generate the forecast ensemble by via the model $\mathcal{M}$:
     \begin{equation}
\xb_k{(e)} = \mathcal{M}_{t_{k-1} \rightarrow t_k} \left( \xa_{k-1}{(e)} \right),\quad e=1,2,\ldots,\nens.
     \end{equation}

    \STATE \textbf{Analysis step:} given the observation vector $\y_k$ at time point $t_k$, follow the steps:
     \begin{enumerate}
  \item[i-  ] Set the initial state
  $ \xtin$ of the Markov Chain to be to the best estimate available, e.g., the mean of the forecast ensemble. One can use the EnKF analysis if the  cost is acceptable, and this choice is expected to result in a faster convergence of the chain to the stationary distribution.
  \item[ii- ] Calculate the ensemble-based forecast error covariance matrix $\mathbf{B}_k$ (and possibly balance it by a fixed (or frequently updated) covariance matrix $\mathbf{B}_0$), and apply localization as in
              equation \eqref{eq:EnKF_localization}. 
              It is important to emphasize that building  the full background error covariance matrix is not necessary for the current algorithm to work.
  \item[iii-] Choose a positive definite diagonal  mass matrix $\mathbf{M}$. 
                              One choice that favors the performance of the sampling algorithm is the diagonal of the matrix 
                              $\mathbf{B}^{-1}_k$ ~\citep{neal2011mcmc} which scales the components of the state vector vary. 
                              Ideally, $\mathbf{M}$ should be set to the diagonal of the inverse posterior covariance matrix.
  \item[iv- ] Apply Algorithm \ref{alg:HMC_sampling} with initial state $ \xtin$ and generate $\nens$ ensemble members. 
  In practice one starts accepting samples after a warm-up phase (of, say, $30$ steps), to guarantee
  that selected members explore the entire state space.
  \item[v-  ] Use the generated samples $\{ \xa_k(e)\}_{e=1,2,\ldots ,\nens}$ as an analysis ensemble and calculate the best estimate of the state 
  (e.g. the mean), and the analysis error covariance matrix.
     \end{enumerate}
    \STATE Increase time $k := k+1$ and repeat steps 1 and 2.
     \end{algorithmic}
 \caption{Sampling Filter}
 \label{alg:sampling_filter}
 \end{algorithm}
As stated in step $ii$ of Algorithm \ref{alg:sampling_filter}, the explicit representation of the matrix $\mathbf{B}_k$ is not necessary - one only needs to apply its inverse to a vector in (\ref{eqn:J-sampling-filter}), (\ref{eqn:J-gradient-sampling-filter}). Typically  $\mathbf{B}_k$ is formed as a linear a combination between a fixed matrix $\mathbf{B}_0$ and the ensemble covariance.  The calculation requires to evaluate the products
  \begin{align}
          u &= \mathbf{B}^{-1}_k\, ( \x - \xb_k ) \\
\nonumber
            &= \left( \gamma\, \mathbf{B}_0 + \frac{1-\gamma}{\nens-1} \sum_{e=1}^{\nens} \Delta \x(e) \left(\Delta\x(e)\right)^T \right)^{-1}  ( \x - \xb_k ) \, ,
  \end{align}
  where $\Delta\x(e)$ is the deviation of the ensemble member $\x(e)$ from the mean of the ensemble.
  The linear system
  \begin{equation}
     \left( \gamma \, \mathbf{B}_0 + \frac{1-\gamma}{\nens-1} \sum_{e=1}^{\nens} \Delta \x(e) \left(\Delta\x(e)\right)^T \right)\cdot u =  \x - \xb_k \, ,
  \end{equation}
  can be solved without having to build the full matrix $\mathbf{B}_k$ as discussed in~\citep{NinoSandu_2014}. 
  
In our numerical experiments we build flow-dependent background error covariance matrices $\mathbf{B}_k$ at each time step. 
   We set $\mathbf{M}$ to be equal to the diagonal of $\mathbf{B}_k$ in case of Lorenz-96 model following (\citep{beskos2011hybrid,liu2008monte}.  Taking $\mathbf{M}$ equal to the diagonal of $\mathbf{B}^{-1}_k$ lead to similar results for the Lorenz-96 model. For the shallow-water model on the sphere we set $\mathbf{M}$ to be equal to the diagonal of $\mathbf{B}^{-1}_k$.
  %

\section{Numerical Results}\label{sec:Results}
%
\subsection{The Lorenz-96 model}\label{subsec:Lorenz}

 Numerical tests are primarily performed using the 40-variables Lorenz-96 model~\citep{lorenz1996predictability} which is described by the equations:
 \begin{equation}
 \label{eqn:Lorenz96}
    \frac{dx_i}{dt} = x_{i-1} \left( x_{i+1} - x_{i-2} \right) - x_i + F \,,
 \end{equation}
 where $\x=(x_1,x_2,\ldots,x_{40})^T\in \mathbb{R}^{40}$ is the state vector. The indices work in a circular fashion, e.g., $x_{0} \equiv x_{40}$. The forcing parameter is
 set to $F=8$ in our experiment. These settings make the system chaotic~\citep{lorenz1998optimal}. 
The initial condition is obtained by integrating a vector of equidistant components ranging from $-2 \text{ to } 2$ for 10 time units before the beginning of the experiment time interval.
The simulation time interval is $\left[ 0 ,10 \right]$ units with observations available at time points $\{t_k = 0.1\times k\}_{k=1,2,\ldots,101} $. 
To study the behavior of the sampling algorithm with small ensemble, the number of ensemble members is chosen to be $30$.
All observations are synthetic, created by  applying the observation operator 
to the reference trajectory (by applying the corresponding observation operator) and adding Gaussian noise with a standard deviation equal to $5\%$ of the average magnitude of the corresponding observation 
along the reference trajectory. 
The background error is Gaussian with a diagonal covariance matrix $\mathbf{B}_0$; the standard deviation of each component is $8\%$ of the average magnitude of the initial condition of the system.                 
 %

\subsection{Observations and observation operators}
\label{subsec:obs_operators}

We choose six different observation operators of different complexities and varying levels of non-linearity to test the performance of the sampling filter. All the six operator were used with the Lorenz-96 model.
Both quadratic and cubic observation operators used here were employed by Zupanski~\citep{zupanski2005maximum,zupanski2008maximum} in the simple case of one dimensional state space. 
Synthetic observations are obtained by applying them to a reference trajectory and adding Gaussian random noise with a standard deviation of $5\%$ of the magnitude of the reference observation values.

 \paragraph{Linear observation operator}
The first observation operator is a linear operator  that selects a specific subset of the components of the state vector. 
This operator makes $\mathcal{J}(\x)$ differentiable. In our experiments we observe each third component of the state, starting with the first component
\begin{equation}
\label{eqn:Linear_H}
\mathcal{H}(\x) =\mathbf{H}\x =(x_1,\ x_4,\ x_7,\ \dots,\ x_{37},\ x_{40})^T \in \mathbb{R}^{14} \,.
\end{equation}
%

 \paragraph{Quadratic observation operator} 
This is a non-linear but differentiable observation operator that squares selected components \eqref{eqn:Linear_H} of the state.
In our experiments we use:
\begin{equation}
\label{eqn:Quadratic_H}
 \mathcal{H}(\x) =(x_1^2,\ x_4^2,\ x_7^2,\ \dots,\ x_{37}^2,\ x_{40}^2)^T \in \mathbb{R}^{14}.
\end{equation}
%

 \paragraph{Cubic observation operator} 

This is another non-linear but differentiable observation operator that squares selected components \eqref{eqn:Linear_H} of the state.
In our experiments we use:
\begin{equation}
\label{eqn:Cubic_H}
 \mathcal{H}(\x) =(x_1^3,\ x_4^3,\ x_7^3,\ \dots,\ x_{37}^3,\ x_{40}^3)^T \in \mathbb{R}^{14}.
\end{equation}
%

 \paragraph{Magnitude observation operator} 
This non-differentiable observation operator returns the absolute values of selected components  \eqref{eqn:Linear_H}.
The observation vector reads: 
\begin{equation}
\label{eqn:Abs_H}
 \mathcal{H}(\x) =(|x_1|,\ |x_4|,\ |x_7|,\ \dots,\ |x_{37}|,\ |x_{40}|)^T \in \mathbb{R}^{14}.
\end{equation}
%

 \paragraph{Quadratic  observation operator with a threshold}
This observation operator is similar to the simple version used by Zupanski et al in~\citep{zupanski2008maximum}. 
The observation vector is:
\begin{equation}
 \label{eqn:Quad_Thresh_H}
 \mathcal{H}(\x) =(x'_1,\ x'_4,\ x'_7,\ \dots,\ x'_{37},\ x'_{40})^T \in \mathbb{R}^{14} \,,
\end{equation}
where
\[
  x'_i = \left\{
    \begin{array}{ll}
 \ \ \ x_i^2  & :  x_i \geq 0.5  \\
 -x_i^2 & :  x_i < 0.5     \,,
    \end{array}  
   \right.
\]
This operator is non-linear and discontinuous.

 \paragraph{Exponential observation operator}
This is a highly nonlinear, differentiable observation operator: 
\begin{equation}
\label{eqn:Exponential_H}
\mathcal{H}(\x) =(e^{r\cdot x_1},\ e^{r\cdot x_4},\ e^{r\cdot x_7},\ \dots,\ e^{r\cdot x_{37}},\ e^{r\cdot x_{40}})^T \in \mathbb{R}^{14} \,,
\end{equation}
where $r \in \mathbb{R}$ is a scaling factor that controls the degree of nonlinearity.

%
\subsection{Experimental setting}
\label{subsec:setting}

We perform two sampling filter data assimilation experiments with each observation operator described in Section \ref{subsec:obs_operators}. Both share the same model parameters but use different step sizes of the symplectic integration during the HMC sampling. This is found to have a great impact on the performance of the sampling filter.
     
In the first experiment a time $T=0.1$ with $h=0.01$ and $m=10$ is used for all integrators tested. This choice guarantees that the standard position Verlet integrator yields satisfactory results with the linear observation operator, but the performance on nonlinear observation operators remains to be checked. The second experiment analyzes the performance of the sampling filter when all time integrators take roughly the same computational cost.  The parameters, $m$, $h$, are tuned by trial and error such as to make the Verlet integrator successful, if possible, with the nonlinear observation operators. The other time integrators use the same total time step $T$ as the Verlet integrator; the values of $m$\ and $h$ are chosen for each method such that the number of gradient calculations done by all integrators is the same.  In general, however, the time-stepping parameters of each symplectic integrator should be set individually to get the best performance of the sampling filter.

Each numerical experiment performs $100$ realizations of the sampling filter. Each realization uses the same settings  but the sequence of random number generated by the  sampling filter, for both the potential variable and the acceptance/rejection rule, was different.  The root mean squared error (RMSE) metric is used to compare the analyses against the reference solution at observation time points:
                      \[
                           \mathbf{RMSE} = \sqrt{\frac{1}{\nvar} \sum_{i=1}^{\nvar}{(x_i - x_i^{\rm true})^2} } \, , \label{eqn:RMSE_Formula}
                      \]
where $\x^{\rm true}$ is the reference state of the system. The RMSE is calculated at all assimilation time points along the trajectory over the time span of the experiment.
     
To guarantee that the Markov chain reaches the stationary distribution before starting the sampling process a set of $200$ steps are perform as burn-in stage. We noticed that the chain always converges in a small number ($10-20$) burn-in steps. Stationarity tests will be given special attention in our future work.

After the burn-in stage an ensemble member is selected after each $30$ generated states; this choice decreases correlation between generated ensemble members since the chain is not memoryless. The number of ensembles that are not retained will be referred to as the number of inter-chain steps. In our experiments the acceptance probability is high (usually over $0.9$) with this sampling strategy. The number of inter-chain steps is a parameter that can be tuned by the user to control the performance of the sampling filter.

Stability requirements impose tight upper bounds on the step size $h$ of the Verlet integrator. The step size should be decreased with the increasing dimension of the system in order to maintain $\mathcal{O}(1)$ acceptance probability~\citep{beskos2013optimal}. On the other hand large steps of the symplectic integrator are needed in order to explore the space efficiently. There is no precise rule available to select the optimal step size values~\citep{sanz2014Markov} and consequently $h$ should be tuned for each problem. 
The higher-order integrators \eqref{eqn:two_stage}, \eqref{eqn:three_stage}, \eqref{eqn:four_stage} are expected to be more stable than Verlet for larger time steps~\citep{blanes2013numerical,neal2011mcmc}.

To guarantee ergodicity of the Markov chain, which is a property required for the chain to converge to its invariant distribution, we follow~\citep{blanes2013numerical,neal2011mcmc} 
and change the step length at the beginning of each Markov step (once at the beginning of the Hamiltonian trajectory) to $h = (1+r)\, h_{\rm ref}$ where $h_{\rm ref}$ is
a reference step size and $r\sim \mathcal{U}(-0.2,0.2)$ is a uniformly distributed random variable. 
Randomizing the step size of the symplectic integrator, in addition to other benefits, ensures that the results obtained are not entrusted with specific choice of the step size~\citep{neal2011mcmc}. 

\FloatBarrier    
     \subsection{Linear observation operator experiments}
     \label{subsec:Linear_H_Results}   
  
\begin{figure}[H]
    \centering
    \subfigure[Position Verlet integrator \eqref{eqn:Verlet}]{%
    \includegraphics[width=0.44\linewidth]{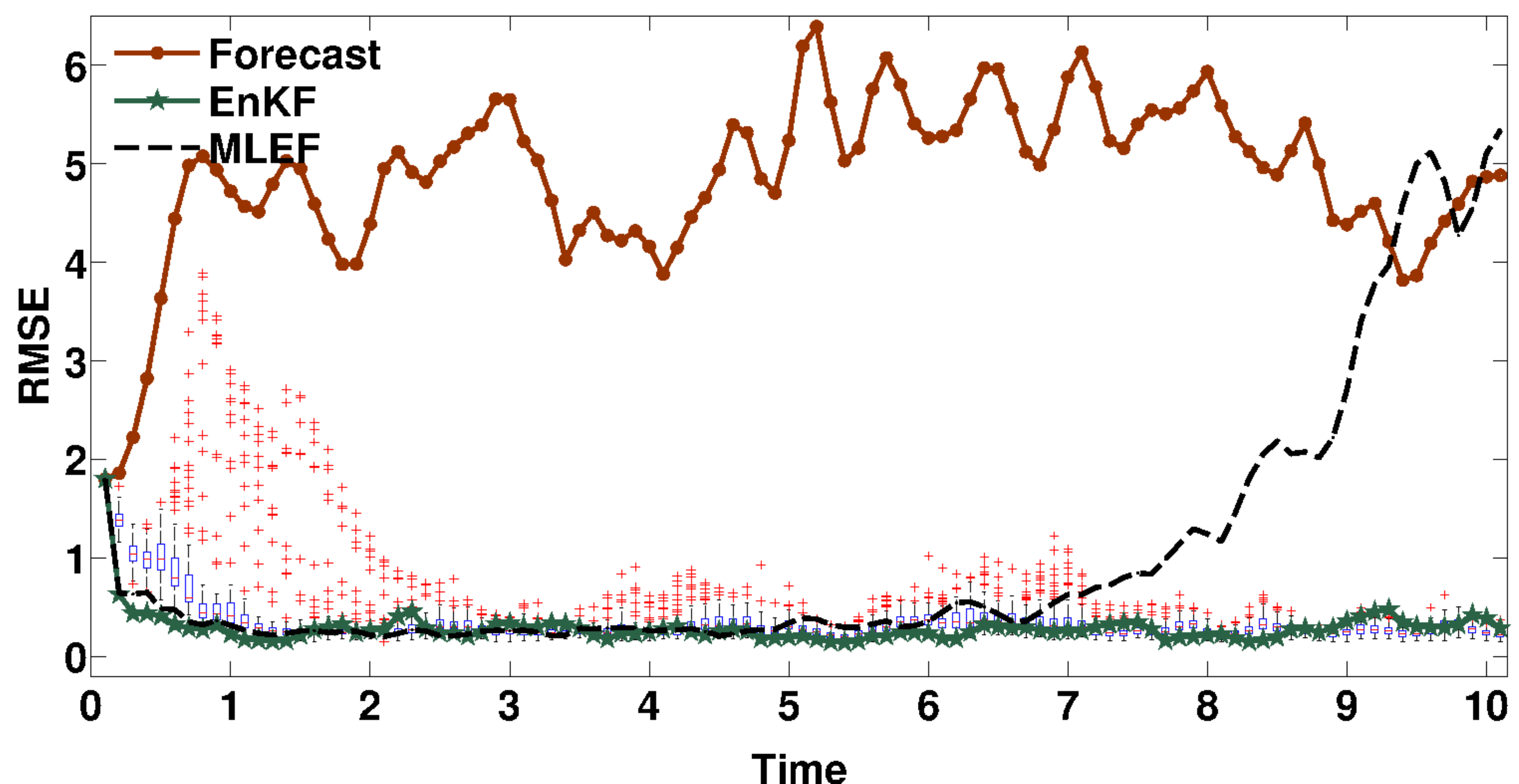}   
    \label{fig:Linear_Exper1_Verlet}}
    \quad  
    \subfigure[Two-stage integrator \eqref{eqn:two_stage}]{%
    \includegraphics[width=0.44\linewidth]{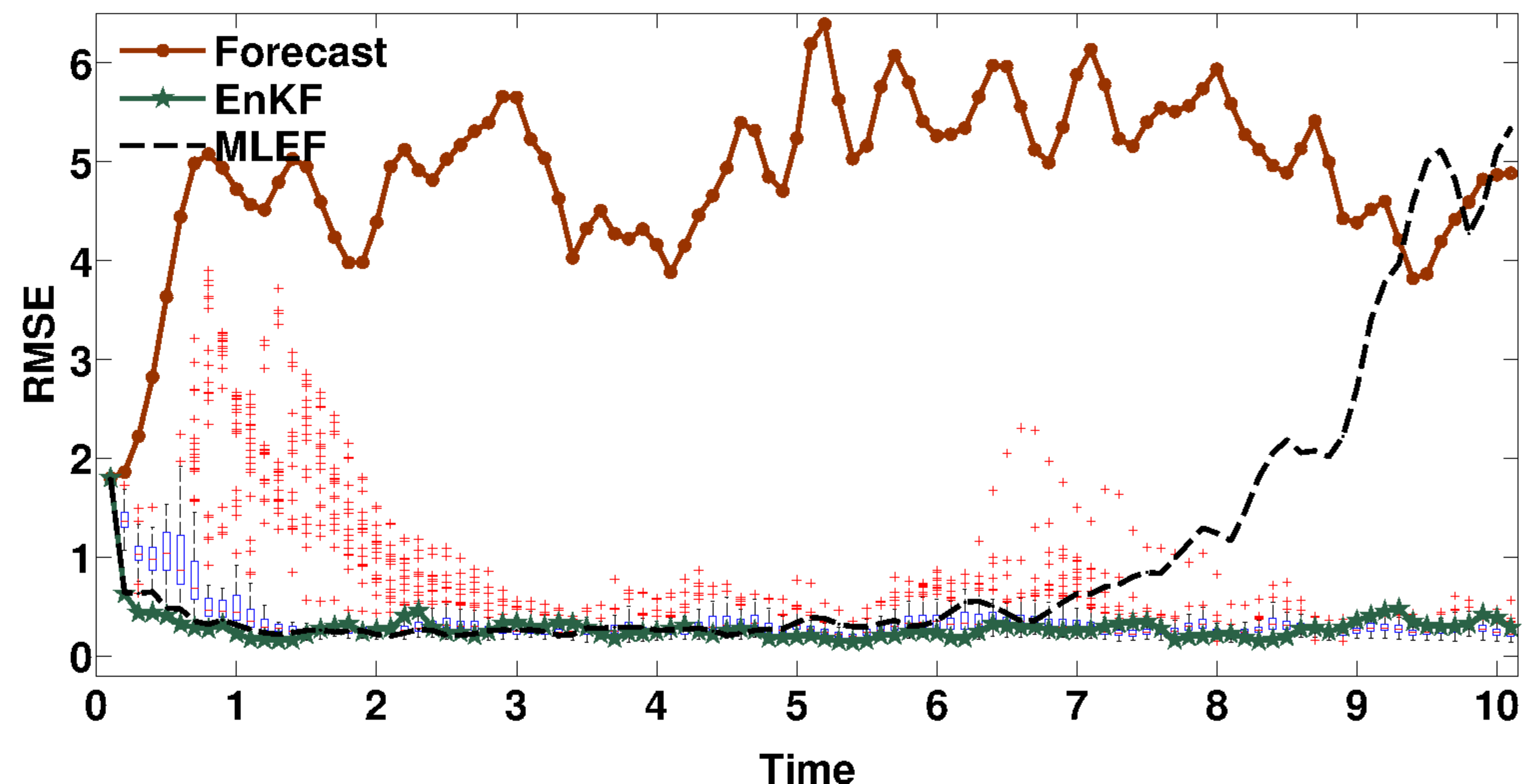}   
    \label{fig:Linear_Exper1_2Stage}}
    \subfigure[Three-stage integrator \eqref{eqn:three_stage}]{%
    \includegraphics[width=0.44\linewidth]{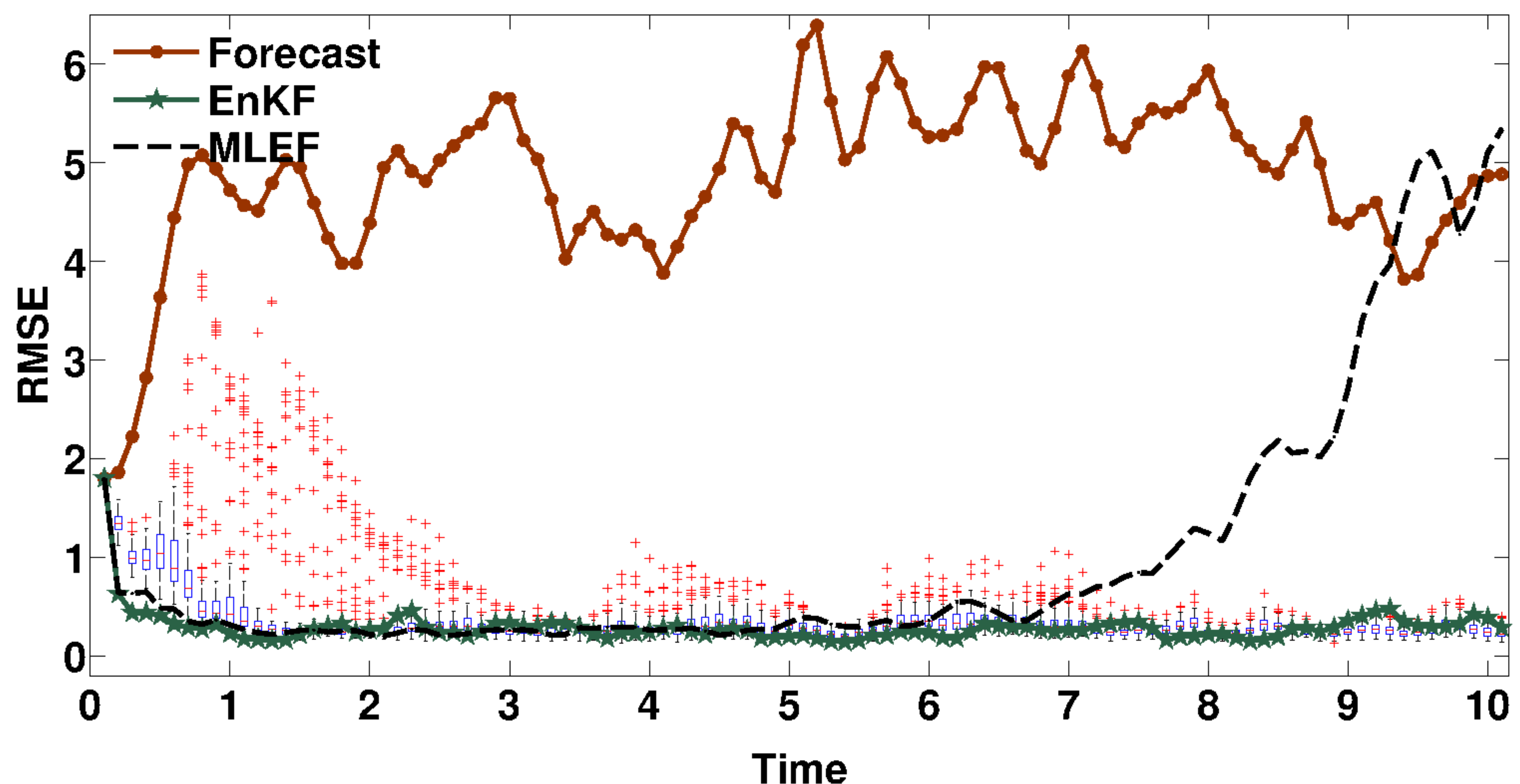}   
    \label{fig:Linear_Exper1_3Stage}}
    \quad
    \subfigure[Four-stage integrator \eqref{eqn:four_stage}]{%
    \includegraphics[width=0.44\linewidth]{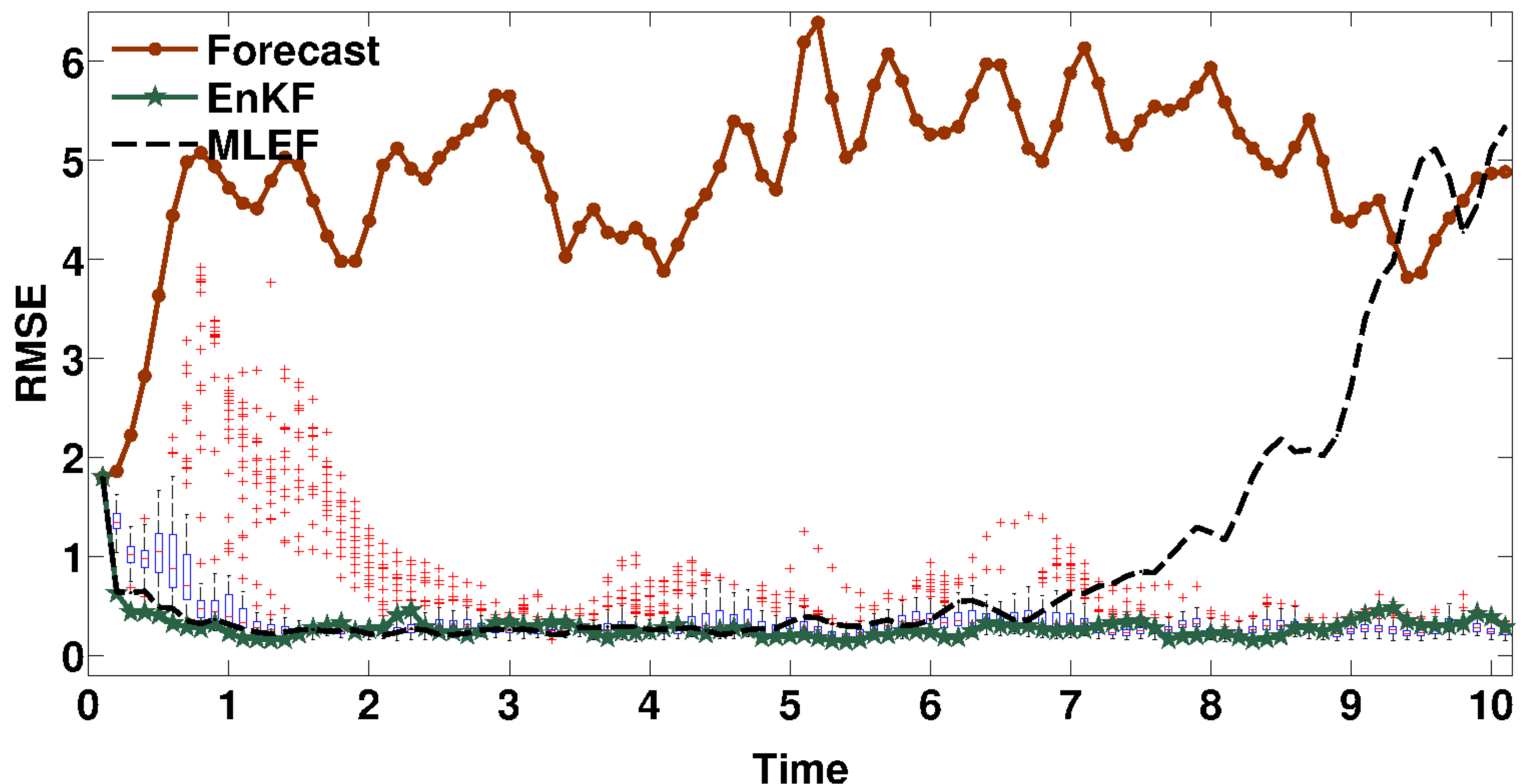}   
    \label{fig:Linear_Exper1_4Stage}}
    \subfigure[Integrator defined on Hilbert space \eqref{eqn:hilbert_integrator}]{%
    \includegraphics[width=0.44\linewidth]{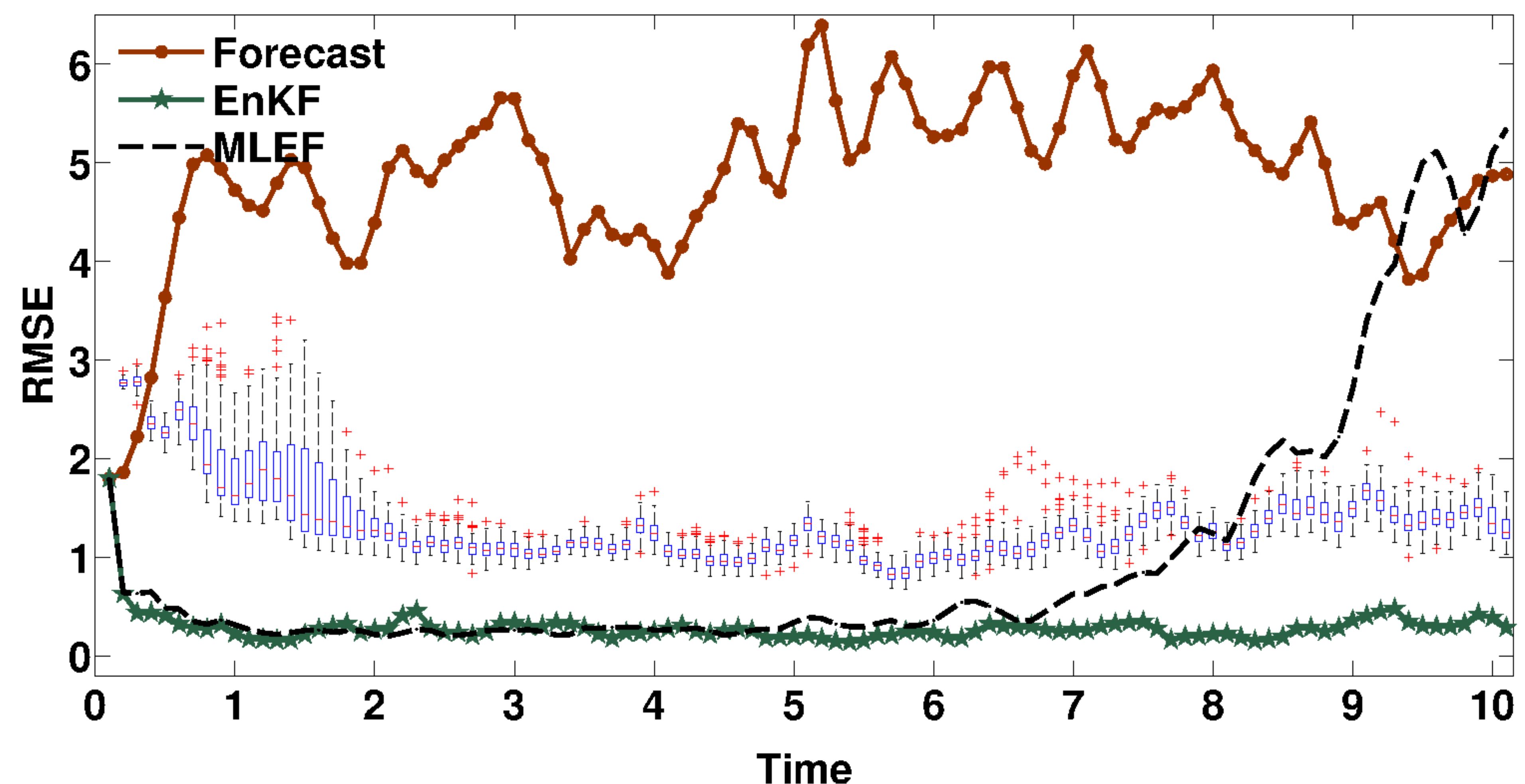}   
    \label{fig:Linear_Exper1_Hilbert}}
  \caption{Data assimilation results with the linear observation operator \eqref{eqn:Linear_H}. The  symplectic integrator used is indicated under each panel.
     The time step for all integrators is $T=0.1$ with $h=0.01$, $m=10$, and $30$ inter-chain steps.
     The RMSE for $100$ instances of the sampling filter results are shown as box plots. The  red line represents the median RMSE values across all instances and the central blue box represents the variance. The two vertical lines (whiskers) extend up to $1.5$ times the height of the central box. The values exceeding the length of the whiskers are considered outliers (extremes) and are plotted as red crosses.} 
  \label{fig:Linear_Exper1}
  \end{figure}

Figure \ref{fig:Linear_Exper1} shows the analysis results of different filters when the system uses linear observation operators \eqref{eqn:Linear_H}. The accuracy of the analyses provided by different filters is plotted  at different time moments. 
Results are reported for 100 instances of the sampling filter. 
The  red line represents the median RMSE values across all instances and the central blue box represents the variance. The two vertical lines (whiskers) extend up to $1.5$ times the height of the central box. The values exceeding the length of the whiskers are considered outliers (extremes) and are plotted as red crosses.
All symplectic integrators show outliers (the red crosses) with the exception of the Hilbert space integrator. The RMSE errors of the sampling integrator are larger than those of EnKF, however they remain small overall. The analysis follows closely the reference trajectory as seen in Figure \ref{fig:ExponentialH_2_State_Variables_Hilbert}. 

\begin{figure}[H]
\centering
\subfigure[Position Verlet integrator \eqref{eqn:Verlet}; $h=0.01,\ m=24$]{%
\includegraphics[width=0.44\linewidth]{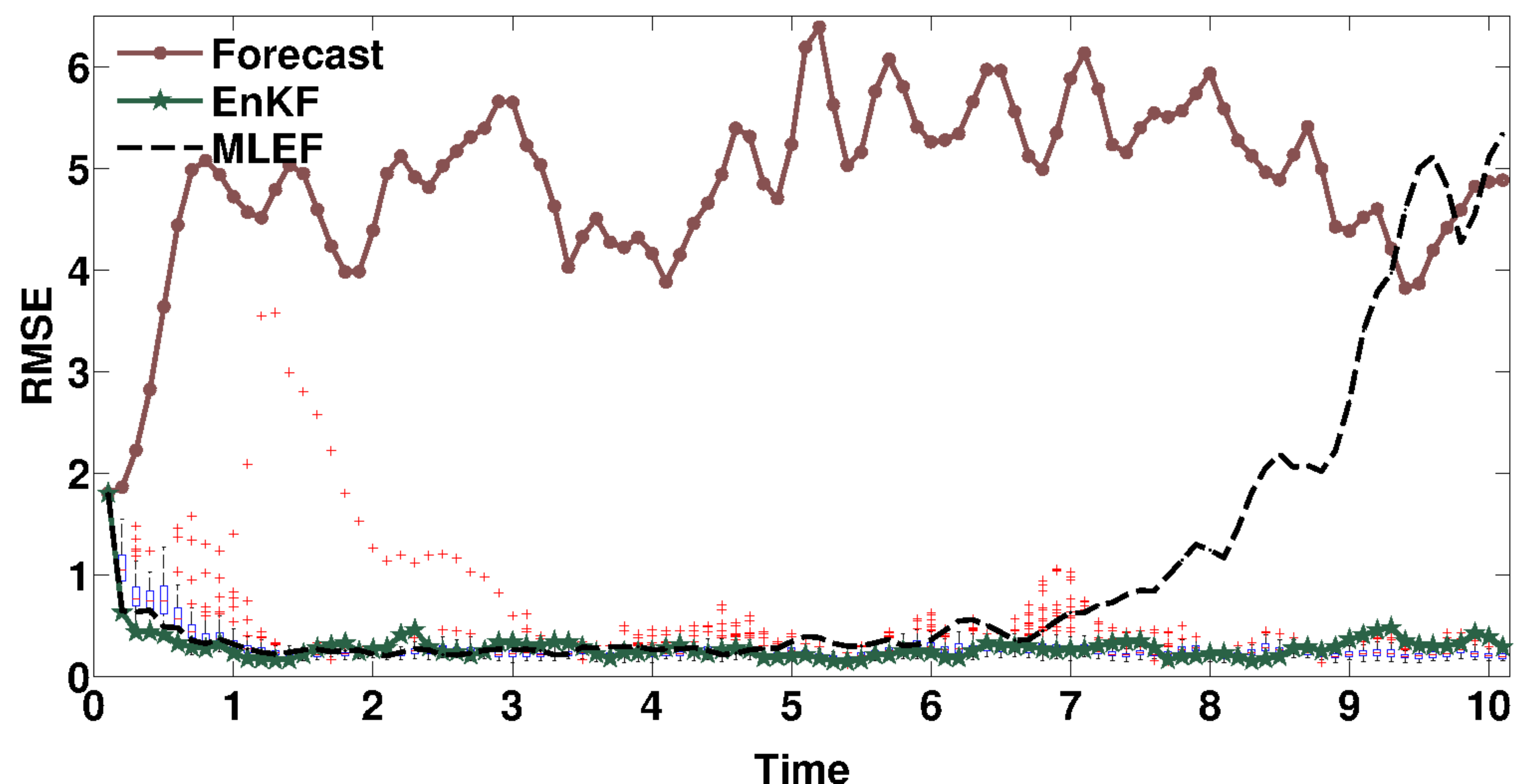}   
\label{fig:Linear_Exper2_Verlet}}
\quad  
\subfigure[Two-stage integrator \eqref{eqn:two_stage}; $h=0.02,\ m=12$]{%
\includegraphics[width=0.44\linewidth]{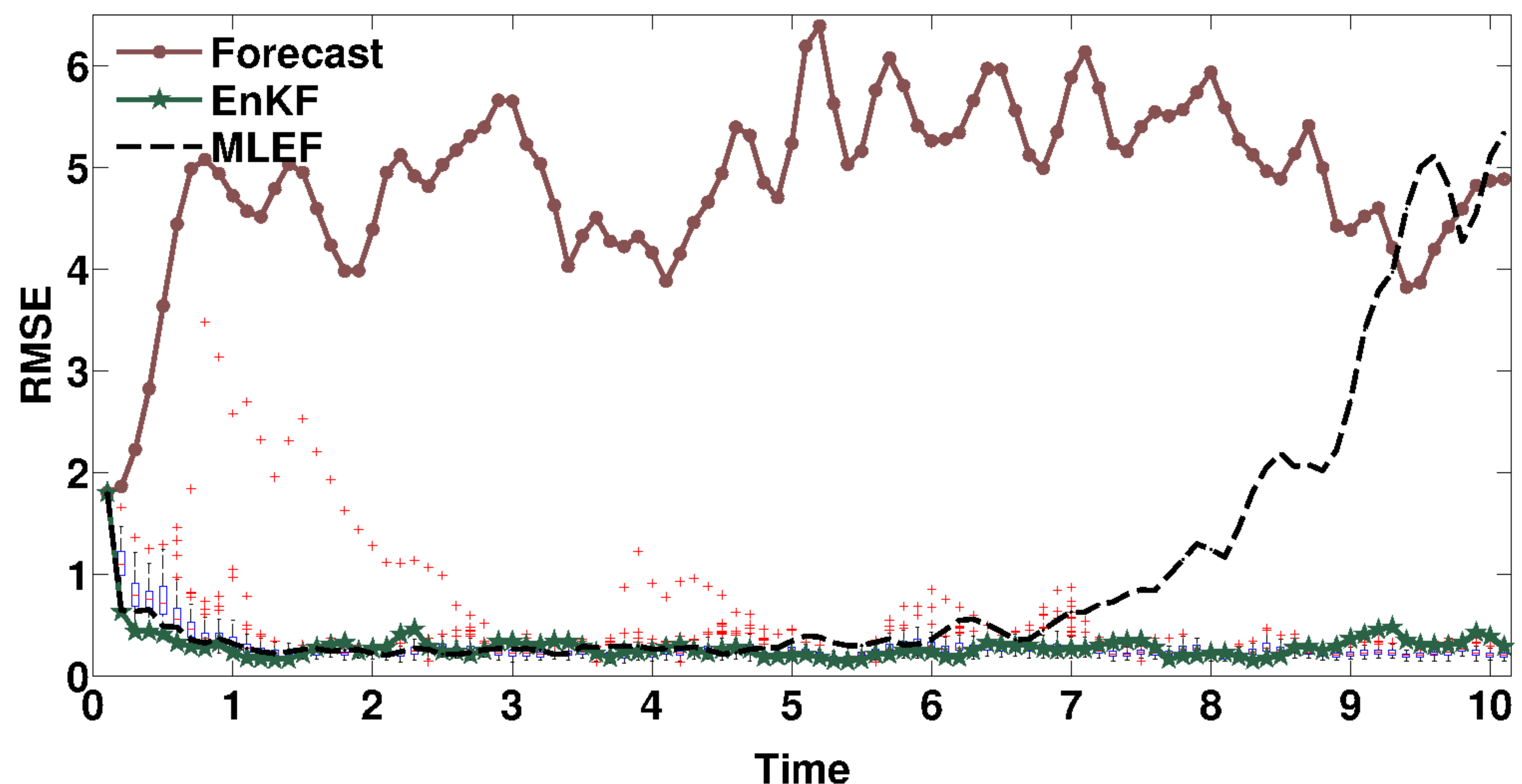}   
\label{fig:Linear_Exper2`_2Stage}}
\subfigure[Three-stage integrator \eqref{eqn:three_stage}; $h=0.03,\ m=8$]{%
\includegraphics[width=0.44\linewidth]{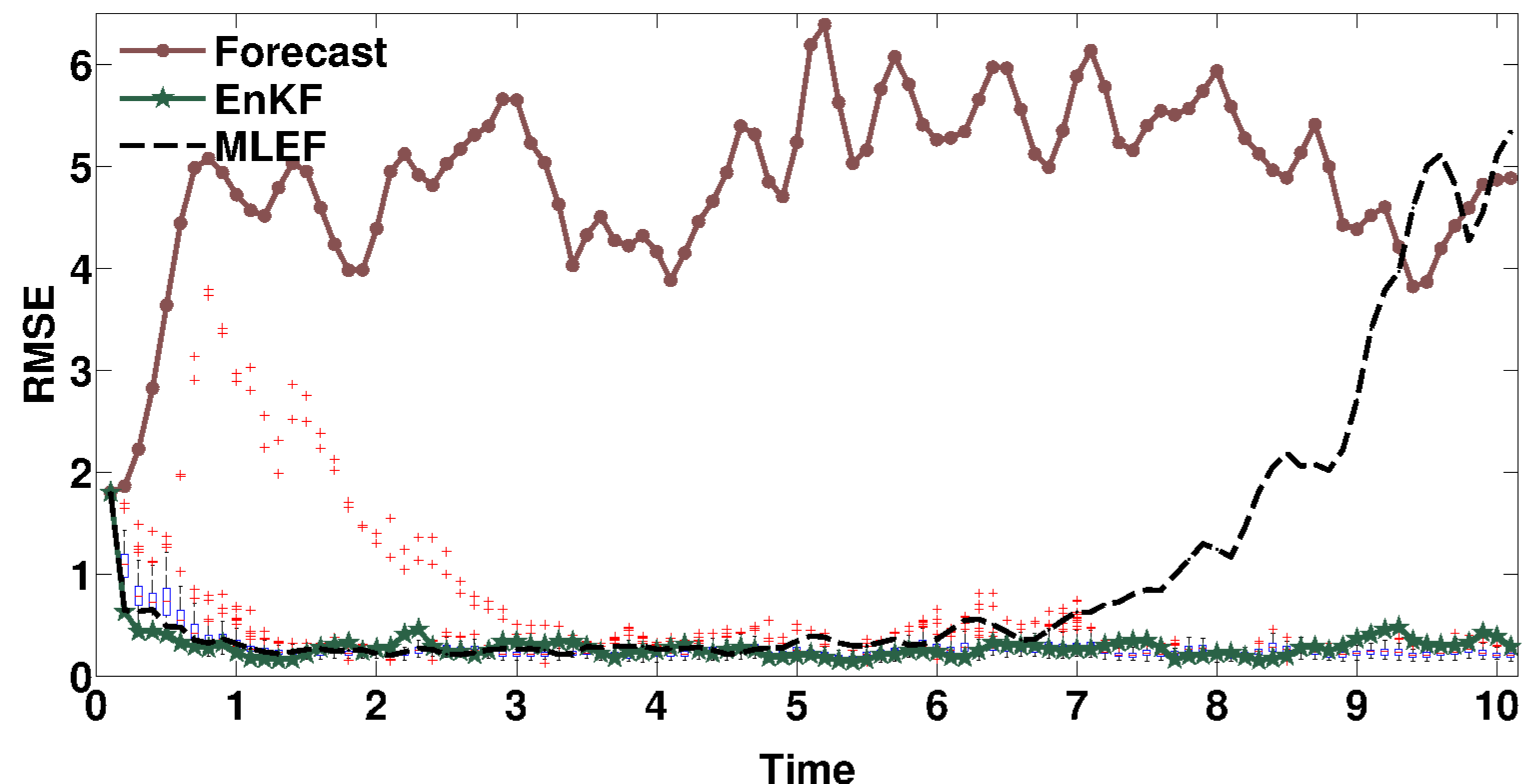}   
\label{fig:Linear_Exper2_3Stage}}
\quad
\subfigure[Four-stage integrator \eqref{eqn:four_stage}; $h=0.04,\ m=6$]{%
\includegraphics[width=0.44\linewidth]{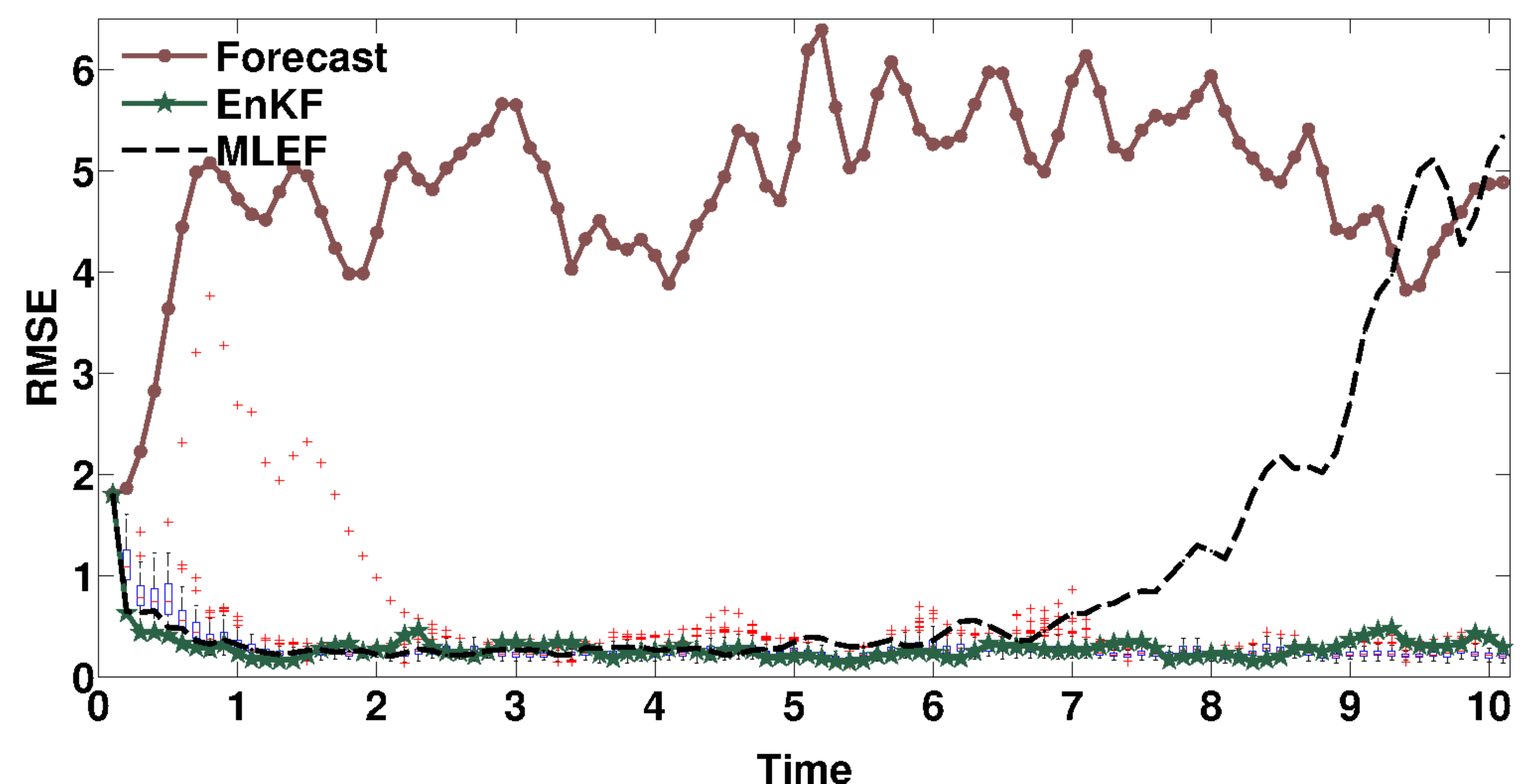}   
\label{fig:Linear_Exper2_4Stage}}
\subfigure[Integrator defined on Hilbert space \eqref{eqn:hilbert_integrator}; $h=0.01,\ m=24$]{%
\includegraphics[width=0.44\linewidth]{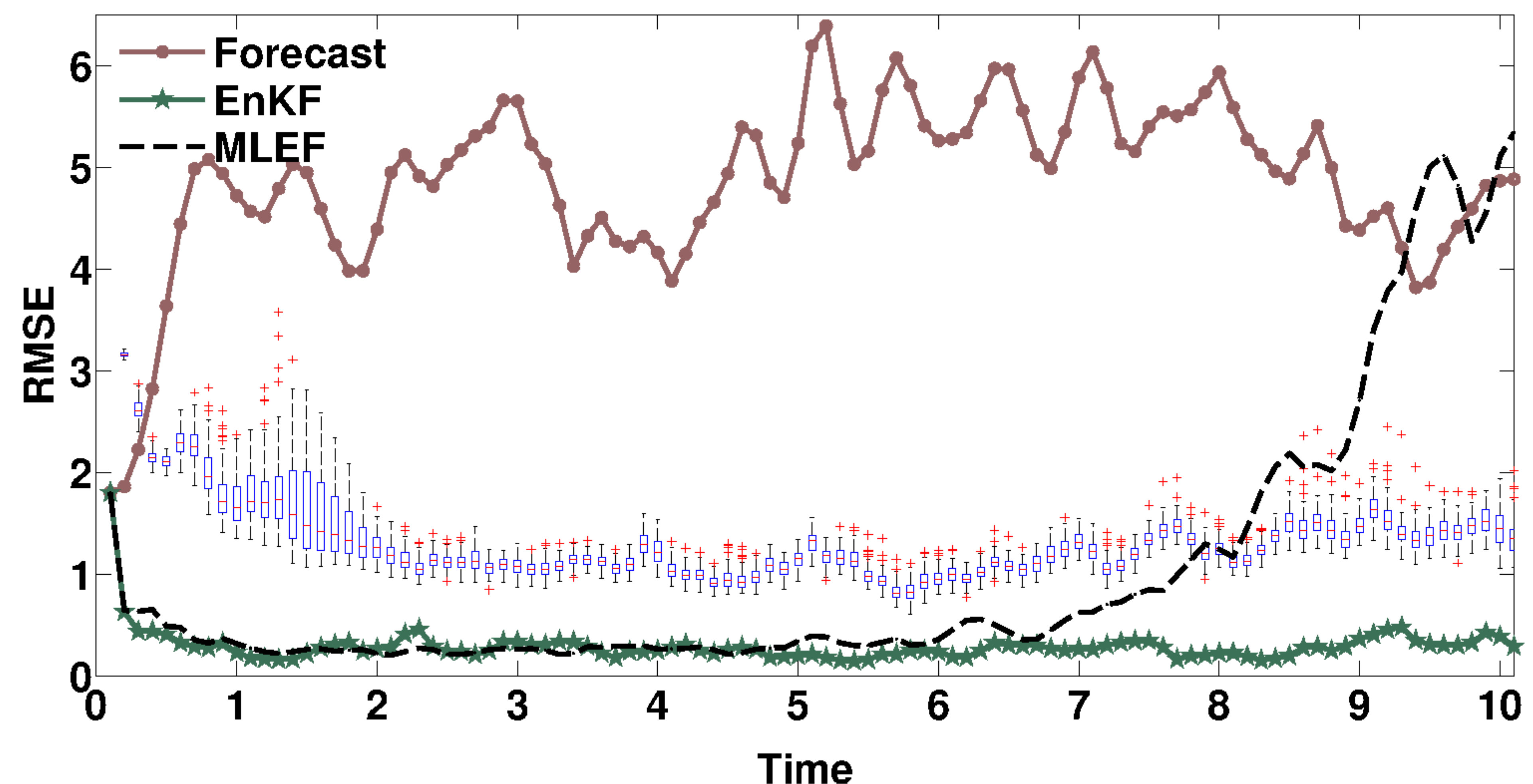}   
\label{fig:Linear_Exper2_Hilbert}}
    \caption{Data assimilation results with the linear observation operator \eqref{eqn:Linear_H}. The symplectic integrator used is indicated under each panel.
     The time step for all integrators is $T=0.24$ (units), and $h$ and $m$ are chosen such as to equalize the computational effort. The number of inter-chain steps is $30$. 
     The RMSE for $100$ instances of the sampling filter results are shown as box plots. The  red line represents the median RMSE values across all instances and the central blue box represents the variance. The two vertical lines (whiskers) extend up to $1.5$ times the height of the central box. The values exceeding the length of the whiskers are considered outliers (extremes) and are plotted as red crosses.
     } 
    \label{fig:Linear_Exper2}
   \end{figure}

Figure \ref{fig:Linear_Exper2} shows results with the time step parameters of symplectic integrators tuned to provide equalized work. The number of steps $m$ for Verlet is increased compared to the tests in Figure \ref{fig:Linear_Exper1} for two reasons: to test the capabilities of position Verlet with different step sizes, and to allow for more accuracy using Verlet integrators. The sampling filters perform well and are as accurate as EnKF with  any choice of integrator, except for the Hilbert space one which yields larger RMSE errors.
   %
   %
     \subsection{Quadratic observation operator experiments}
     \label{subsec:Quadratic_H_Results}
   
\begin{figure}[H]
\centering
\subfigure[Position Verlet integrator \eqref{eqn:Verlet}]{%
\includegraphics[width=0.44\linewidth]{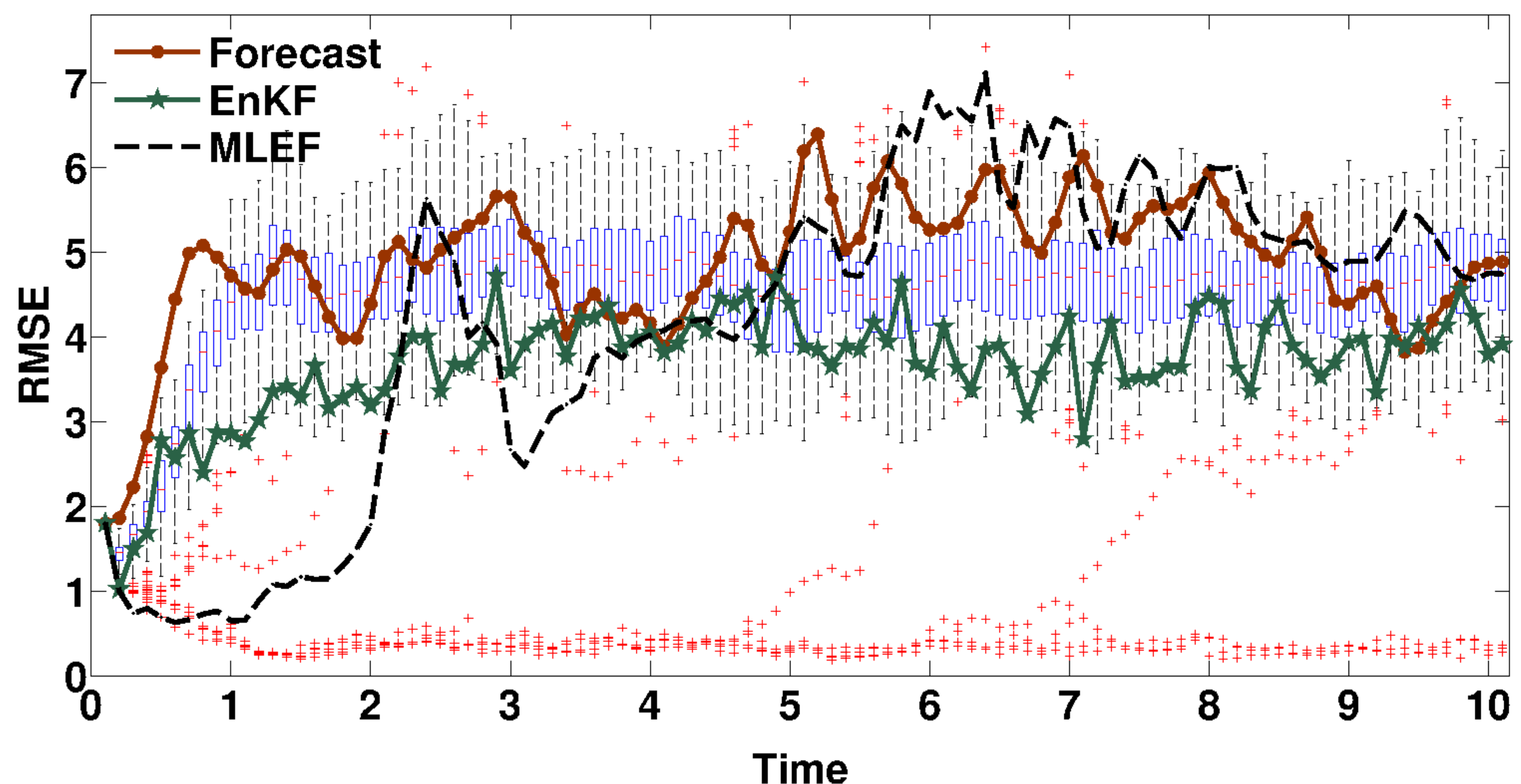}   
\label{fig:Quadratic_Exper1_Verlet}}
\quad  
\subfigure[Two-stage integrator \eqref{eqn:two_stage}]{%
\includegraphics[width=0.44\linewidth]{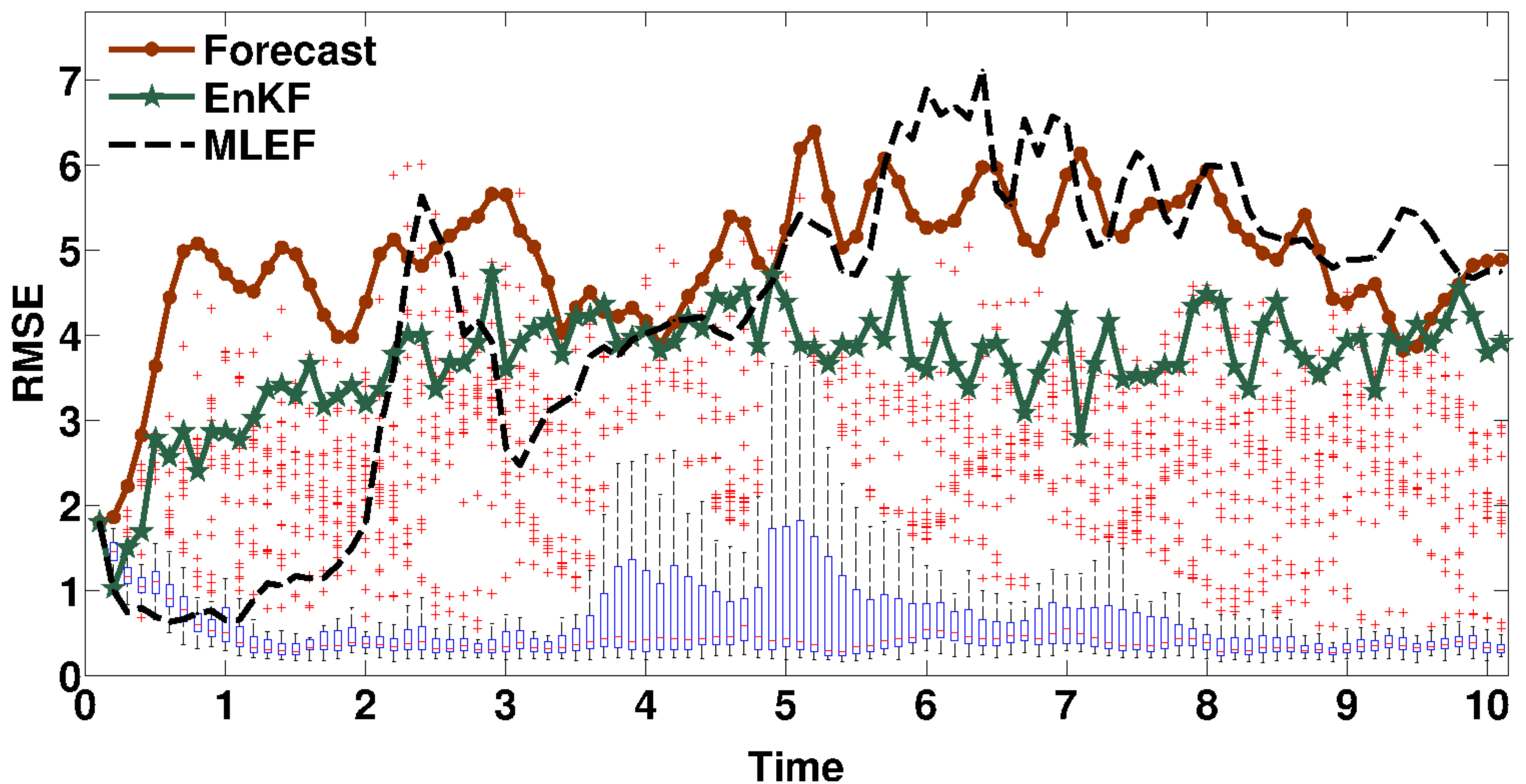}   
\label{fig:Quadratic_Exper1_2Stage}}
\subfigure[Three-stage integrator \eqref{eqn:three_stage}]{%
\includegraphics[width=0.44\linewidth]{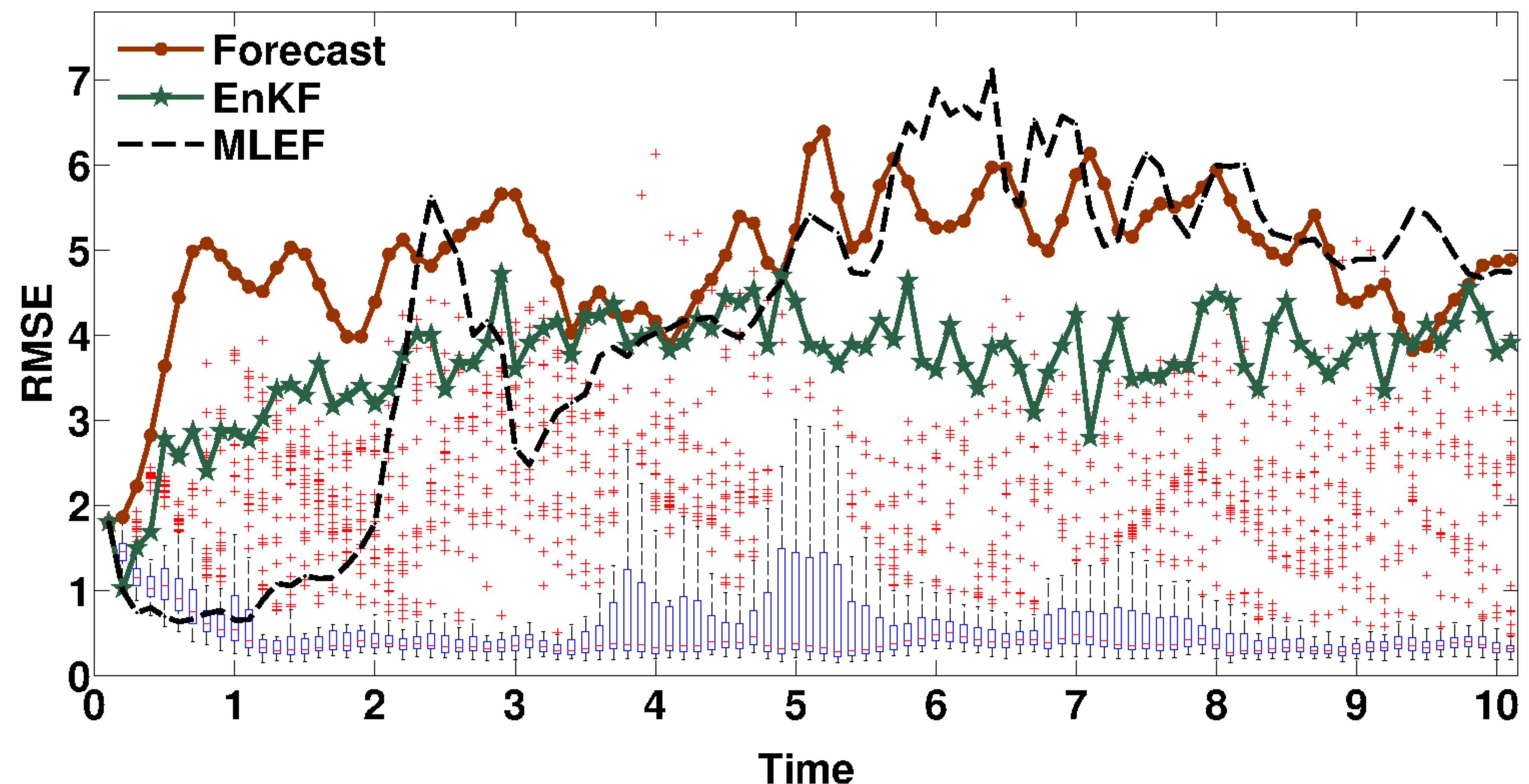}   
\label{fig:Quadratic_Exper1_3Stage}}
\quad
\subfigure[Four-stage integrator \eqref{eqn:four_stage}]{%
\includegraphics[width=0.44\linewidth]{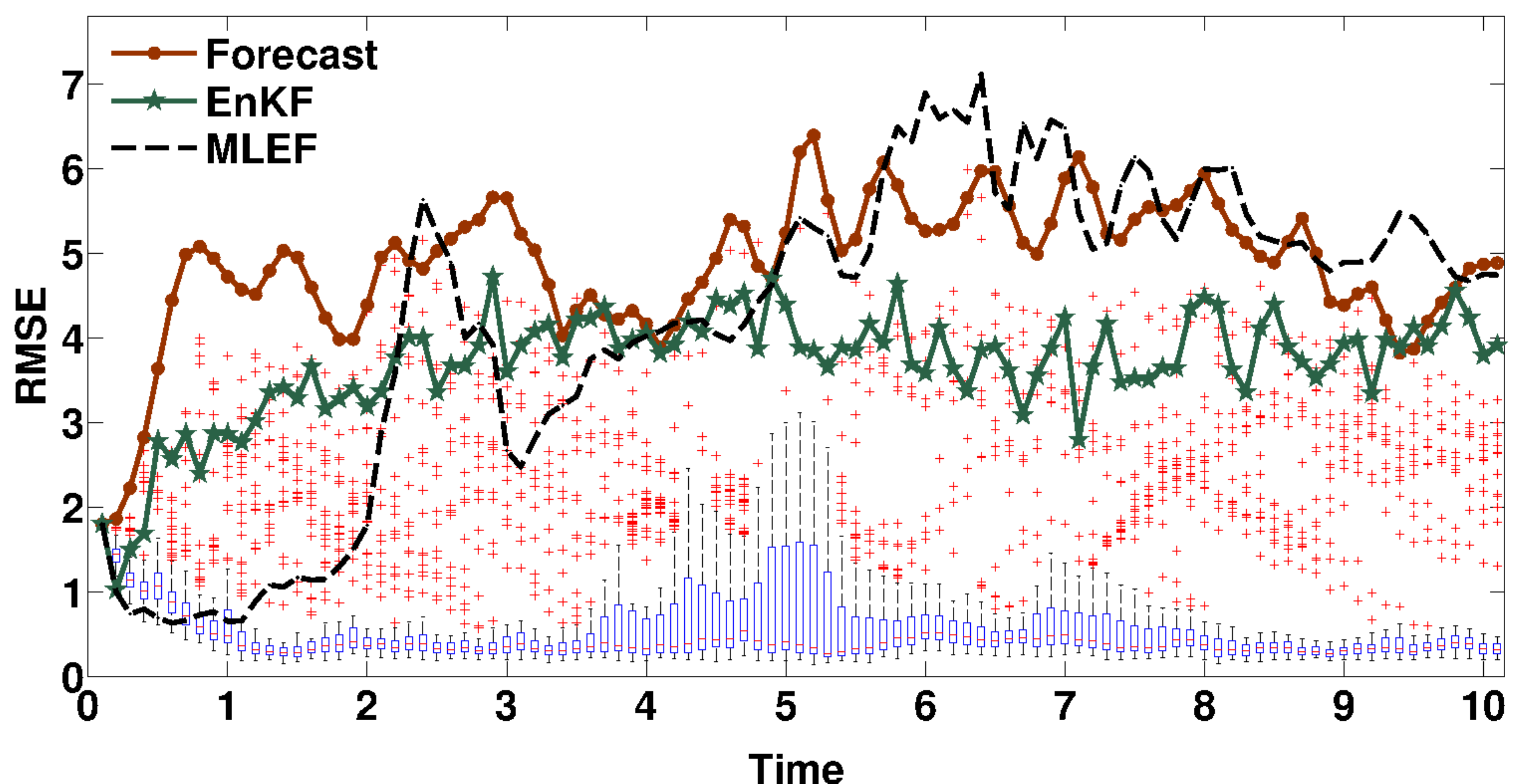}   
\label{fig:Quadratic_Exper1_4Stage}}
\subfigure[Integrator defined on Hilbert space \eqref{eqn:hilbert_integrator}]{%
\includegraphics[width=0.44\linewidth]{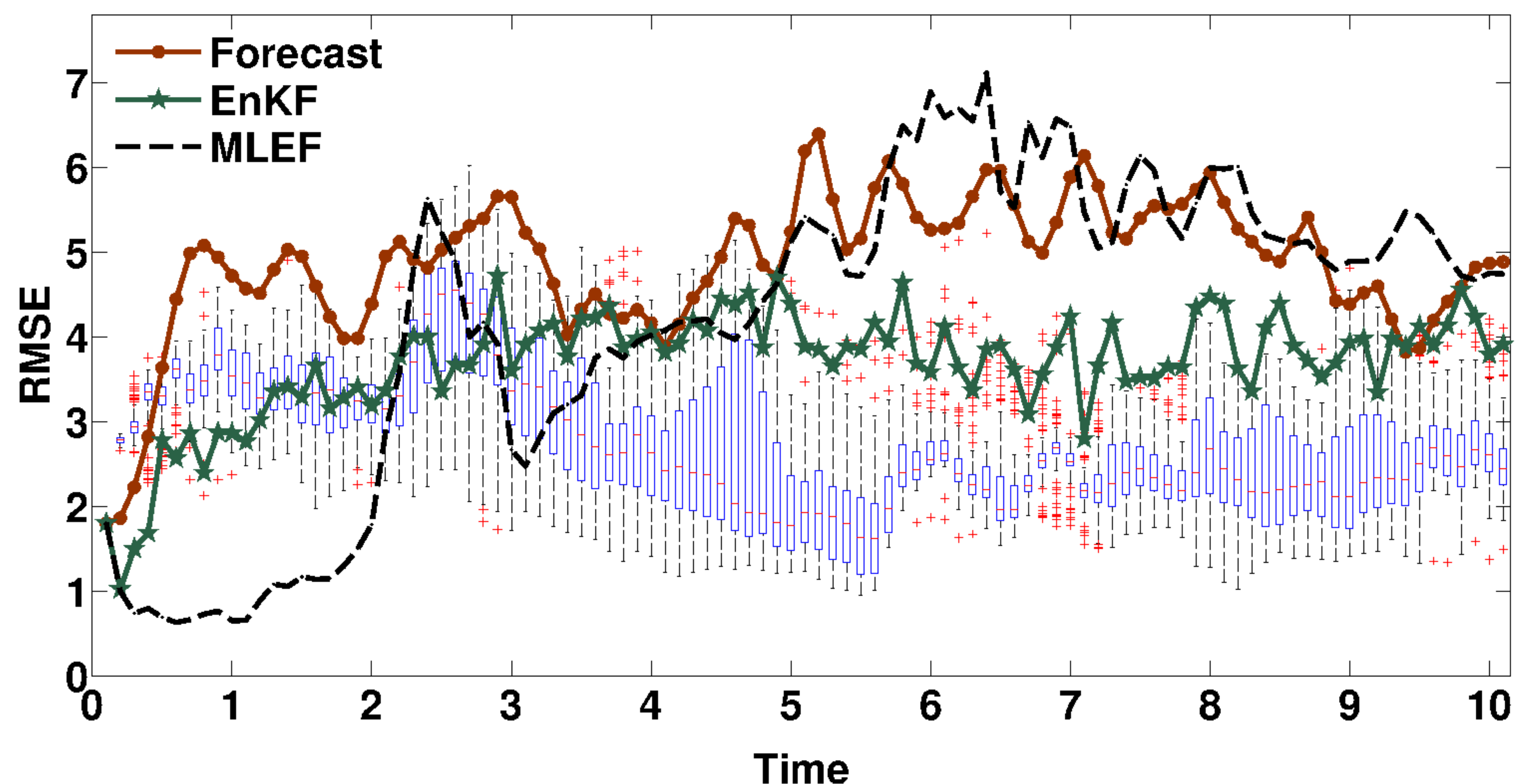}   
\label{fig:Quadratic_Exper1_Hilbert}}
    \caption{Data assimilation results with the quadratic observation operator \eqref{eqn:Quadratic_H}. The symplectic integrator used is indicated under each panel.
     The time step for all integrators is $T=0.1$ with $h=0.01$, $m=10$, and $30$ inter-chain steps.
     The RMSE for $100$ instances of the sampling filter results are shown as box plots. The  red line represents the median RMSE values across all instances and the central blue box represents the variance. The two vertical lines (whiskers) extend up to $1.5$ times the height of the central box. The values exceeding the length of the whiskers are considered outliers (extremes) and are plotted as red crosses.
     } 
    \label{fig:Quadratic_Exper1}
    \end{figure}

 Figure \ref{fig:Quadratic_Exper1} shows the results with quadratic observation operator \eqref{eqn:Quadratic_H}. All symplectic integrators use the parameters $h=0.01$ and $m=10$. The sampling filter with Verlet integrator fails to converge and produce representative samples from the analysis PDF. When high-order integrators are used the sampling filter gives satisfactory analysis RMSE, comparable to that obtained by MLEF, except for occasional failures represented in the plots as outliers (red crosses). Section \ref{subsec:MC_Steps_Effect} will discuss strategies to handle possible failures and avoid these outliers. The filter with the Hilbert space integrator has a larger RMSE error than EnKF, however it does not suffer from outliers as much as the other integrators.
%
%
%
    %
    %

\begin{figure}[H]
\centering
\subfigure[Position Verlet integrator \eqref{eqn:Verlet}; $h=0.01,\ m=40$]{%
\includegraphics[width=0.44\linewidth]{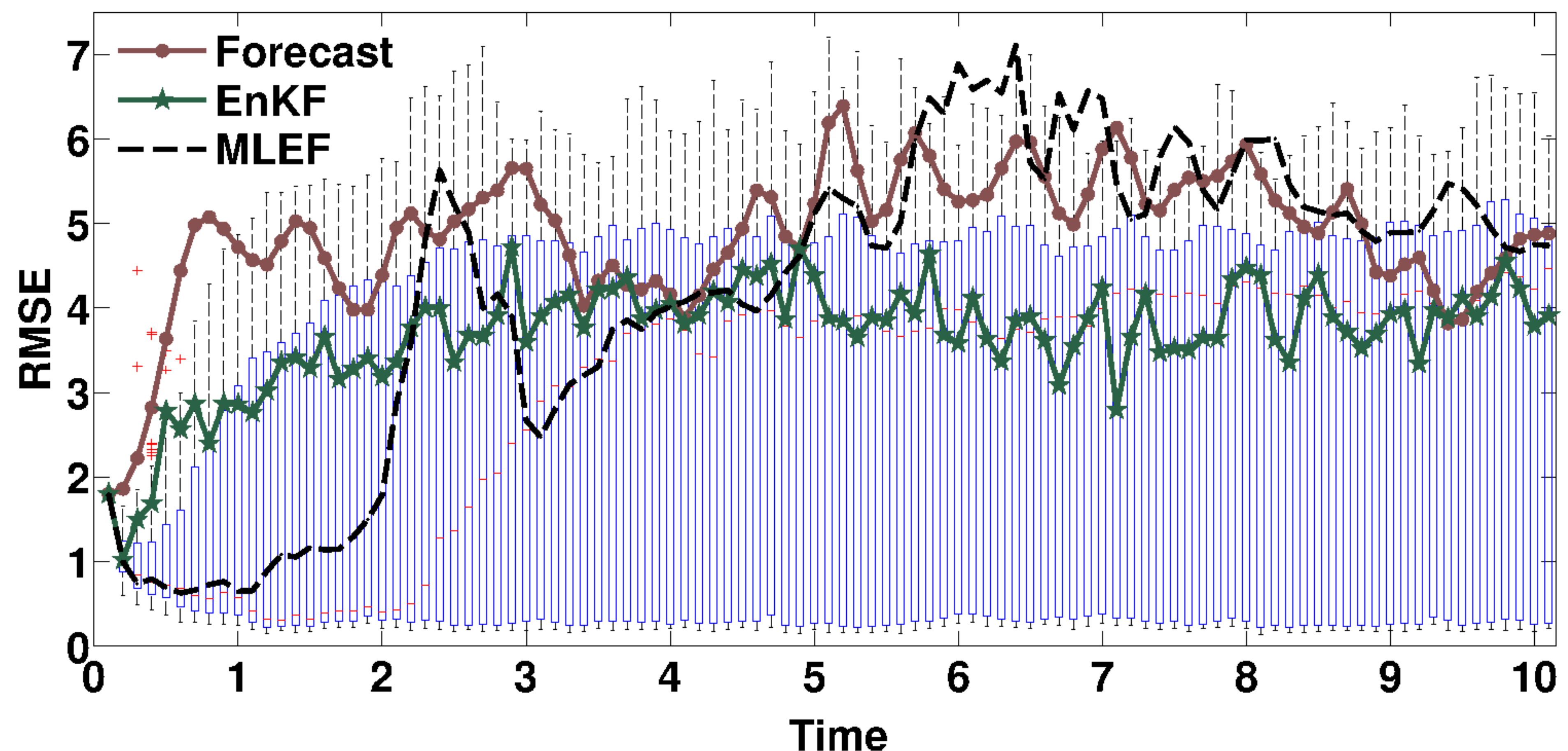}   
\label{fig:Quadratic_Exper2_Verlet}}
\quad  
\subfigure[Two-stage integrator \eqref{eqn:two_stage}; $h=0.02,\ m=20$]{%
\includegraphics[width=0.44\linewidth]{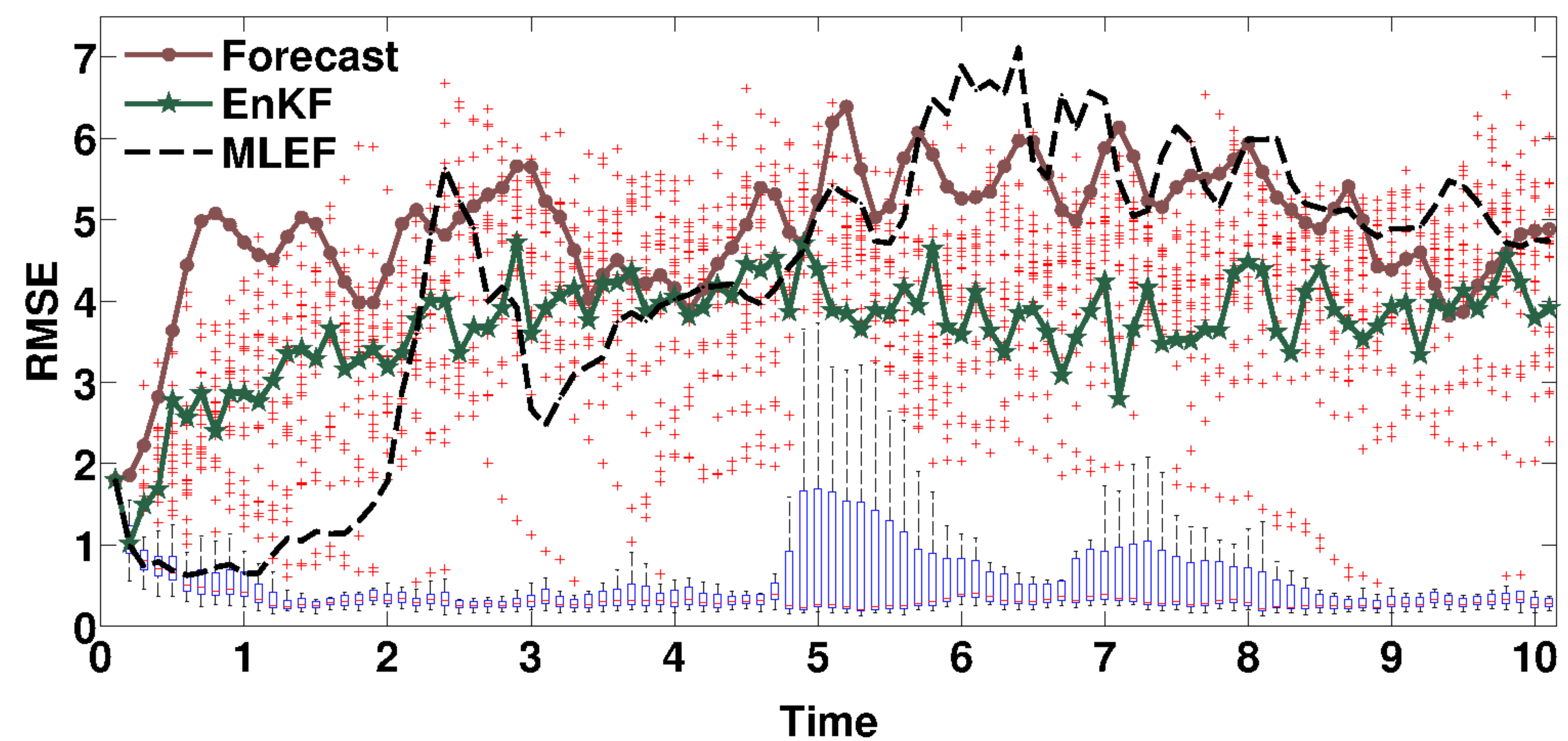}   
\label{fig:Quadratic_Exper2_2Stage}}
\subfigure[Three-stage integrator \eqref{eqn:three_stage}; $h=0.03,\ m=13$]{%
\includegraphics[width=0.44\linewidth]{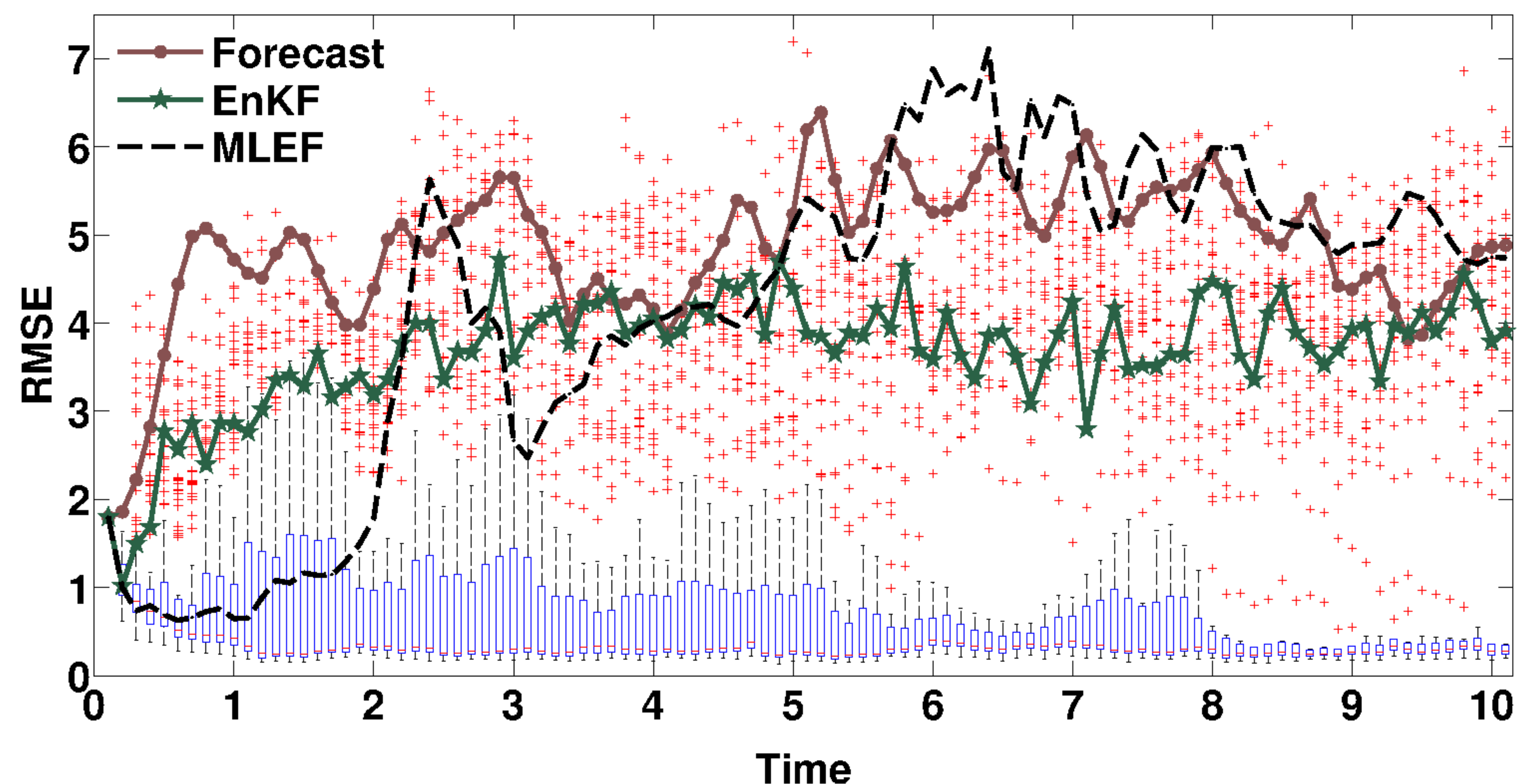}   
\label{fig:Quadratic_Exper2_3Stage}}
\quad
\subfigure[Four-stage integrator \eqref{eqn:four_stage}; $h=0.04,\ m=10$]{%
\includegraphics[width=0.44\linewidth]{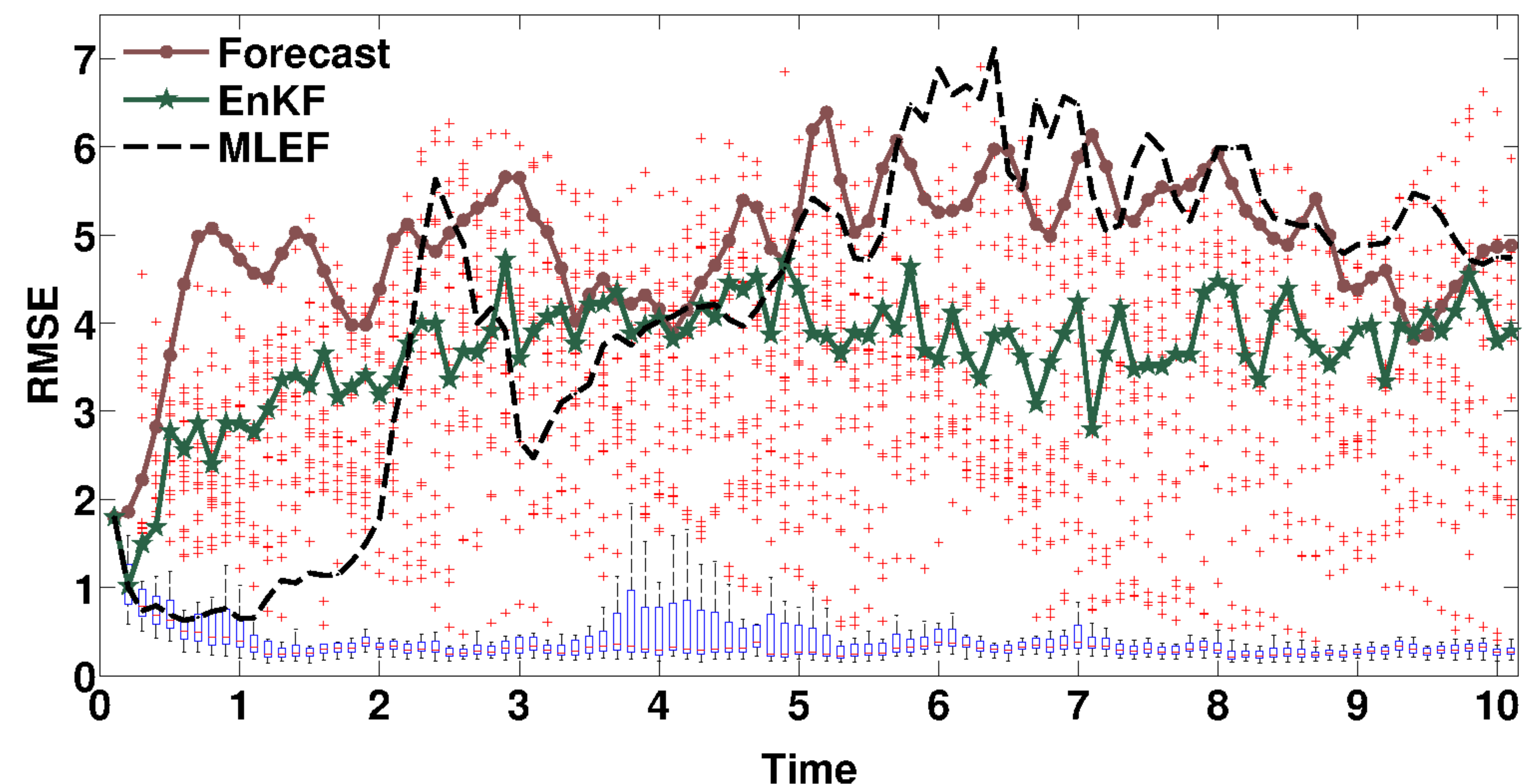}   
\label{fig:Quadratic_Exper2_4Stage}}
\subfigure[Integrator defined on Hilbert space \eqref{eqn:hilbert_integrator}; $h=0.01,\ m=40$]{%
\includegraphics[width=0.44\linewidth]{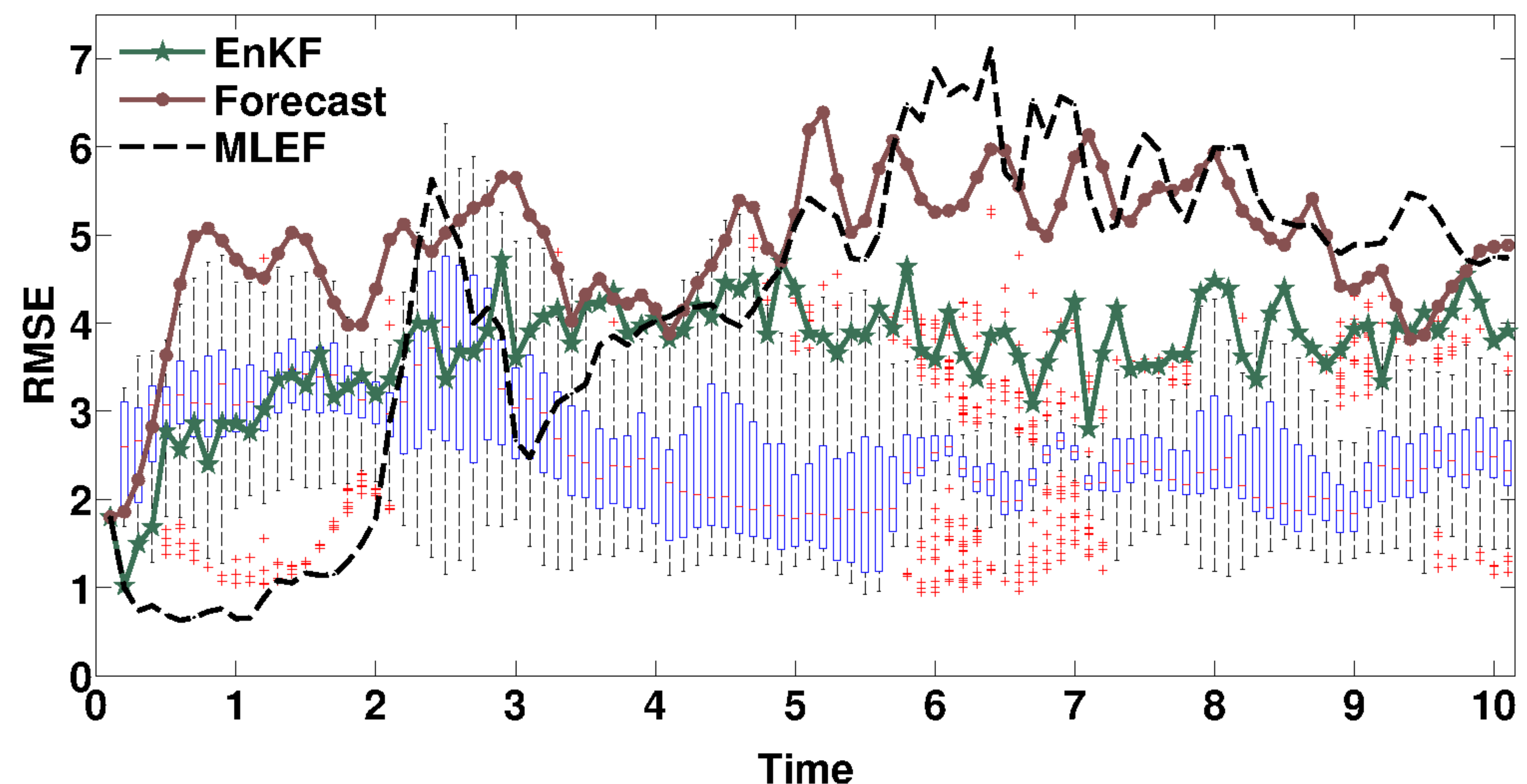}   
\label{fig:Quadratic_Exper2_Hilbert}}
    \caption{Data assimilation results with the quadratic observation operator \eqref{eqn:Quadratic_H}. The symplectic integrator used is indicated under each panel.
     The time step for all integrators is $T=mh$, with $h,\, m$, as indicated under each panel. The number of inter-chain steps is $30$. 
     The RMSE for $100$ instances of the sampling filter results are shown as box plots. The  red line represents the median RMSE values across all instances and the central blue box represents the variance. The two vertical lines (whiskers) extend up to $1.5$ times the height of the central box. The values exceeding the length of the whiskers are considered outliers (extremes) and are plotted as red crosses.
     } 
    \label{fig:Quadratic_Exper2}
    \end{figure}

Figure \ref{fig:Quadratic_Exper2} shows results with the time parameters tuned such as to obtain the best possible results with the Verlet scheme; the step sizes of other integrators are chosen such that their work per step is equal to Verlet's work per step.  The Verlet integrator results in high uncertainty in the RMSE which makes the divergence of the filter very likely. High-order integrators continue to give good results but the number outliers seem to increase. The integrator defined on Hilbert space fails completely and yields large RMSE  in many cases. We conclude that the step sizes should be tuned independently for each integrator. The experiments indicate that each of the integrators, except perhaps Verlet, can be tuned to give very satisfactory filtering results.             
           
    %
    %
    %
    %
     
     \subsection{Cubic observation operator experiments}
     \label{subsec:Cubic_H_Results}
   
\begin{figure}[H]
    \centering
\subfigure[Position Verlet integrator \eqref{eqn:Verlet}]{%
\includegraphics[width=0.44\linewidth]{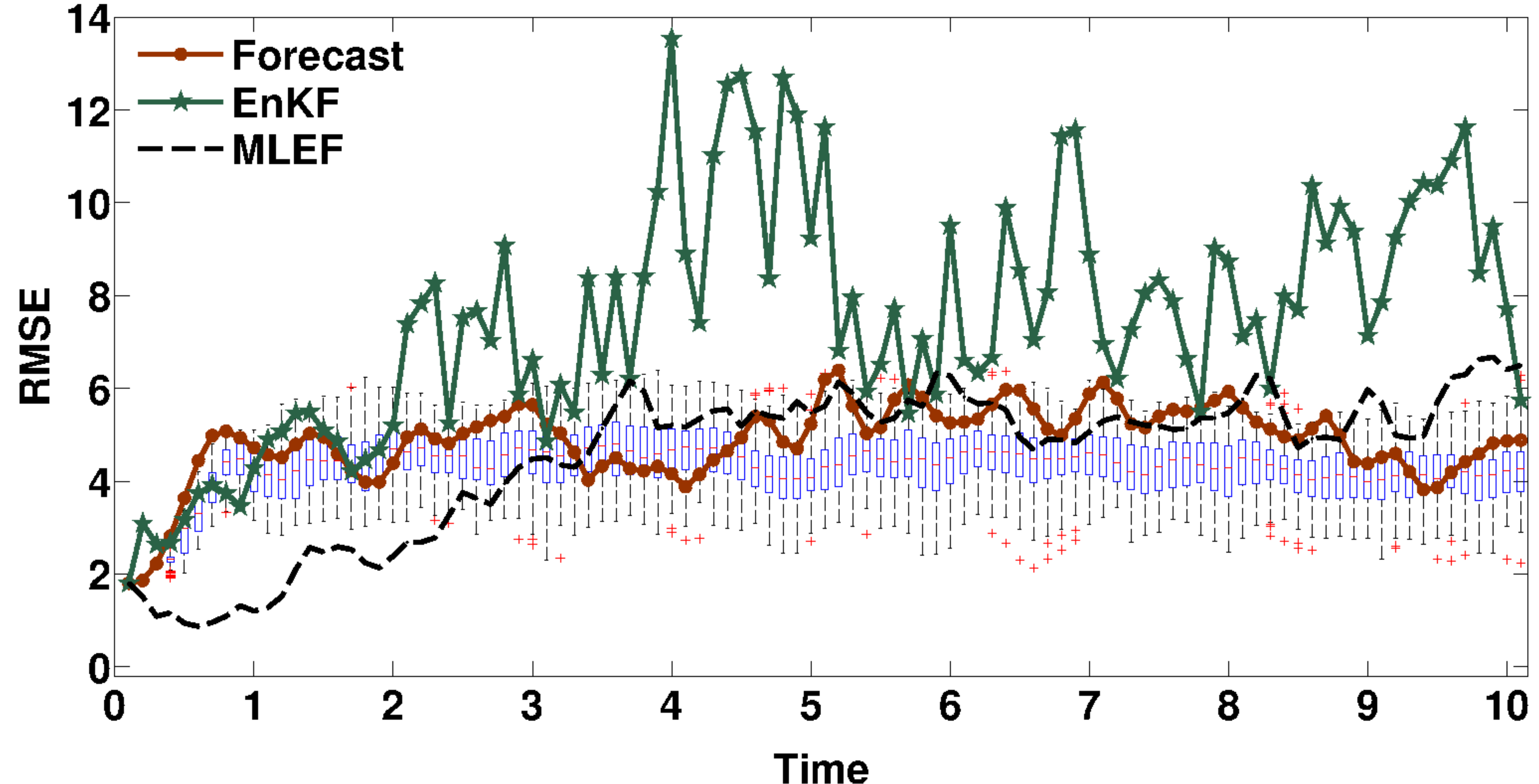}   
\label{fig:Cubic_Exper1_Verlet}}
\quad  
\subfigure[Two-stage integrator \eqref{eqn:two_stage}]{%
\includegraphics[width=0.44\linewidth]{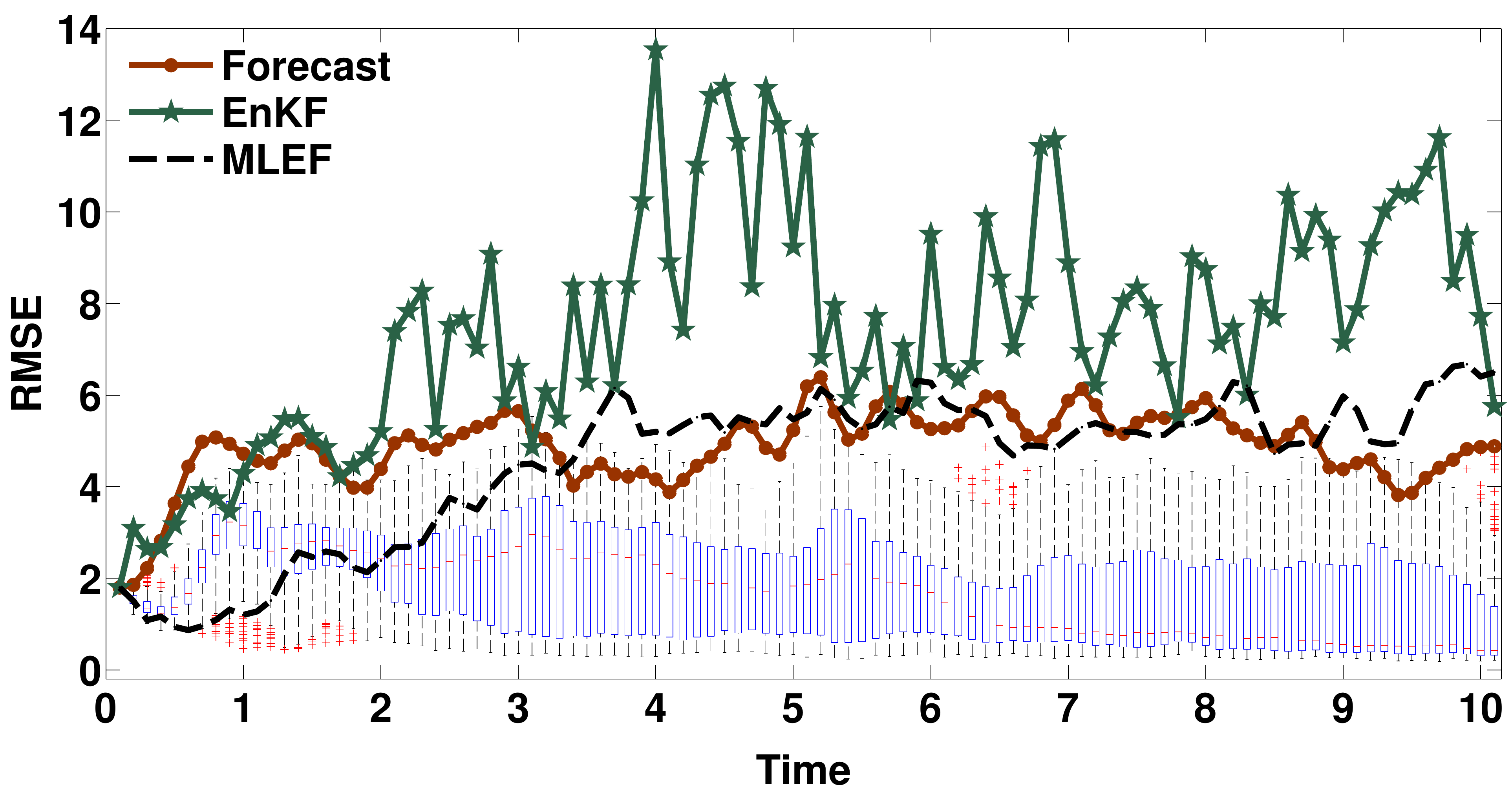}   
\label{fig:Cubic_Exper1_2Stage}}
\subfigure[Three-stage integrator \eqref{eqn:three_stage}]{%
\includegraphics[width=0.44\linewidth]{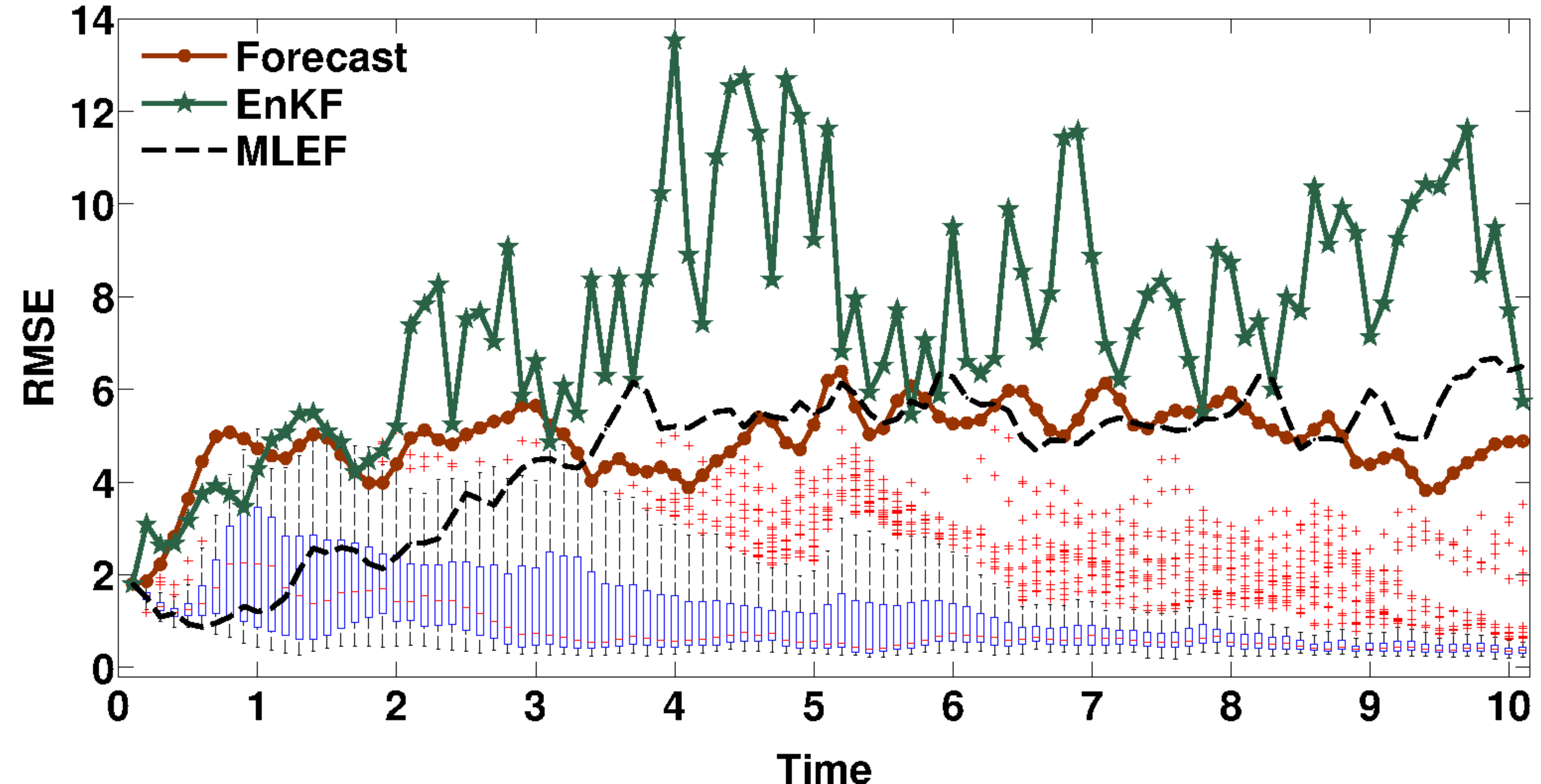}   
\label{fig:Cubic_Exper1_3Stage}}
\quad
\subfigure[Four-stage integrator \eqref{eqn:four_stage}]{%
\includegraphics[width=0.44\linewidth]{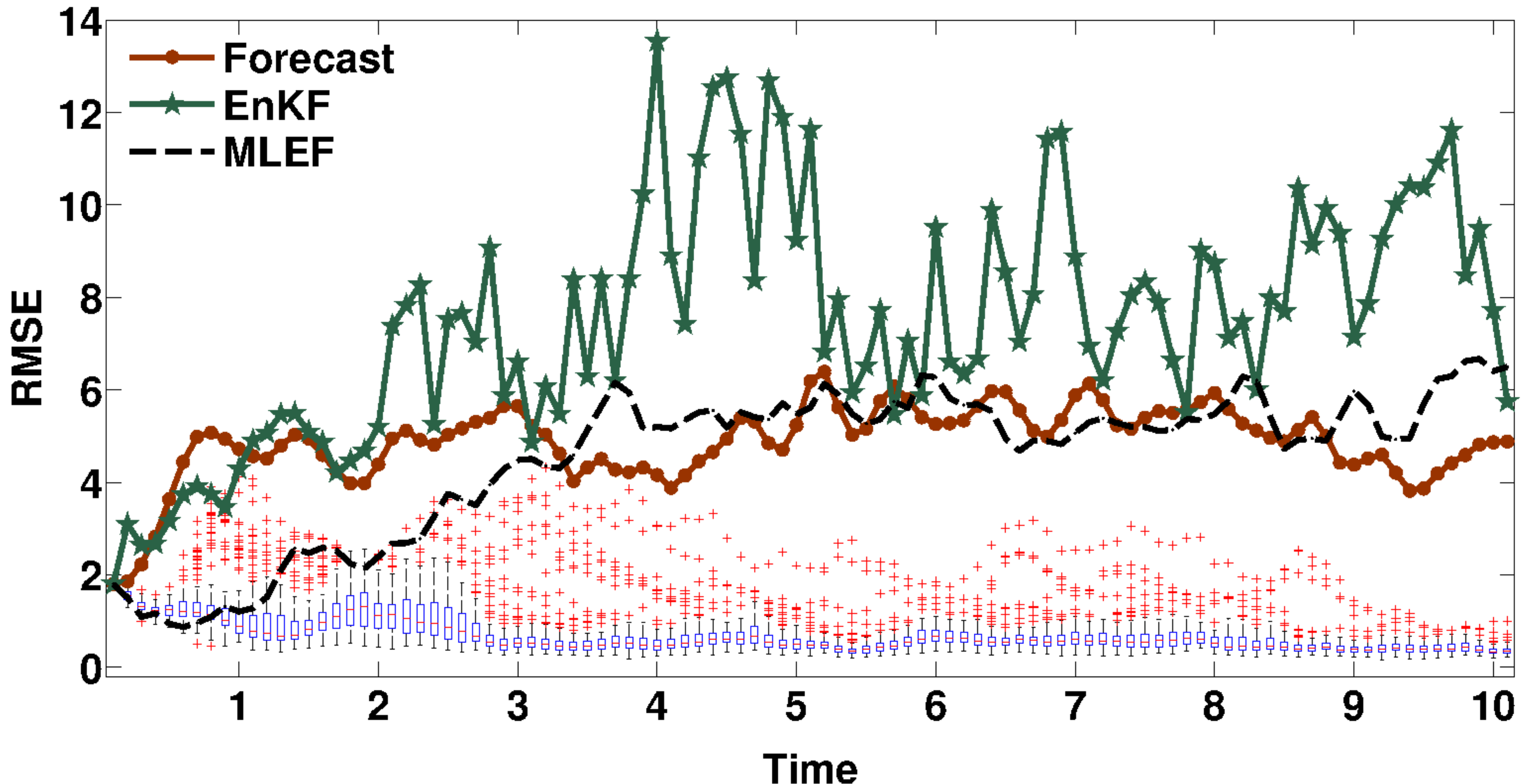}   
\label{fig:Cubic_Exper1_4Stage}}
\subfigure[Integrator defined on Hilbert space \eqref{eqn:hilbert_integrator}]{%
\includegraphics[width=0.44\linewidth]{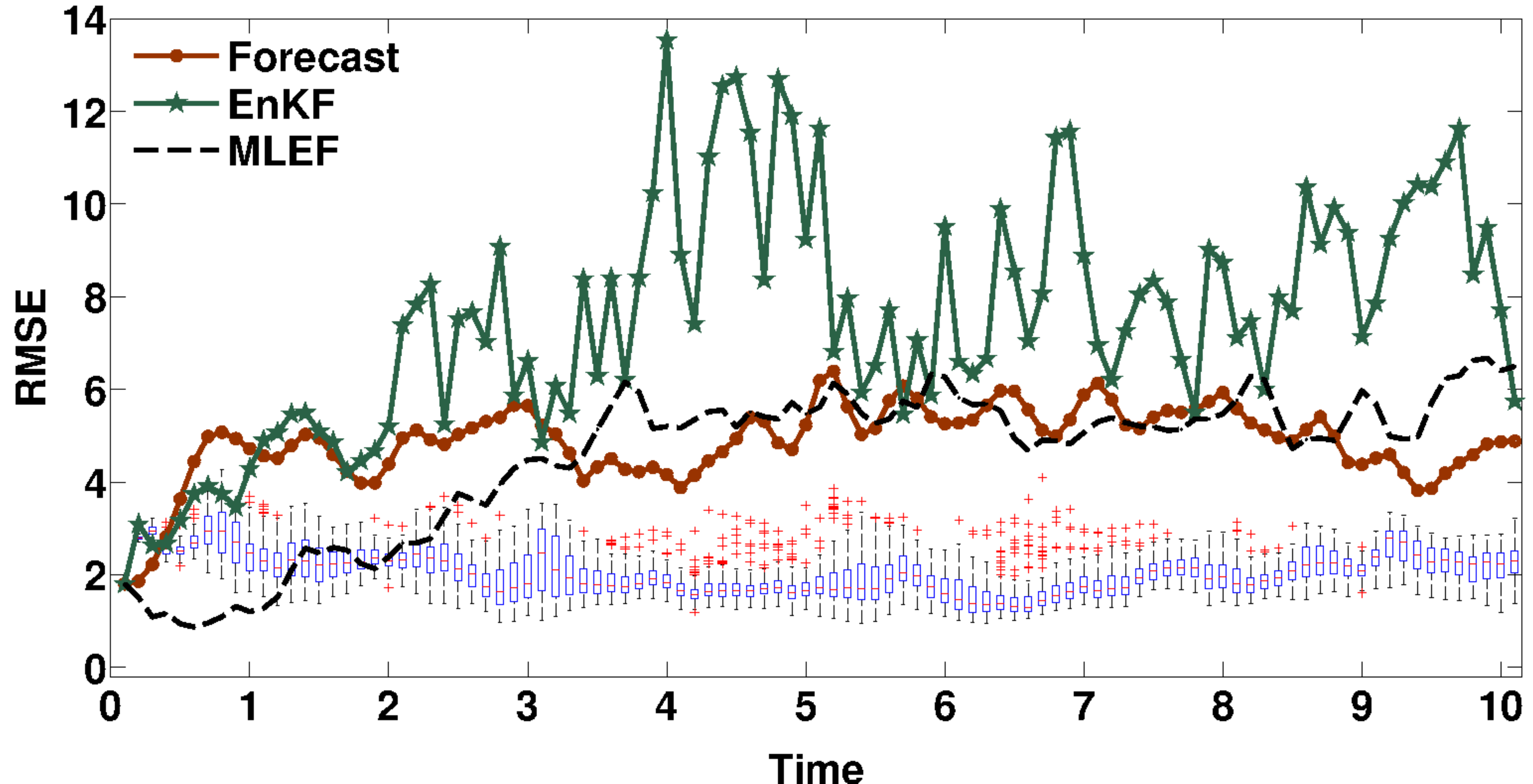}   
\label{fig:Cubic_Exper1_Hilbert}}
    \caption{Data assimilation results with the cubic observation operator \eqref{eqn:Cubic_H}. The symplectic integrator used is indicated under each panel.
     The time step for all integrators is $T=0.1$ with $h=0.01$, $m=10$, and $30$ inter-chain steps.
     The RMSE for $100$ instances of the sampling filter results are shown as box plots. The  red line represents the median RMSE values across all instances and the central blue box represents the variance. The two vertical lines (whiskers) extend up to $1.5$ times the height of the central box. The values exceeding the length of the whiskers are considered outliers (extremes) and are plotted as red crosses.
     } 
   \label{fig:Cubic_Exper1}
 \end{figure}

Figure \ref{fig:Cubic_Exper1} shows the results with cubic observation operator \eqref{eqn:Cubic_H}. Both EnKF and MLEF fail to converge due to high non-linearity of the observation operator. The MLEF failure was unexpected and may be due to its sensitivity to the uncertainty levels of either the background, or the observations or both. The level of nonlinearity of the observation operator has a major impact on the success of the MLEF filter as shown in \ref{subsec:More_on_MLEF_results}. 
The sampling filter with position Verlet integrator fails to converge. The results are better for high-stage integrators, and the four-stage integrator provides satisfactory results that are similar to those obtained with linear observation operators.  

\begin{figure}[H]
\centering
\subfigure[Position Verlet integrator \eqref{eqn:Verlet}; $h=0.001,\ m=40$]{%
\includegraphics[width=0.44\linewidth]{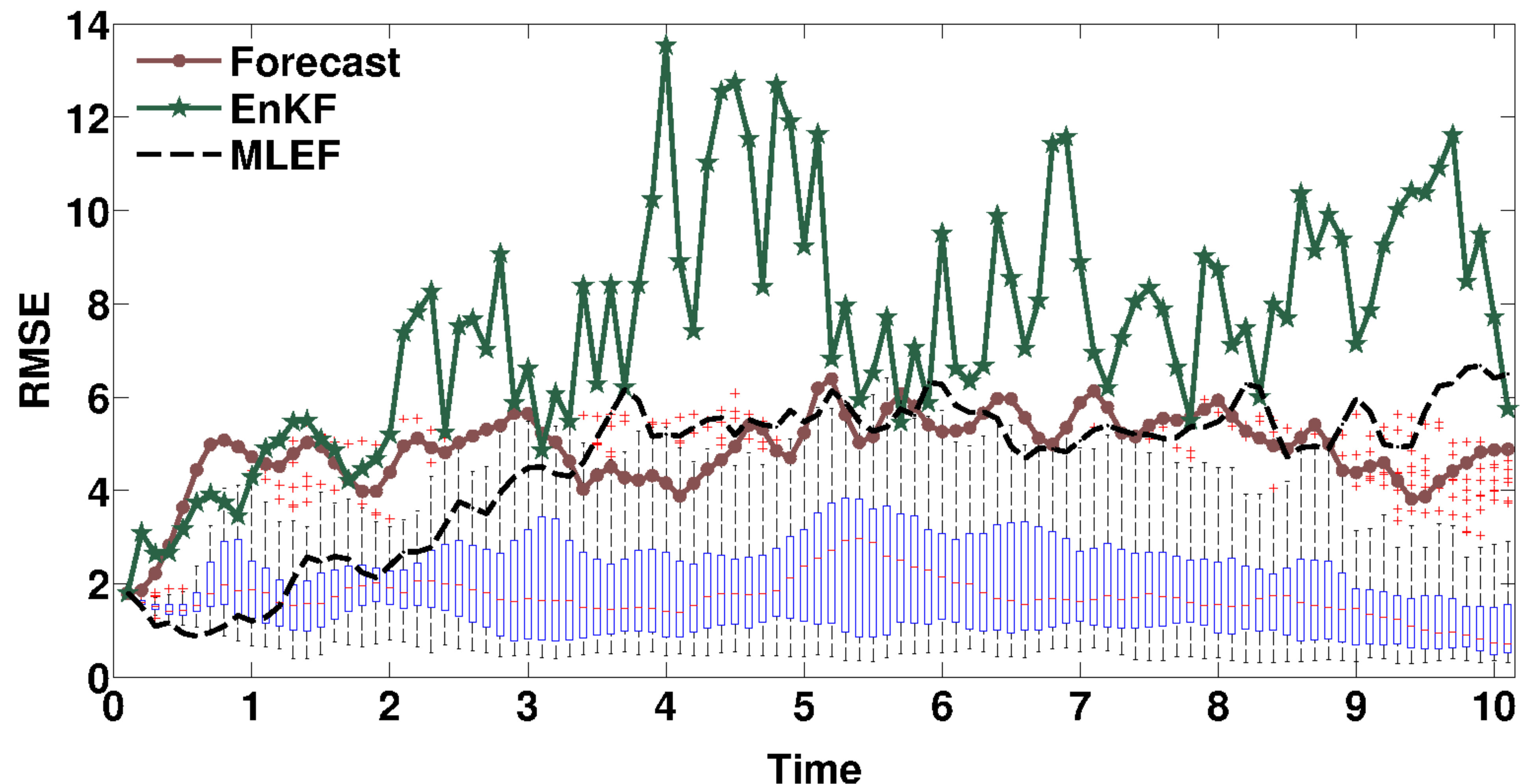}   
\label{fig:Cubic_Exper2_Verlet}}
\quad  
\subfigure[Two-stage integrator \eqref{eqn:two_stage}; $h=0.002,\ m=20$]{%
\includegraphics[width=0.44\linewidth]{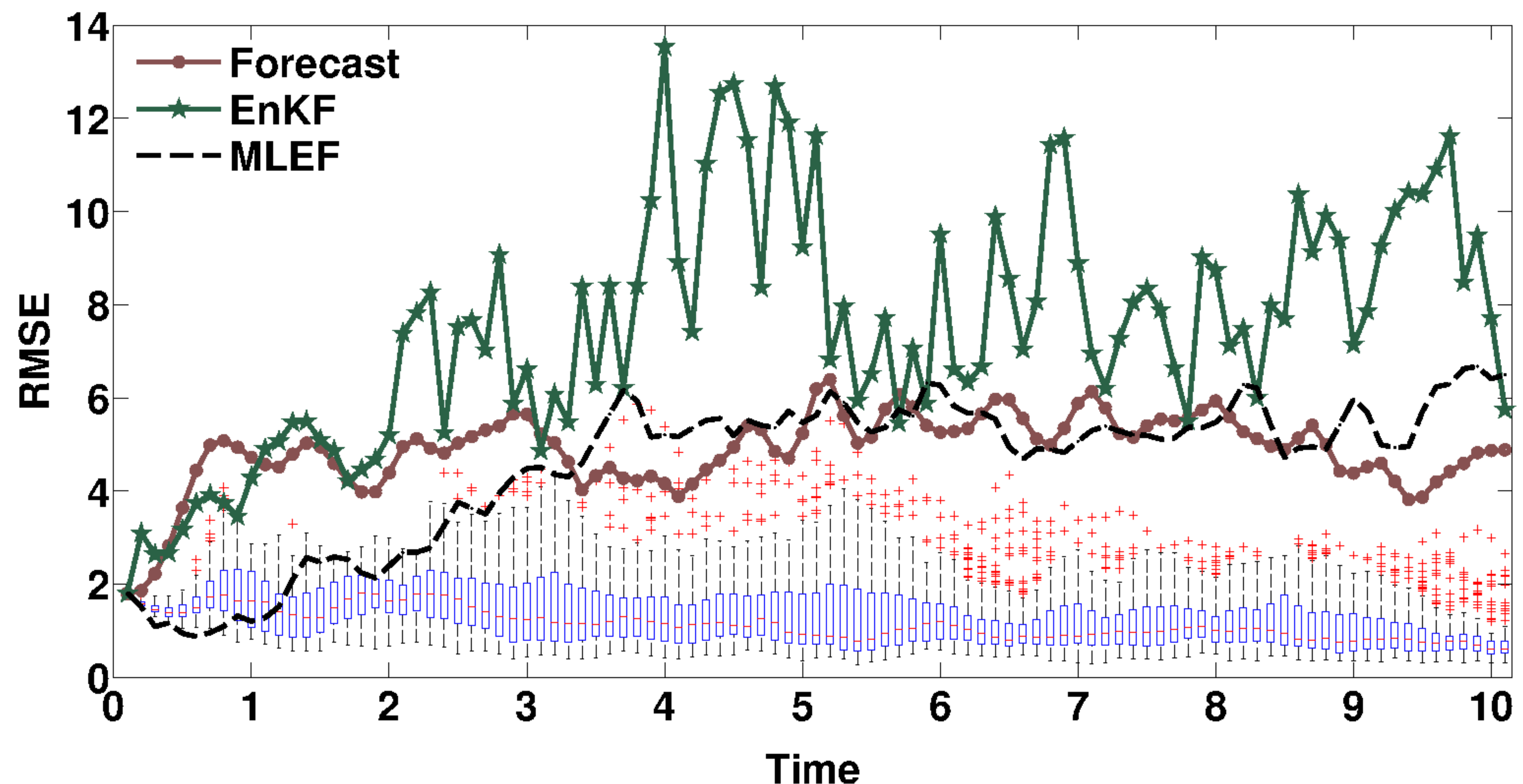}   
\label{fig:Cubic_Exper2_2Stage}}
\subfigure[Three-stage integrator \eqref{eqn:three_stage}; $h=0.003,\ m=13$]{%
\includegraphics[width=0.44\linewidth]{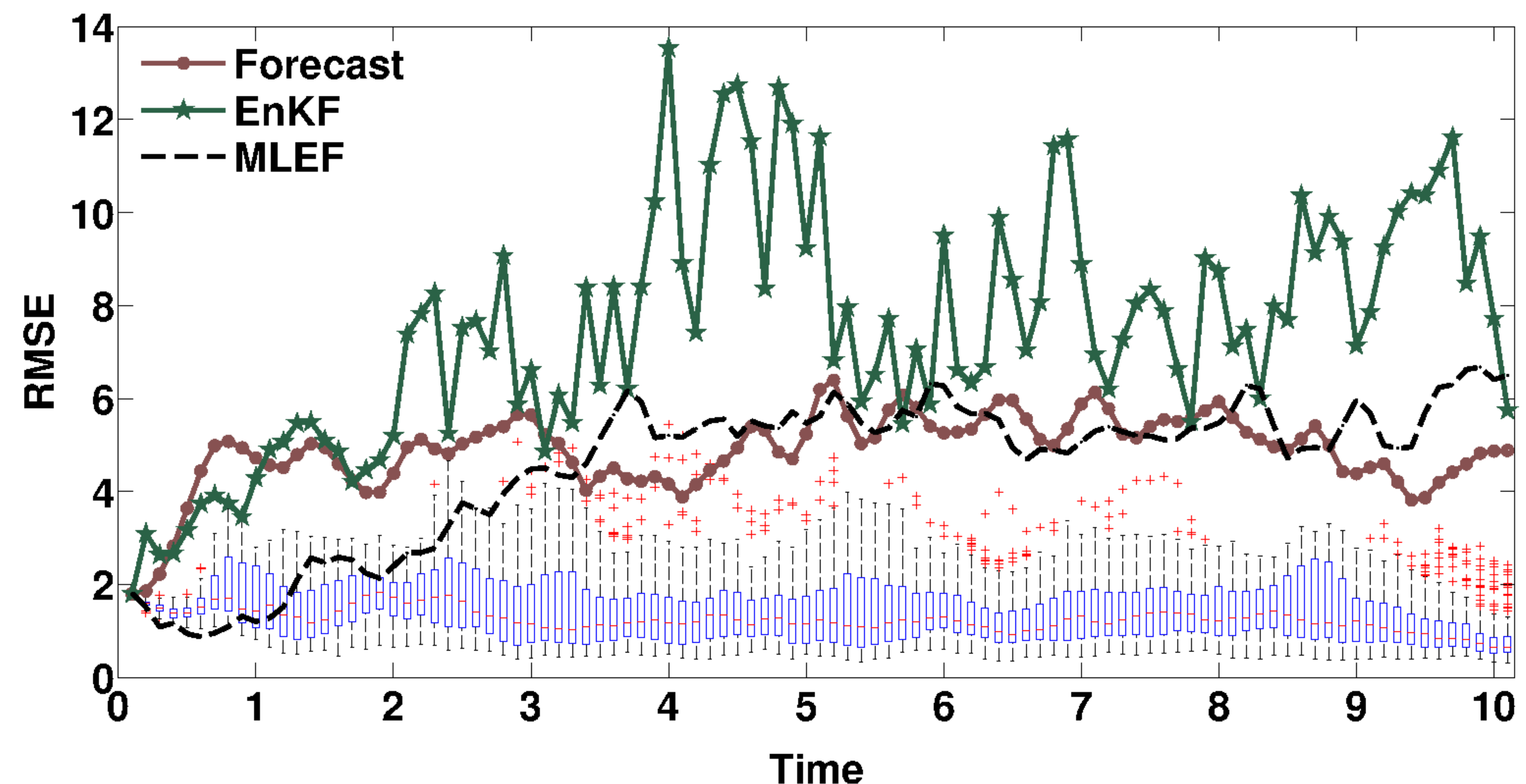}   
\label{fig:Cubic_Exper2_3Stage}}
\quad
\subfigure[Four-stage integrator \eqref{eqn:four_stage}; $h=0.004,\ m=10$]{%
\includegraphics[width=0.44\linewidth]{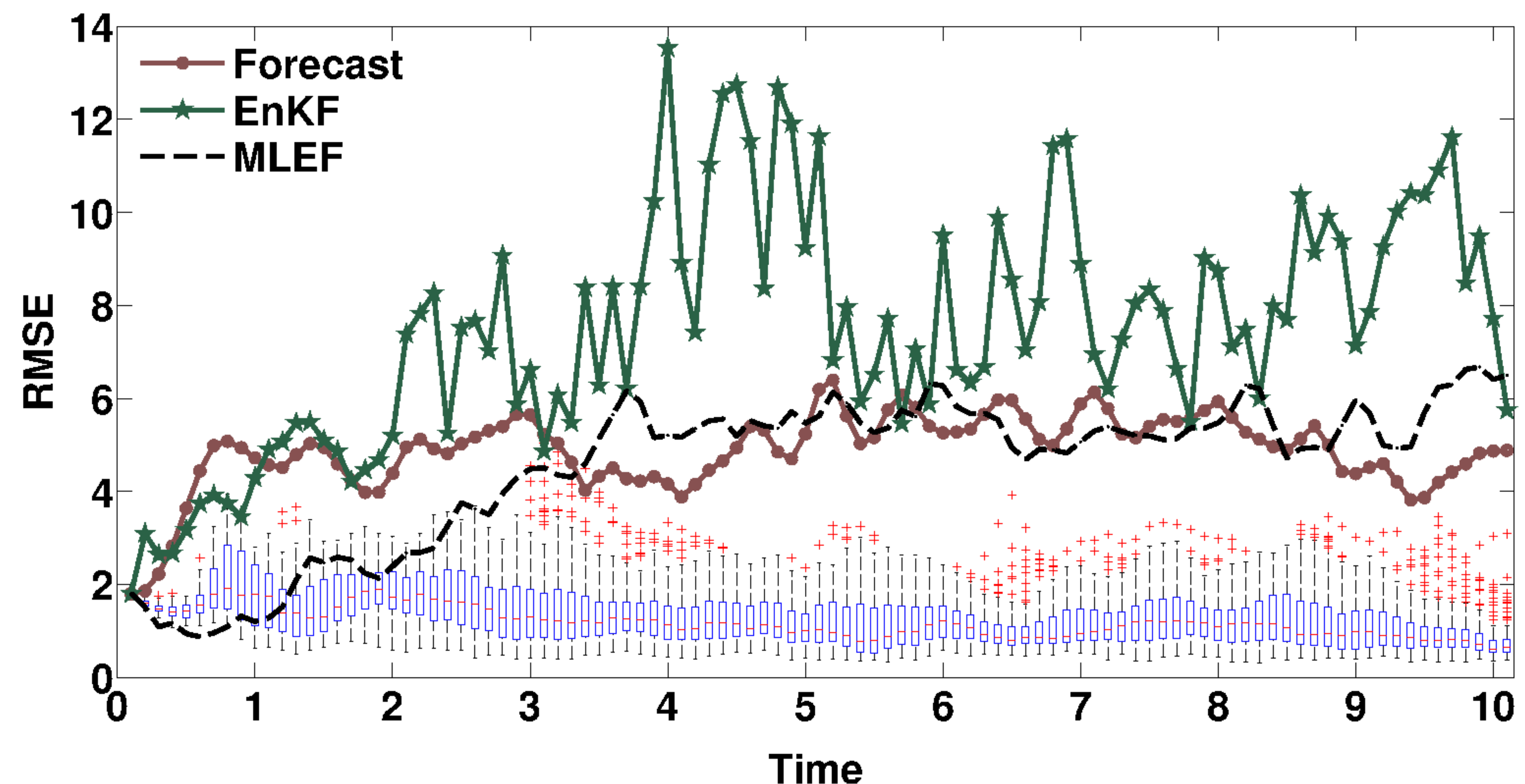}   
\label{fig:Cubic_Exper2_4Stage}}
\subfigure[Integrator defined on Hilbert space \eqref{eqn:hilbert_integrator}; $h=0.001,\ m=40$]{%
\includegraphics[width=0.44\linewidth]{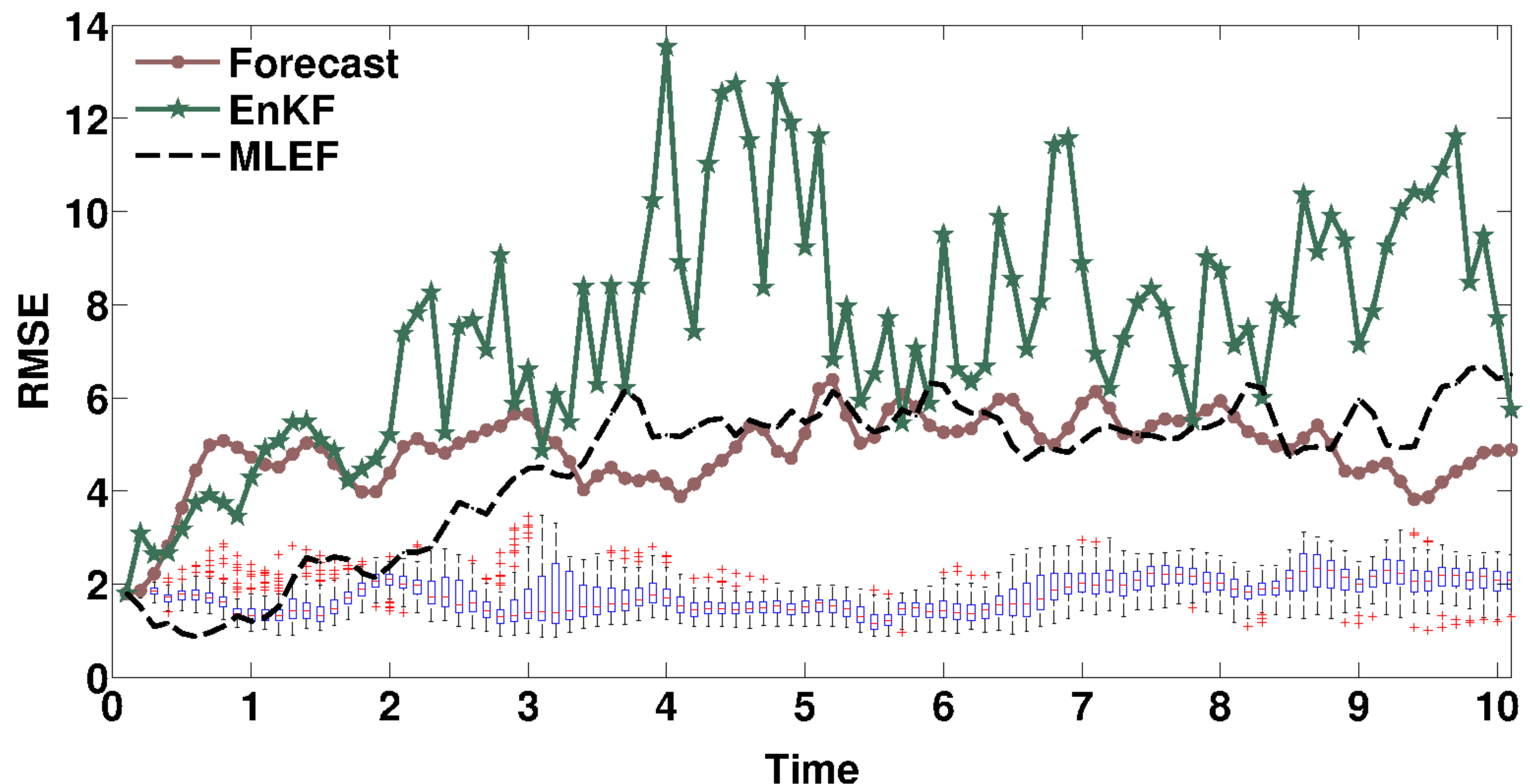}   
\label{fig:Cubic_Exper2_Hilbert}}
    \caption{Data assimilation results with the cubic observation operator \eqref{eqn:Cubic_H}. The symplectic integrator used is indicated under each panel.
     The time step for all integrators is $T=mh$, with $h,\, m$, as indicated under each panel. The number of inter-chain steps is $30$. 
     The RMSE for $100$ instances of the sampling filter results are shown as box plots. The  red line represents the median RMSE values across all instances and the central blue box represents the variance. The two vertical lines (whiskers) extend up to $1.5$ times the height of the central box. The values exceeding the length of the whiskers are considered outliers (extremes) and are plotted as red crosses.
     } 
    \label{fig:Cubic_Exper2}
    \end{figure}

Figure \ref{fig:Cubic_Exper2} shows results with tuned parameters such that the work is equal for all integrators. Position Verlet requires more work and finer step sizes to provide convergence of the sampling filter, and even in this case there are many outliers that show divergence, as seen in Figure \ref{fig:Cubic_Exper2_Verlet}. The high-order integrators give good results, but reducing the step size increases their computational costs. As shown in Figure \ref{fig:Cubic_Exper2_Hilbert}, the Hilbert integrator leads to large RMSE with this setting of step size. Again, it is advisable to tune the step size of this integrator independently.

     \subsection{Absolute value observation operator experiments}
     \label{subsec:Abs_H_Results}
   
\begin{figure}[H]
\centering
\subfigure[Position Verlet integrator \eqref{eqn:Verlet}]{%
\includegraphics[width=0.44\linewidth]{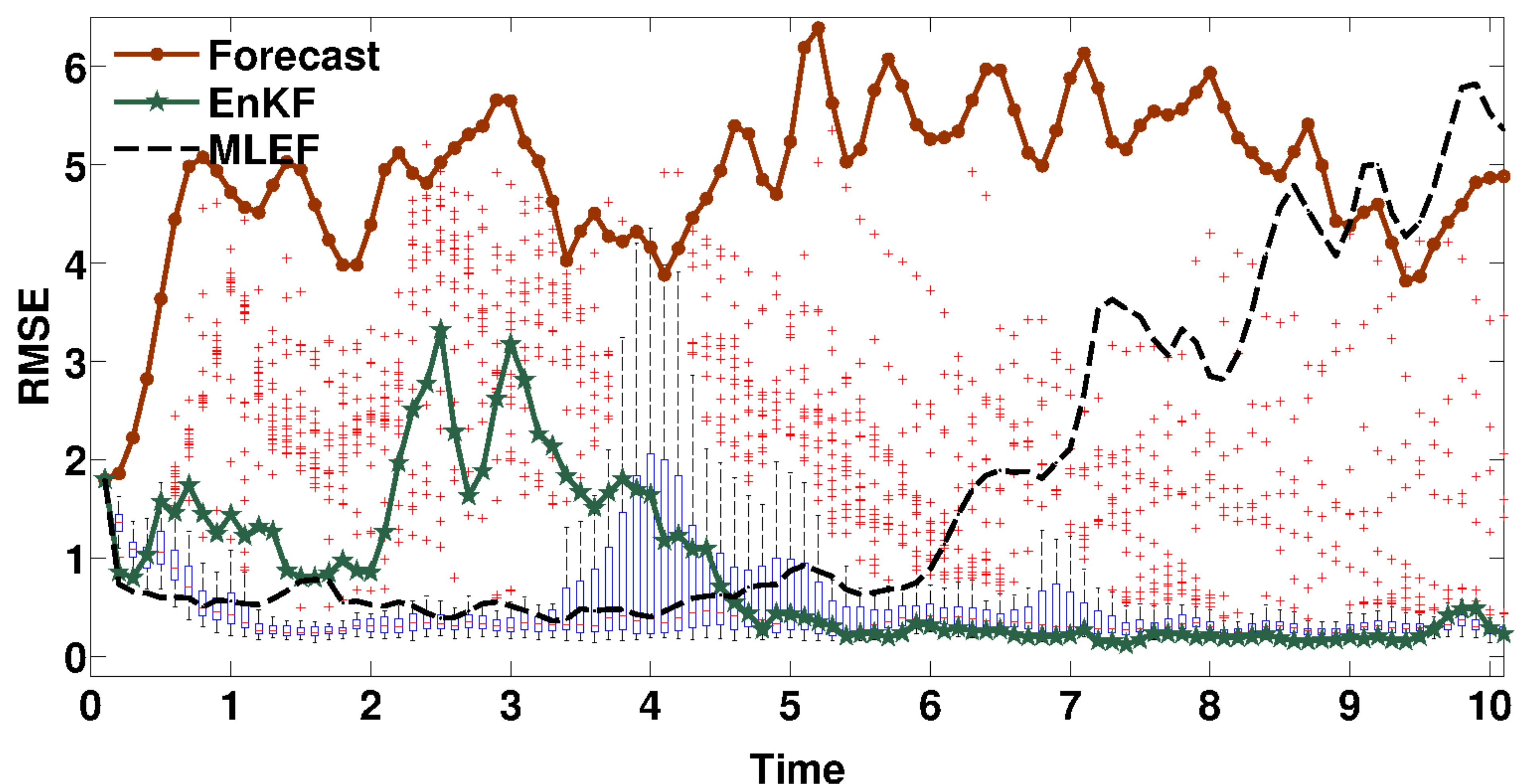}   
\label{fig:Abs_Exper1_Verlet}}
\quad  
\subfigure[Two-stage integrator \eqref{eqn:two_stage}]{%
\includegraphics[width=0.44\linewidth]{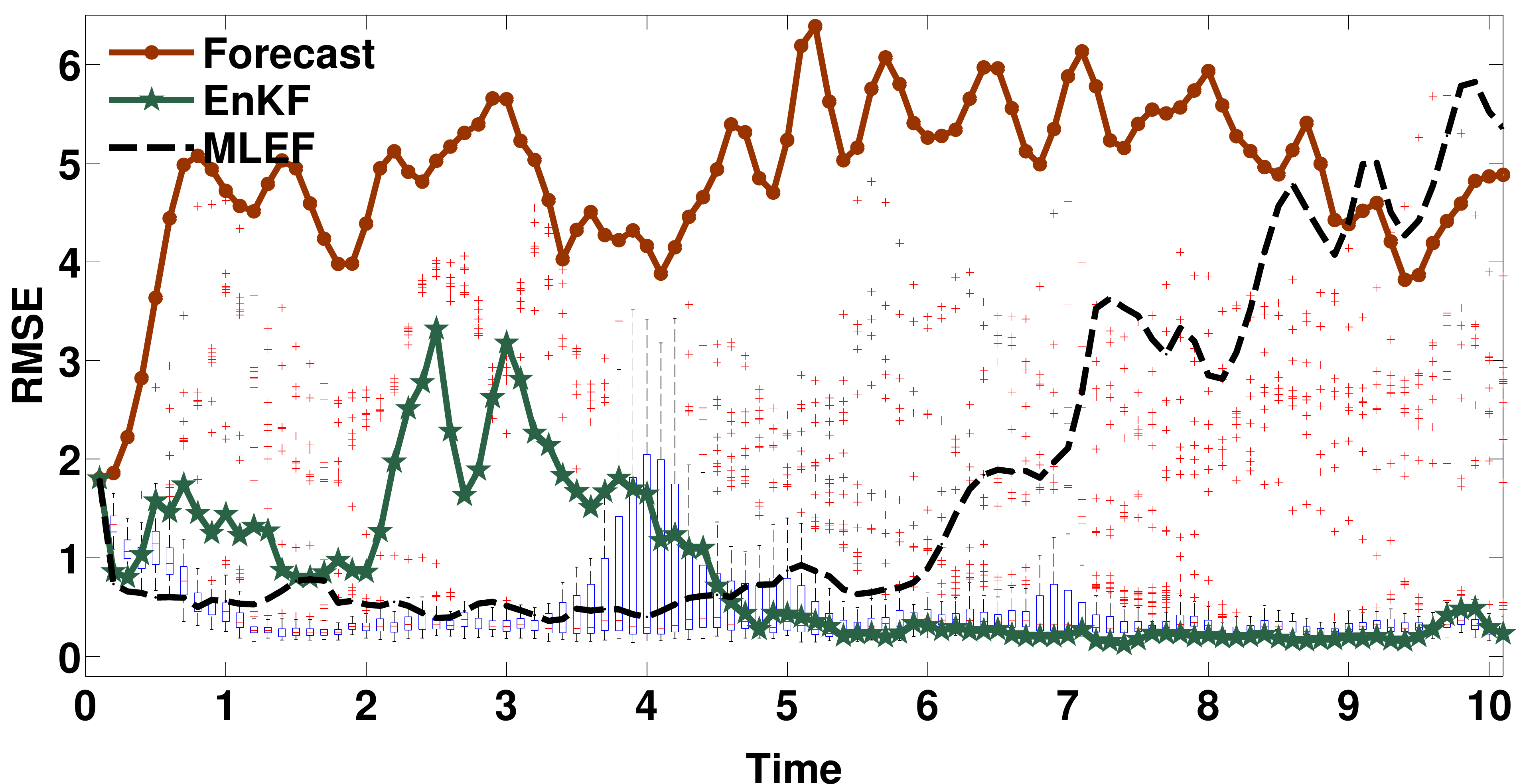}   
\label{fig:Abs_Exper1_2Stage}}
\subfigure[Three-stage integrator \eqref{eqn:three_stage}]{%
\includegraphics[width=0.44\linewidth]{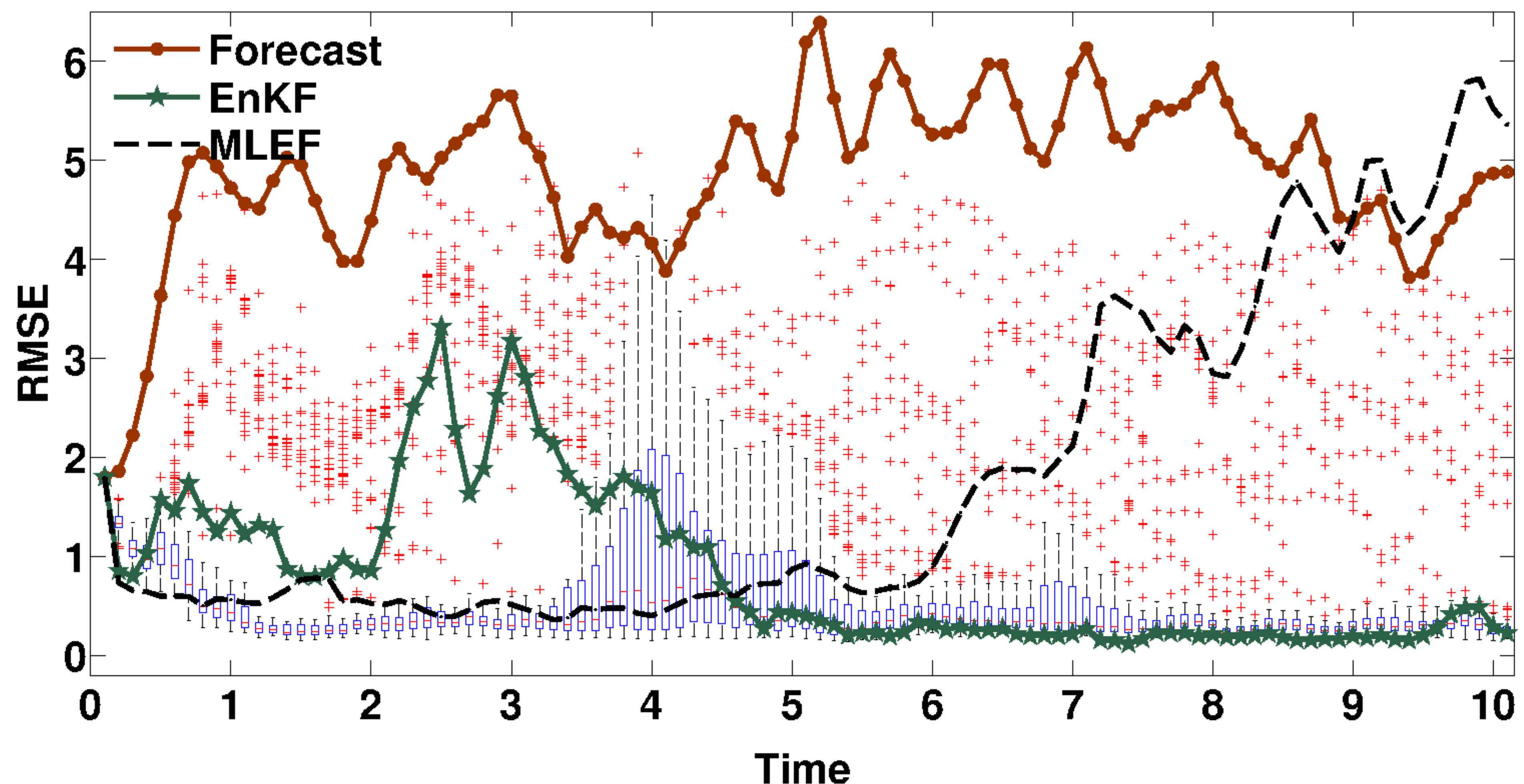}   
\label{fig:Abs_Exper1_3Stage}}
\quad
\subfigure[Four-stage integrator \eqref{eqn:four_stage}]{%
\includegraphics[width=0.44\linewidth]{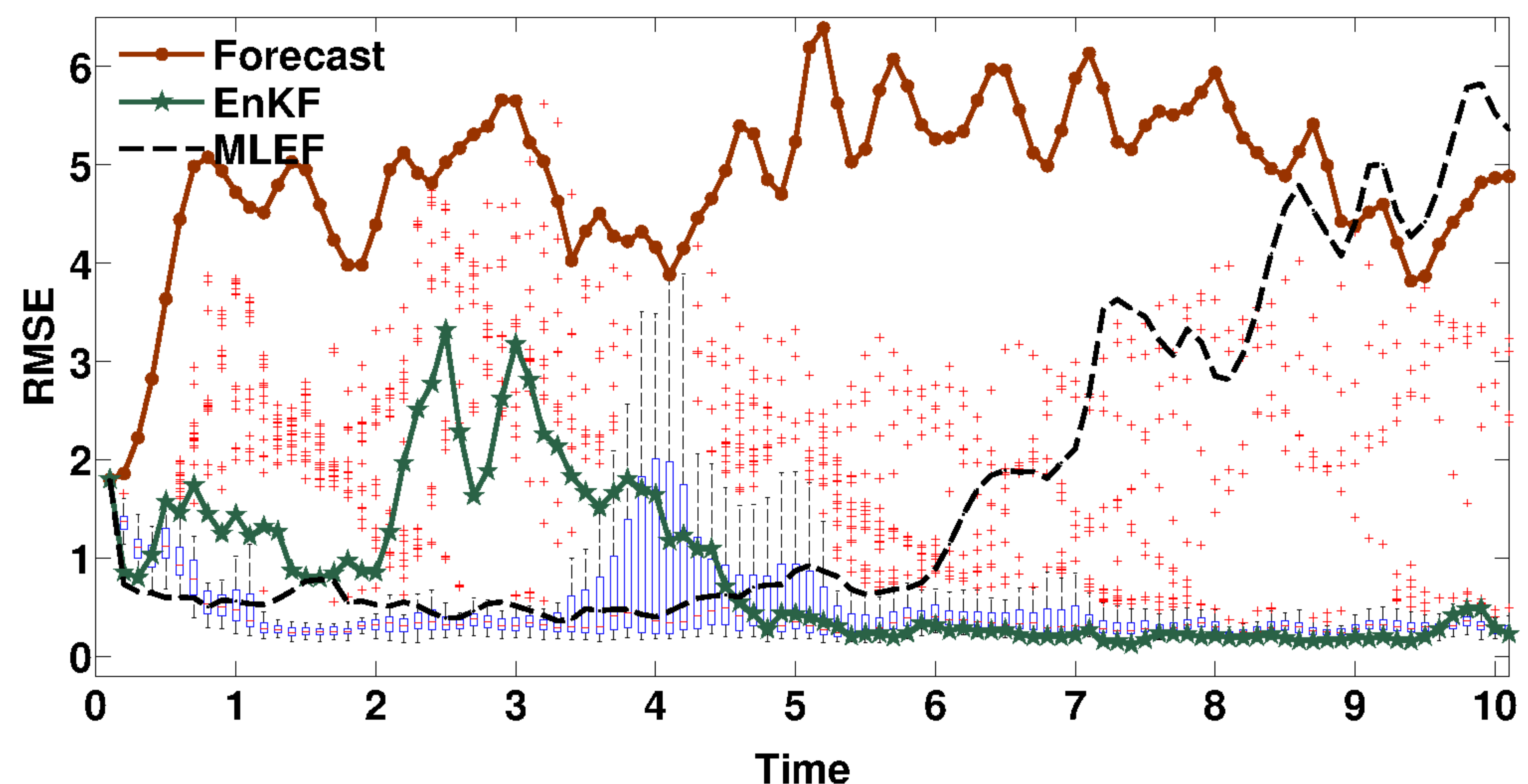}   
\label{fig:Abs_Exper    1_4Stage}}
\subfigure[Integrator defined on Hilbert space \eqref{eqn:hilbert_integrator}]{%
\includegraphics[width=0.44\linewidth]{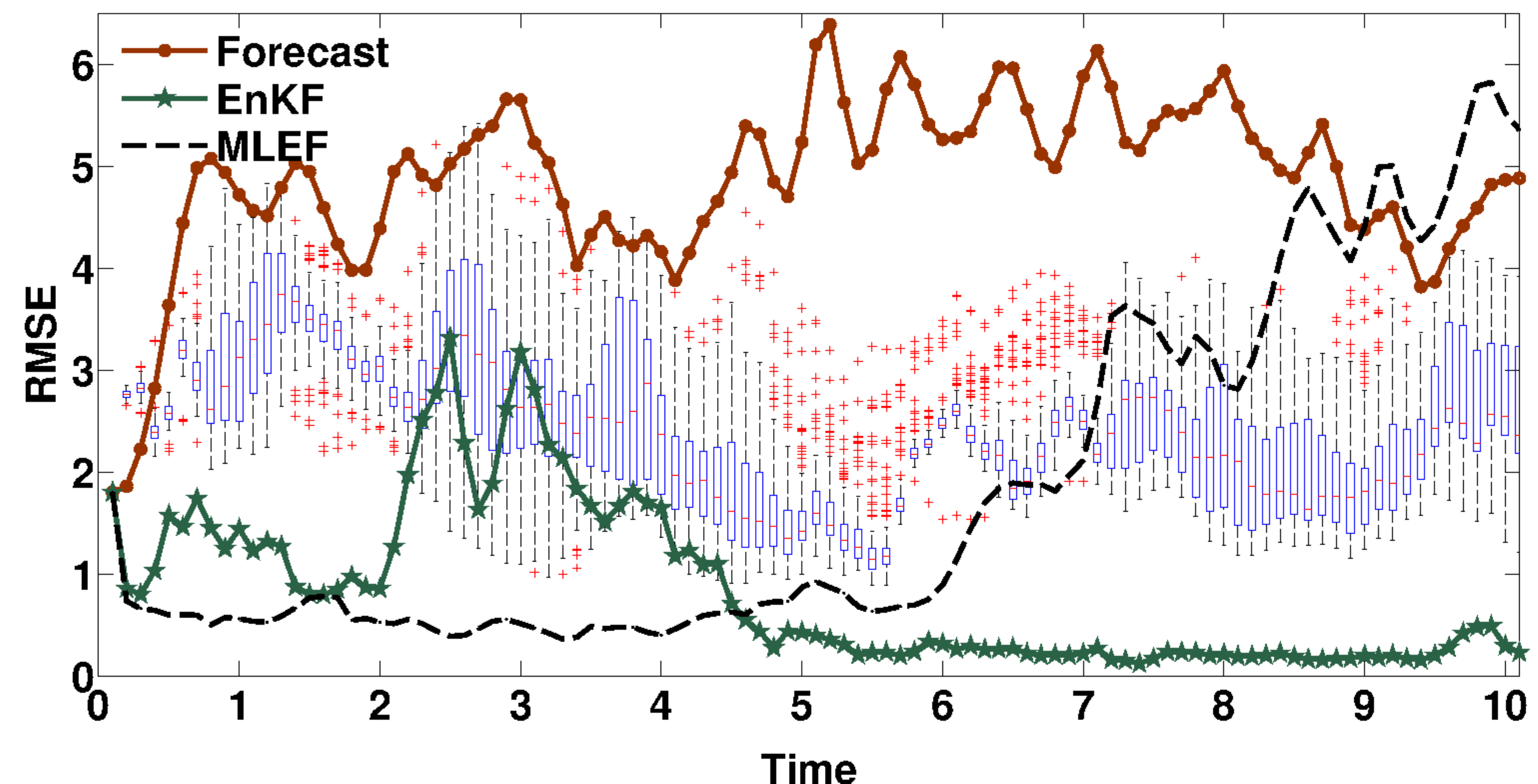}   
\label{fig:Abs_Exper1_Hilbert}}
    \caption{Data assimilation results with the magnitude observation operator \eqref{eqn:Abs_H}. The symplectic integrator used is indicated under each panel.
     The time step for all integrators is $T=0.1$ with $h=0.01$, $m=10$, and $30$ inter-chain steps. 
     The RMSE for $100$ instances of the sampling filter results are shown as box plots. The  red line represents the median RMSE values across all instances and the central blue box represents the variance. The two vertical lines (whiskers) extend up to $1.5$ times the height of the central box. The values exceeding the length of the whiskers are considered outliers (extremes) and are plotted as red crosses.
     } 
    \label{fig:Abs_Exper1}
    \end{figure}

Figure \ref{fig:Abs_Exper1} shows the results with absolute observation operator \eqref{eqn:Abs_H}. The Jacobian of this observation operator is taken as the sign of the measured components of the state vector. Similar to the case of linear observation operator, MLEF converges in the beginning, since the observation operator is weekly non-linear, but diverges later in the experiment,  mostly due to large observation errors and low observation frequency. The sampling filter using Hilbert integrator shows improvement over the forecast, but its analysis is less accurate than MLEF or EnKF analyses (when they converge). Verlet and the high-order integrators behave almost identically. The distribution of outliers is similar to that for quadratic observation operator. We will discuss how to deal with the occasional filter divergence in Section \ref{subsec:MC_Steps_Effect}. 

\begin{figure}[H]
\centering
\subfigure[Position Verlet integrator \eqref{eqn:Verlet}; $h=0.01,\ m=24$]{%
\includegraphics[width=0.44\linewidth]{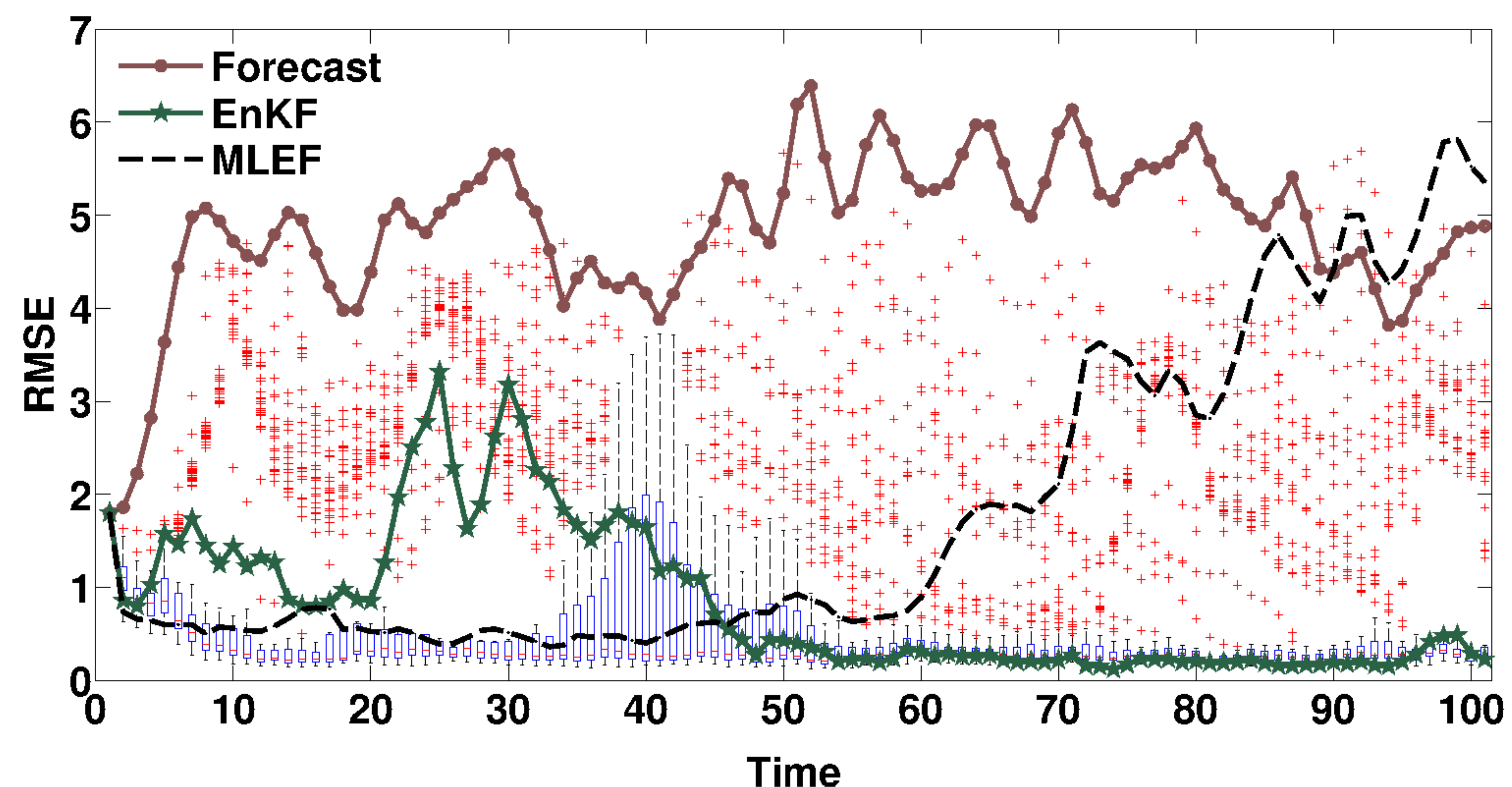}   
\label{fig:AbsH_Exper2_Verlet}}
\quad  
\subfigure[Two-stage integrator \eqref{eqn:two_stage}; $h=0.02,\ m=12$]{%
\includegraphics[width=0.44\linewidth]{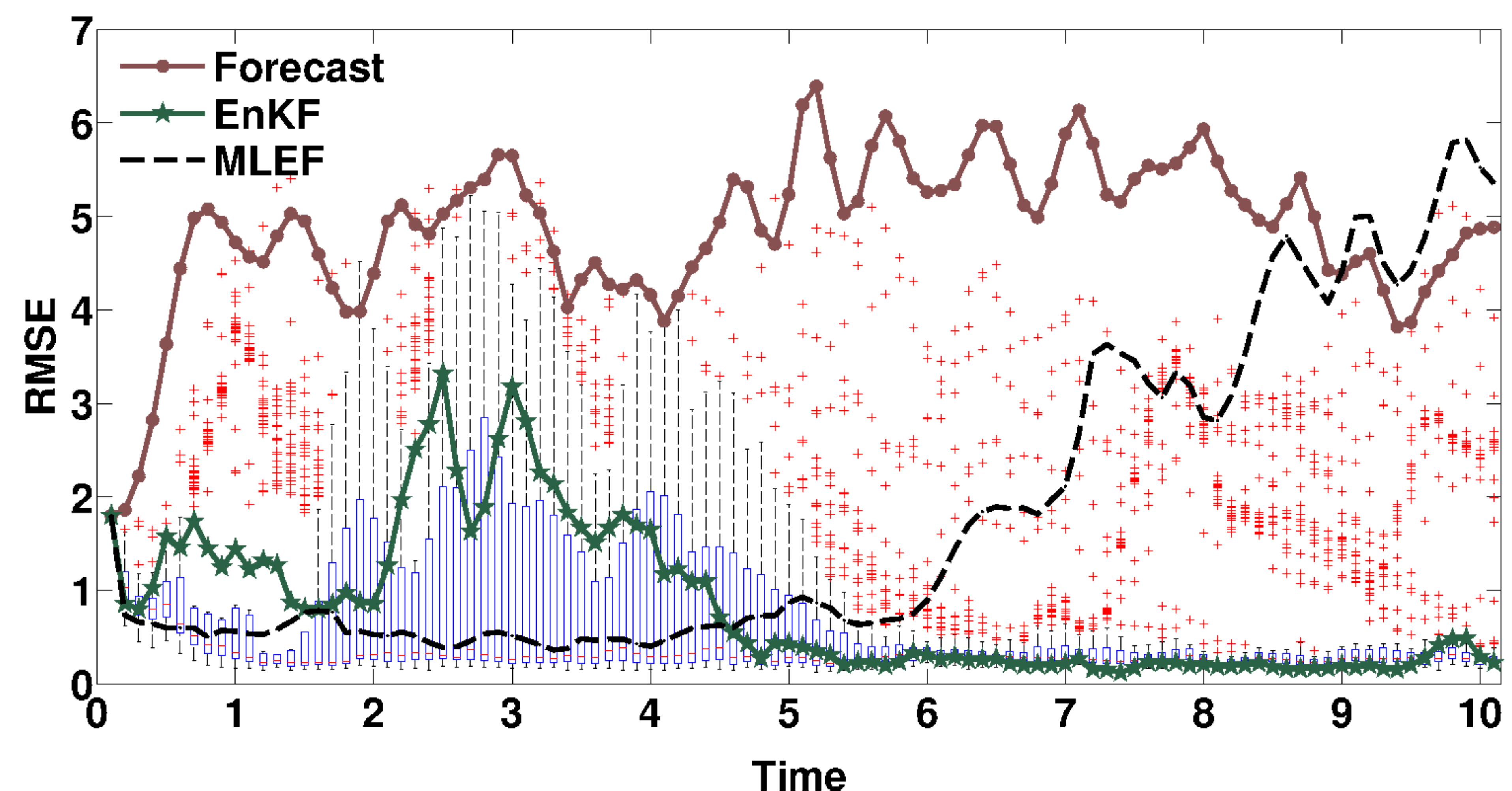}   
\label{fig:AbsH_Exper2`_2Stage}}
\subfigure[Three-stage integrator \eqref{eqn:three_stage}; $h=0.03,\ m=8$]{%
\includegraphics[width=0.44\linewidth]{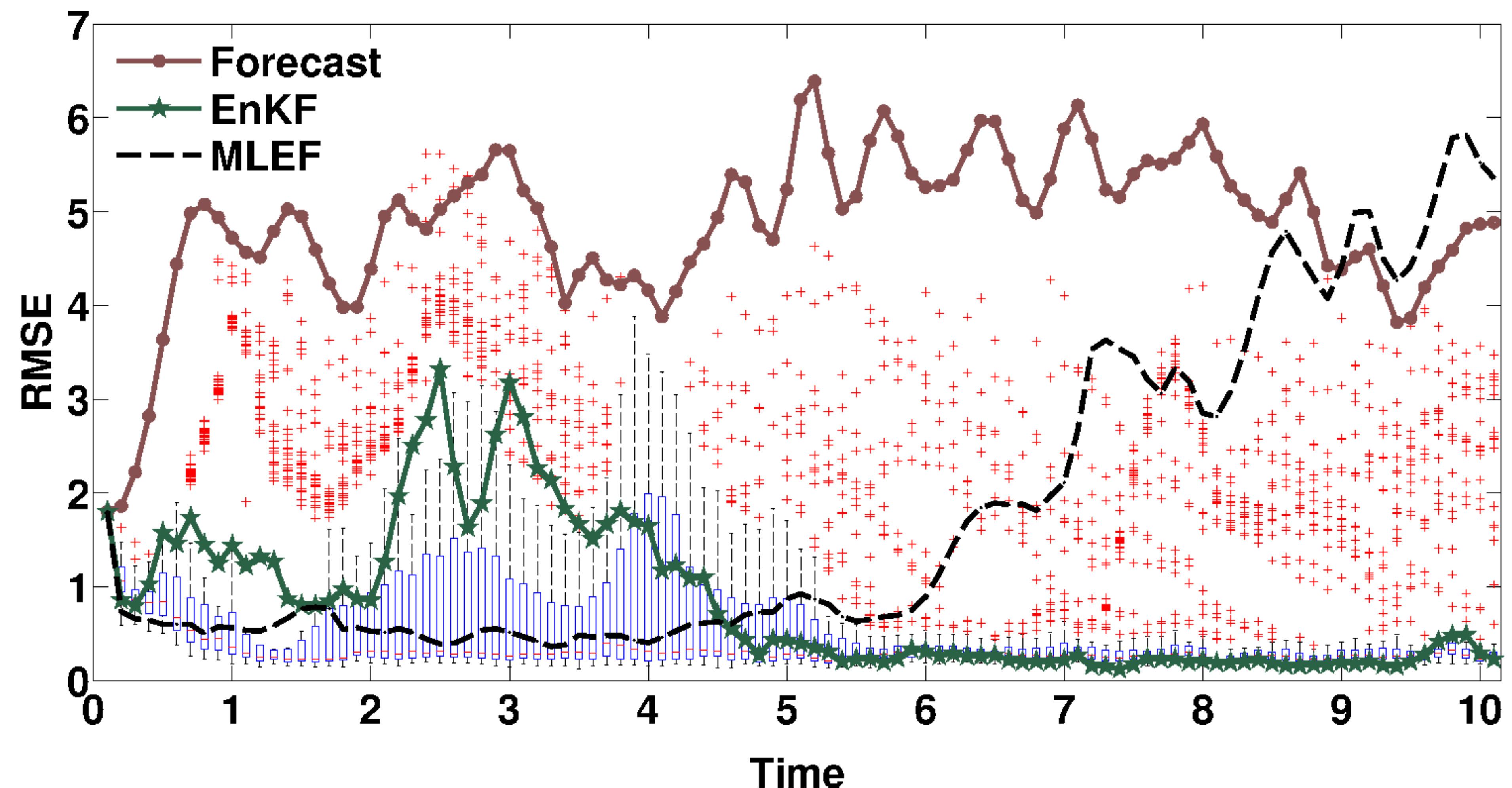}   
\label{fig:AbsH_Exper2_3Stage}}
\quad
\subfigure[Four-stage integrator \eqref{eqn:four_stage}; $h=0.04,\ m=6$]{%
\includegraphics[width=0.44\linewidth]{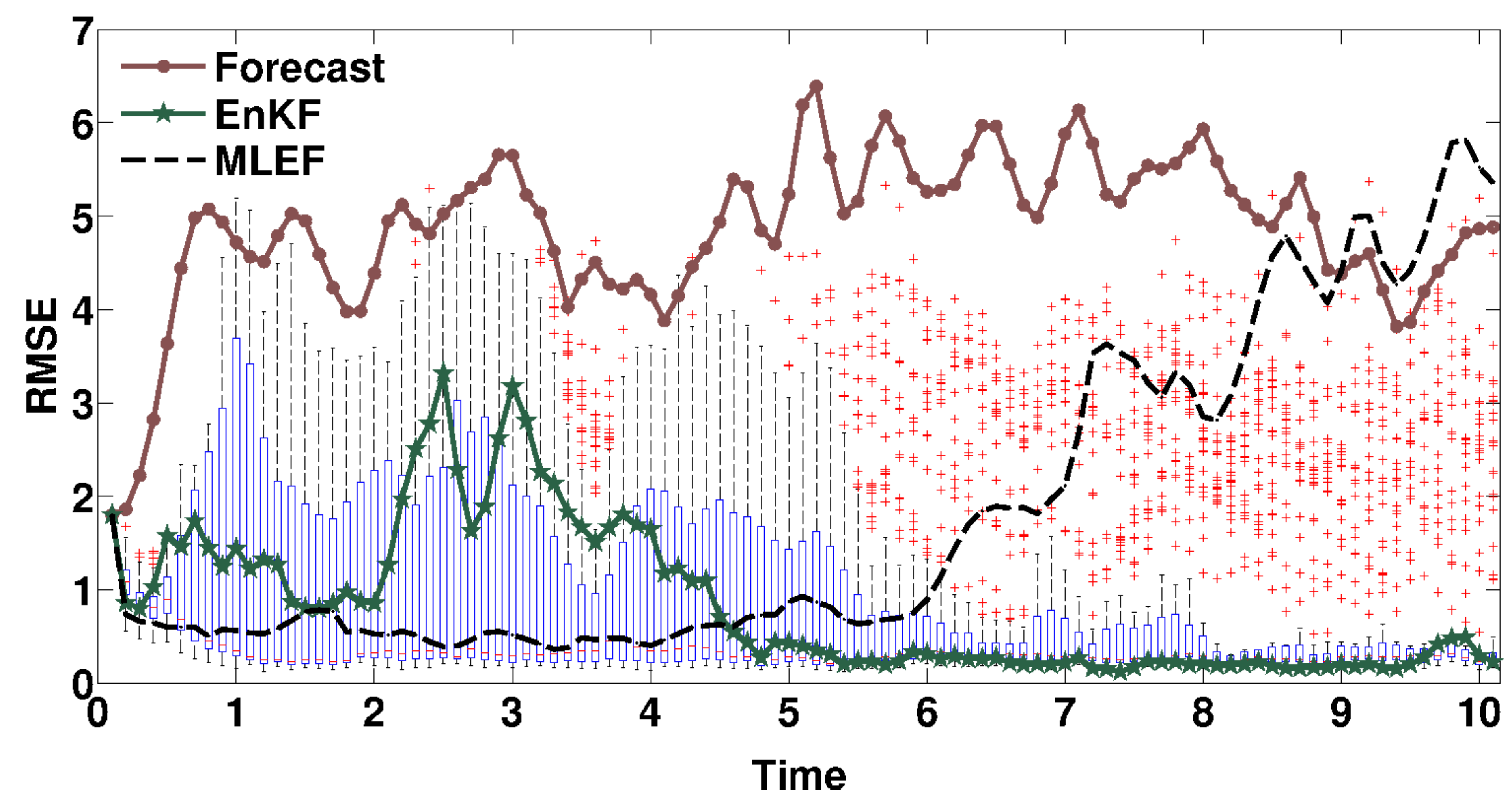}   
\label{fig:AbsH_Exper2_4Stage}}
\subfigure[Integrator defined on Hilbert space \eqref{eqn:hilbert_integrator}; $h=0.01,\ m=24$]{%
\includegraphics[width=0.44\linewidth]{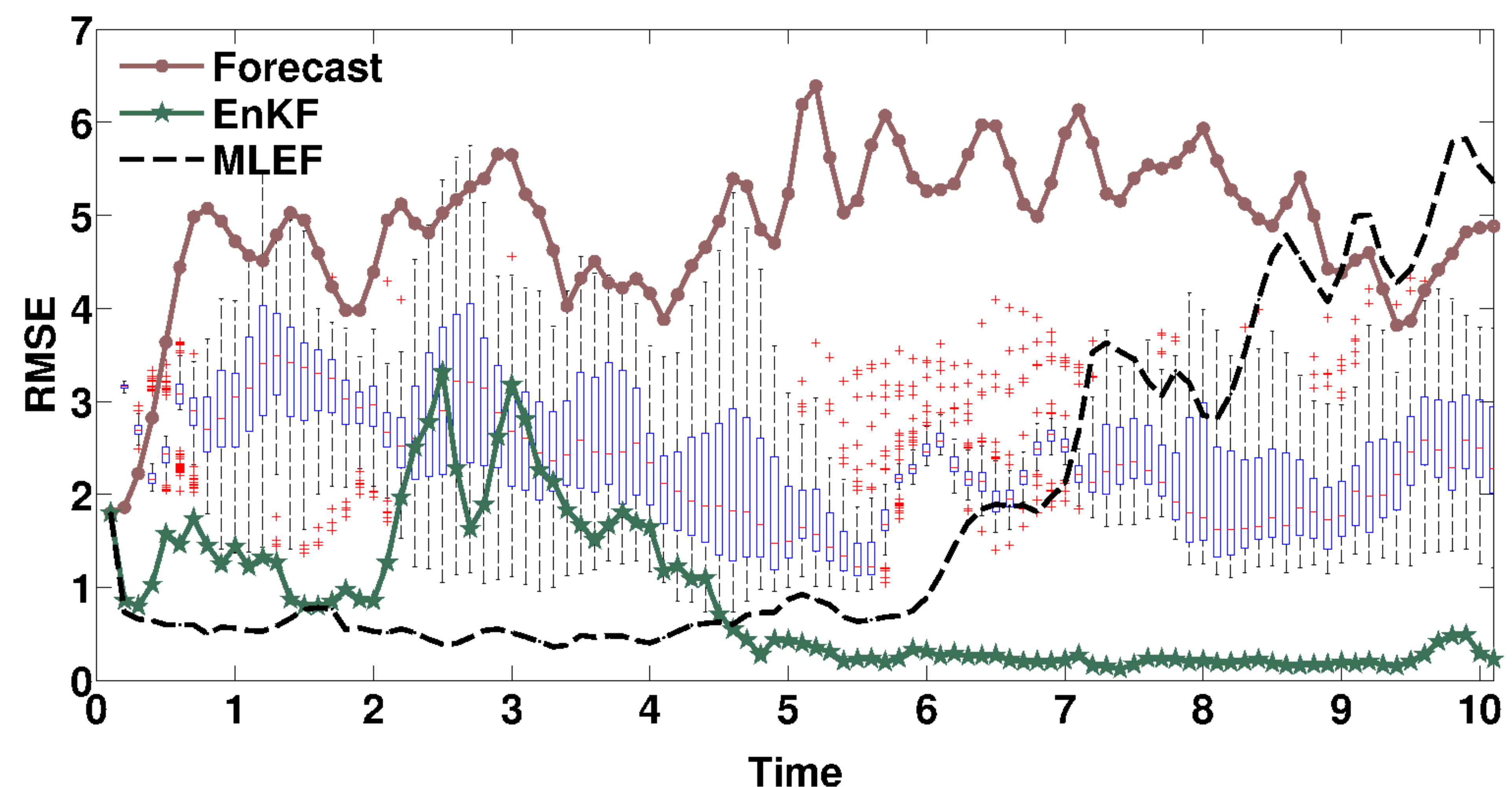}   
\label{fig:AbsH_Exper2_Hilbert}}
    \caption{Data assimilation results with the magnitude observation operator \eqref{eqn:Abs_H}. The symplectic integrator used is indicated under each panel.
     The time step for all integrators is $T=mh$, with $h,\, m$, as indicated under each panel. The number of inter-chain steps is $30$.
     The RMSE for $100$ instances of the sampling filter results are shown as box plots. The  red line represents the median RMSE values across all instances and the central blue box represents the variance. The two vertical lines (whiskers) extend up to $1.5$ times the height of the central box. The values exceeding the length of the whiskers are considered outliers (extremes) and are plotted as red crosses.
     } 
    \label{fig:AbsH_Exper2}
    \end{figure}

Figure \ref{fig:AbsH_Exper2} shows results with a larger step size and with equalized integrator work. The results with Verlet are similar to those reported in Figure \ref{fig:Abs_Exper1},  however the results obtained using high-order integrators are worse than before.  The use of Hilbert space integrator results in large RMSE, however these errors are stable (do not increase) with time.

    %
    %
     
     \subsection{Quadratic observation operator with threshold experiments}
     \label{subsec:Quad_Thresh}

\begin{figure}[H]
\centering
\subfigure[Position Verlet integrator \eqref{eqn:Verlet}]{%
\includegraphics[width=0.44\linewidth]{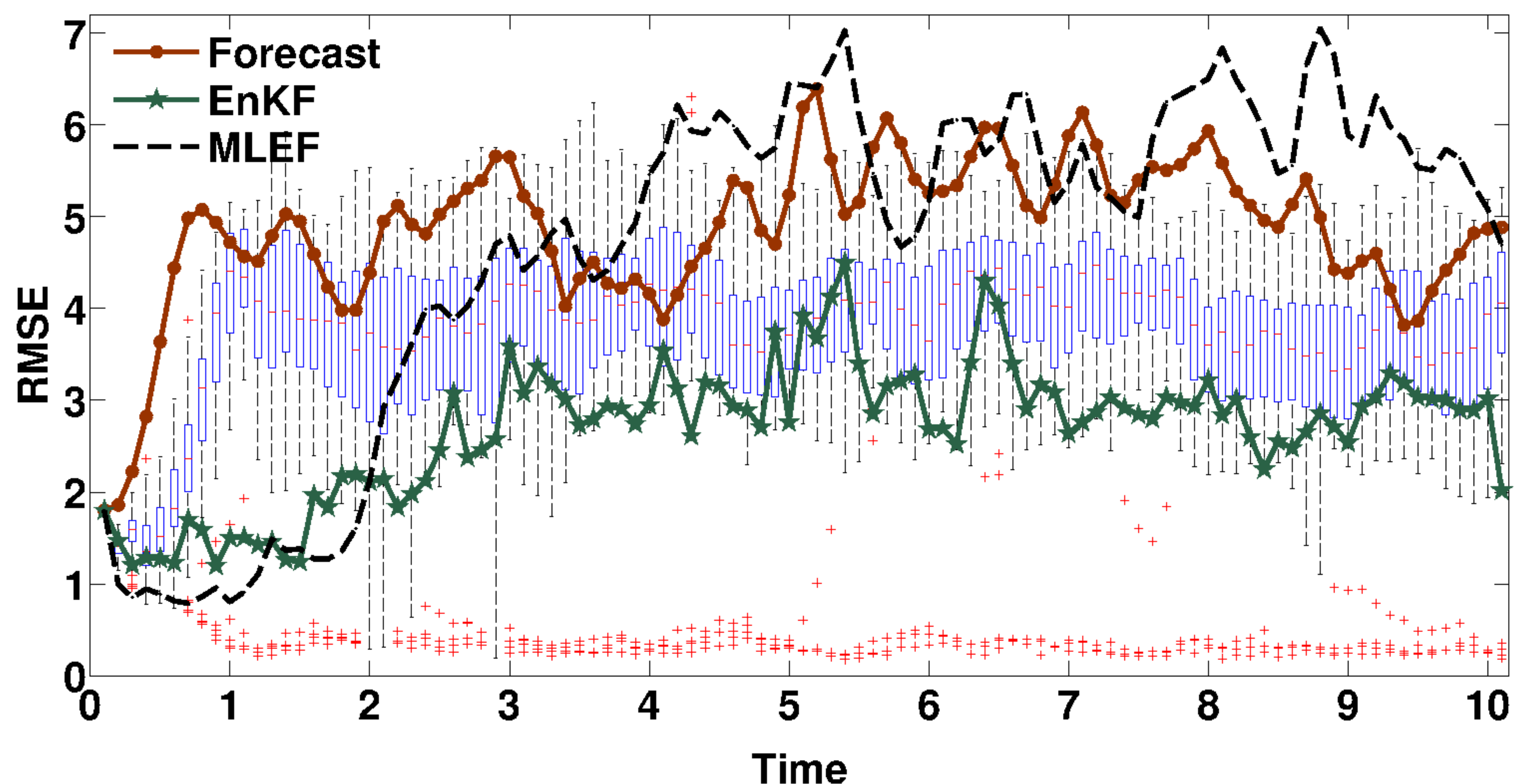}   
\label{fig:Quadratic_thresh_Exper1_Verlet}}
\quad  
\subfigure[Two-stage integrator \eqref{eqn:two_stage}]{%
\includegraphics[width=0.44\linewidth]{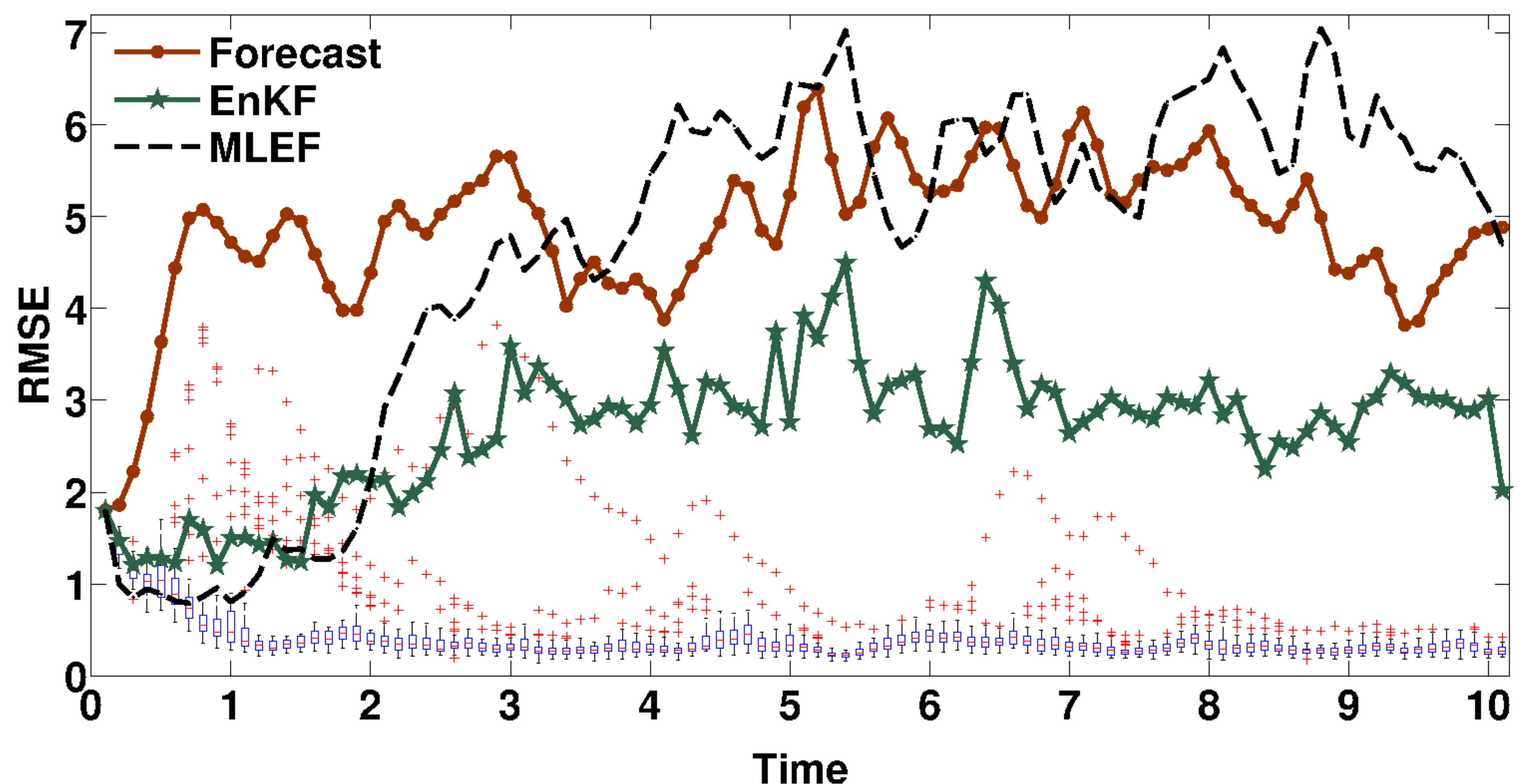}   
\label{fig:Quadratic_thresh_Exper1_2Stage}}
\subfigure[Three-stage  integrator]{%
\includegraphics[width=0.44\linewidth]{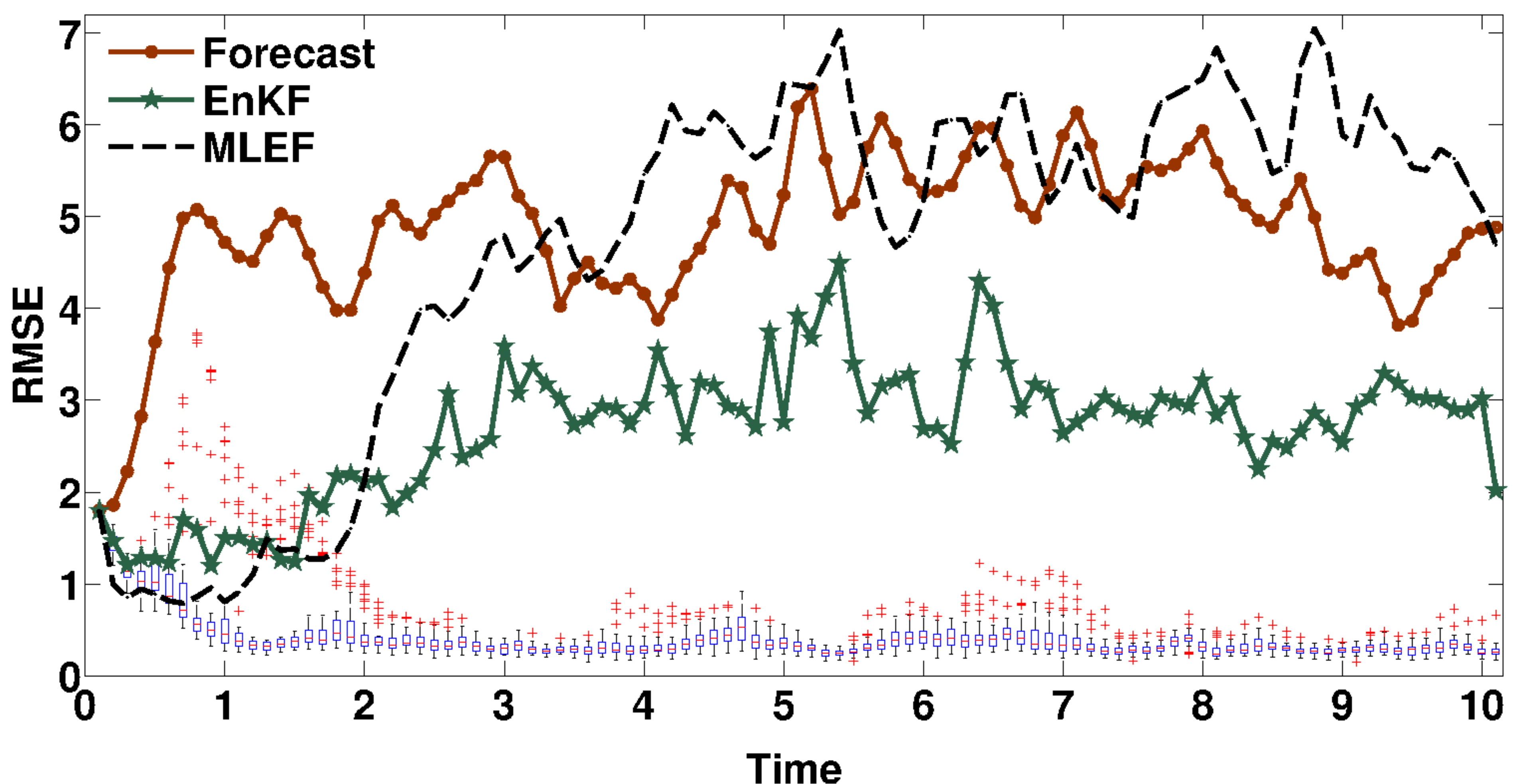}   
\label{fig:Quadratic_thresh_Exper1_3Stage}}
\quad
\subfigure[Four-stage integrator \eqref{eqn:four_stage}]{%
\includegraphics[width=0.44\linewidth]{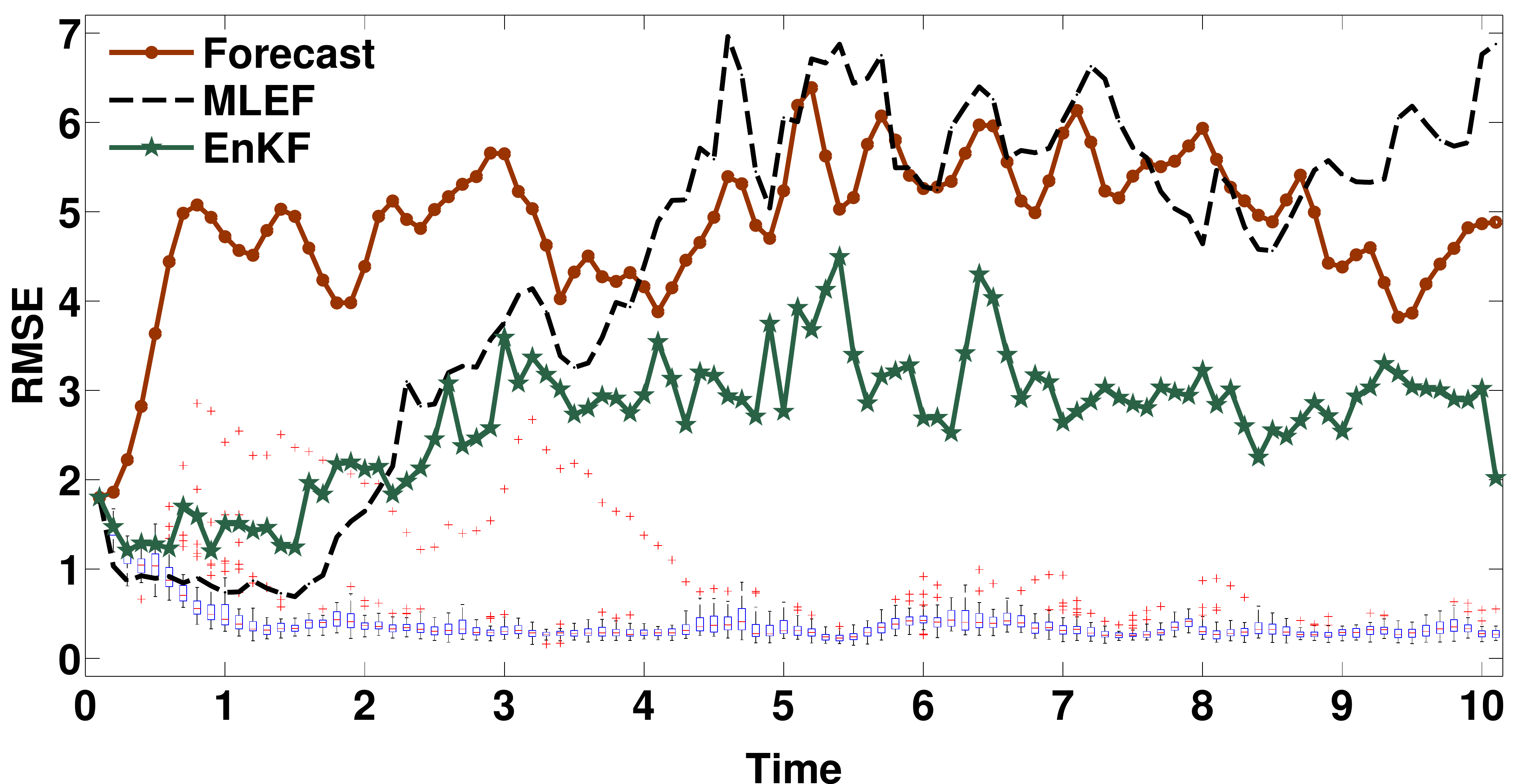}   
\label{fig:Quadratic_thresh_Exper1_4Stage}}
\subfigure[Integrator defined on Hilbert space \eqref{eqn:hilbert_integrator}]{%
\includegraphics[width=0.44\linewidth]{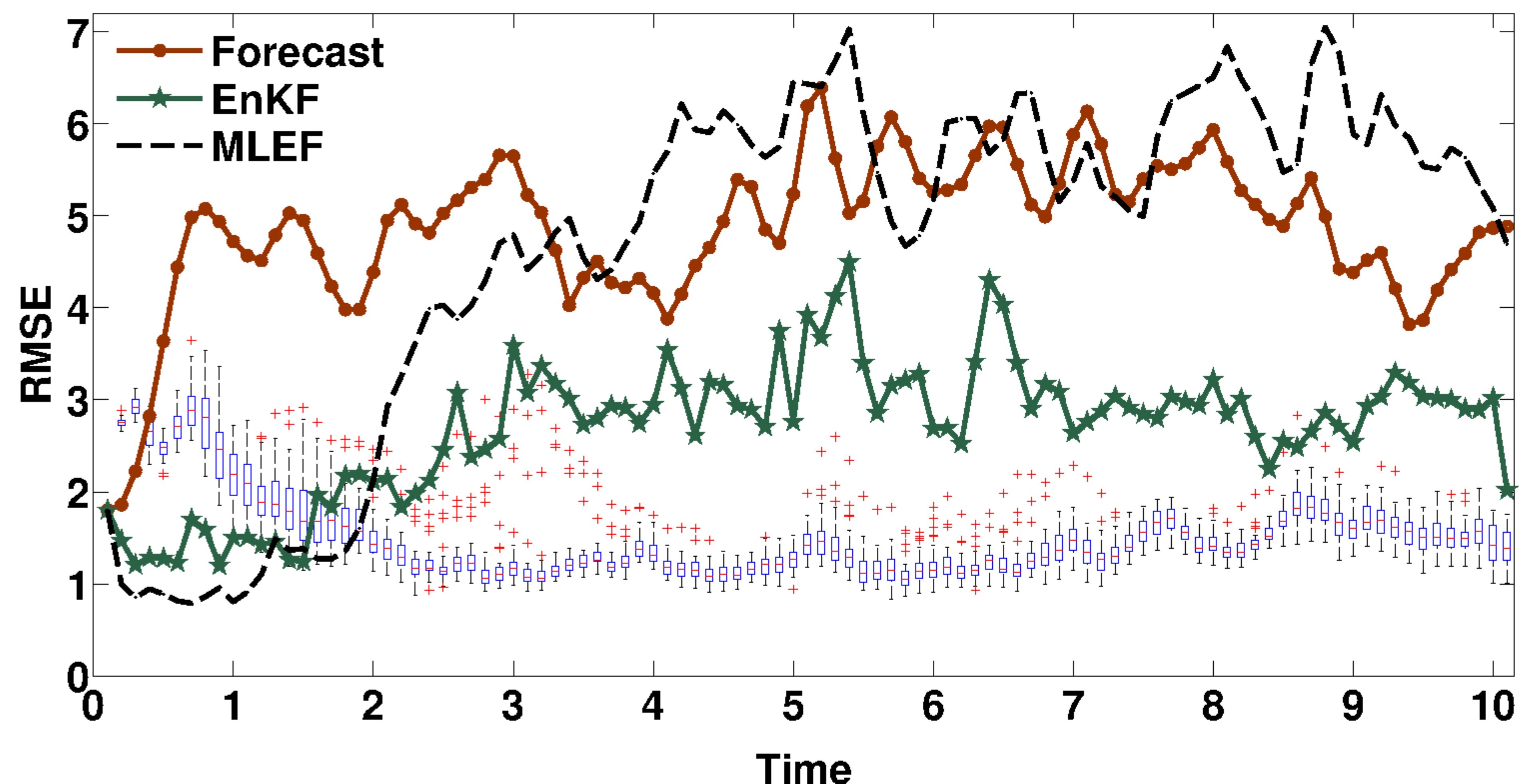}   
\label{fig:Quadratic_thresh_Exper1_Hilbert}}
    \caption{Data assimilation results with the quadratic observation operator \eqref{eqn:Quad_Thresh_H} with a threshold $a=0.5$. The symplectic integrator used is indicated under each panel.
     The time step for all integrators is $T=0.1$ with $h=0.01$, $m=10$, and $30$ inter-chain steps.
     The RMSE for $100$ instances of the sampling filter results are shown as box plots. The  red line represents the median RMSE values across all instances and the central blue box represents the variance. The two vertical lines (whiskers) extend up to $1.5$ times the height of the central box. The values exceeding the length of the whiskers are considered outliers (extremes) and are plotted as red crosses.
} 
    \label{fig:Quadratic_thresh_Exper1}
    \end{figure}

Figure \ref{fig:Quadratic_thresh_Exper1} shows the results with quadratic observation operator \eqref{eqn:Quad_Thresh_H} with threshold $a=0.5$. Even if MLEF was successfully tested with one dimensional models with this version of observation operator~\citep{zupanski2008maximum},  it does not perform well with the Lorenz model. The sampling filter using Verlet integrator fails due to the high non-linearity of the observation operator and/or the uncertainty levels. The high order integrators show good results and the analysis RMSE has the level obtained in case of linear observation operator. We can conclude that the ensemble produced by the filter is representative to the posterior PDF as both the mean and the covariance are incorporated in the analysis steps.The likelihood of outliers is small and decreases using higher-order integrators. The Hilbert integrator gives reasonable results. 

\begin{figure}[H]
\centering
\subfigure[Position Verlet integrator \eqref{eqn:Verlet}; $h=0.01,\ m=24$]{%
\includegraphics[width=0.44\linewidth]{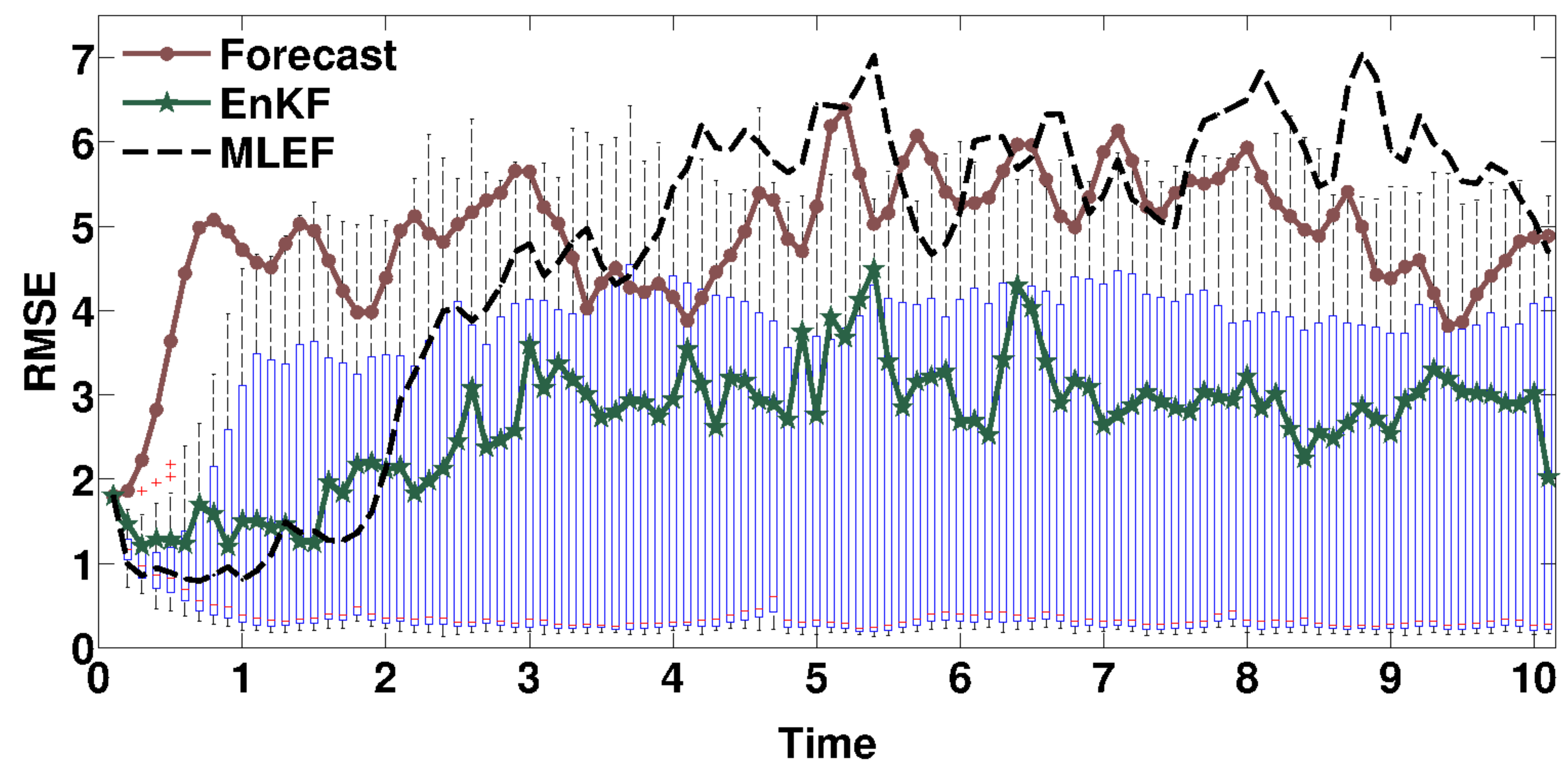}   
\label{fig:QuadThreshH_Exper2_Verlet}}
\quad  
\subfigure[Two-stage integrator \eqref{eqn:two_stage}; $h=0.02,\ m=12$]{%
\includegraphics[width=0.44\linewidth]{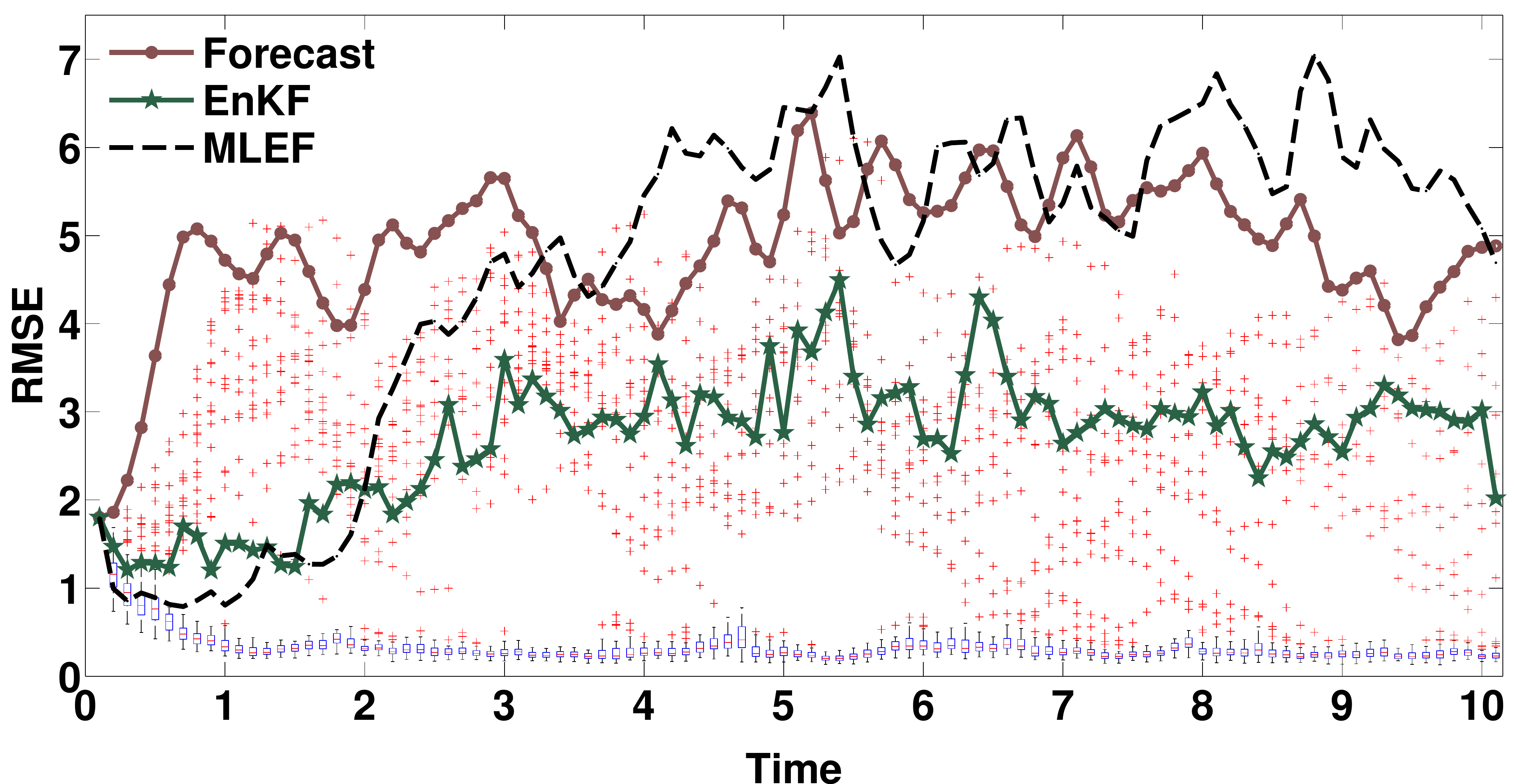}   
\label{fig:QuadThreshH_Exper2`_2Stage}}
\subfigure[Three-stage integrator \eqref{eqn:three_stage}; $h=0.03,\ m=8$]{%
\includegraphics[width=0.44\linewidth]{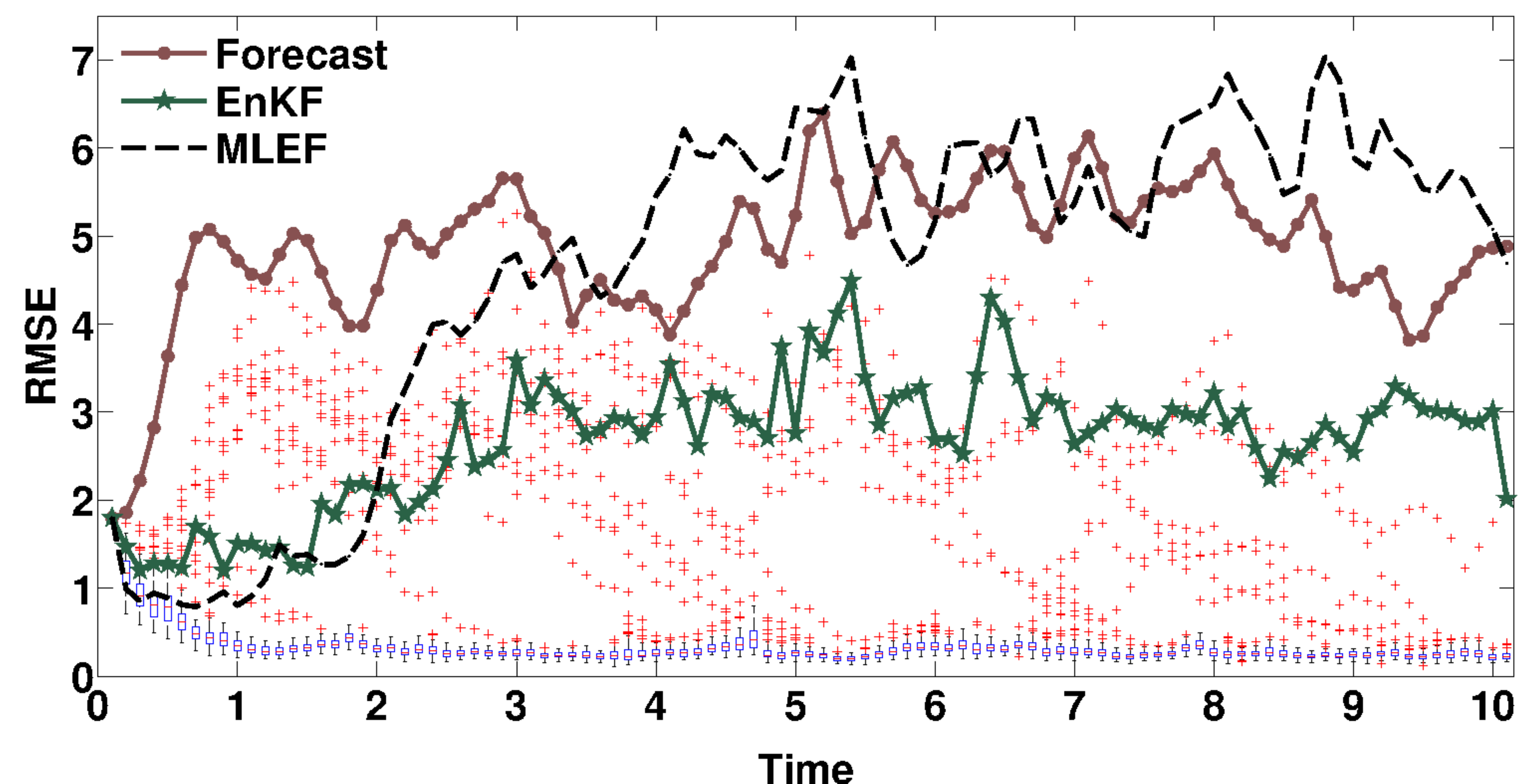}   
\label{fig:QuadThreshH_Exper2_3Stage}}
\quad
\subfigure[Four-stage integrator \eqref{eqn:four_stage}; $h=0.04,\ m=6$]{%
\includegraphics[width=0.44\linewidth]{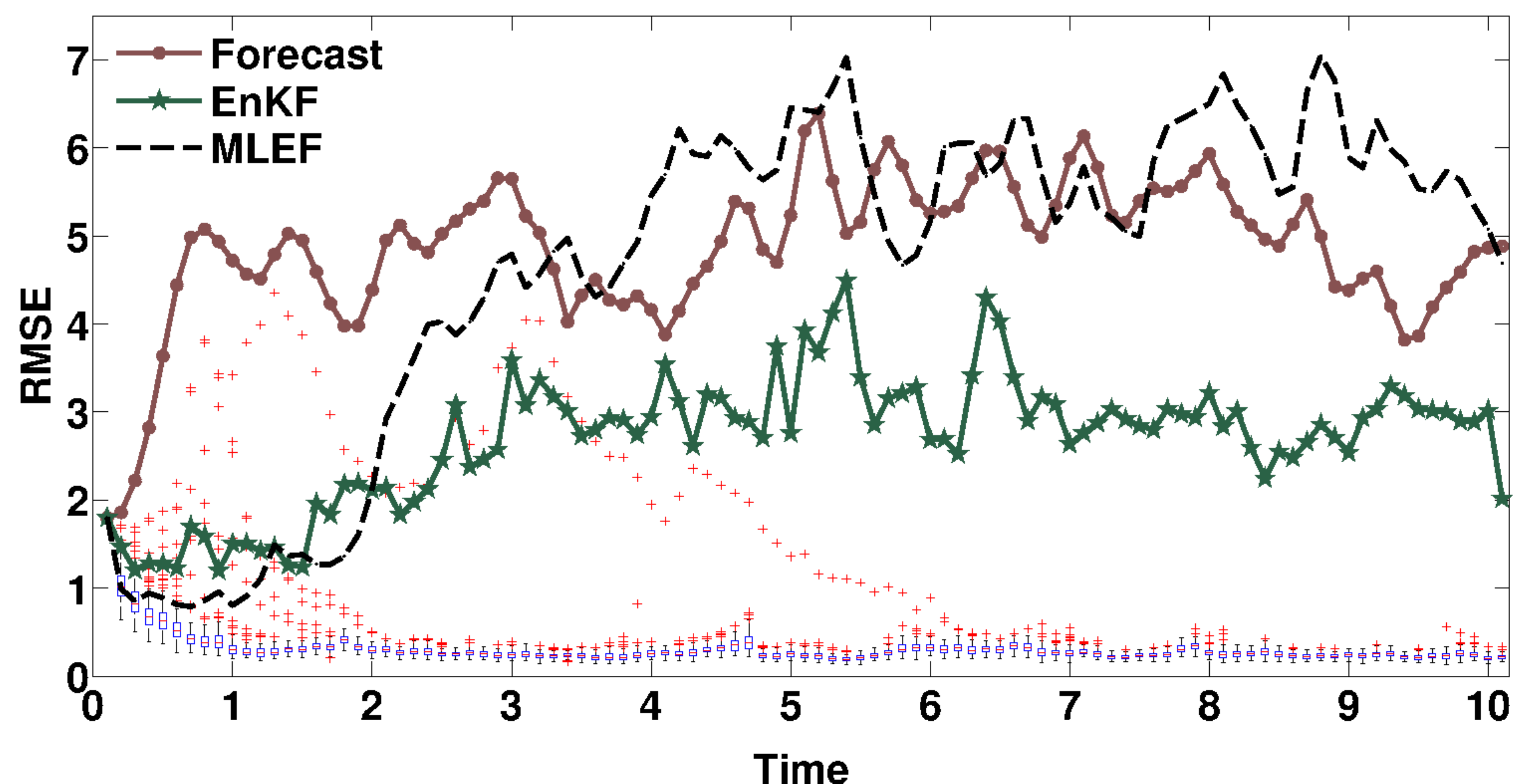}   
\label{fig:QuadThreshH_Exper2_4Stage}}
\subfigure[Integrator defined on Hilbert space \eqref{eqn:hilbert_integrator}; $h=0.01,\ m=24$]{%
\includegraphics[width=0.44\linewidth]{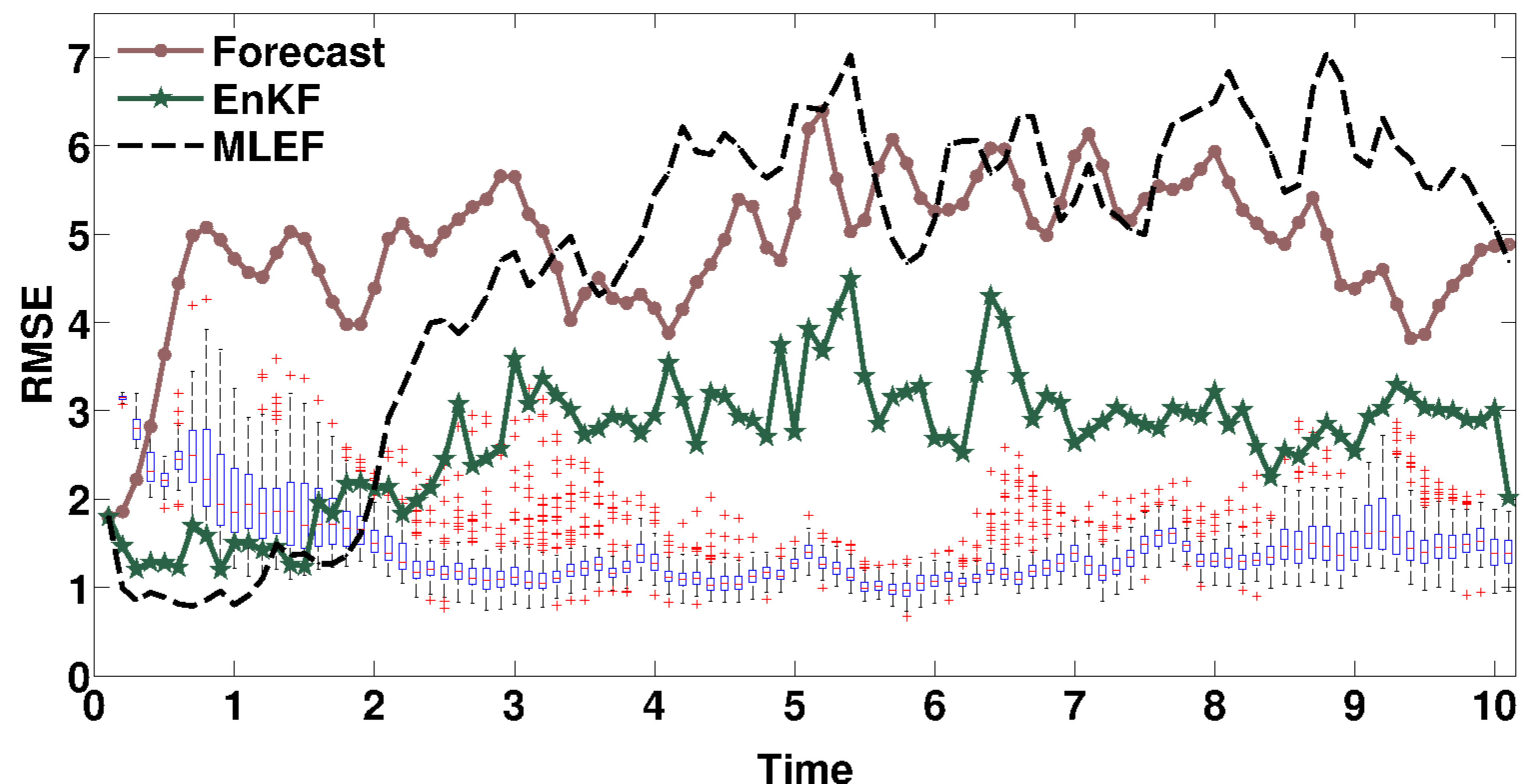}   
\label{fig:QuadThreshH_Exper2_Hilbert}}
    \caption{Data assimilation results with the quadratic observation operator \eqref{eqn:Quad_Thresh_H} with a threshold $a=0.5$. The symplectic integrator used is indicated under each panel.
     The time step for all integrators is $T=mh$, with $h,\, m$, as indicated under each panel. The number of inter-chain steps is $30$.
     The RMSE for $100$ instances of the sampling filter results are shown as box plots. The  red line represents the median RMSE values across all instances and the central blue box represents the variance. The two vertical lines (whiskers) extend up to $1.5$ times the height of the central box. The values exceeding the length of the whiskers are considered outliers (extremes) and are plotted as red crosses.
     } 
    \label{fig:Quadratic_thresh_Exper2}
    \end{figure}

Figure \ref{fig:Quadratic_thresh_Exper2} shows results with  obtained with step sizes and equalized work.The filter with four stage integrator is superior with such level of non-linearity as it suffers the least from outliers and gives a small RMSE. The Verlet integrator also gives satisfactory results. The Hilbert integrator results in large analysis RMSE but it performs robustly even if the analysis is very far from the true solution and large step sizes are selected.
    
    The Jacobian of this observation operator is approximated using finite differences. Alternatives will be considered in the future.
     
     \subsection{Exponential observation operator (with factor $r=0.2$) experiments}
     \label{subsec:Exponential_2_Results}

\begin{figure}[H]
\centering
\subfigure[Position Verlet integrator \eqref{eqn:Verlet}]{%
\includegraphics[width=0.44\linewidth]{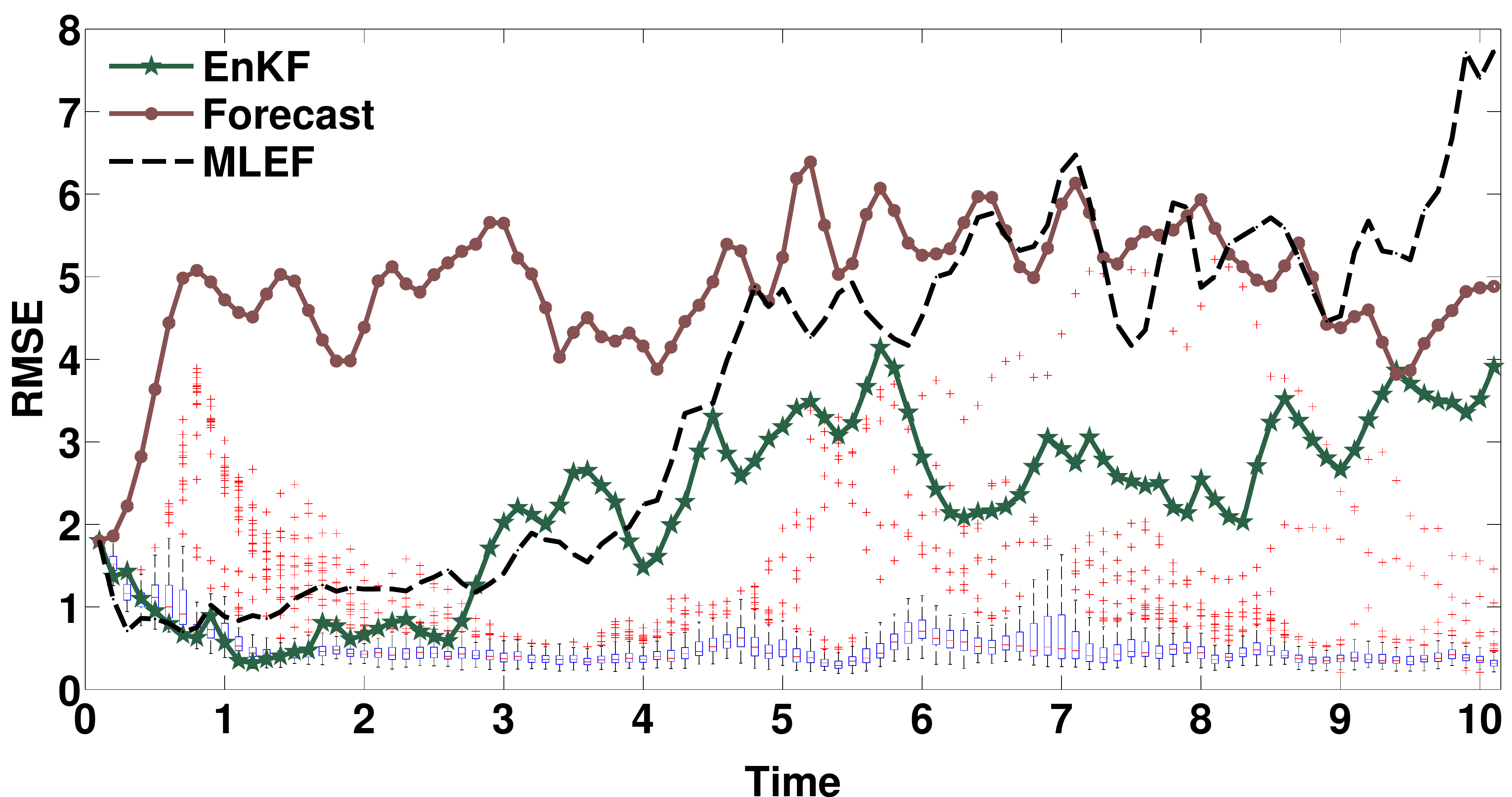}   
\label{fig:Exponential1_Exper1_Verlet}}
\quad  
\subfigure[Two-stage integrator \eqref{eqn:two_stage}]{%
\includegraphics[width=0.44\linewidth]{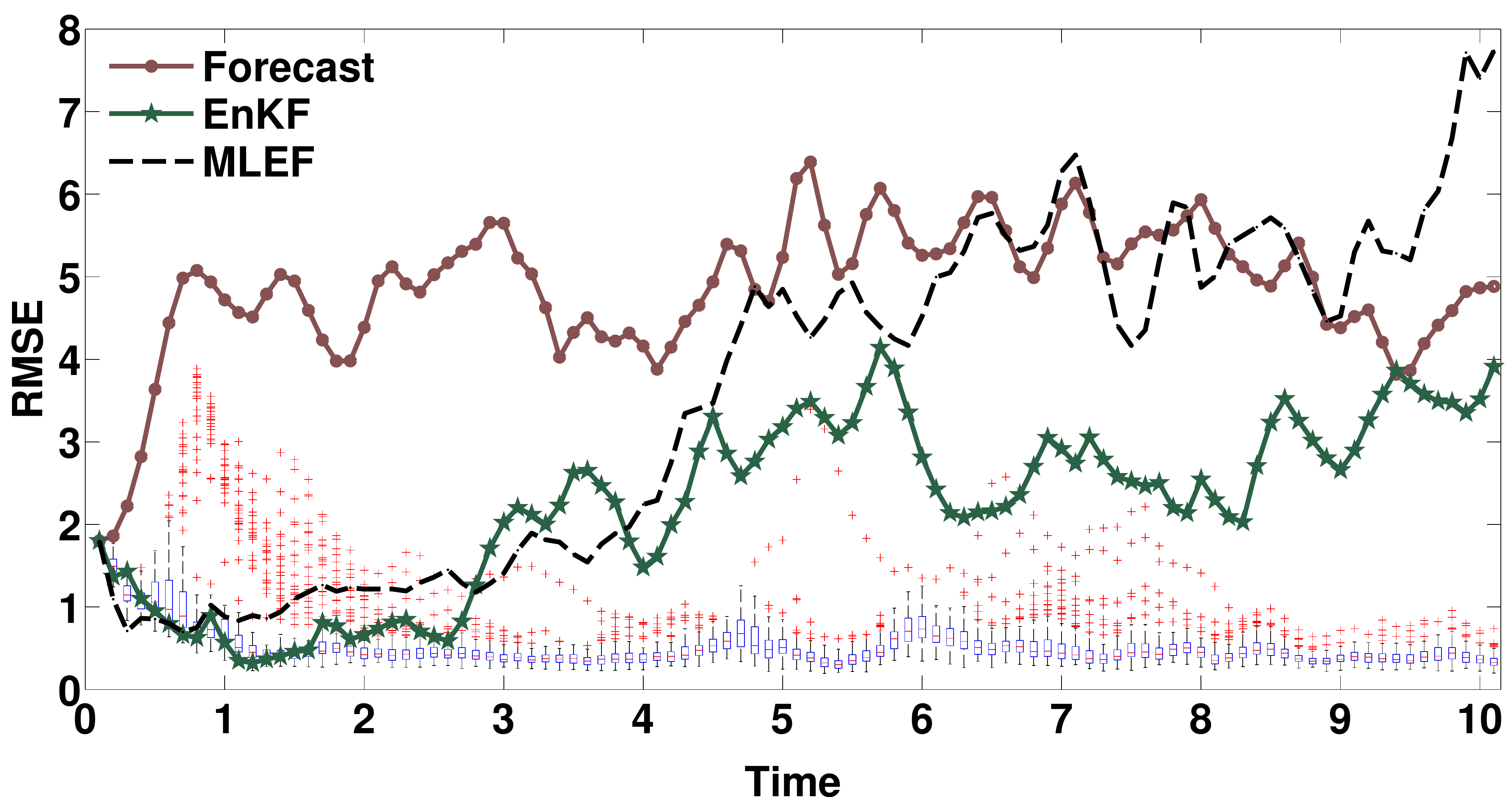}   
\label{fig:Exponential1_Exper1_2Stage}}
\subfigure[Three-stage integrator \eqref{eqn:three_stage}]{%
\includegraphics[width=0.44\linewidth]{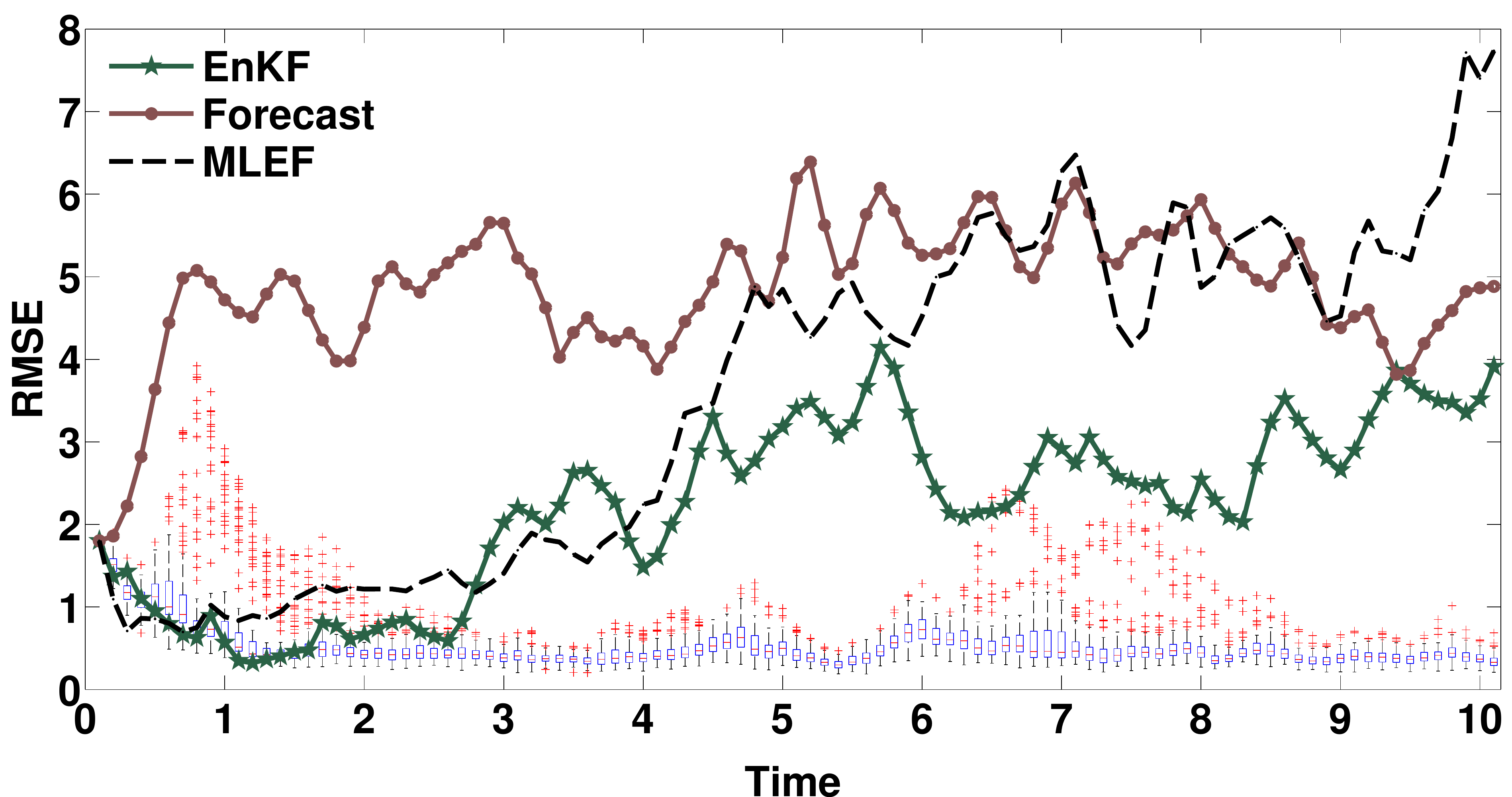}   
\label{fig:Exponential1_Exper1_3Stage}}
\quad
\subfigure[Four-stage integrator \eqref{eqn:four_stage}]{%
\includegraphics[width=0.44\linewidth]{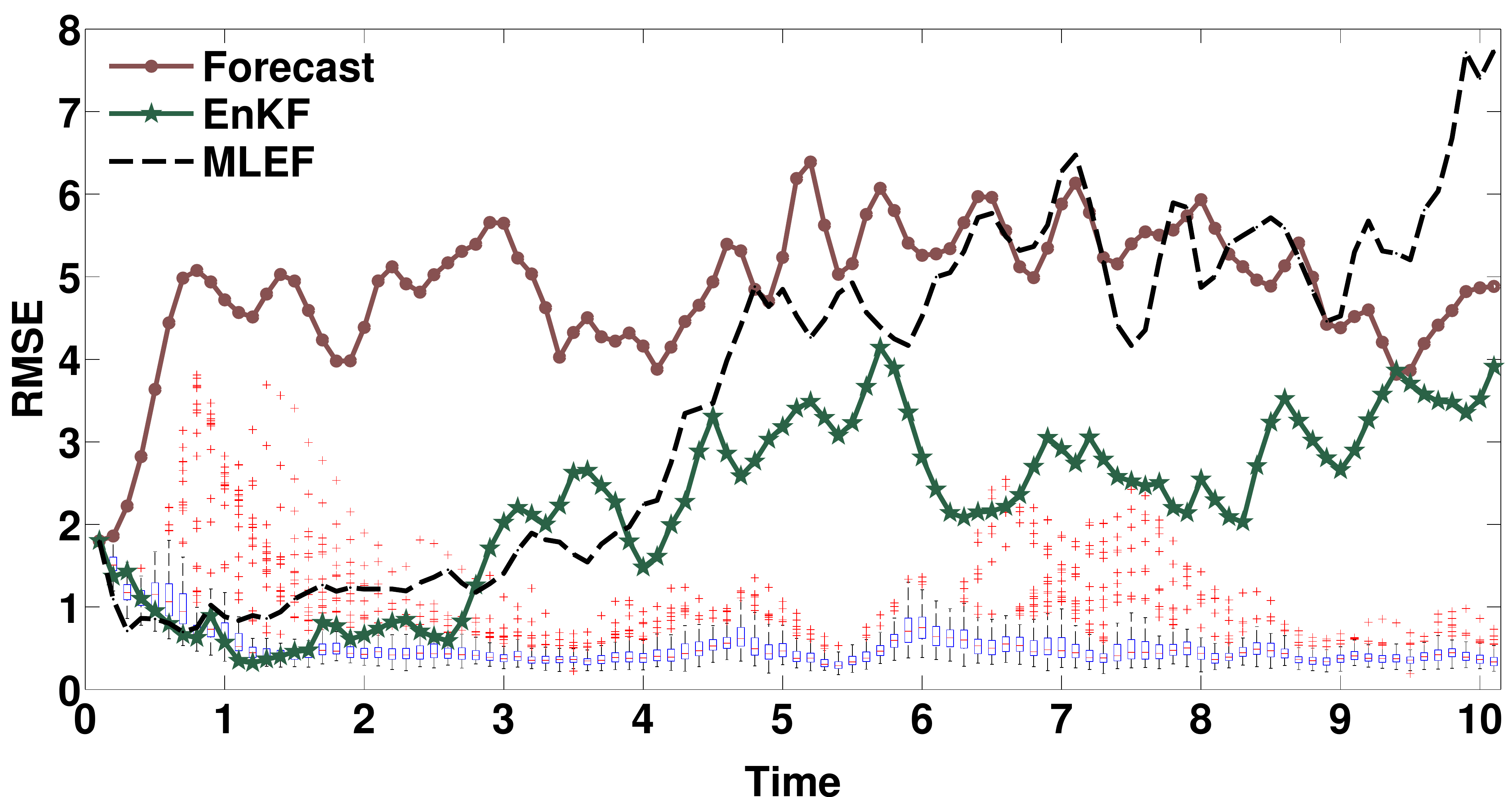}   
\label{fig:Exponential1_Exper1_4Stage}}
\subfigure[Integrator defined on Hilbert space \eqref{eqn:hilbert_integrator}]{%
\includegraphics[width=0.44\linewidth]{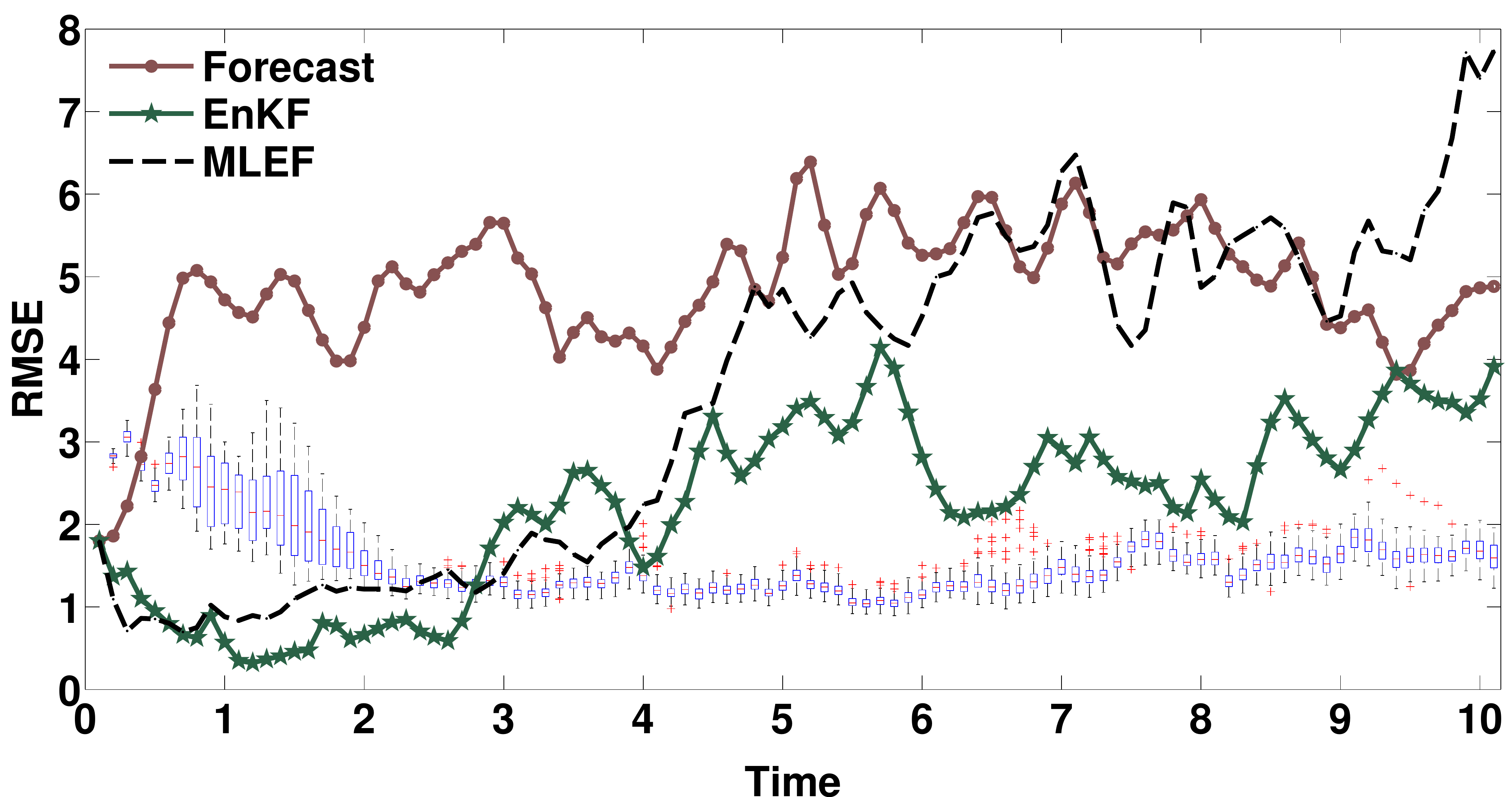}   
\label{fig:Exponential1_Exper1_Hilbert}}
    \caption{Data assimilation results with the exponential observation operator \eqref{eqn:Exponential_H} with a factor $r=0.2$. The symplectic integrator used is indicated under each panel.
     The time step for all integrators is $T=0.1$ with $h=0.01$, $m=10$, and $30$ inter-chain steps.
     The RMSE for $100$ instances of the sampling filter results are shown as box plots. The  red line represents the median RMSE values across all instances and the central blue box represents the variance. The two vertical lines (whiskers) extend up to $1.5$ times the height of the central box. The values exceeding the length of the whiskers are considered outliers (extremes) and are plotted as red crosses.
     } 
    \label{fig:Exponential1_Exper1}
    \end{figure}

Figure \ref{fig:Exponential1_Exper1} shows the results with the exponential observation operator \eqref{eqn:Exponential_H} with factor $r=0.2$. This observation operator is differentiable, however small perturbations in the state might result in relatively large changes in the measured vales. Under strongly nonlinear conditions the sampling filter performs better than either MLEF and EnKF. The performance of the sampling filter in this experiment is similar to its performance in case of linear observation operators.

\begin{figure}[H]
\centering
\subfigure[Position Verlet integrator \eqref{eqn:Verlet}; $h=0.01,\ m=16$]{%
\includegraphics[width=0.44\linewidth]{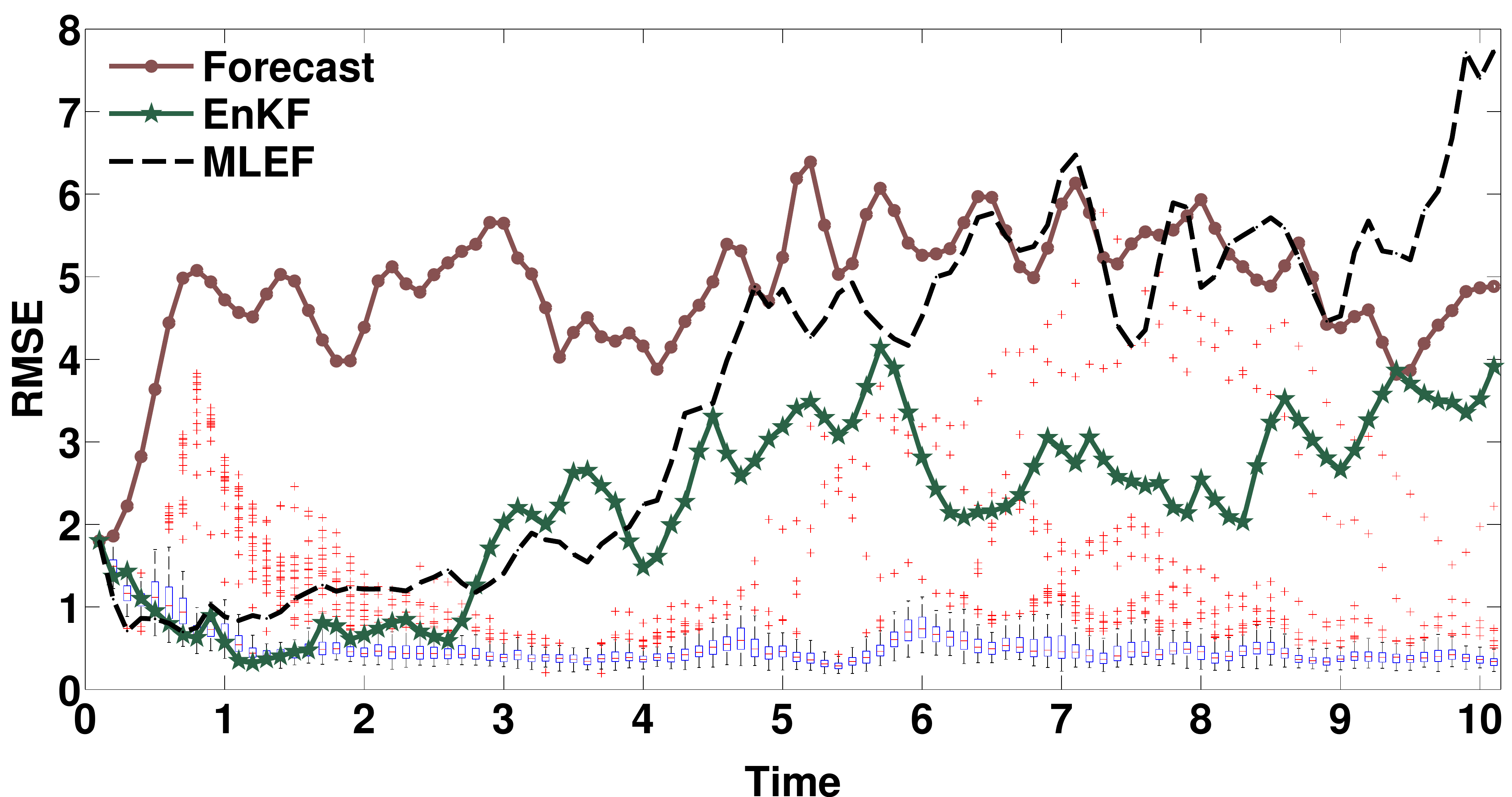}   
\label{fig:Exponential1_Exper2_Verlet}}
\quad  
\subfigure[Two-stage integrator \eqref{eqn:two_stage}; $h=0.02,\ m=8$]{%
\includegraphics[width=0.44\linewidth]{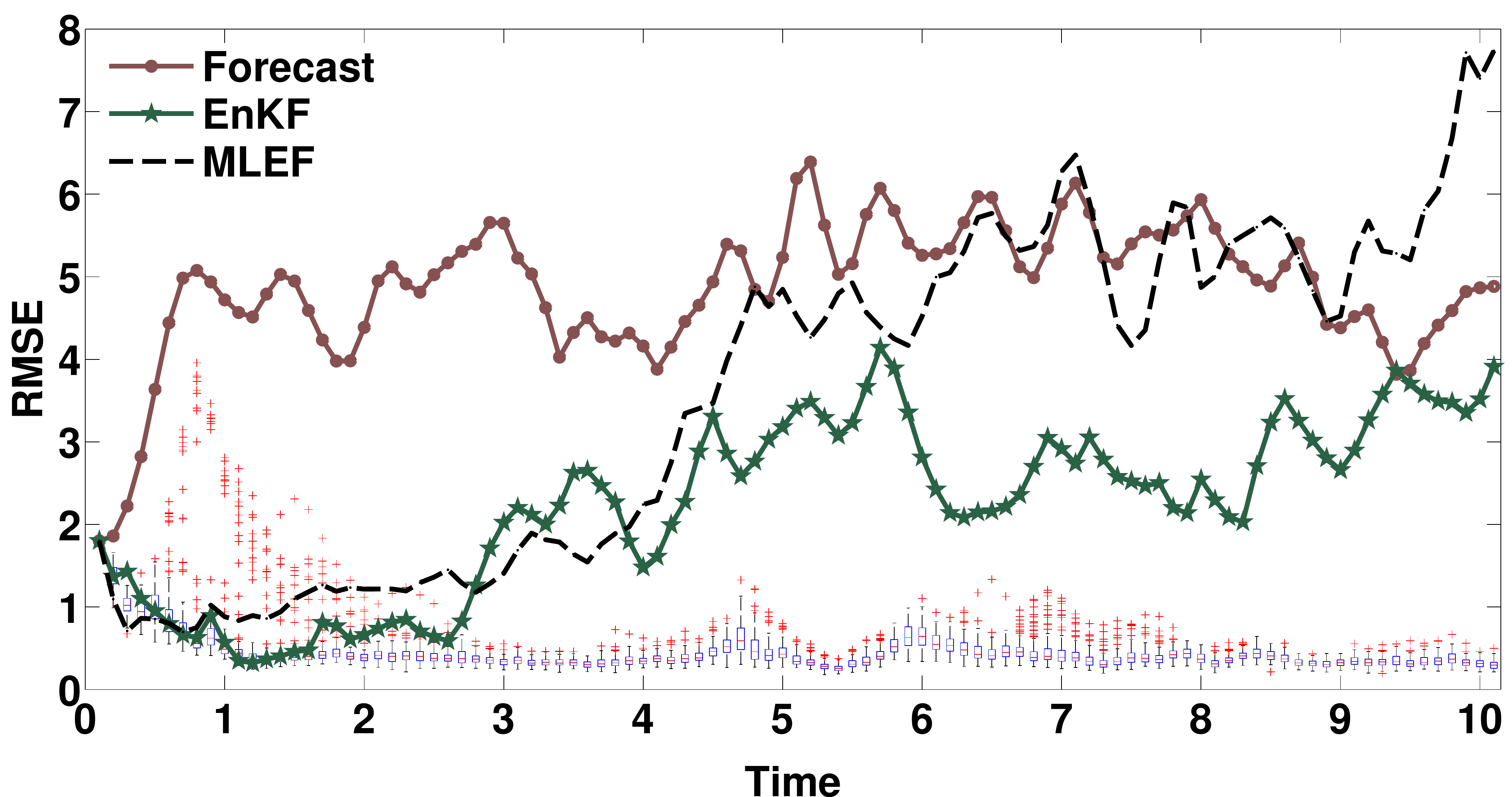}   
\label{fig:Exponential1_Exper2_2Stage}}
\subfigure[Three-stage integrator \eqref{eqn:three_stage}; $h=0.03,\ m=6$]{%
\includegraphics[width=0.44\linewidth]{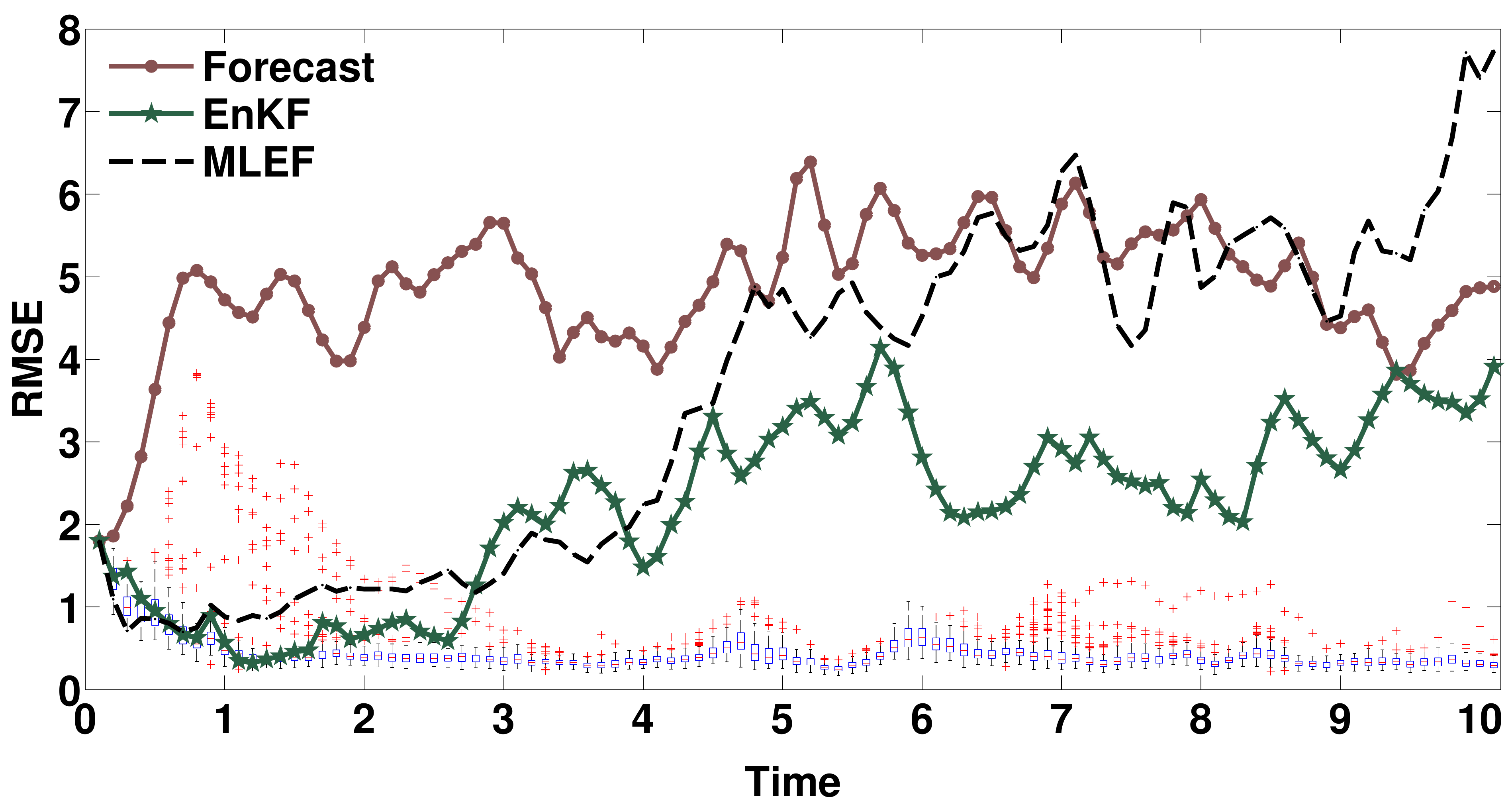}   
\label{fig:Exponential1_Exper2_3Stage}}
\quad
\subfigure[Four-stage integrator \eqref{eqn:four_stage}; $h=0.04,\ m=4$]{%
\includegraphics[width=0.44\linewidth]{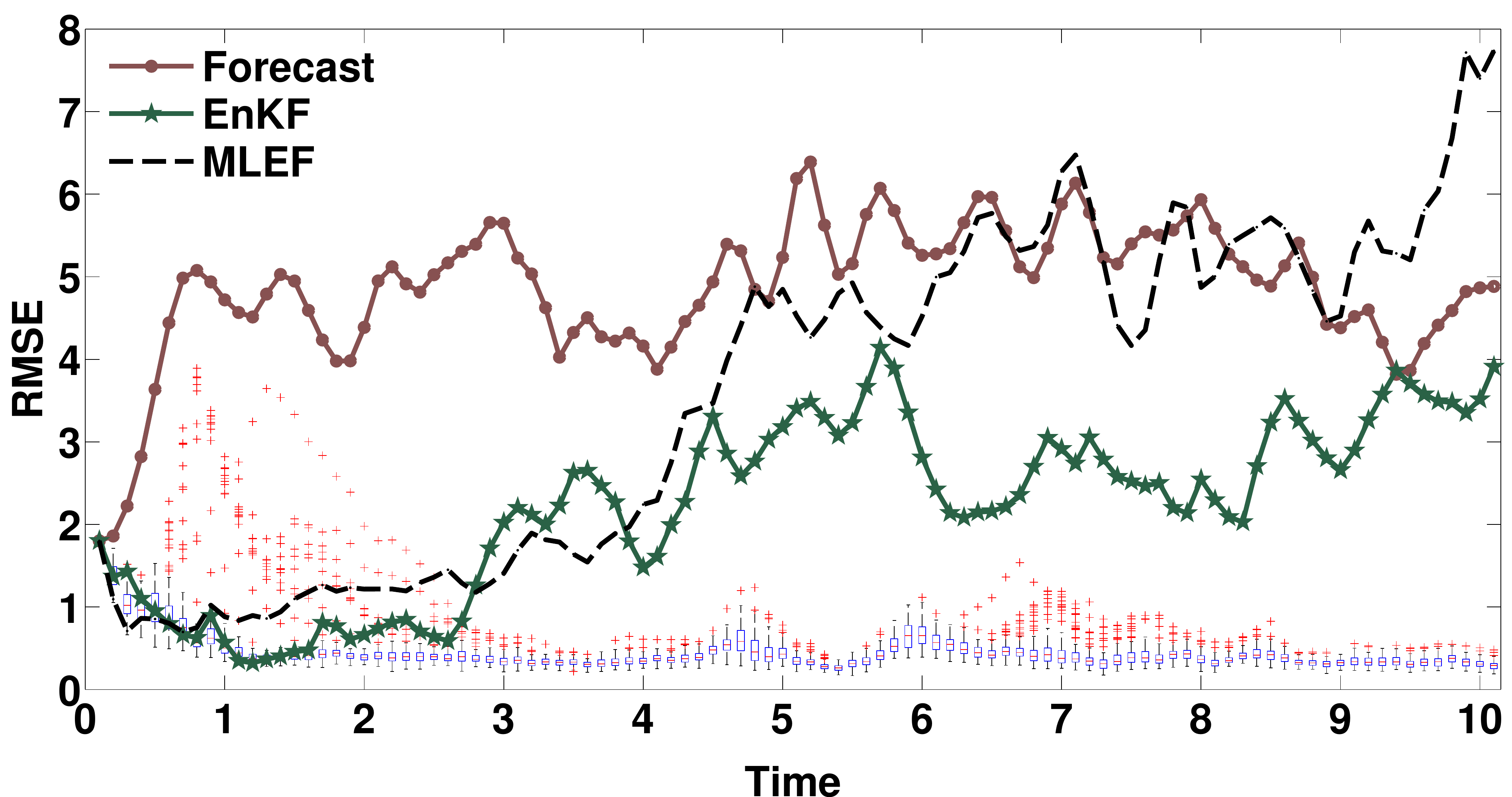}   
\label{fig:Exponential1_Exper2_4Stage}}
\subfigure[Integrator defined on Hilbert space \eqref{eqn:hilbert_integrator}; $h=0.01,\ m=16$]{%
\includegraphics[width=0.44\linewidth]{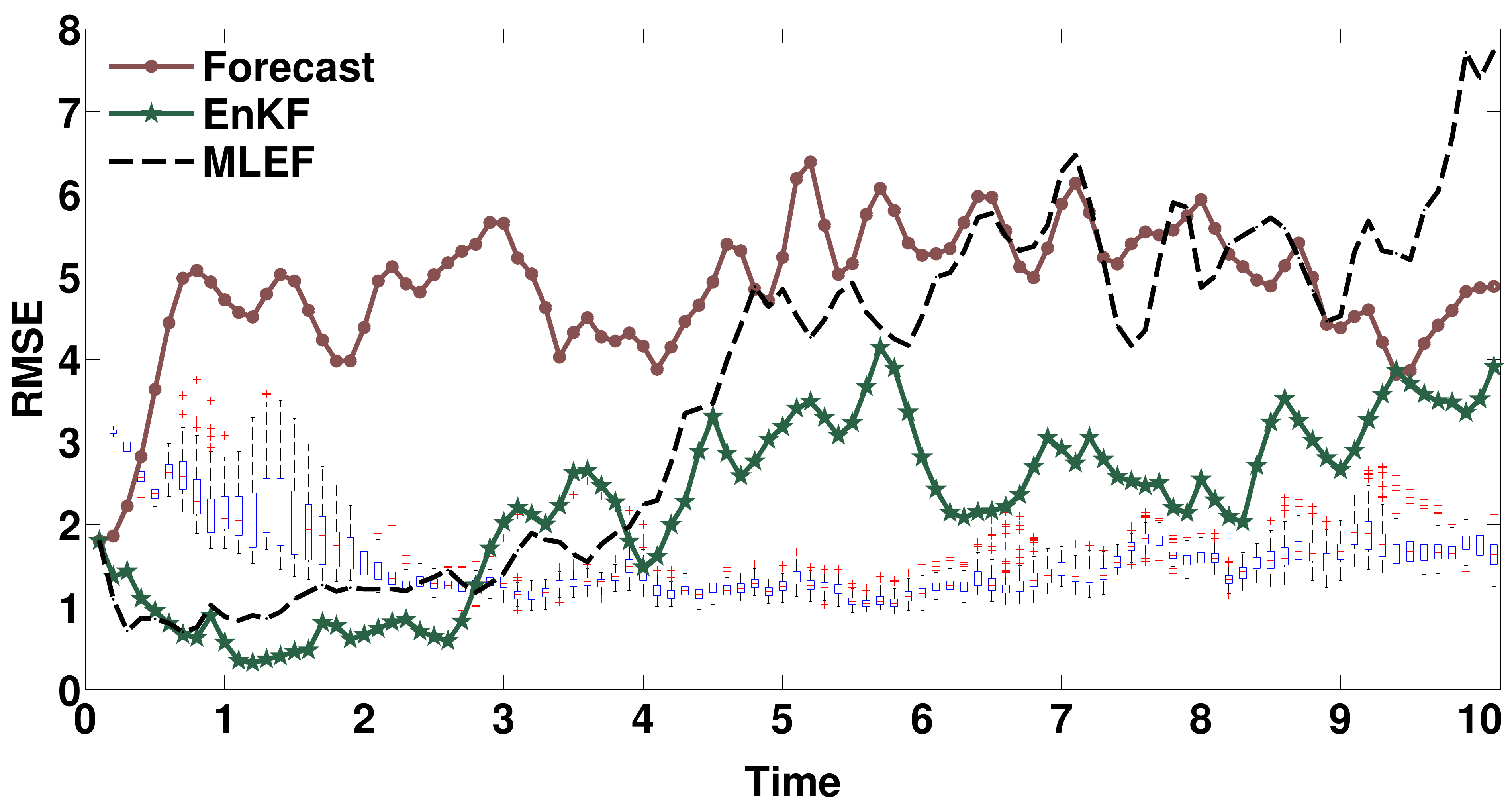}   
\label{fig:Exponential1_Exper2_Hilbert}}
    \caption{Data assimilation results with the exponential observation operator \eqref{eqn:Exponential_H}. The symplectic integrator used is indicated under each panel.
     The time step for all integrators is $T=mh$, with $h,\, m$, as indicated under each panel. The number of inter-chain steps is $30$.
     The RMSE for $100$ instances of the sampling filter results are shown as box plots. The  red line represents the median RMSE values across all instances and the central blue box represents the variance. The two vertical lines (whiskers) extend up to $1.5$ times the height of the central box. The values exceeding the length of the whiskers are considered outliers (extremes) and are plotted as red crosses.
     } 
    \label{fig:Exponential1_Exper2}
    \end{figure}

As shown in Figure \ref{fig:Exponential1_Exper2_Verlet}, further tuning of the step size for Verlet integrator does not result in notable improvements over the results in Figure \ref{fig:Exponential1_Exper1_Verlet}. 
Figures \ref{fig:Exponential1_Exper2_2Stage},  \ref{fig:Exponential1_Exper2_3Stage}, and \ref{fig:Exponential1_Exper2_4Stage} show that the two-stage, three-stage, and four-stage integrators behave similarly, and give slightly better results than those reported in Figures \ref{fig:Exponential1_Exper1_2Stage}, \ref{fig:Exponential1_Exper1_3Stage}, and \ref{fig:Exponential1_Exper1_4Stage}, respectively. The infinite dimensional integrator performance does not change with the change in step size, as can be seen in Figures \ref{fig:Exponential1_Exper1_Hilbert}, and \ref{fig:Exponential1_Exper2_Hilbert}.

\FloatBarrier     
     %
     \subsection{MLEF performance}
     \label{subsec:More_on_MLEF_results}
\begin{figure}[!htbp]
\centering
\subfigure[All components are observed]{%
\includegraphics[width=0.44\linewidth]{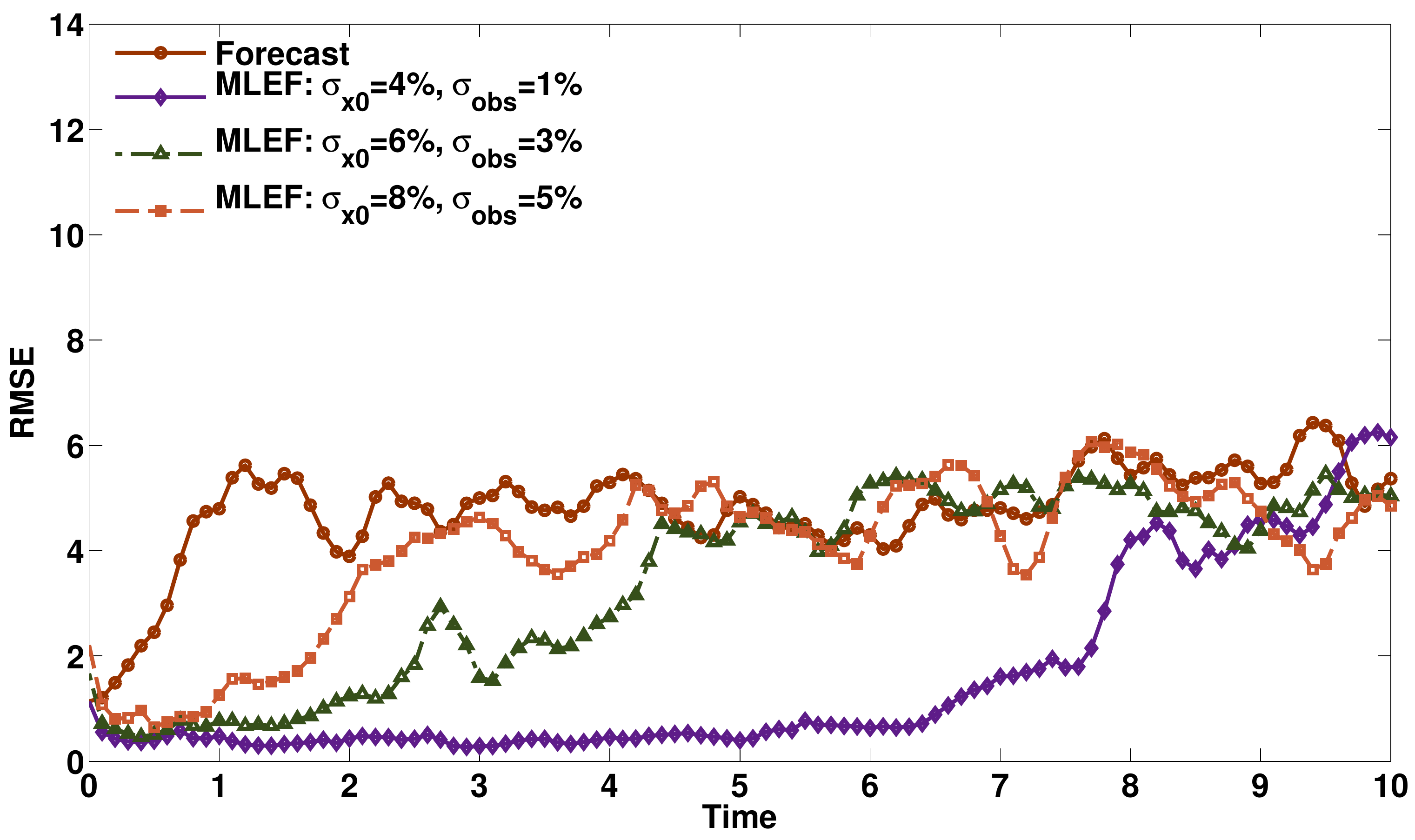}   
\label{fig:More_MLEF_Cubic_All}}
\quad  
\subfigure[Each second component is observed]{%
\includegraphics[width=0.44\linewidth]{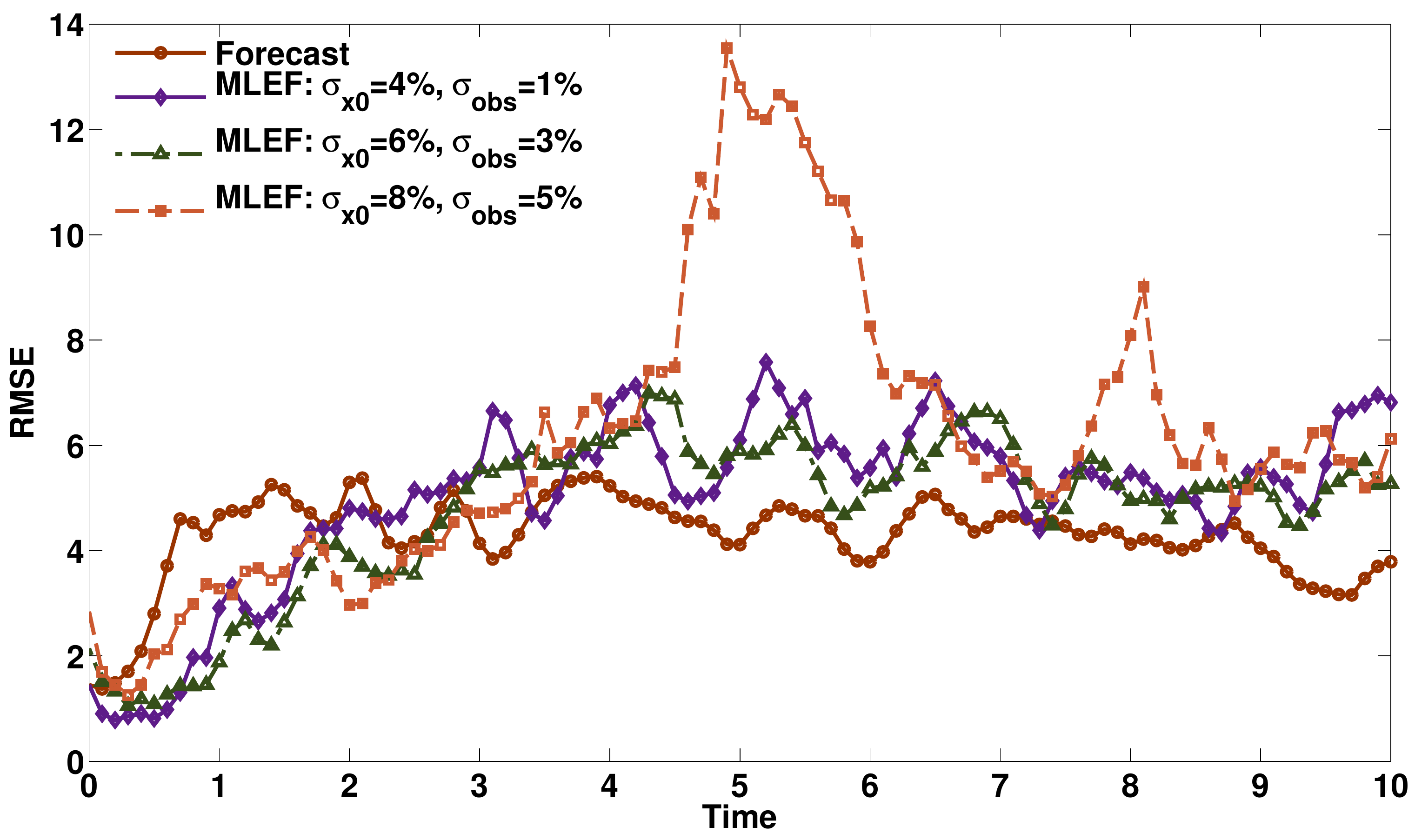}   
\label{fig:More_MLEF_Cubic_2nds}}
\subfigure[Each third component is observed]{%
\includegraphics[width=0.44\linewidth]{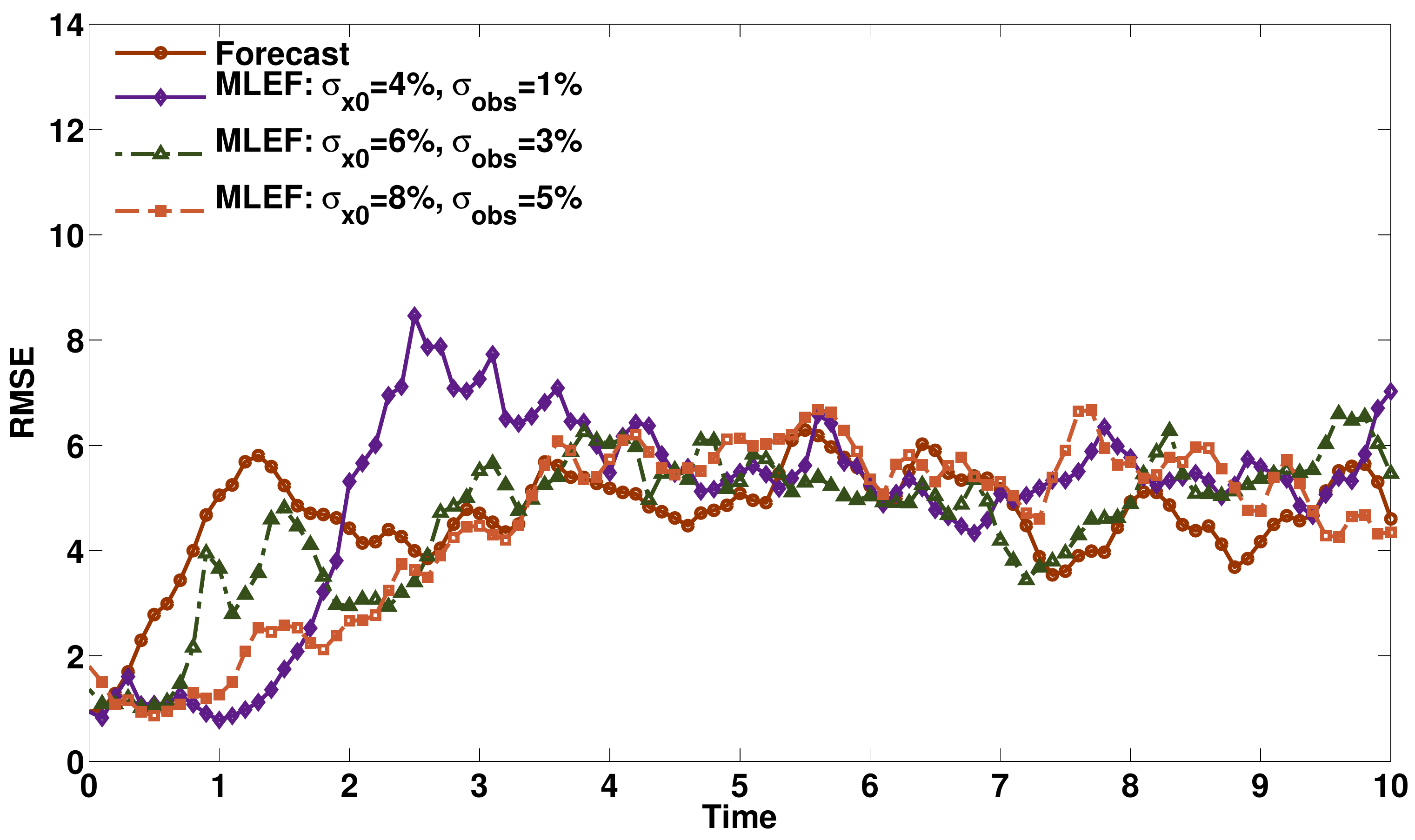}   
\label{fig:More_MLEF_Cubic_3rds}}
    \caption{MLEF data assimilation results with the cubic observation operator \eqref{eqn:Cubic_H}. 
             MLEF is applied with varying observation frequencies, and with different noise levels for both observations and the background state.
             The frequency of observations is indicated under each panel. The background error standard deviation is $\sigma_{x0}$, and the observation  error standard deviation is $\sigma_{\rm obs}$.} 
    \label{fig:More_MLEF_Cubic}
    \end{figure}

\begin{figure}[!htbp]
\centering
\subfigure[All components are observed]{%
\includegraphics[width=0.44\linewidth]{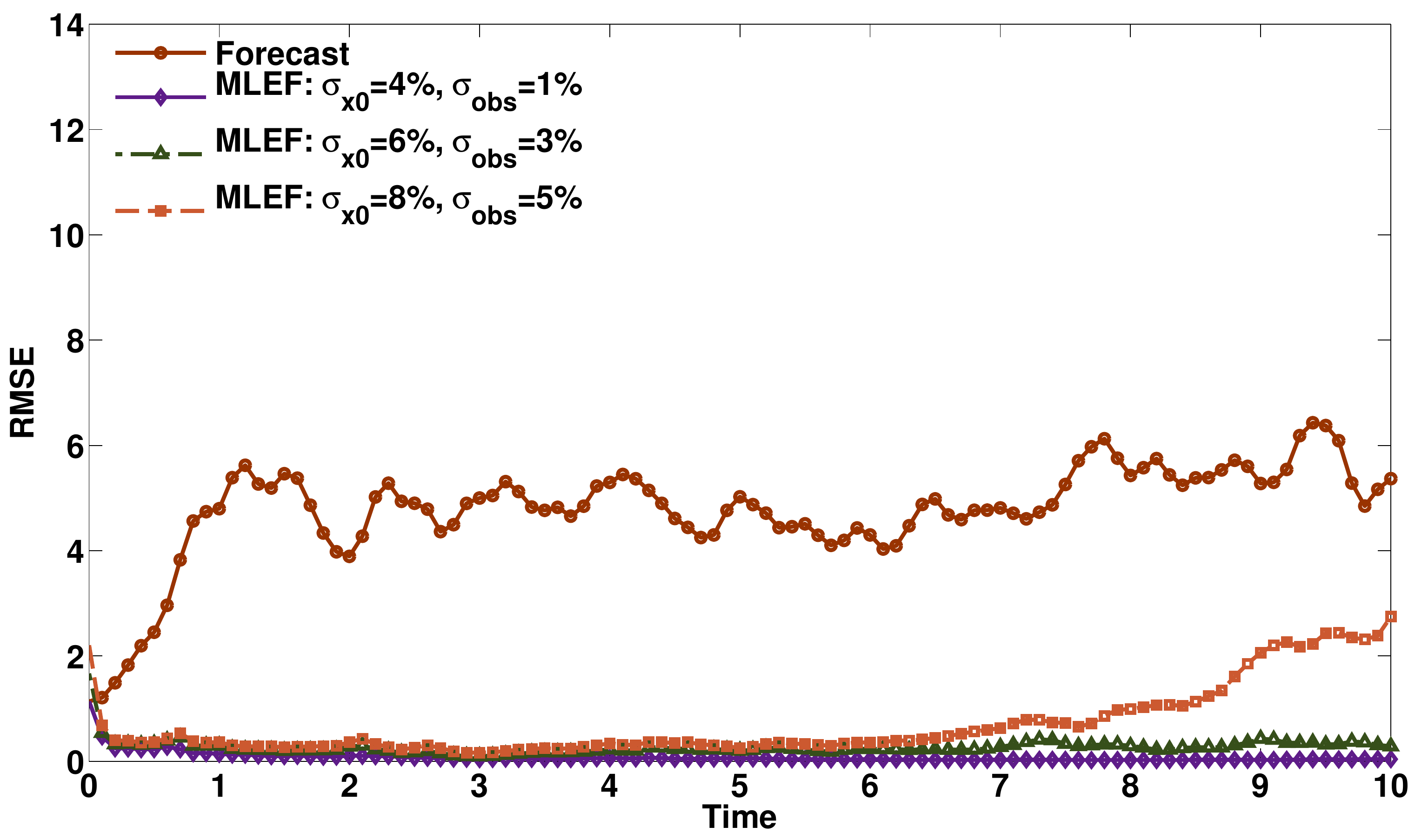}   
\label{fig:More_MLEF_Quad_Thresh_All}}
\quad  
\subfigure[Each second component is observed]{%
\includegraphics[width=0.44\linewidth]{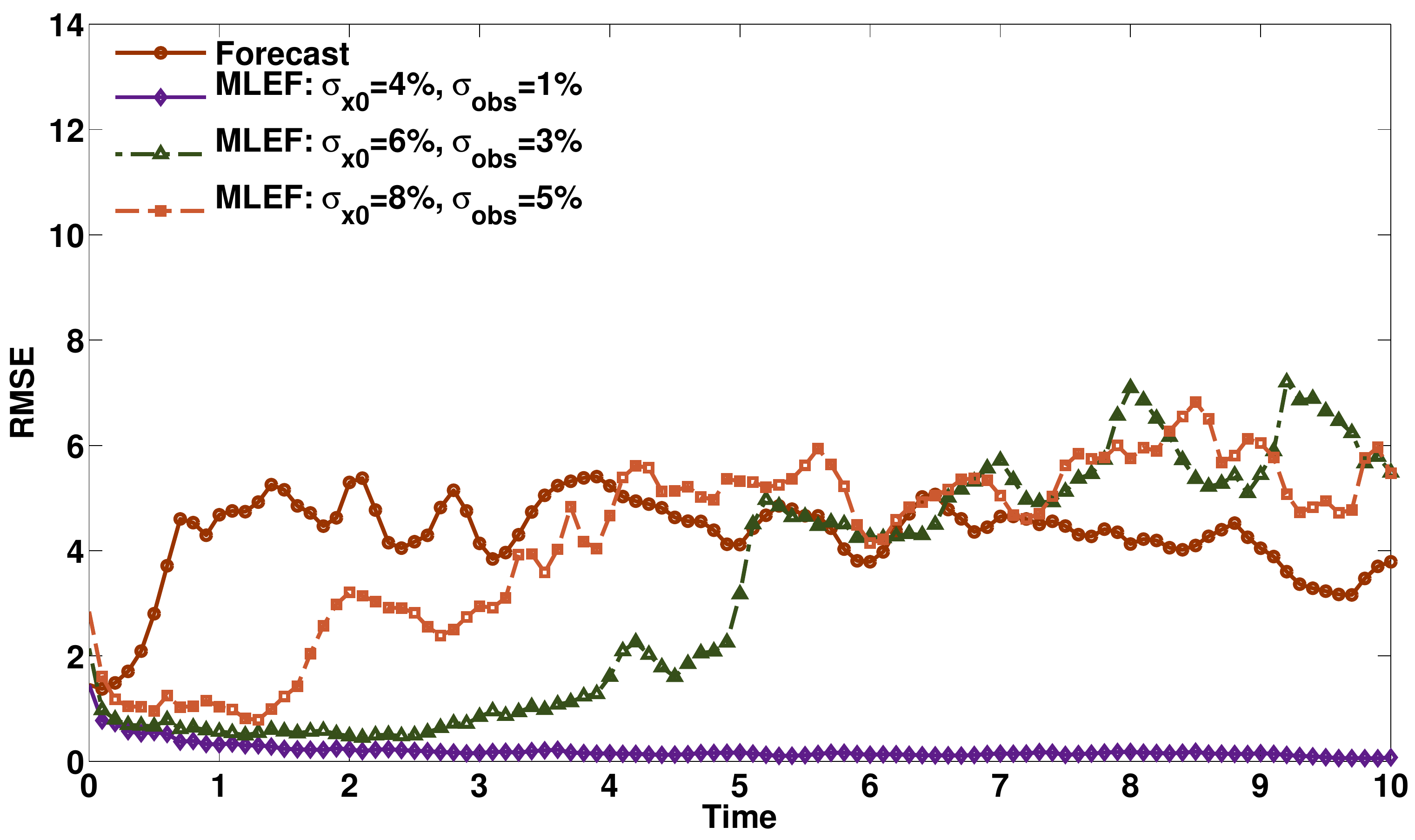}
\label{fig:More_MLEF_Quad_Thresh_2nds}}
\subfigure[Each third component is observed]{%
\includegraphics[width=0.44\linewidth]{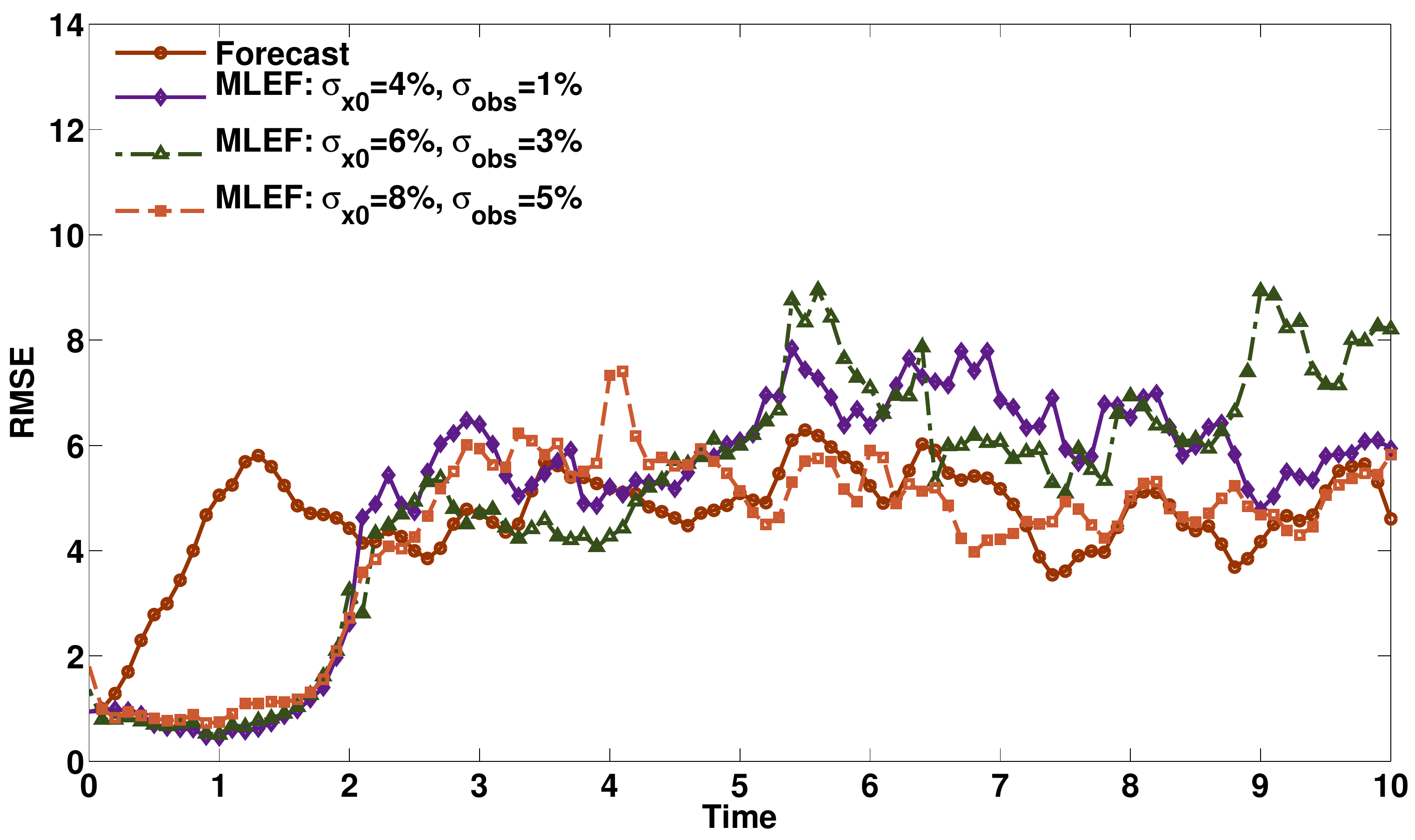}   
\label{fig:More_MLEF_Quad_Thresh_3rds}}
    \caption{MLEF data assimilation results with the quadratic observation operator \eqref{eqn:Quad_Thresh_H} with a threshold $a=0.5$. 
             MLEF is applied with varying observation frequencies, and with different noise levels for both observations and the background state.
             The frequency of observations is indicated under each panel. The background error standard deviation is $\sigma_{x0}$, and the observation  error standard deviation is $\sigma_{\rm obs}$.} 
    \label{fig:More_MLEF_Quad_Thresh}
    \end{figure}

\begin{figure}[!htbp]
\centering
\subfigure[All components are observed]{%
\includegraphics[width=0.44\linewidth]{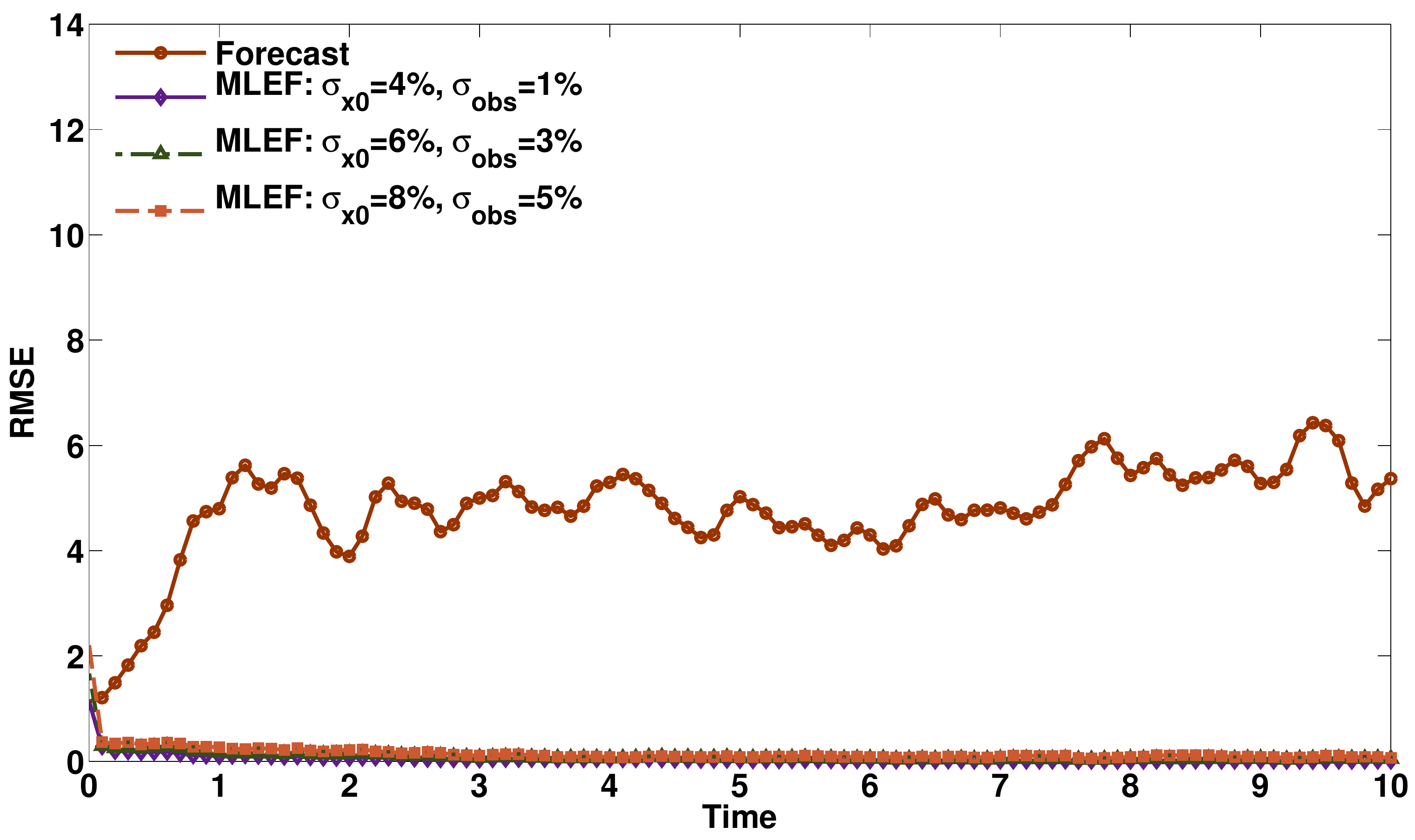}   
\label{fig:More_MLEF_Exponential2_All}}
\quad  
\subfigure[Each second component is observed]{%
\includegraphics[width=0.44\linewidth]{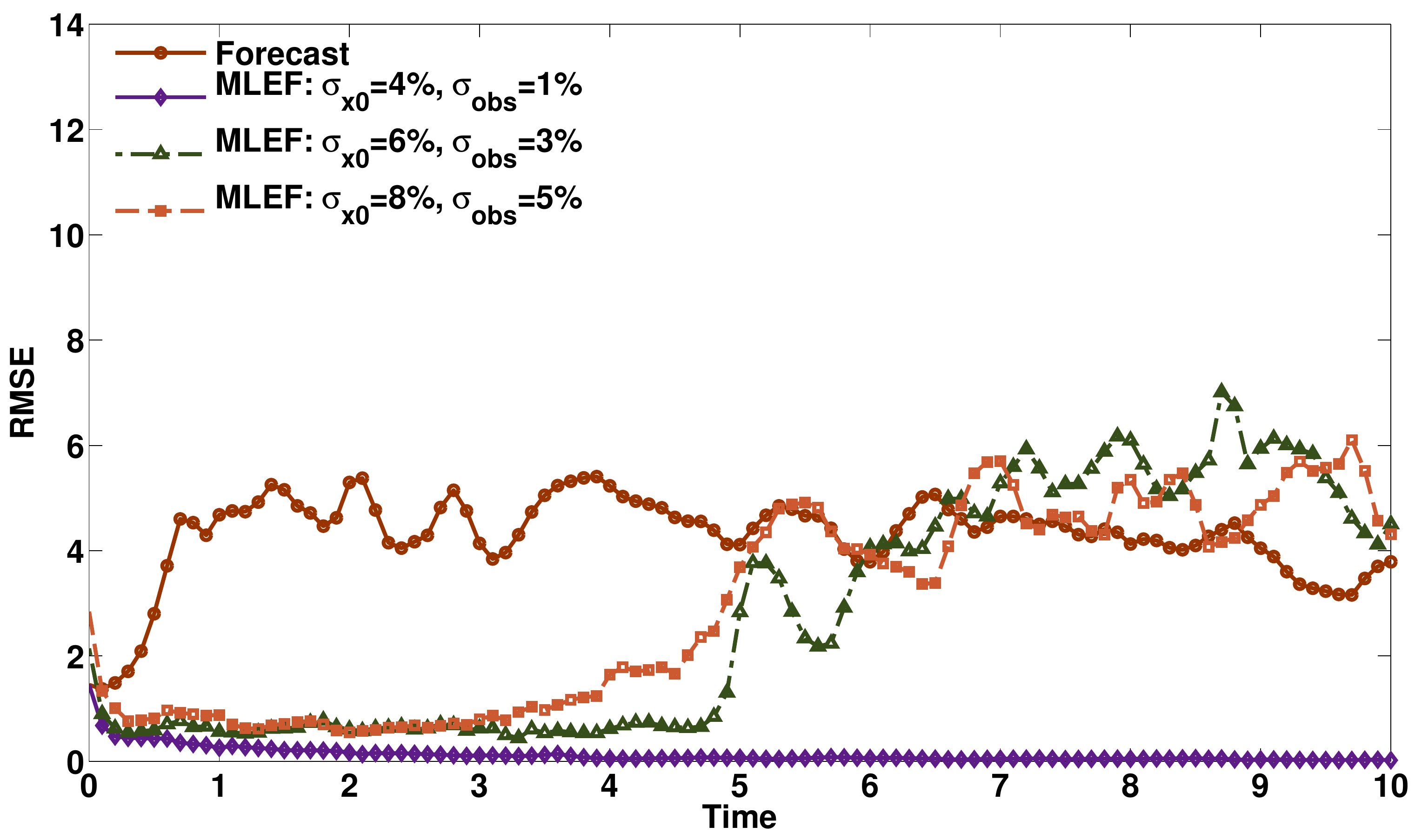}   
\label{fig:More_MLEF_Exponential2_2nds}}
\subfigure[Each third component is observed]{%
\includegraphics[width=0.44\linewidth]{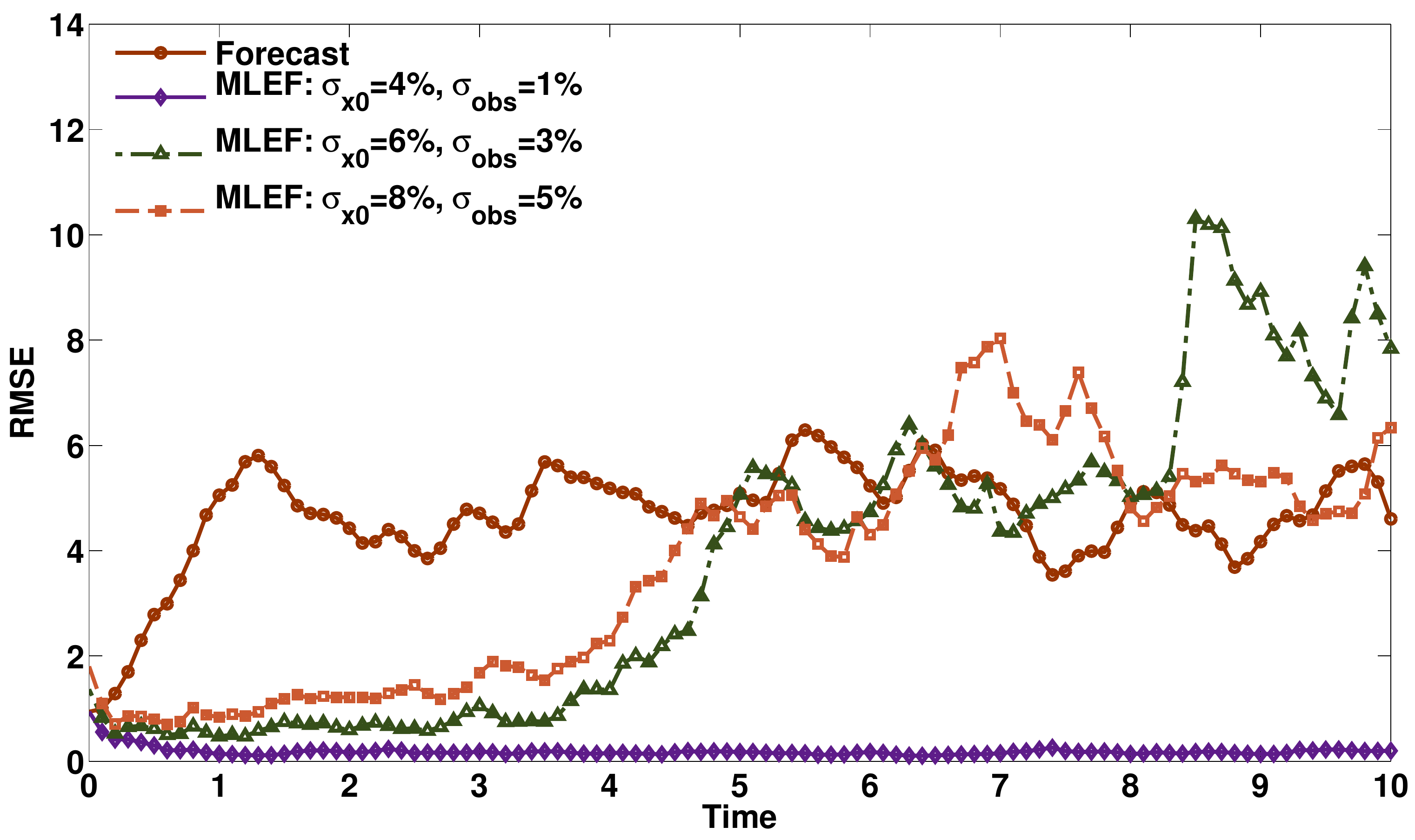}   
\label{fig:More_MLEF_Exponential2_3rds}}
    \caption{MLEF data assimilation results with the exponential observation operator \eqref{eqn:Exponential_H}. 
             MLEF is applied with varying observation frequencies, and with different noise levels for both observations and the background state.
             The frequency of observations is indicated under each panel. The background error standard deviation is $\sigma_{x0}$, and the observation  error standard deviation is $\sigma_{\rm obs}$.} 
    \label{fig:More_MLEF_Exponential2}
    \end{figure}

 The standard version of MLEF seems to be sensitive to the level of uncertainties in observations and background state, and to the degree of nonlinearity of the observation operator. Figures \ref{fig:More_MLEF_Cubic}, \ref{fig:More_MLEF_Quad_Thresh}, and \ref{fig:More_MLEF_Exponential2}  show the results of MLEF applied to the tests with cubic, quadratic with a threshold, and exponential (with a factor $r=0.2$) observation operators, respectively. In these tests all variables of the model are observed (unlike observing only each third component of the state vector as in the previous tests). Also, several uncertainty levels are considered. 
The results indicate that the MLEF performance degrades considerably when the observations are sparser (when only each second or third variables are observed). Also, the performance degrades  for higher uncertainty levels and for higher degrees of nonlinearity of the observation operators.
    \begin{figure}[!htbp]
\centering
\subfigure[All components are observed]{%
\includegraphics[width=0.44\linewidth]{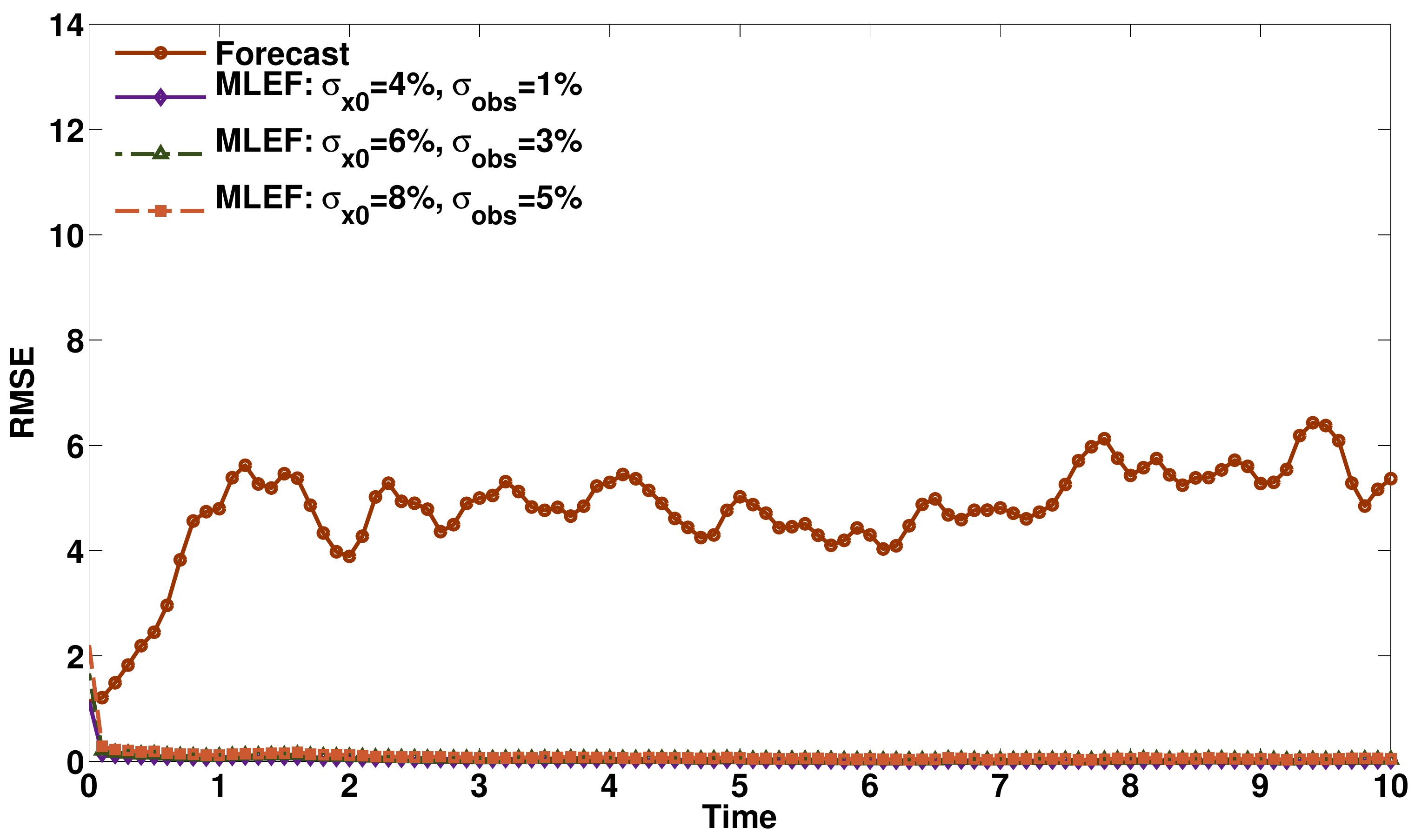}   
\label{fig:More_MLEF_Linear_All}}
\quad  
\subfigure[Each second component is observed]{%
\includegraphics[width=0.44\linewidth]{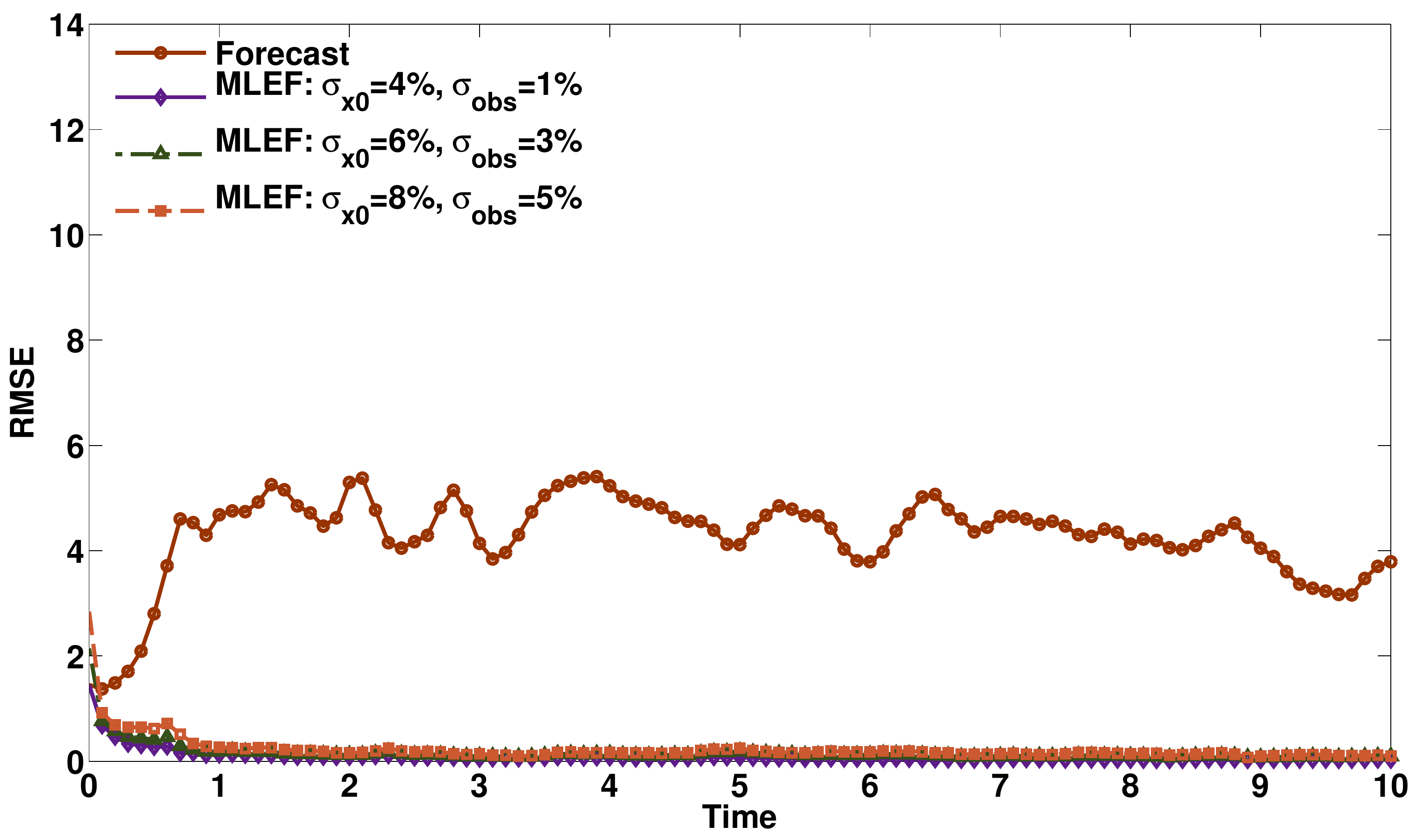}   
\label{fig:More_MLEF_Linear_2nds}}
\subfigure[Each third component is observed]{%
\includegraphics[width=0.44\linewidth]{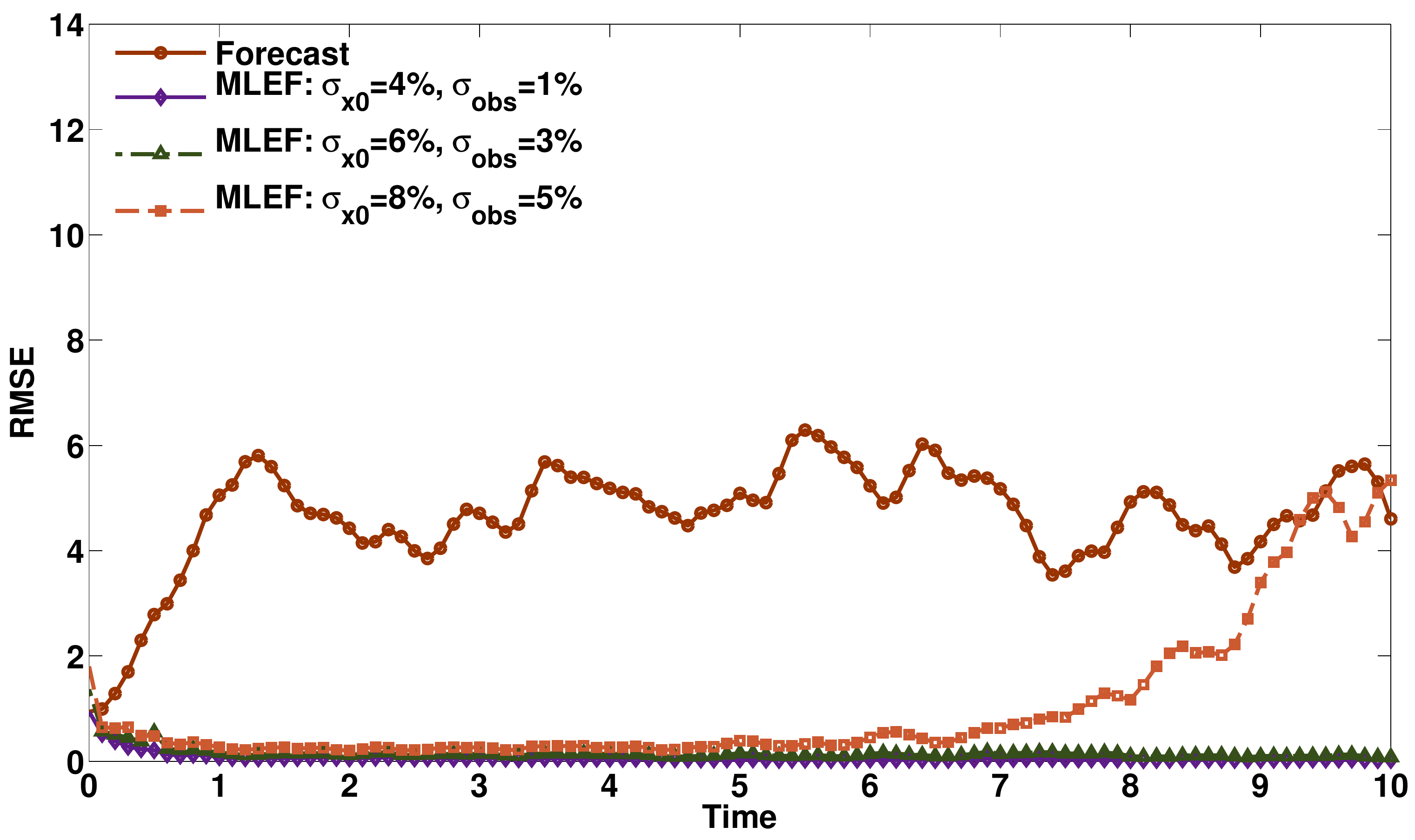}   
\label{fig:More_MLEF_Linear_3rds}}
    \caption{MLEF data assimilation results with the linear observation operator \eqref{eqn:Linear_H}. 
             MLEF is applied with varying observation frequencies, and with different noise levels for both observations and the background state.
             The frequency of observations is indicated under each panel. The background error standard deviation is $\sigma_{x0}$, and the observation  error standard deviation is $\sigma_{\rm obs}$.} 
    \label{fig:More_MLEF_Linear}
    \end{figure}
    \begin{figure}[!htbp]
\centering
\subfigure[All components are observed]{%
\includegraphics[width=0.44\linewidth]{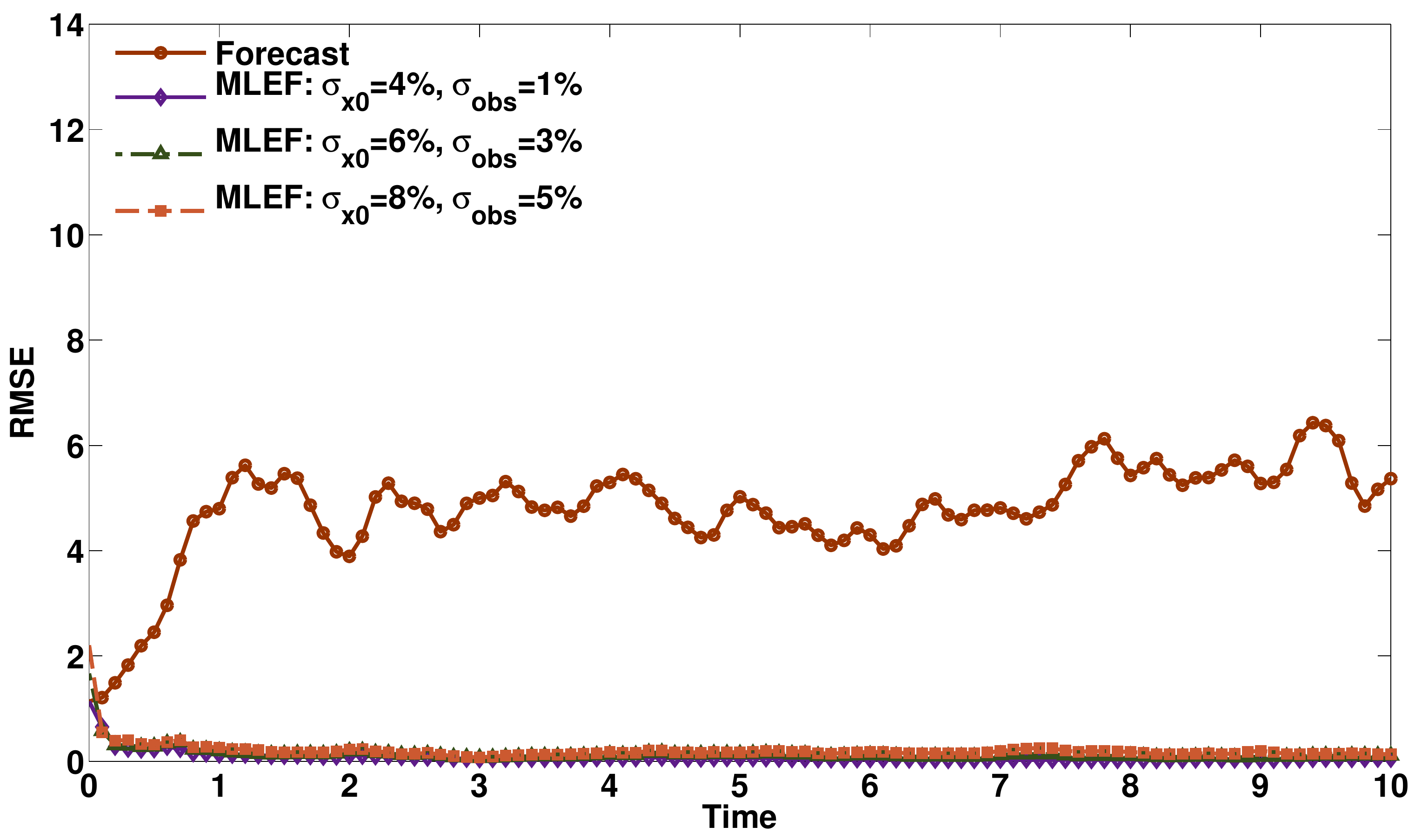}   
\label{fig:More_MLEF_Quadratic_All}}
\quad  
\subfigure[Each second component is observed]{%
\includegraphics[width=0.44\linewidth]{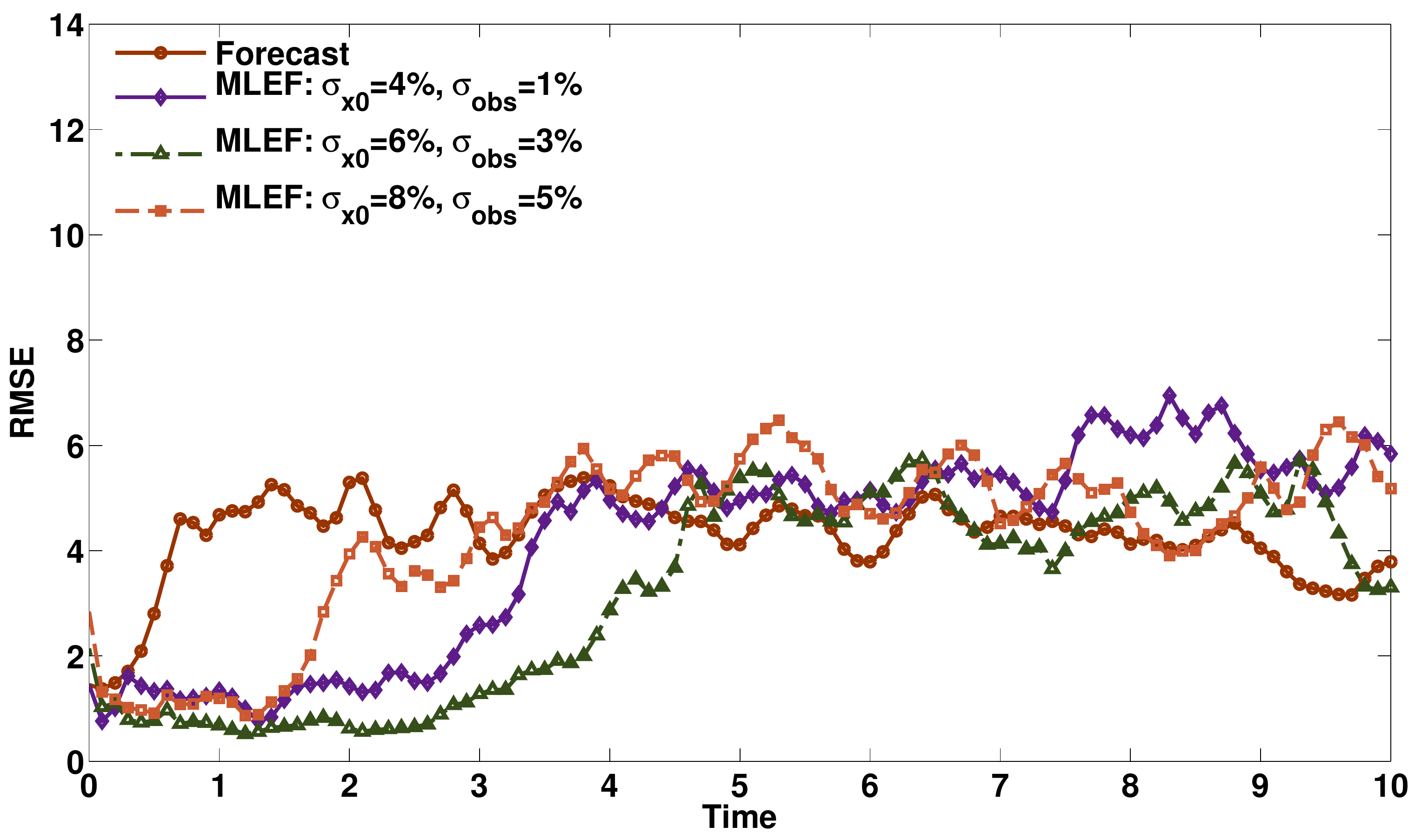}   
\label{fig:More_MLEF_Quadratic_2nds}}
\subfigure[Each third component is observed]{%
\includegraphics[width=0.44\linewidth]{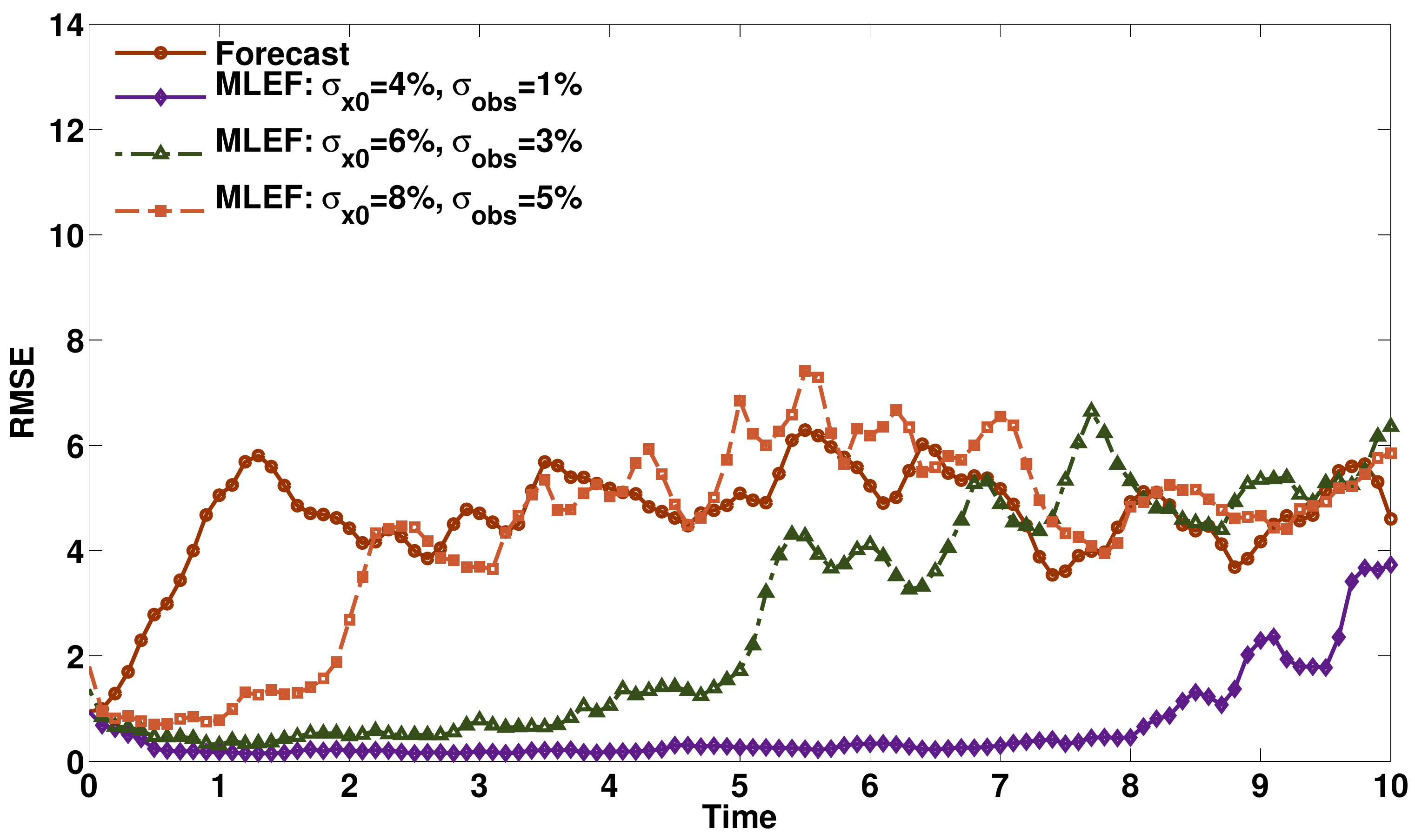}   
\label{fig:More_MLEF_Quadratic_3rds}}
    \caption{MLEF data assimilation results with the quadratic observation operator \eqref{eqn:Quadratic_H}. 
             MLEF is applied with varying observation frequencies, and with different noise levels for both observations and the background state.
             The frequency of observations is indicated under each panel. The background error standard deviation is $\sigma_{x0}$, and the observation  error standard deviation is $\sigma_{\rm obs}$.} 
    \label{fig:More_MLEF_Quadratic}
    \end{figure}
    \begin{figure}[!htbp]
\centering
\subfigure[All components are observed]{%
\includegraphics[width=0.44\linewidth]{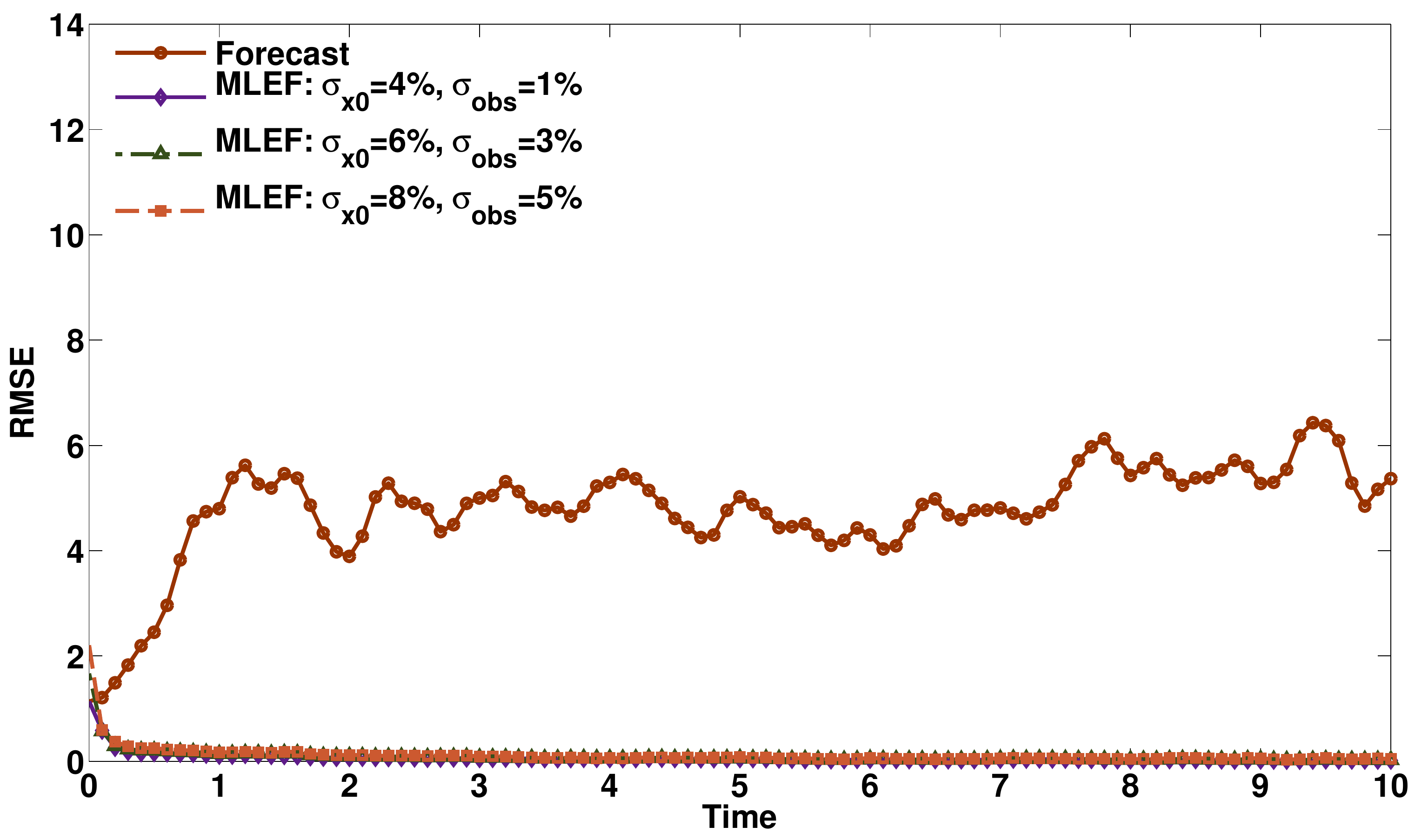}   
\label{fig:More_MLEF_Abs_All}}
\quad  
\subfigure[Each second component is observed]{%
\includegraphics[width=0.44\linewidth]{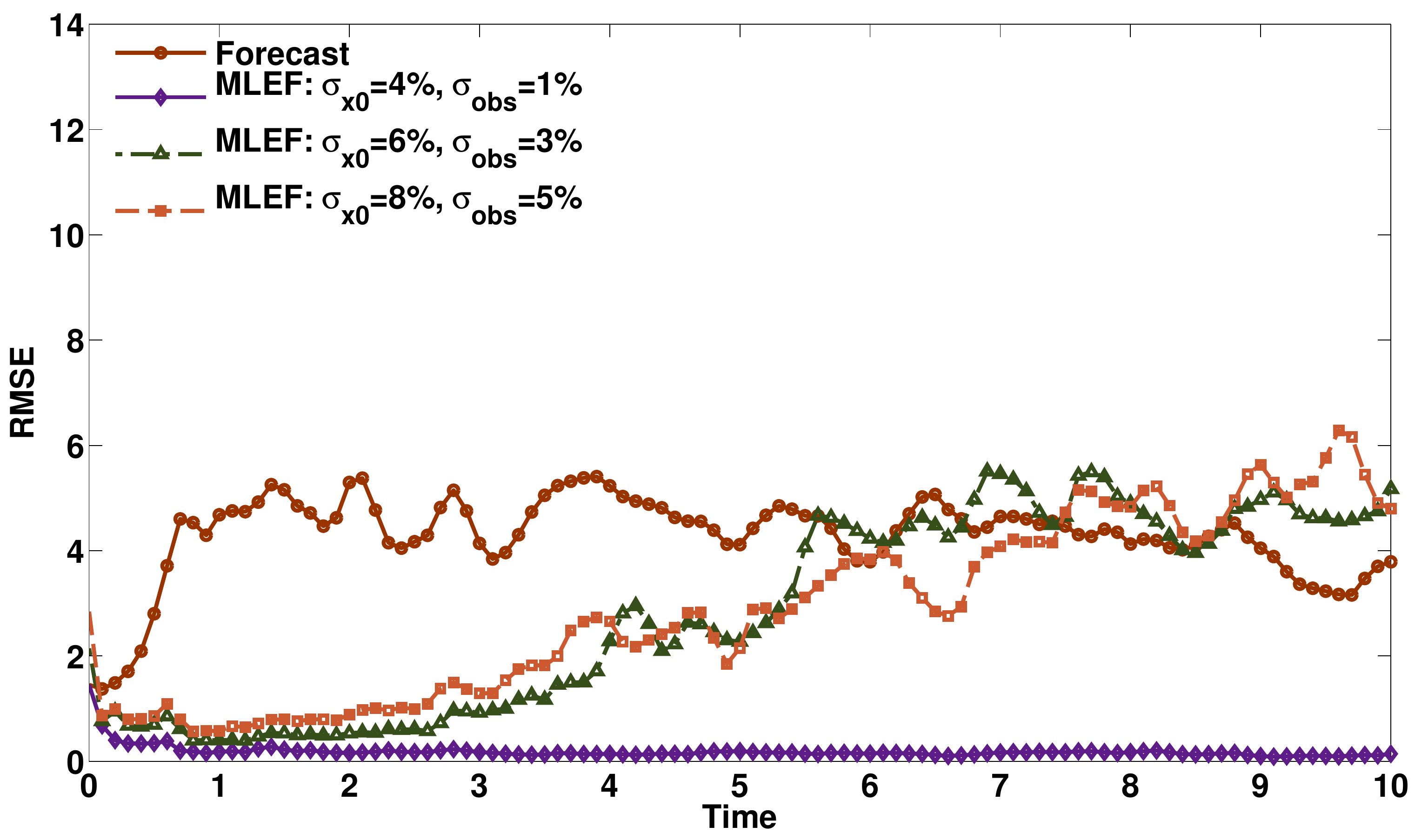}   
\label{fig:More_MLEF_Abs_2nds}}
\subfigure[Each third component is observed]{%
\includegraphics[width=0.44\linewidth]{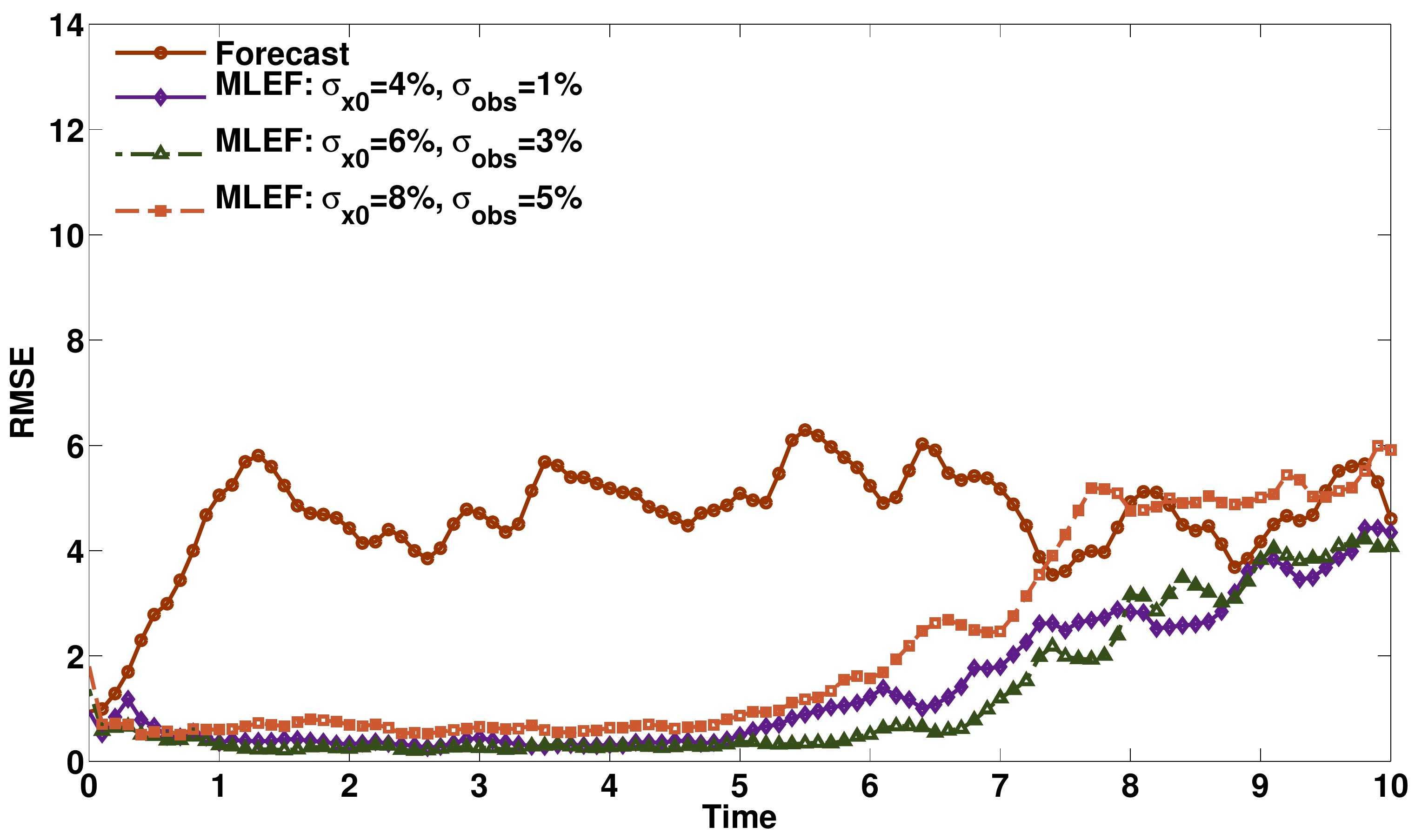}   
\label{fig:More_MLEF_Abs_3rds}}
    \caption{MLEF data assimilation results with the magnitude observation operator \eqref{eqn:Abs_H}. 
             MLEF is applied with varying observation frequencies, and with different noise levels for both observations and the background state.
             The frequency of observations is indicated under each panel. The background error standard deviation is $\sigma_{x0}$, and the observation  error standard deviation is $\sigma_{\rm obs}$.} 
    \label{fig:More_MLEF_Abs}
    \end{figure}

   \FloatBarrier
     %
     %
     \subsection{Tuning the number of MC steps between successive state selections}
     \label{subsec:MC_Steps_Effect}
\begin{figure}[H]
\centering
\subfigure[RMSE mean: two-stage integrator; $h=2/\nvar,\ m=\nvar$]{%
\includegraphics[width=0.44\linewidth]{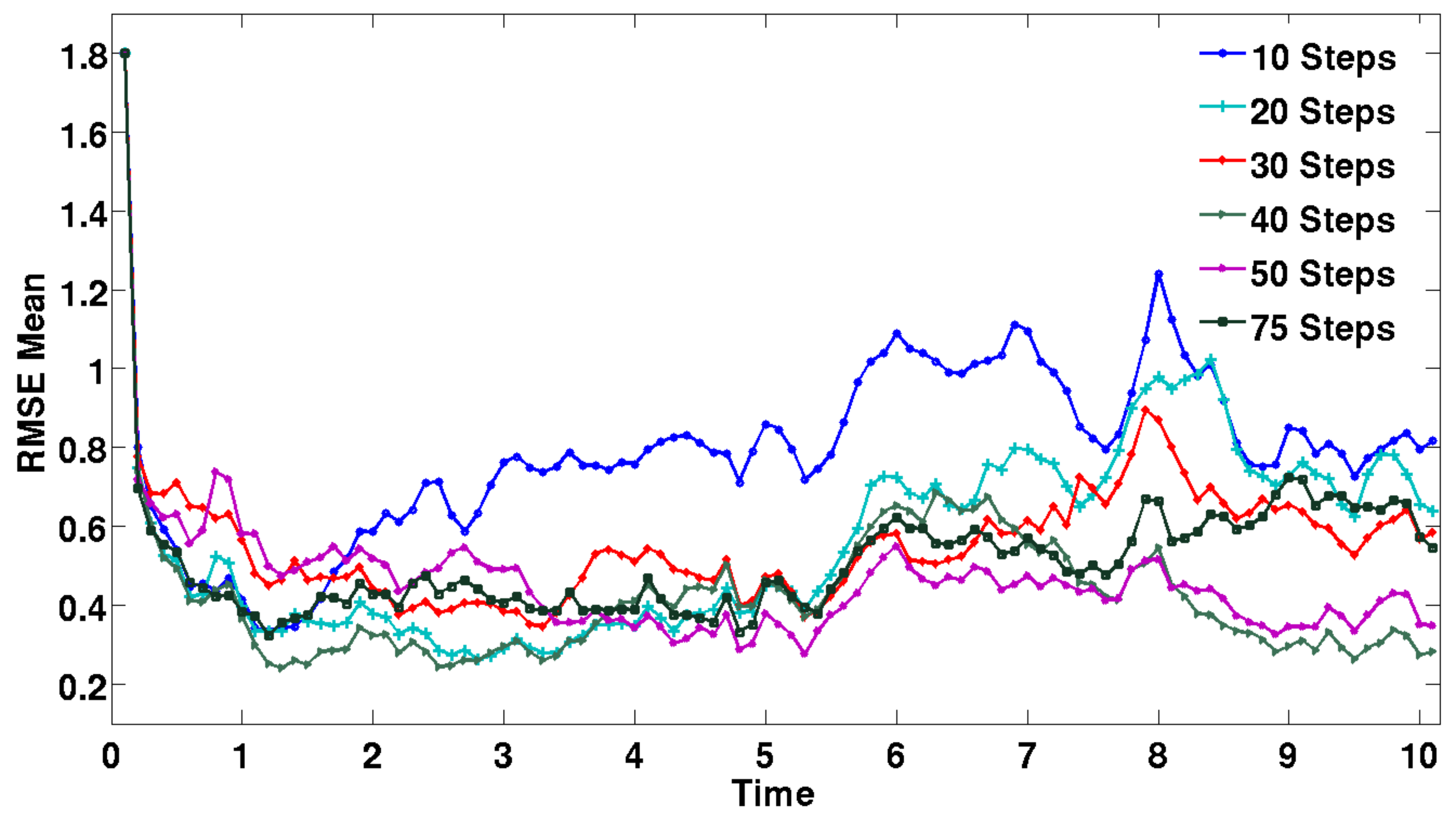}   
\label{fig:Theo_MCSteps_2Stage_RMSE-Mean}}
\quad  
\subfigure[RMSE standard deviation: two-stage integrator; $h=2/\nvar,\ m=\nvar$]{%
\includegraphics[width=0.44\linewidth]{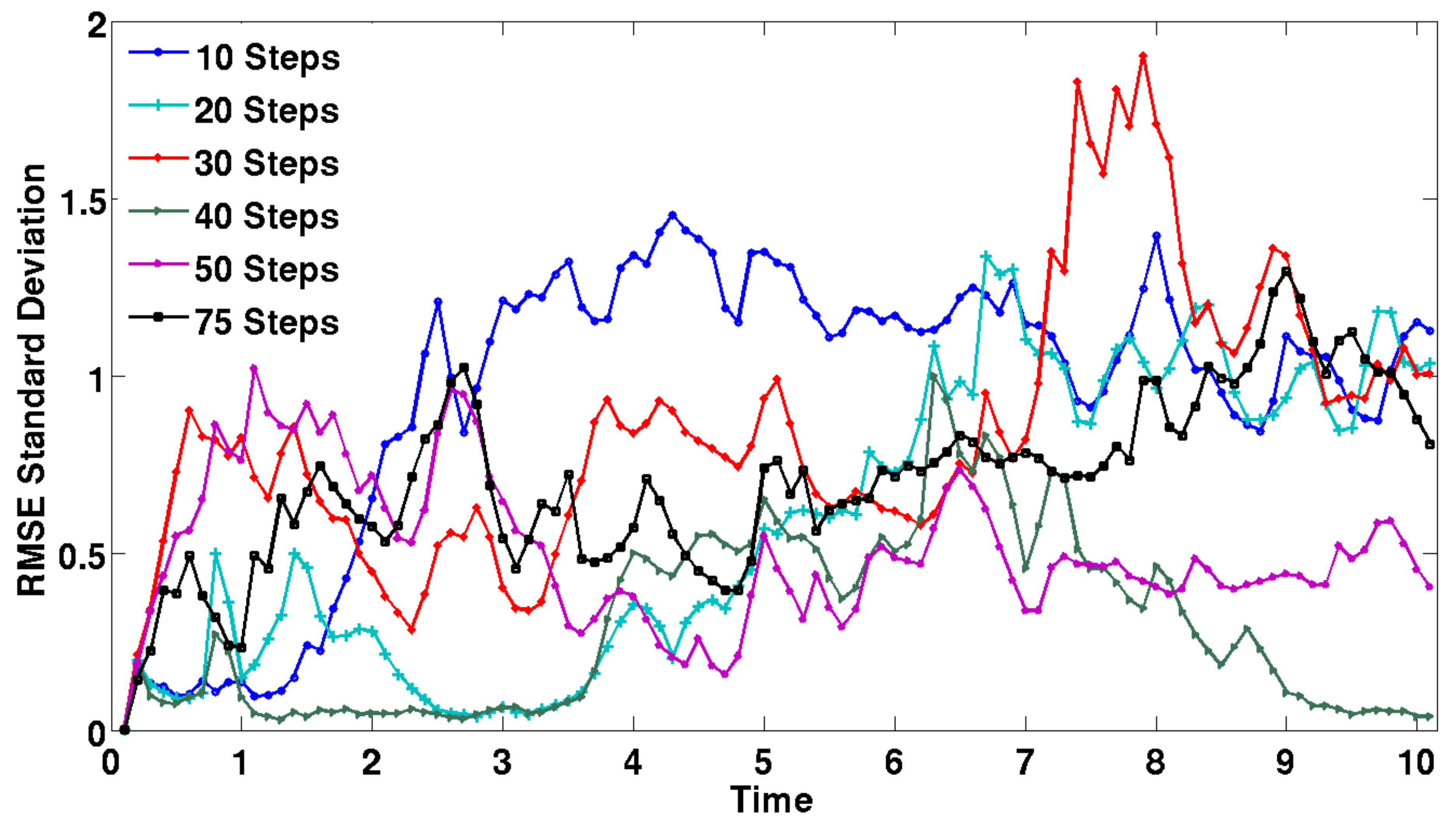}   
\label{fig:Theo_MCSteps_2Stage_RMSE-StDev}}
\subfigure[RMSE mean: three-stage integrator; $h=3/\nvar,\ m=\nvar/2$]{%
\includegraphics[width=0.44\linewidth]{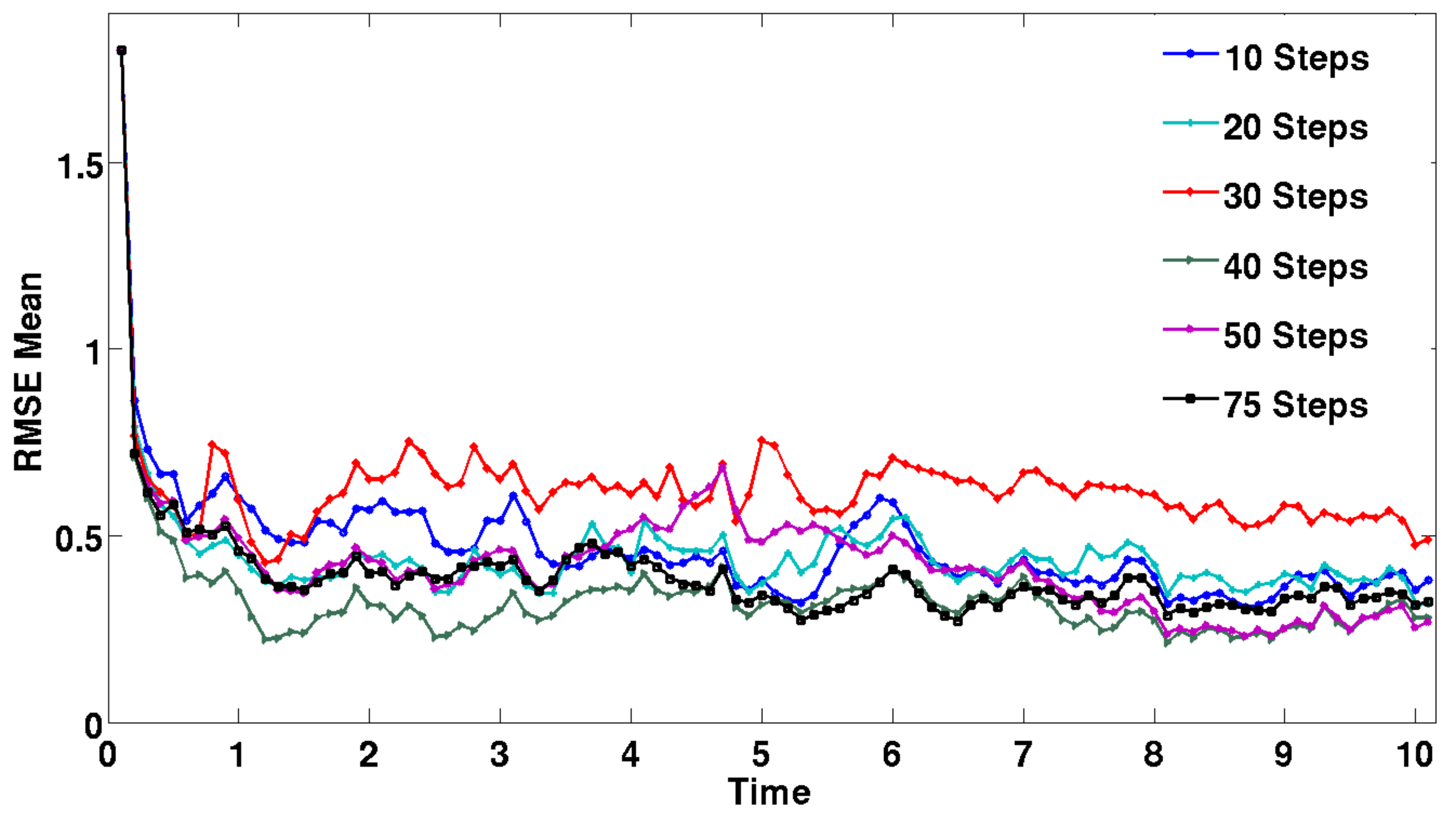}   
\label{fig:Theo_MCSteps_3Stage_RMSE-Mean}}
\quad  
\subfigure[RMSE standard deviation: three-stage integrator; $h=3/\nvar,\ m=\nvar/2$]{%
\includegraphics[width=0.44\linewidth]{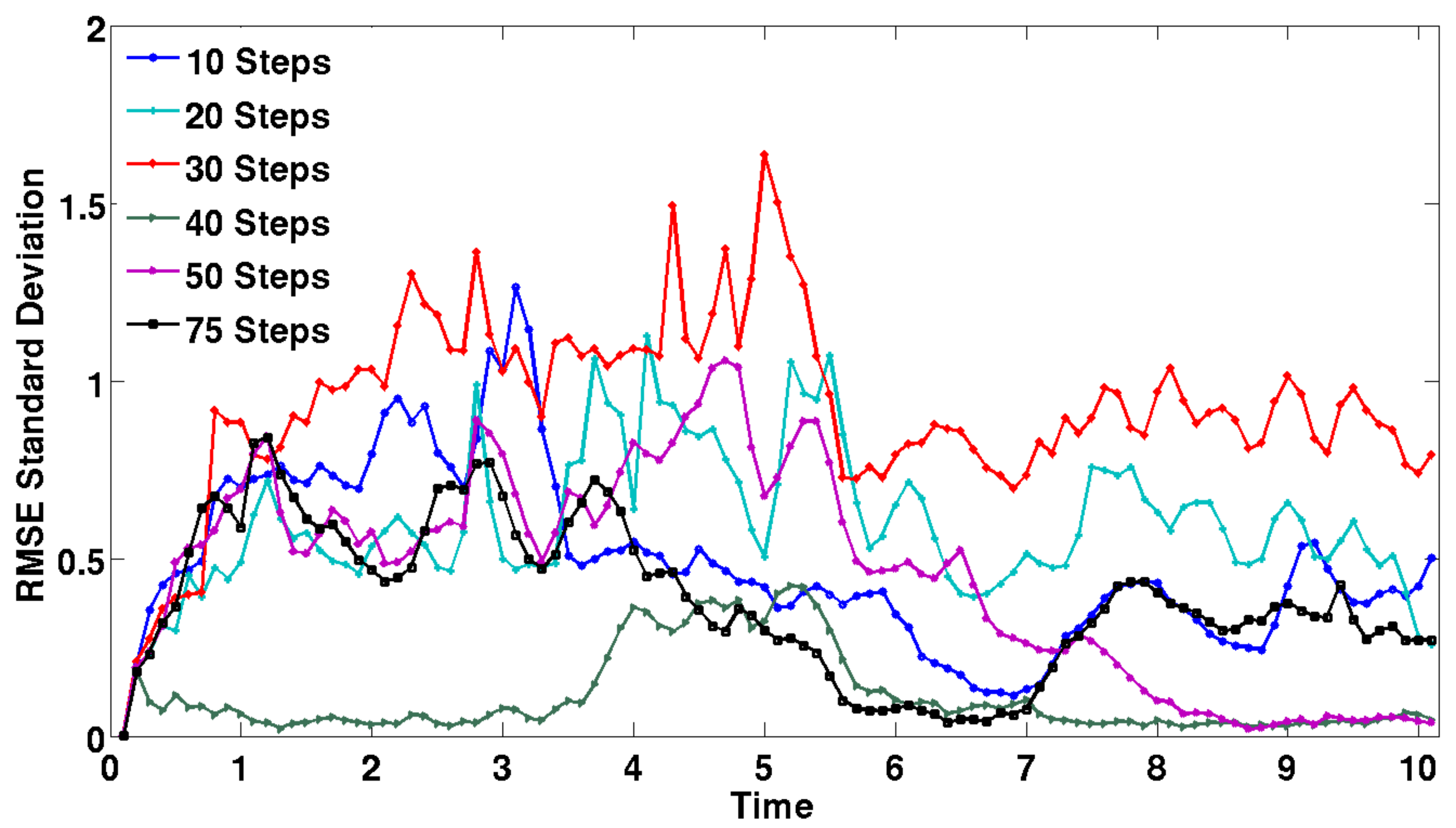}   
\label{fig:Theo_MCSteps_3Stage_RMSE-StDev}}
\subfigure[RMSE mean: four-stage integrator; $h=4/\nvar,\ m=\nvar/2$]{%
\includegraphics[width=0.44\linewidth]{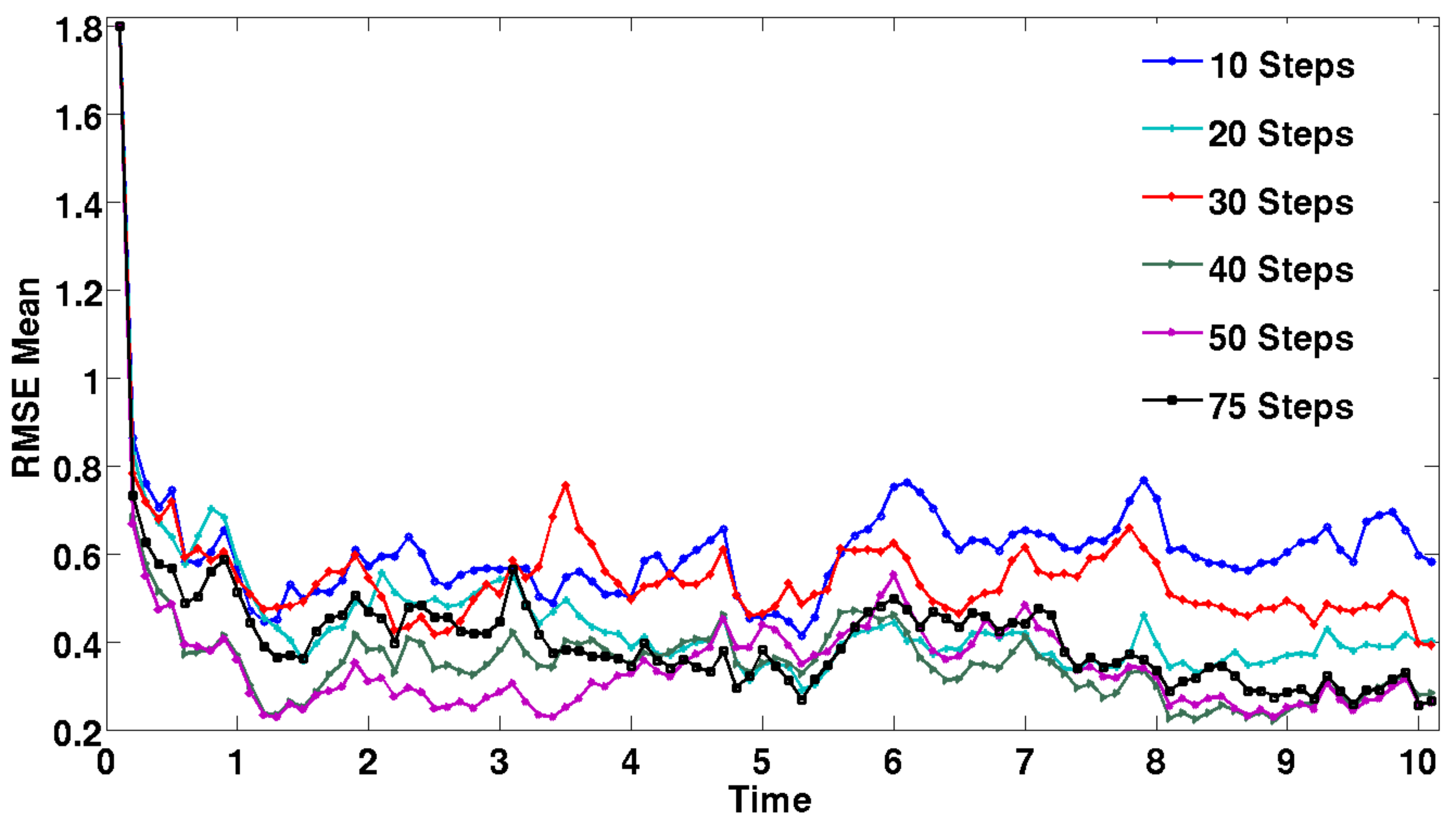}   
\label{fig:Theo_MCSteps_4Stage_RMSE-Mean}}
\quad  
\subfigure[RMSE standard deviation: two-stage integrator; $h=4/\nvar,\ m=\nvar/2$]{%
\includegraphics[width=0.44\linewidth]{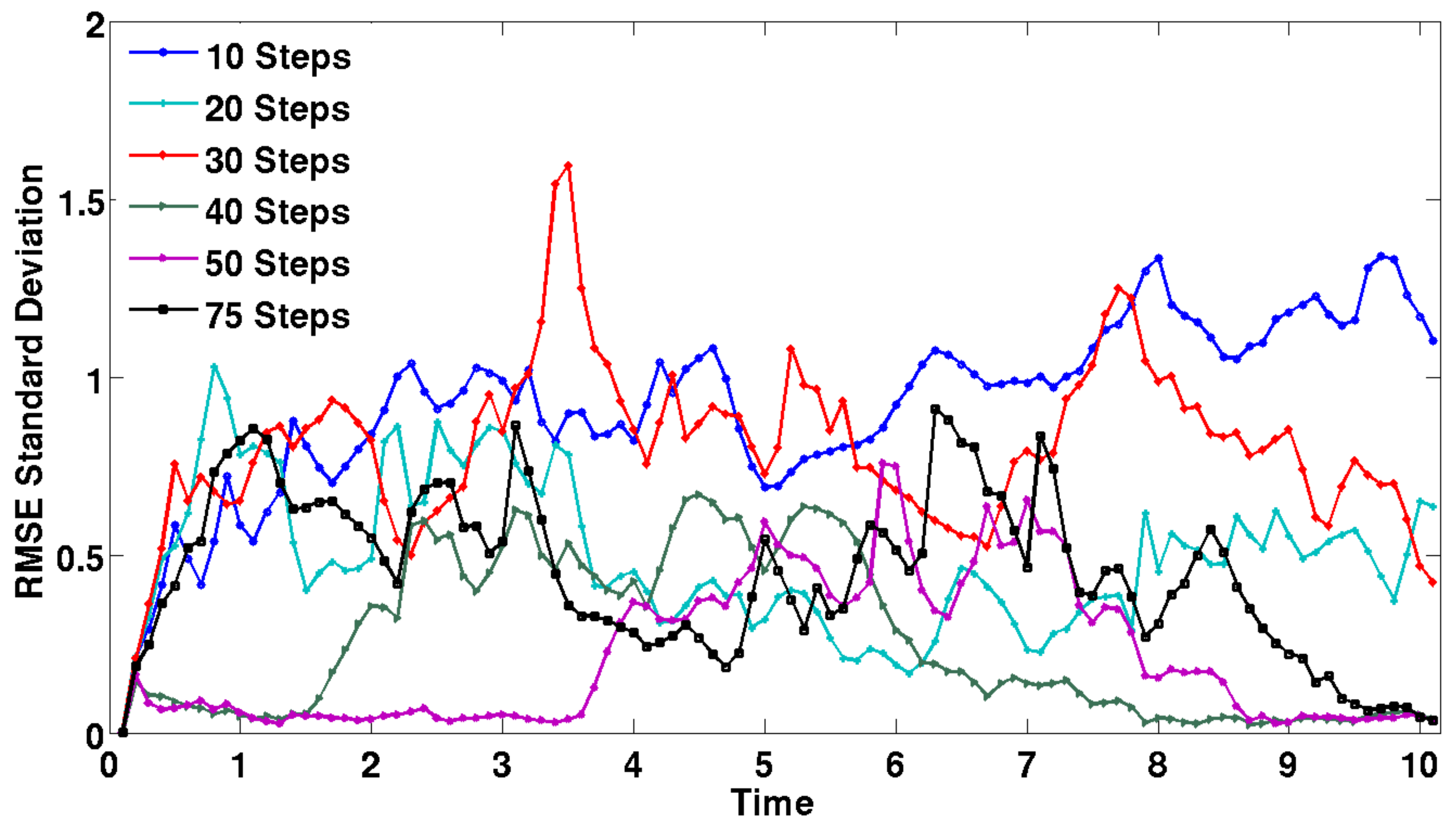}   
\label{fig:Theo_MCSteps_4Stage_RMSE-StDev}}
\caption{Data assimilation results with the quadratic observation operator \eqref{eqn:Quadratic_H}. The sampling filter is applied with several settings of the number of inter-chain steps.
         The average and standard deviation of RMS errors are both evaluated, at each observation time point, over the $25$ realization of the sampling filter. 
         The step size  $h$ and the number of steps $m$ for the symplectic integrator are indicated under each panel, where $\nvar=40$ is the dimension of the state vector.} 
    \label{fig:Theo_MCSteps}
  \end{figure}

This section discusses the prevention of outliers (filter divergence) that can happen for nonlinear observation operators (e.g., in case of quadratic observation operator in our experiments). This is done by tuning integration parameters. In addition to selecting the mass matrix $\mathbf{M}$ and the number of burn-in steps, there two more parameters to be tuned. They are the step size of the symplectic integrator as discussed before, and the number of steps skipped between selected states at stationarity (referred to as inter-chain steps). To study the effect of tuning the last two parameters the quadratic observation operator is re-tested with the high-order integrators and with the optimal step sizes suggested by Blanes~\citep{blanes2013numerical}. Figure \ref{fig:Theo_MCSteps} shows the average and the standard deviation of analysis RMSE for  $25$ realizations of the sampling filter. Various settings of the number of inter-chain steps are used to study its effect on the performance of the proposed filter. Tuning the step size of the symplectic integration results in a notable reduction in the average RMSE compared to the results obtained with the empirical settings and presented in Section \ref{subsec:Quadratic_H_Results}. Outliers are still present as inferred from Figures \ref{fig:Theo_MCSteps_2Stage_RMSE-StDev}, \ref{fig:Theo_MCSteps_3Stage_RMSE-StDev}, and \ref{fig:Theo_MCSteps_4Stage_RMSE-StDev}.  

\begin{figure}[H]
\centering
   \includegraphics[width=0.44\linewidth]{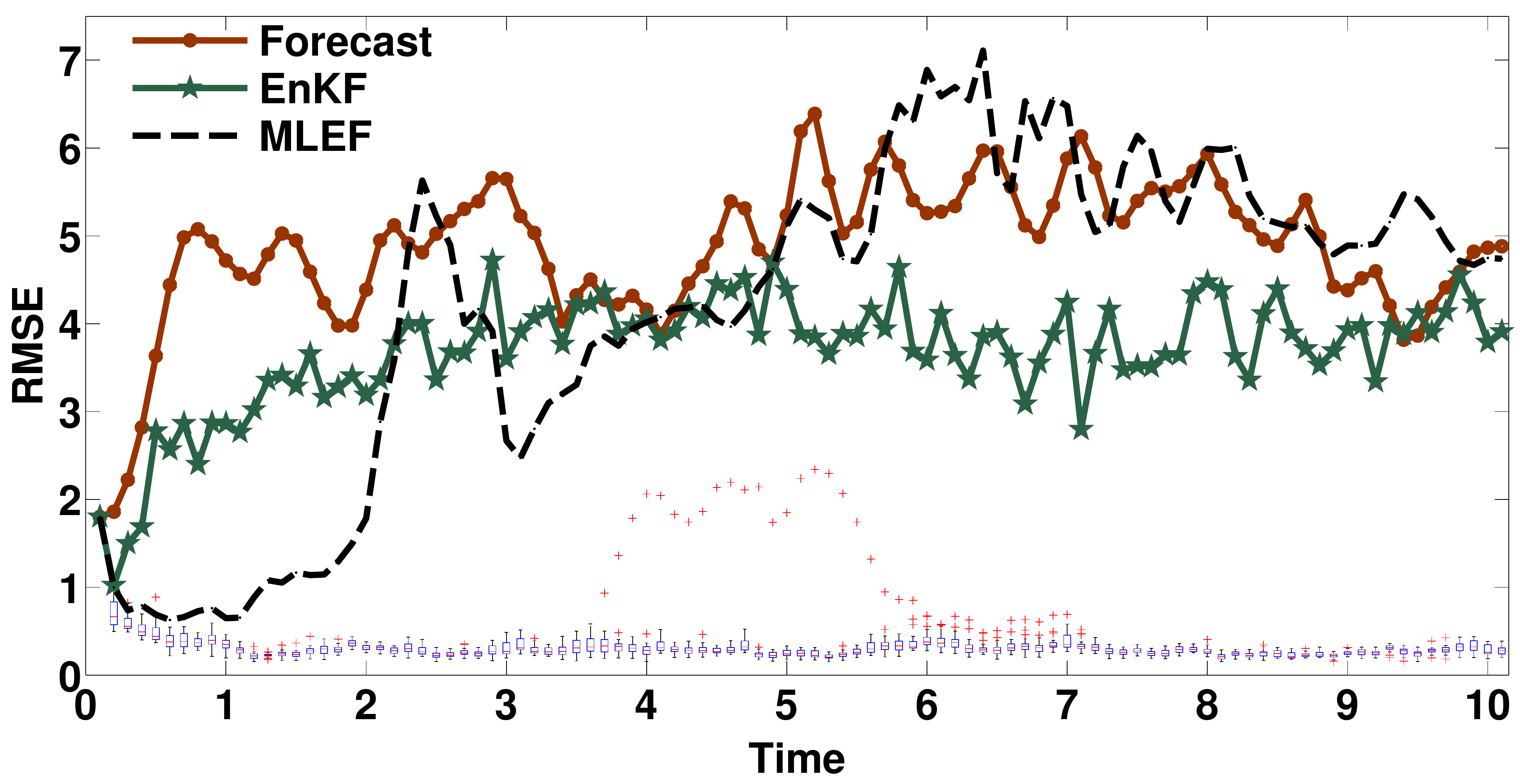}   
\caption{Data assimilation results with the quadratic observation operator \eqref{eqn:Quadratic_H}. The symplectic integrator used is the three-stage integrator \eqref{eqn:three_stage}.
         The time step for the integrator is $T=3/2$ with $h=3/\nvar,\ m=\nvar/2$, and $40$ inter-chain steps. 
         The RMSE for $100$ instances of the sampling filter results are shown as box plots. The  red line represents the median RMSE values across all instances and the central blue box represents the variance. The two vertical lines (whiskers) extend up to $1.5$ times the height of the central box. The values exceeding the length of the whiskers are considered outliers (extremes) and are plotted as red crosses.
         } 
\label{fig:Theo_RMSE_Quadratic_H_3Stage}
    \end{figure}

Tuning the number of inter-chain steps can in principle greatly enhance both the performance of the filter and the 
reliability of the results. Setting the number of inter-chain steps to $30$ is not optimal for the quadratic observation operator, and better results can be obtained with $40$ steps, as seen in the results reported in Figure \ref{fig:Theo_RMSE_Quadratic_H_3Stage}. These results indicate that a careful tuning of both the step size and the number of inter-steps in the chain may overcome the problem of outliers and lead to the desired performance of the filter. 

\FloatBarrier
  \FloatBarrier
 
    %
    %
    %
In addition to controlling the time step settings of the integrator, and tuning the number of steps of the chain, we can use the Hilbert integrator (with tuned step size) to periodically validate the ensembles obtained using other integrators, since the Hilbert integrator suffers less from outliers. A simple solution is to run the assimilation process several times and exclude outlier states by creating a combined ensemble. Care must be exercised, however, to not change the probability density.  These alternatives will be inspected in depth in future work in the context of more complex models. 
    %
     \subsection{A highly nonlinear observation operator}
     \label{subsec:Second_Exponential_Oper_Results}
We have also tested the sampling filter capabilities in a very challenging setting: the exponential observation operator \eqref{eqn:Exponential_H} is considered with a factor of $r=0.5$. This factor leads to a large range of observation values (from $e^{-3.7}$ to $e^{6.2}$). In addition, small perturbations in the state variables cause very large changes in the corresponding measurements. Both traditional methods EnKF and MLEF diverge when applied to this test, and consequently their results are not reported here.

\begin{figure}[H]
   \centering
\subfigure[Three-stage integrator \eqref{eqn:three_stage}; $h=0.01,\ m=60$]{%
\includegraphics[width=0.44\linewidth]{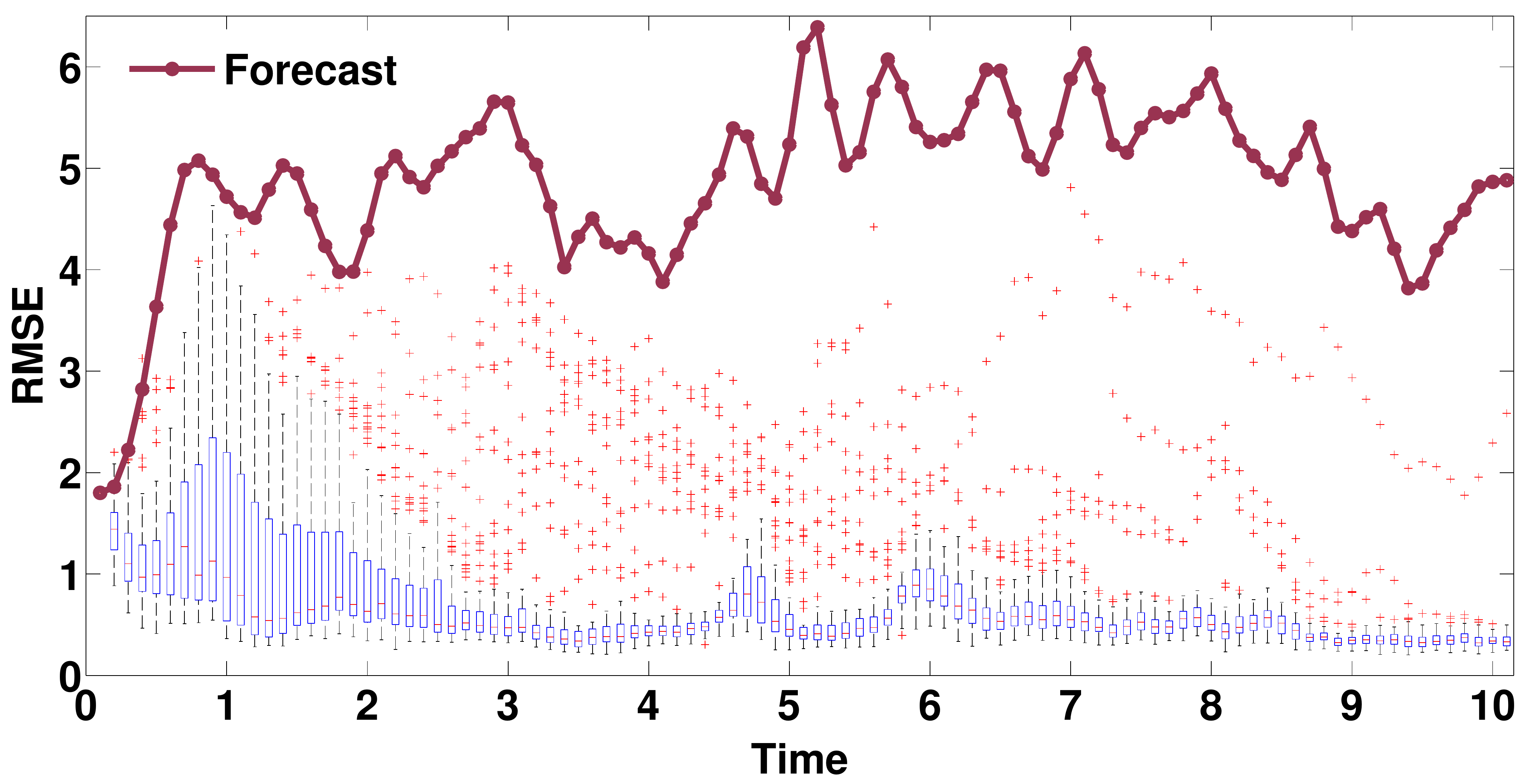}   
\label{fig:ExponentialH_2_3Stage}}
\quad
\subfigure[Four-stage integrator \eqref{eqn:four_stage}; $h=0.001,\ m=30$]{%
\includegraphics[width=0.44\linewidth]{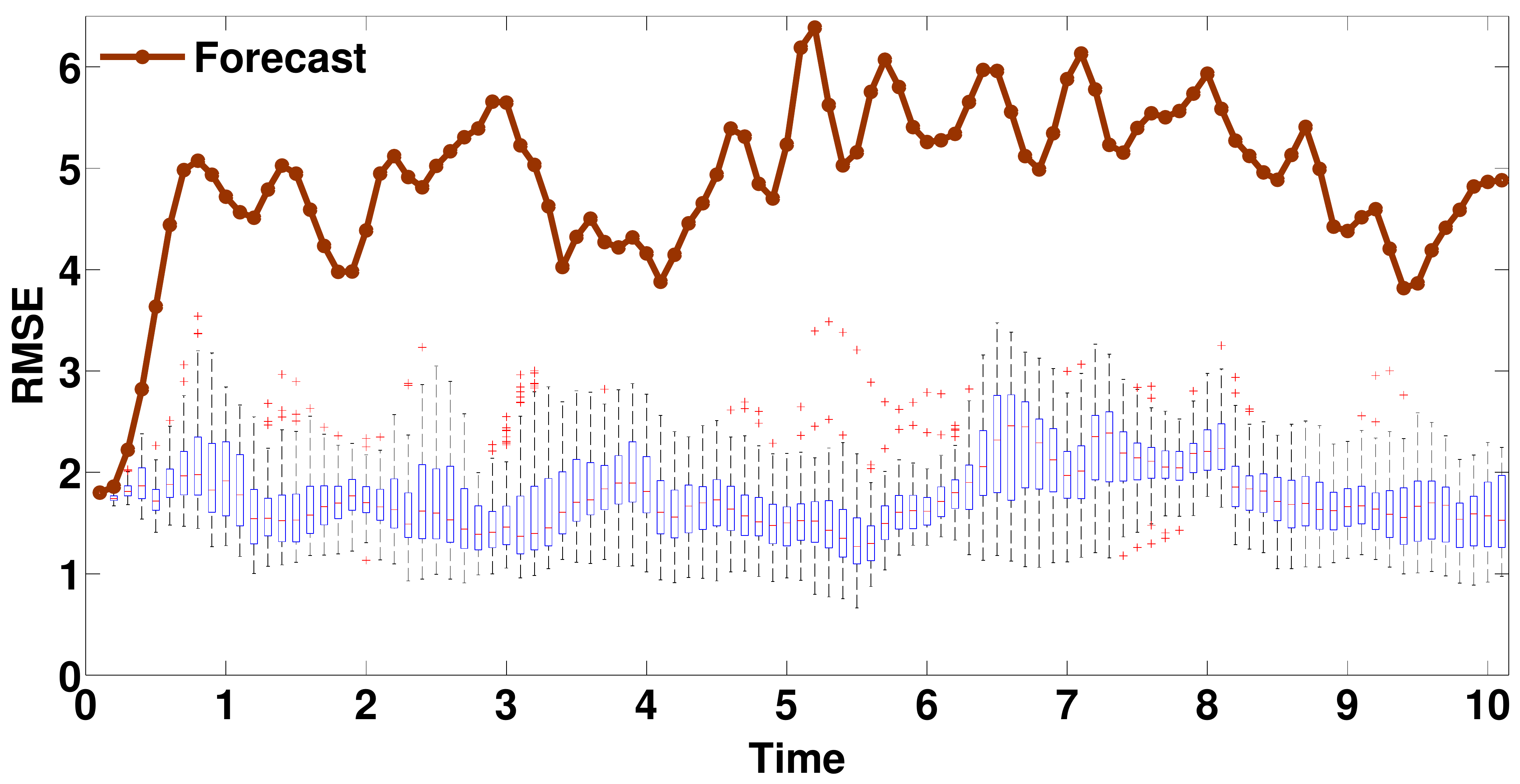}   
\label{fig:ExponentialH_2_Hilbert}}
\caption{Data assimilation results with the exponential observation operator \eqref{eqn:Exponential_H} with a factor $r=0.2$. The symplectic integrator used is indicated under each panel.
         The step size  $h$ and the number of steps $m$ are indicated under each panel. The number of inter-chain steps is $30$.
         The RMSE for $100$ instances of the sampling filter results are shown as box plots. The  red line represents the median RMSE values across all instances and the central blue box represents the variance. The two vertical lines (whiskers) extend up to $1.5$ times the height of the central box. The values exceeding the length of the whiskers are considered outliers (extremes) and are plotted as red crosses.
      } 
\label{fig:ExponentialH_2_3Stage_Hilbert}
\end{figure}

This test problem is challenging for the sampling filter as well and the symplectic integration step sizes need to be tuned to achieve convergence. For example, the number of steps taken by Verlet integrator has to be increased to $m=60$ while keeping the step-size fixed to $h=0.01$, to result in good performance. The length of the trajectory of the Hamiltonian system has to be increased as well.
For the three-stage integrator a shorter trajectory of the Hamiltonian system works well if the step size is sufficiently reduced, e.g., $h=0.001$, and $m=30$. Empirical tuning of the Verlet, two-stage, and four-stage integrators proved to be challenging with this observation operator. However, the three-stage integrator produced very satisfactory results with larger step sizes, as shown in Figure \ref{fig:ExponentialH_2_3Stage}.

\begin{figure}[H]
   \centering
\subfigure[$x_1$]{%
\includegraphics[width=0.44\linewidth]{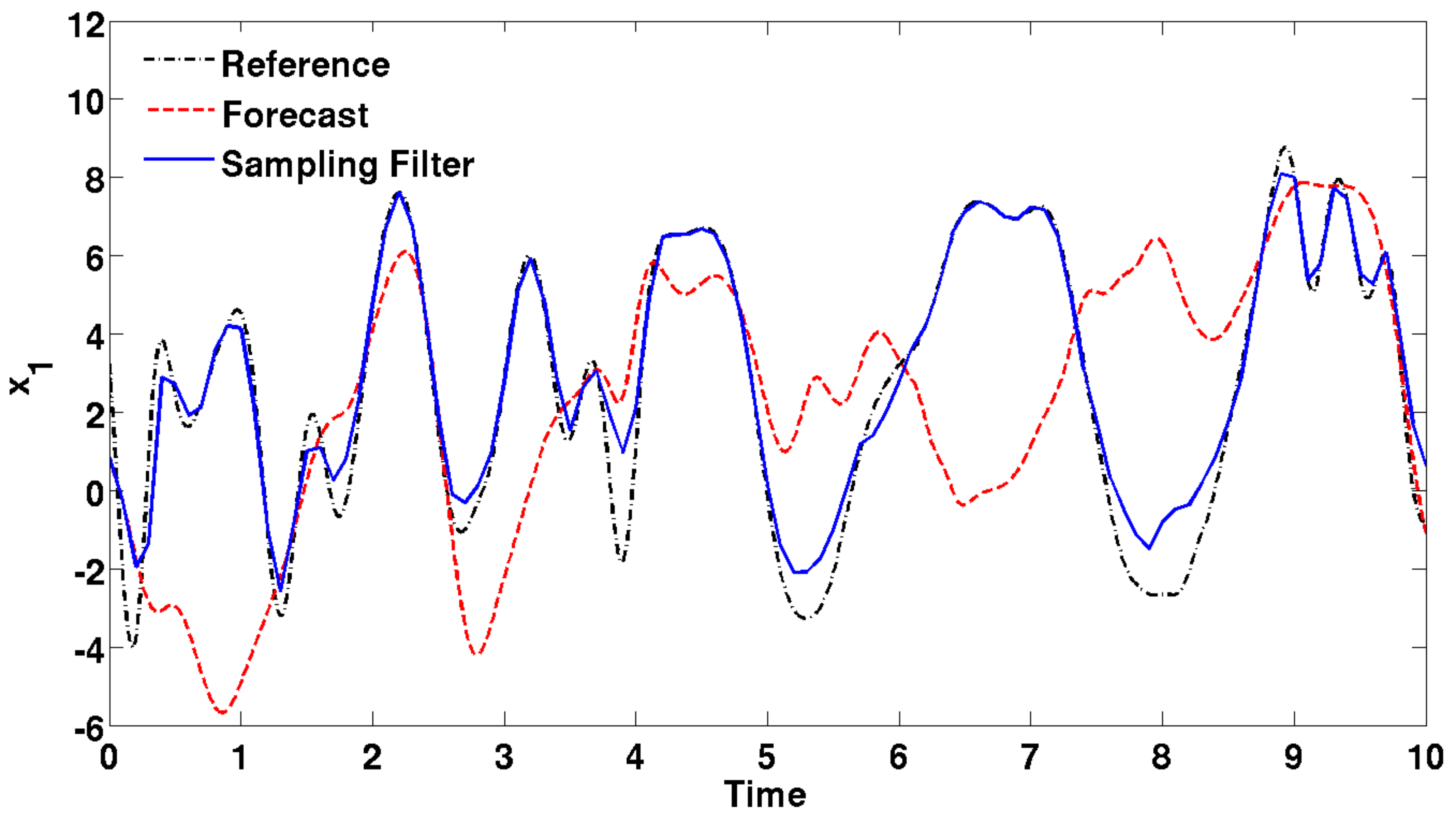}   
\label{fig:ExponentialH_2_Hilbert_X1}}
\quad
\subfigure[$x_4$]{%
\includegraphics[width=0.44\linewidth]{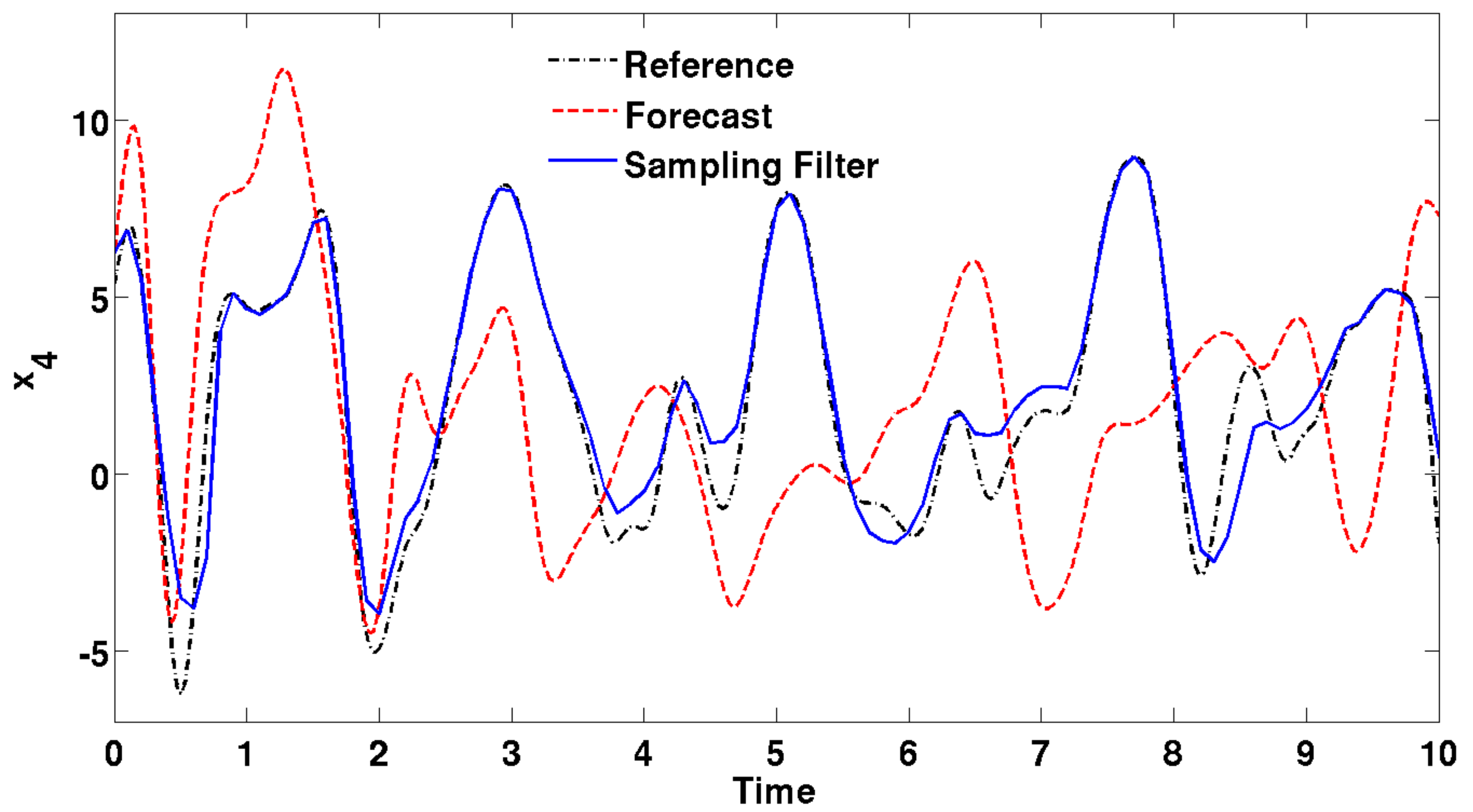}   
\label{fig:ExponentialH_2_Hilbert_X4}}
\subfigure[$x_7$]{%
\includegraphics[width=0.44\linewidth]{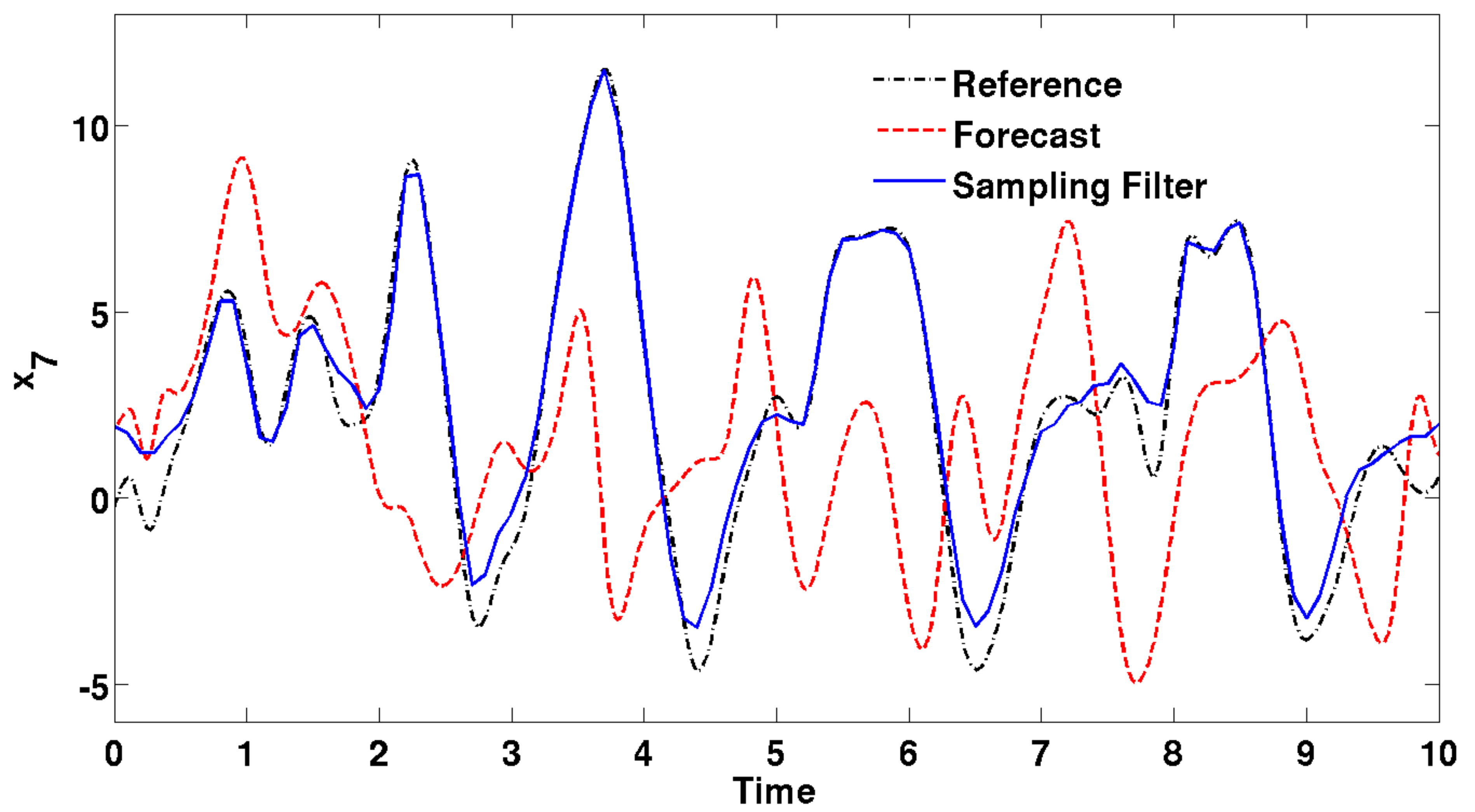}   
\label{fig:ExponentialH_2_Hilbert_X7}}
\quad
\subfigure[$x_{11}$]{%
\includegraphics[width=0.44\linewidth]{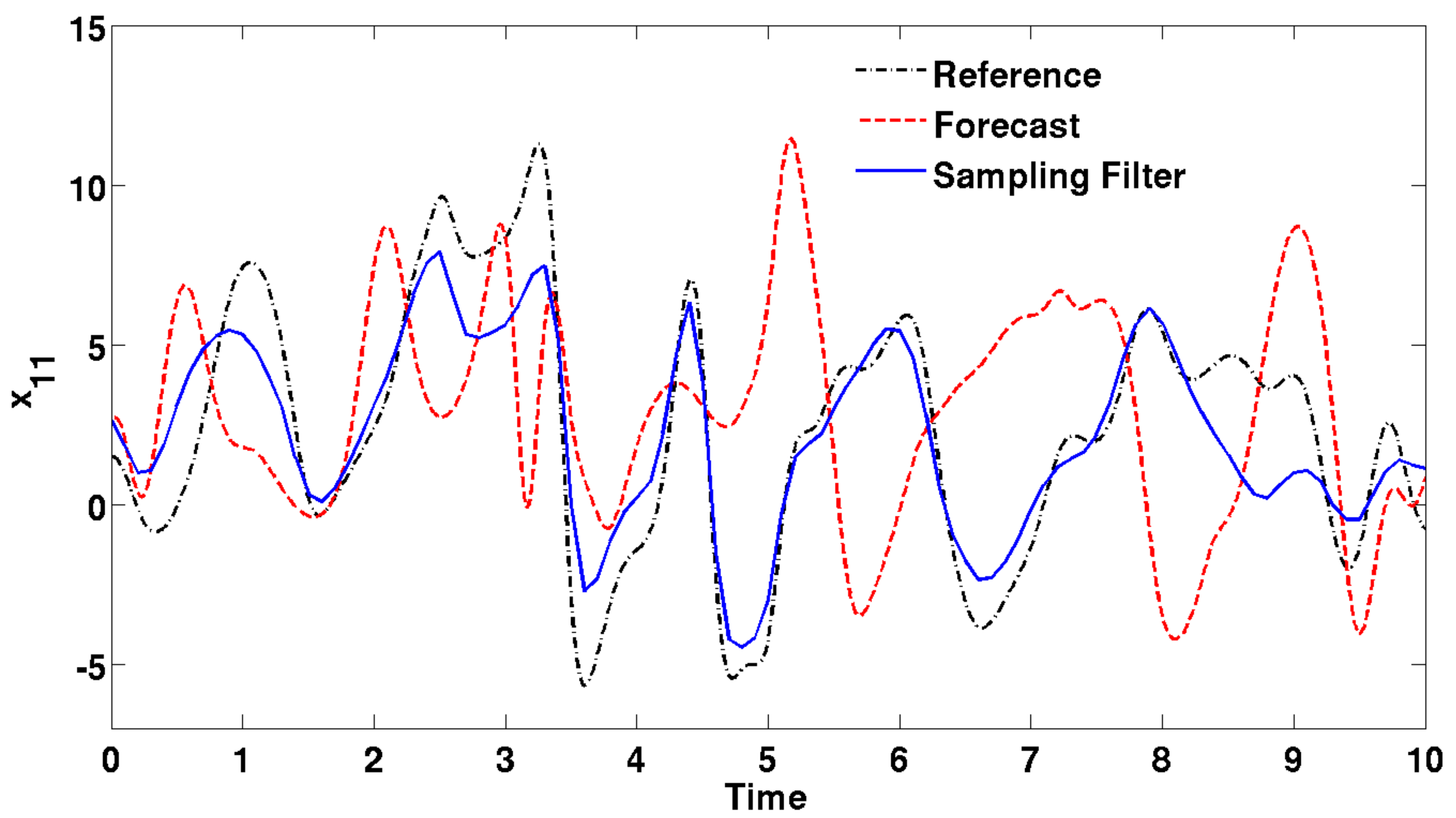}   
\label{fig:ExponentialH_2_Hilbert_X11}}
\caption{Integrator defined on Hilbert space is used; $h=0.001,\ m=30$. The components $x_1,x_4,x_7,x_{11}$ of the state vector $\x$ are plotted.
    The number of inter-chain steps is $30$.} 
\label{fig:ExponentialH_2_State_Variables_Hilbert}
\end{figure}

The Hilbert space integrator performs robustly and yields analyses that are as accurate as the ones for the simpler observation operators; see Figure \ref{fig:ExponentialH_2_State_Variables_Hilbert}. While the RMSE value achieved by the filter using the Hilbert space integrator is relatively large, one can argue that this level is acceptable when dealing with large systems, nonlinear operators where all other filters fail. The results in Figure \ref{fig:ExponentialH_2_State_Variables_Hilbert} show that the analysis (of selected components) follows the truth reasonably closely.

\begin{figure}[H]
   \centering
\subfigure[$x_1$]{%
\includegraphics[width=0.44\linewidth]{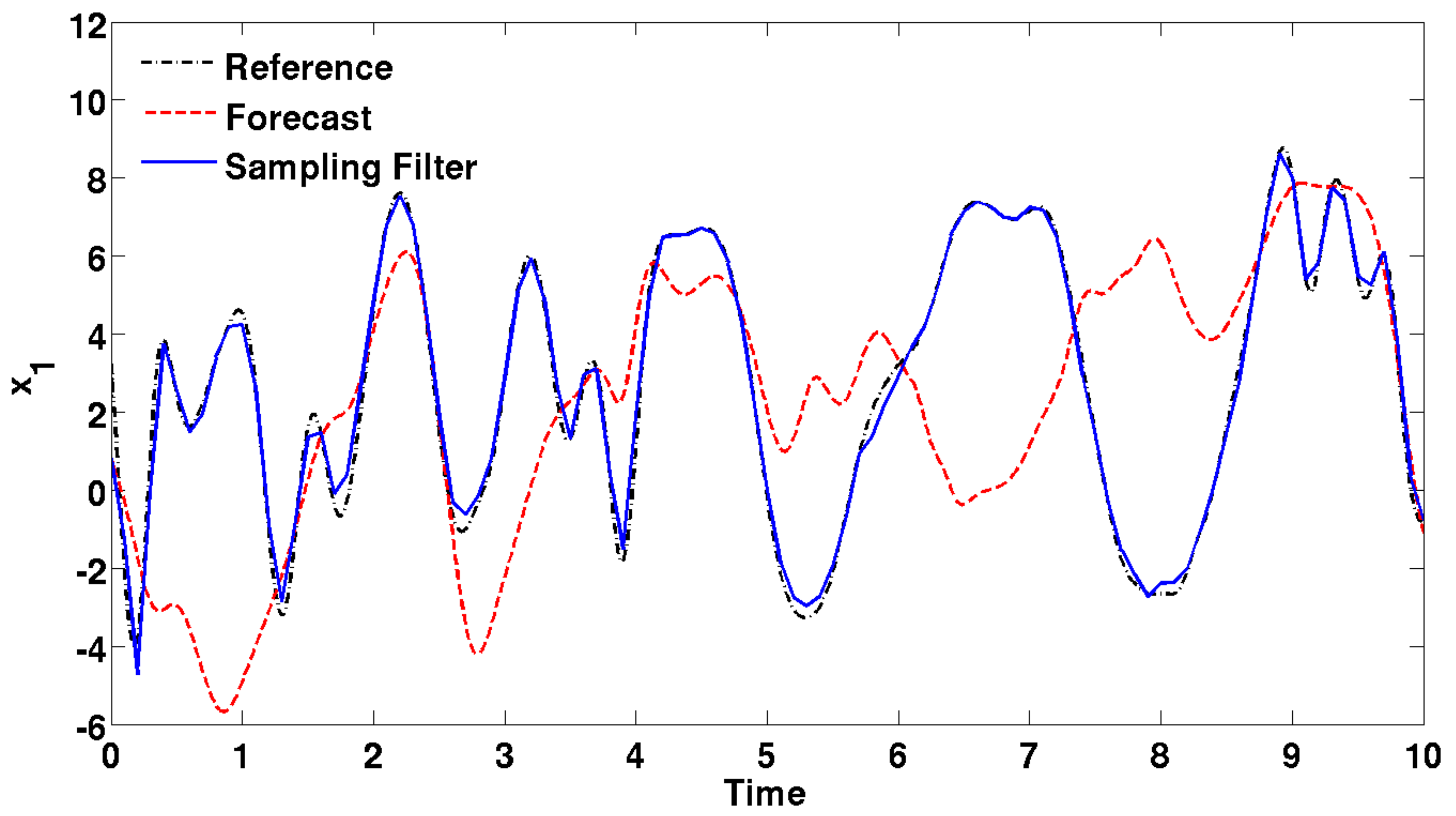}   
\label{fig:ExponentialH_2_3Stage_X1}}
\quad
\subfigure[$x_4$]{%
\includegraphics[width=0.44\linewidth]{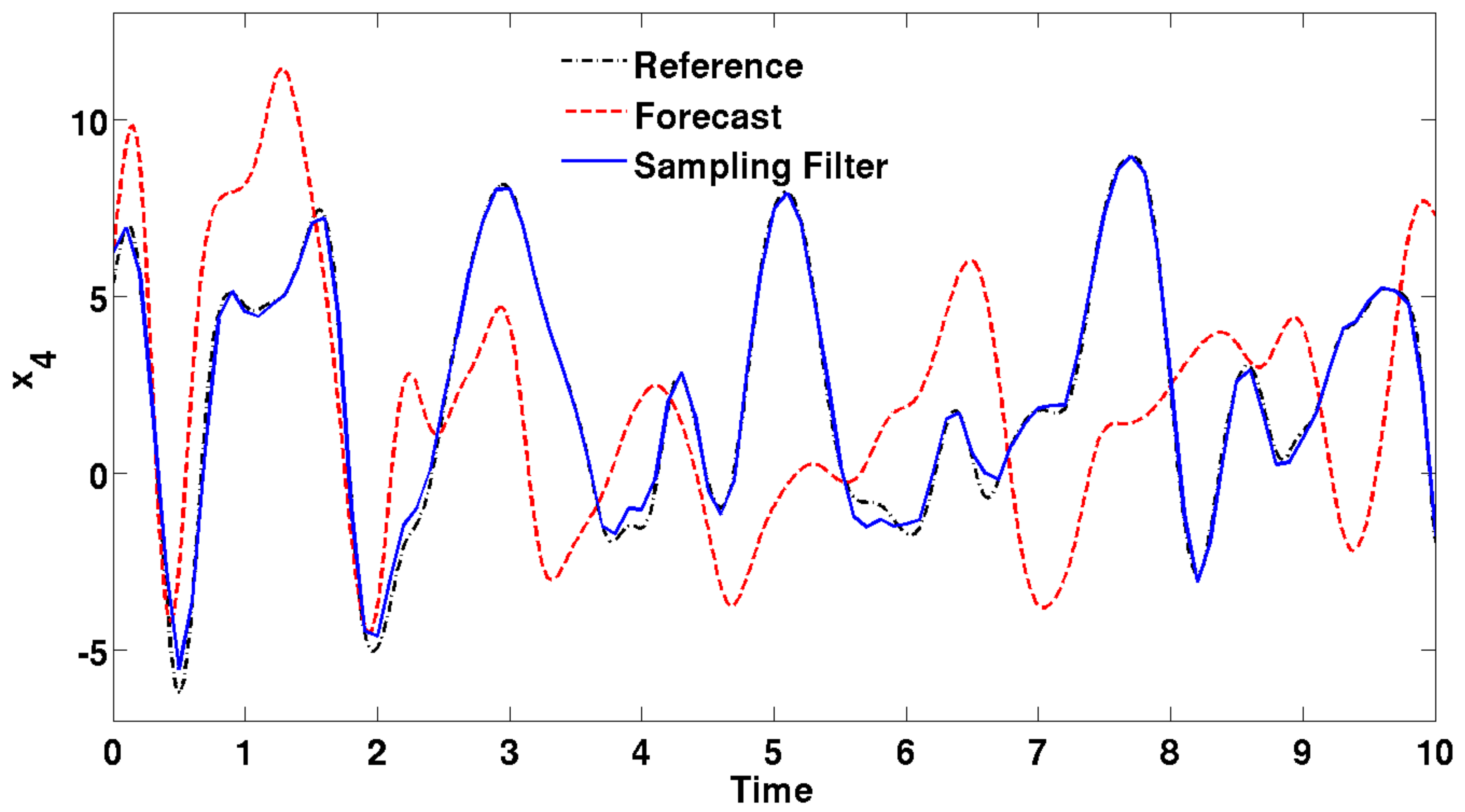}   
\label{fig:ExponentialH_2_3Stage_X4}}
\subfigure[$x_7$]{%
\includegraphics[width=0.44\linewidth]{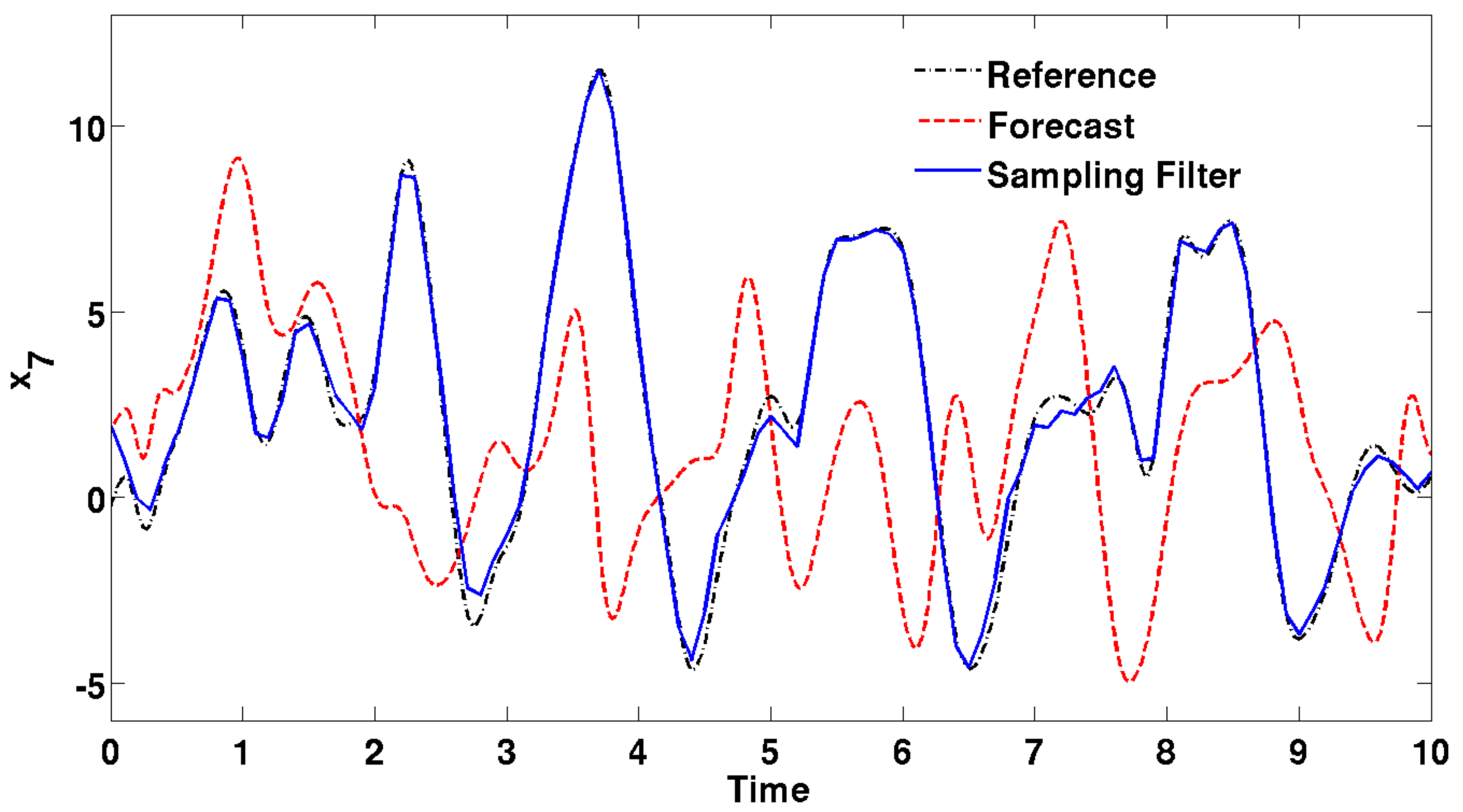}   
\label{fig:ExponentialH_2_3Stage_X7}}
\quad
\subfigure[$x_{11}$]{%
\includegraphics[width=0.44\linewidth]{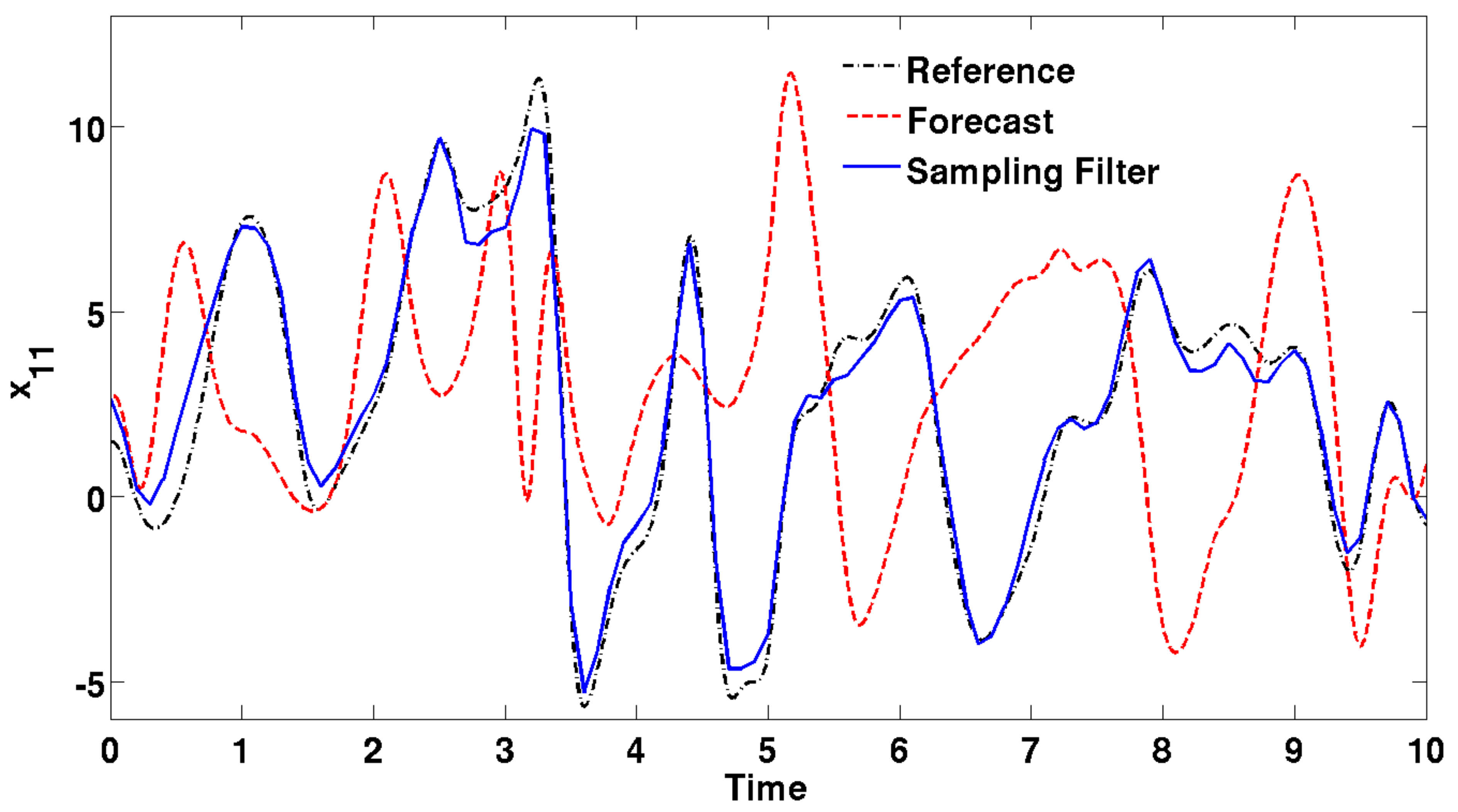}   
\label{fig:ExponentialH_2_3Stage_X11}}
\caption{Three-stage integrator is used; $h=0.01,\ m=60$. The components $x_1,x_4,x_7,x_{11}$ of the state vector $\x$ are plotted.
    The number of inter-chain steps is $30$.} 
\label{fig:ExponentialH_3_State_Variables_3Stage}
\end{figure}

Figure \ref{fig:ExponentialH_3_State_Variables_3Stage} plots the analyses obtained with the three-stage integrator sampling filter. A  large number of steps is required to achieve good results due to the large magnitude of observations. The Hilbert integrator operates at a much lower cost, and can be used to periodically check the results obtained with the three-stage integrator to safeguard against outliers.

The statistics of the results with Lorenz-96 model are summarized in Tables \ref{table:RMSE_Statistics} through \ref{table:RMSE_Statistics_Exponential_divide_by_2_Short}. The results for $100$ instances of the sampling filter, EnKF, and MLEF, over the time interval $[8,10]$, are summarized in Table \ref{table:RMSE_Statistics} and \ref{table:RMSE_Statistics_Exponential_divide_by_2_Full}. The results obtained with the exponential observation operator \eqref{eqn:Exponential_H} with $r=0.5$ are shown in Table \ref{table:RMSE_Statistics_Exponential_divide_by_2_Full}. 
In Table \ref{table:RMSE_Statistics} the columns named ``\textit{Fixed step}'' present statistics obtained from experiments with the fixed step size settings  $h=0.01$ and $m=10$. The columns named ``\textit{Different step}''  report statistics from the experiments where the work was equalized among the symplectic integrators. Tables \ref{table:RMSE_Statistics_Fixed_Step} and \ref{table:RMSE_Statistics_Exponential_divide_by_2_Short} are shorter versions of Tables \ref{table:RMSE_Statistics}, and \ref{table:RMSE_Statistics_Exponential_divide_by_2_Short} respectively; only the results of the sampling filter with fixed time step of the symplectic integrators are included and only the averages and the standard deviations over the time interval $[8,10]$ are summarized. 
%
%

   %
   %

\subsection{Shallow water model on a sphere}\label{subsec:SWE}

As a first step towards large models we test the proposed sampling filter on the shallow water model on a sphere, using linear observation operator where all components are observed.

The shallow water equations provide a simplified  model of the atmosphere which describes the essential wave propagation mechanisms found in general circulation models (GCMs)~\citep{Amik:2007}.
 The shallow water equations in spherical coordinates are given as
\begin{subequations}
\label{eqn:swe}
\begin{align}
 \frac{\partial u}{\partial t} &+ \frac{1}{a\cos \theta} \left( u \frac{\partial u}{\partial \lambda} + v \cos \theta \frac{\partial u}{\partial \theta} \right) 
                               - \left(f + \frac{u \tan \theta}{a} \right) v + \frac{g}{a \cos \theta} \frac{\partial h} {\partial \lambda} = 0, \\
 \frac{\partial v}{\partial t} &+ \frac{1}{a\cos \theta} \left( u \frac{\partial v}{\partial \lambda} + v \cos \theta \frac{\partial v}{\partial \theta} \right) 
                               + \left(f + \frac{u \tan \theta}{a} \right) u + \frac{g}{a} \frac{\partial h} {\partial \theta} = 0, \\
 \frac{\partial h}{\partial \theta} &+ \frac{1}{a \cos \theta} \left(\frac{\partial\left(hu\right)}{\partial \lambda} + \frac{\partial{\left(hv \cos \theta \right)}}{\partial \theta} \right) = 0\,.
\end{align}
\end{subequations}
The Coriolis parameter is  given by $f = 2 \Omega \sin \theta$, where  $\Omega$ is the angular speed of the rotation of the Earth, and $\theta$  is latitudinal direction. The longitudinal direction is $\lambda$. The height of the homogeneous atmosphere is represented by $h$, the zonal and meridional wind components are given by  $u$ and $v$ respectively.
The radius of the earth is $a$, and the gravitational constant is given by $g$.
The space discretization follows the unstaggered Turkel-Zwas scheme~\citep{neta:1997}. The discretization has $nlon=72$ nodes in longitudinal direction and  $nlat=36$ nodes in the latitudinal direction.
The semi-discretization in space results in the following discrete model: 
\begin{align}\label{swe:ode}
 \x_{k+1} &= \mathcal{M}_{{t_k}\rightarrow {t_{k+1}}}\left(\x_k, \theta\right) , \quad k = 0, \dots, N\,, \\
 \x_0 & = \x_0\left(\theta\right)\, .
\end{align}
 The state space vector $\x$ in \eqref{swe:ode} combines the zonal wind, the meridional wind, and the height variables into the vector $\x \in \mathbb{R}^\nvar$ with $\nvar=3\times{\rm nlat}\times{\rm nlon}$.
 The time integration is conducted using an adaptive time-stepping algorithm. 
 A reference initial condition is used to generate a reference trajectory. Synthetic observations are created from the reference trajectory by adding Gaussian noise with zero mean and fixed standard deviation for each of the 
 three components. The level of observation noise for height component is set to $1.5\%$ of the average magnitude of the reference height component in the reference initial condition.
 The level of observation noise for wind components is set to $10\%$ of the average magnitude of the reference wind component in the initial condition. 
 The initial background state is created by perturbing the reference initial condition by a Gaussian noise drawn from a modelled background error covariance matrix $\mathbf{B}_0$.
 The standard deviation of the background errors for the height component is $2\%$ of the average magnitude of the reference height component in the reference initial condition.
 The standard deviation of the background errors for the wind components is $15\%$ of the average magnitude of the reference wind component in the reference initial condition. 
The modeled version of the background error covariance, $\mathbf{B}_0$, that accounts for correlations between state variables is created as follows:
 \begin{itemize}
  \item Start with a diagonal background error covariance matrix with uncertainty levels as mentioned previously.
  \item Apply the ensemble Kalman filter for $48$ hours. Synthetic initial ensemble is created by adding zero-mean Gaussian noise to the reference initial condition with
            covariances set to the initial (diagonal) background error covariance matrix. 
  \item Decorrelate the ensemble-based covariances using a decorrelation matrix $\rho$ with decorrelation distance $L=1000\,km$.
  \item Calculate $\mathbf{B}_0$ by averaging the covariances over the last $6$ hours.
 \end{itemize}

This method of creating a synthetic initial background error covariance matrix is totally empirical, but we found that the resulting background error covariance matrix
performs well for several algorithms including 4DVAR.
Enhancing the quality of this background error covariance matrix can be done by making use of the ensembles generated 
by the sampling filter. In future work, we will investigate the possibility of estimating the background error covariances using the proposed sampling filter.

\subsection{Results for shallow water model with linear observations}\label{subsec:SWE_Results}
The assimilation time interval is 6 hours and there are hourly observations available.  The number of burn-in steps in the Markov chain is set to $50$. We use the two-stage symplectic integrator \eqref{eqn:two_stage} with step size $h=0.01$ and number of steps $m=10$. The number of inter-chain steps is $10$. 

\begin{figure}[H]
    \centering
           \subfigure[EnKF analysis: $h$]{%
           \includegraphics[width=0.44\linewidth]{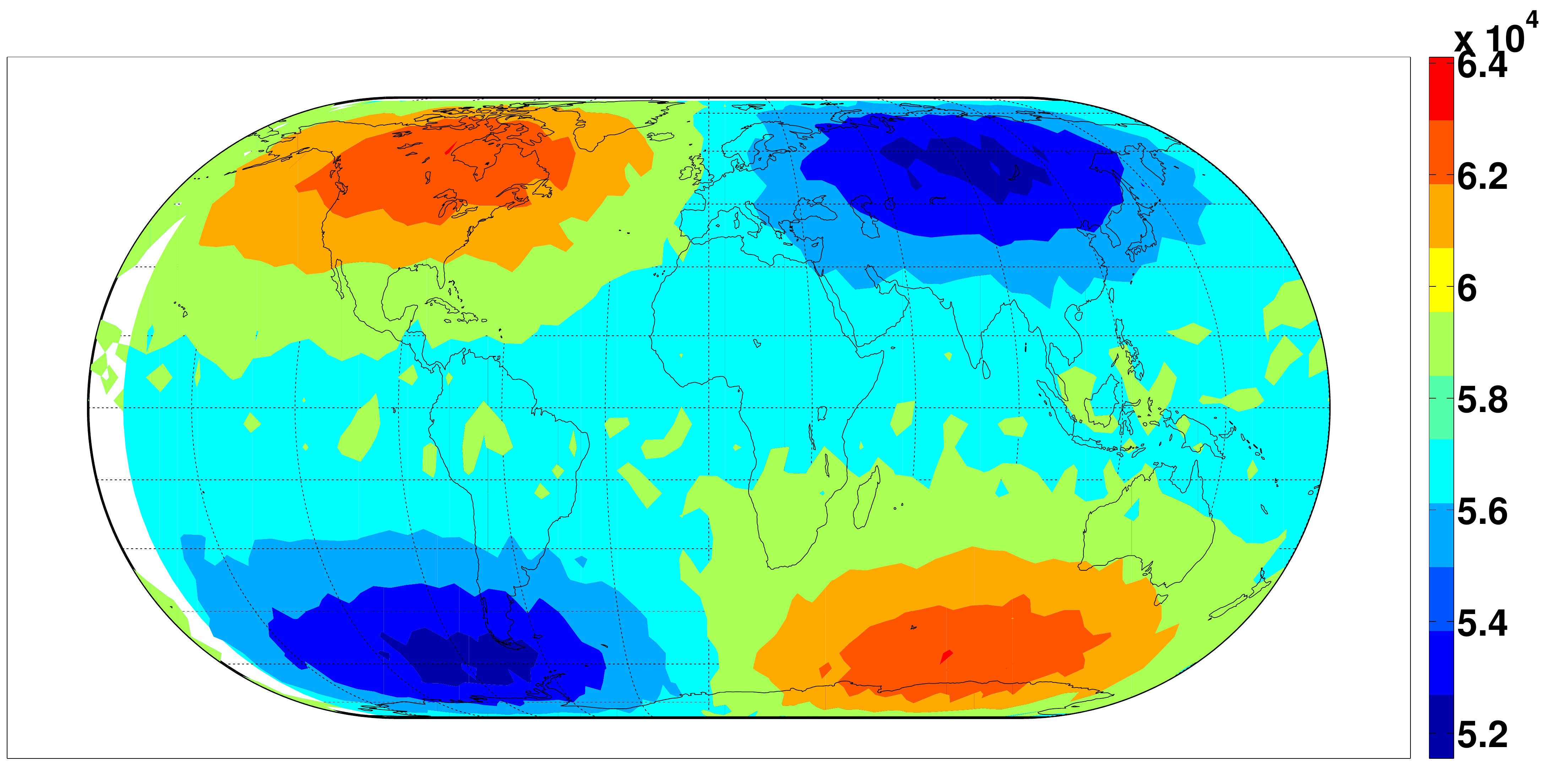}   
           \label{fig:SWE_H_EnKF}}
           \quad
           \subfigure[Sampling filter analysis: $h$]{%
           \includegraphics[width=0.44\linewidth]{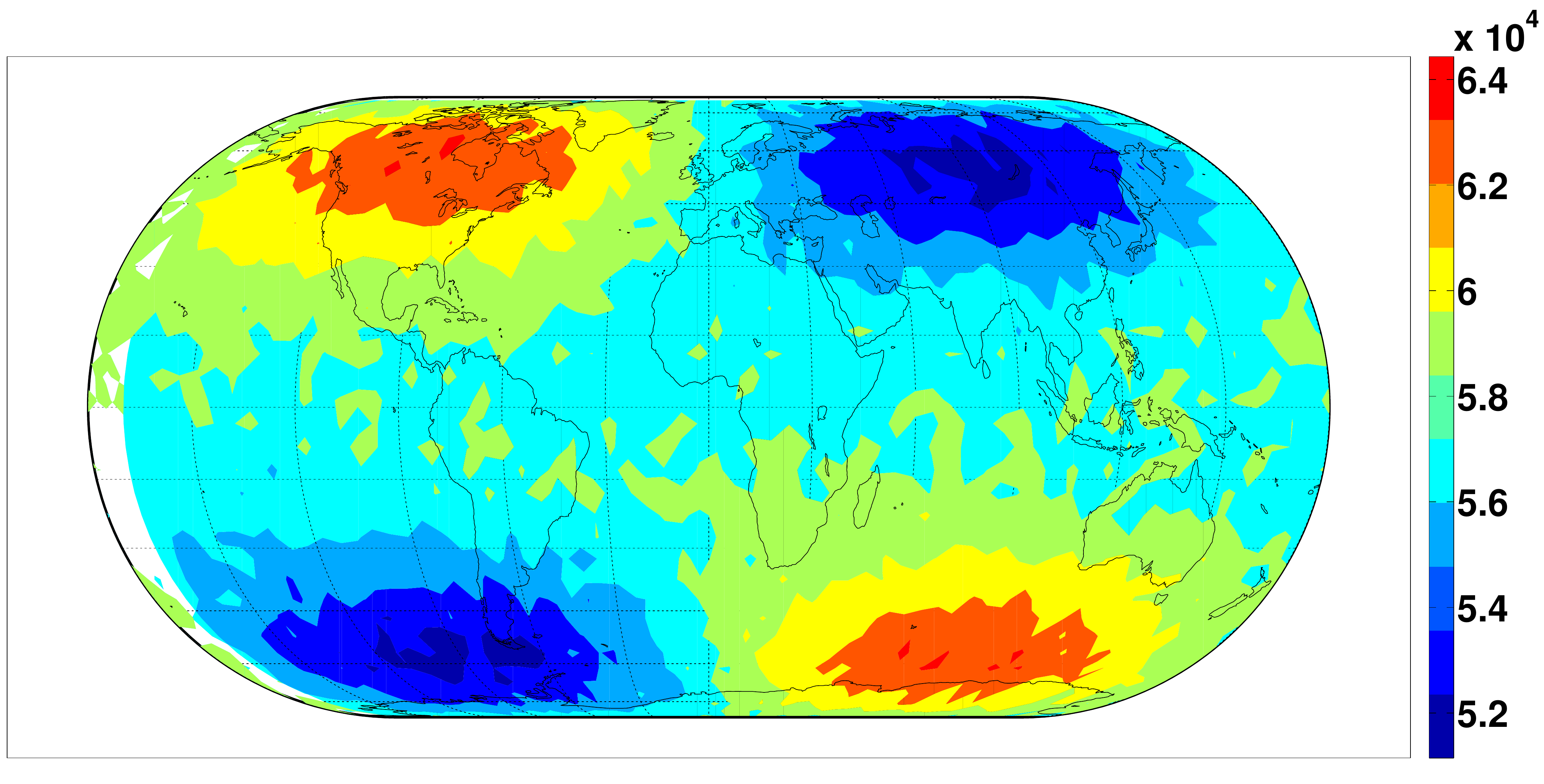}   
           \label{fig:SWE_H_HMC}}
           \subfigure[EnKF analysis: $u$]{%
           \includegraphics[width=0.44\linewidth]{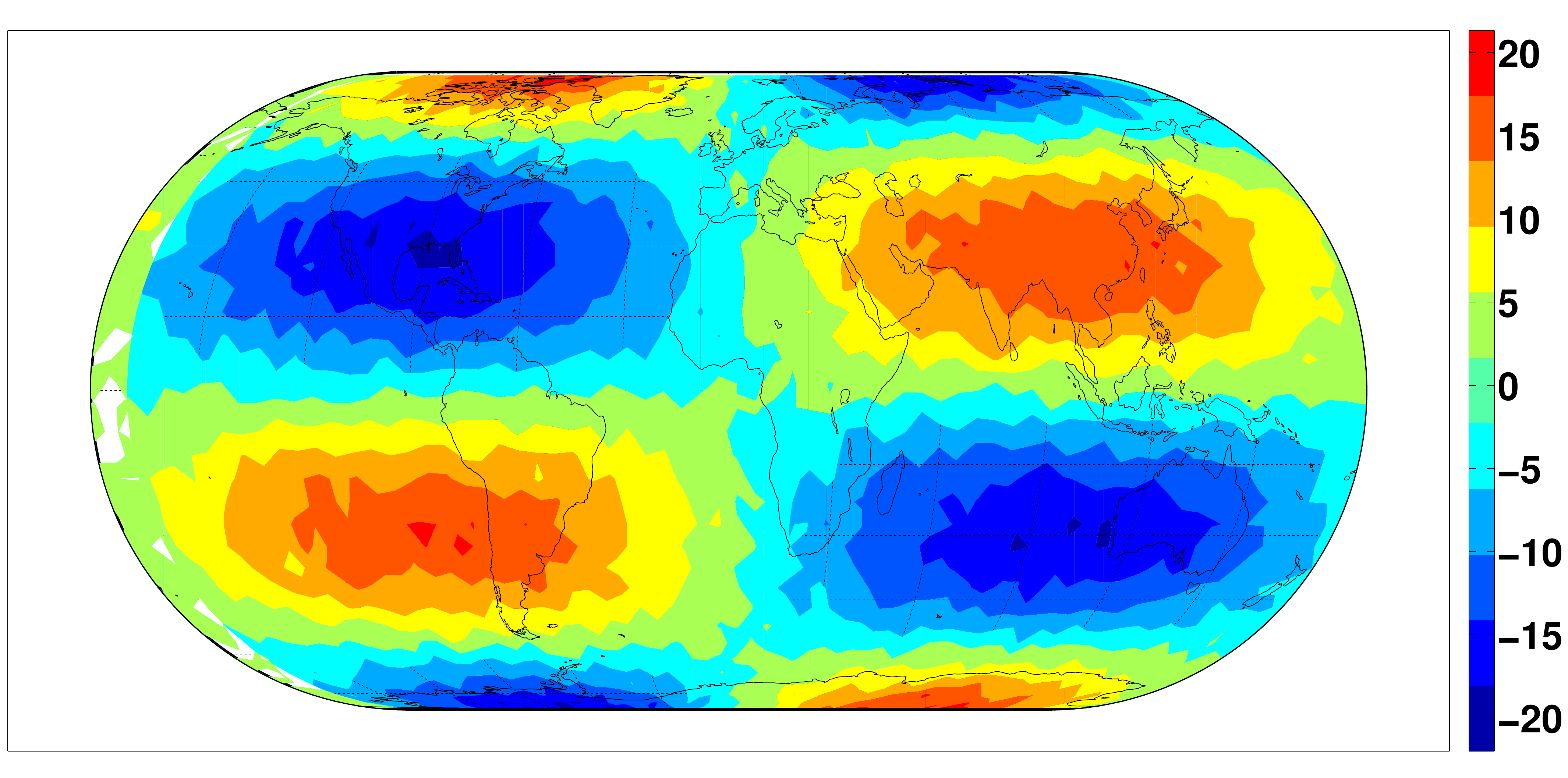}   
           \label{fig:SWE_U_EnKF}}
           \quad
           \subfigure[Sampling filter analysis: $u$]{%
           \includegraphics[width=0.44\linewidth]{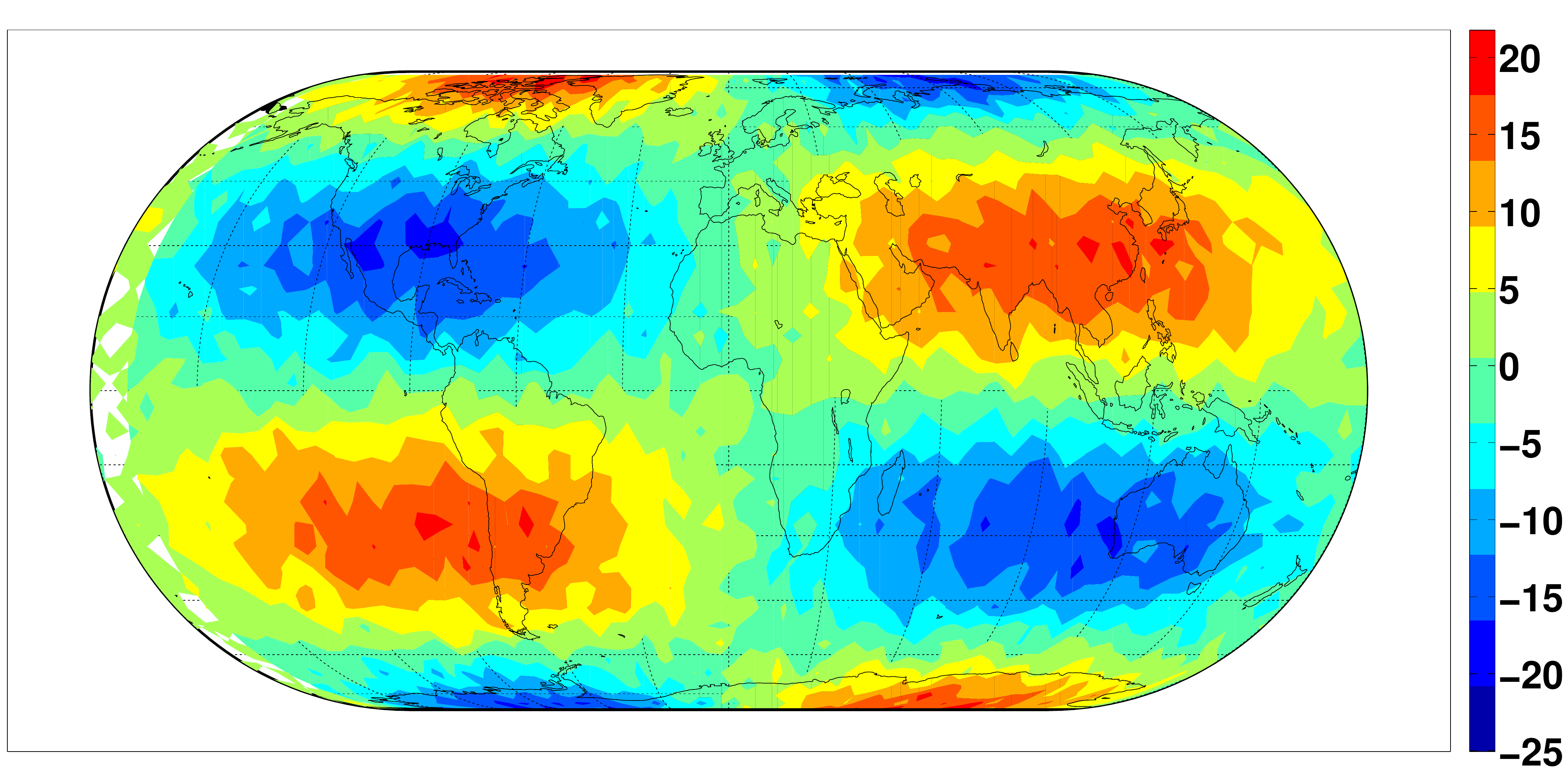}   
           \label{fig:SWE_U_HMC}}
           \subfigure[EnKF analysis: $v$]{%
           \includegraphics[width=0.44\linewidth]{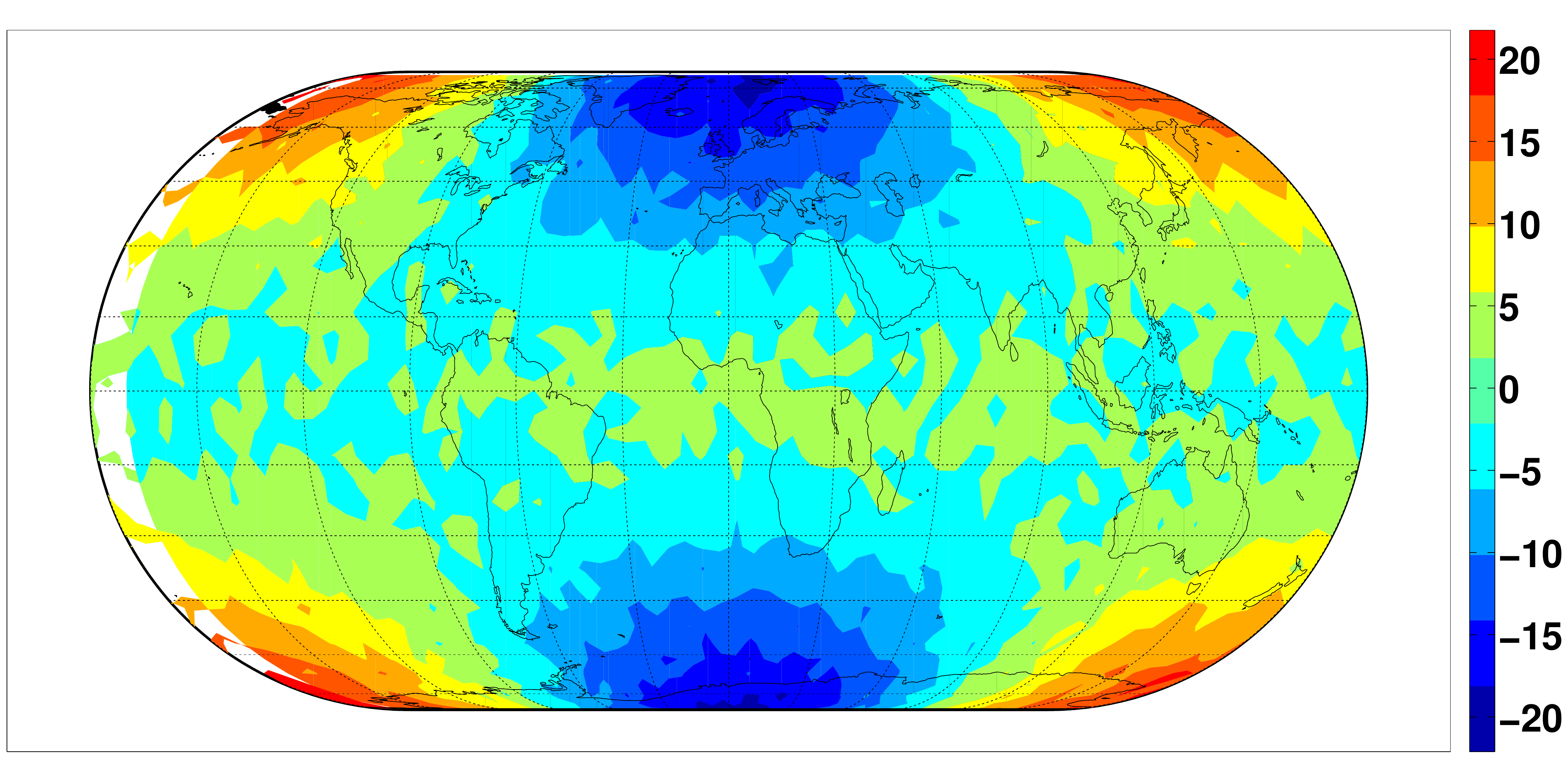}   
           \label{fig:SWE_V_EnKF}}
           \quad
           \subfigure[Sampling filter analysis: $v$]{%
           \includegraphics[width=0.44\linewidth]{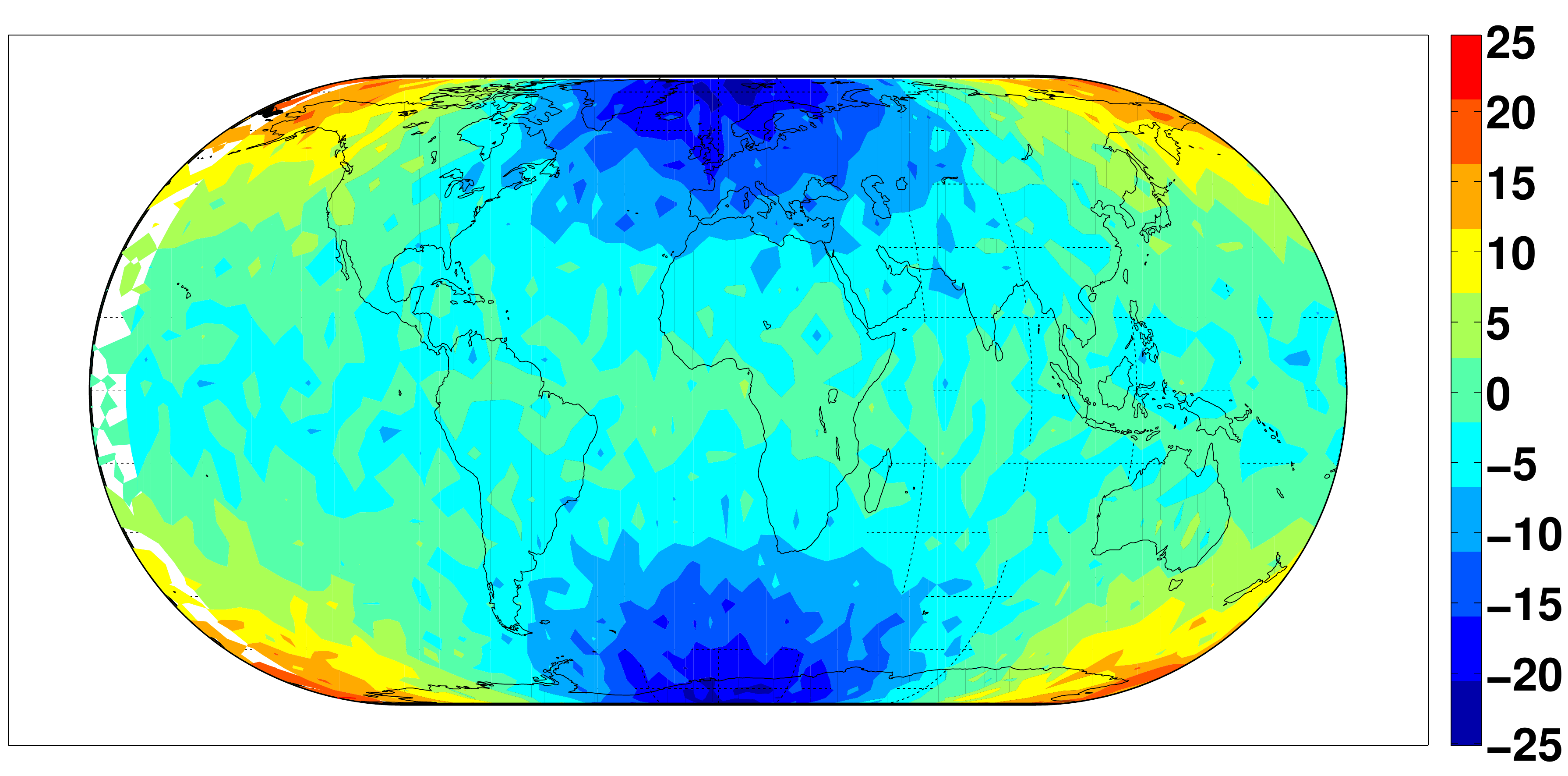}   
           \label{fig:SWE_V_HMC}}
           %
     \caption{Data assimilation results for SWE on the sphere with linear observations where all components are observed. The plotted component of the state vector is indicated under each panel.
              The analysis shown is obtained after sequential assimilation of hourly observations for seven hours. Only state at the seventh hour is plotted. The sampling filter analysis is an overage of analysis 
              states obtained from 50 instances of the sampling filter.
              The symplectic integrator used is the two-stage integrator \eqref{eqn:two_stage}. The length of the Hamiltonian trajectory is $T=mh$, with $h=0.01,\, m=10$. The number of inter-chain steps is $10$.
              } 
     \label{fig:SWE_Results}
     \end{figure}

The EnKF and the sampling filter results are shown in Figure \ref{fig:SWE_Results}. As shown in Figures \ref{fig:SWE_H_HMC}, \ref{fig:SWE_U_HMC}, and \ref{fig:SWE_V_HMC} the analysis is noisy and further tuning of the sampling filter parameters is needed in order to outperform the EnKF analysis. Parameter tuning for the sampling filter with this model will be studied in the future.  Moreover ensemble based forecast covariances need to used for all analyses to improve results.
    %
    %
 \FloatBarrier


   \begin{sidewaystable}[!h]
   \caption{RMS error statistics of experiments for assimilation time points $8\leq t\leq10$ (after filter stabilizes)} 
   \centering
   
   \scalebox{0.65}{ 
   
   \begin{tabular}{|c|c||c|c||c|c||c|c||c|c||c|c||c|c|}
   \hline
   \multirow{3}{2.5cm}{\centering \textbf{Observation Operator}}   & \multirow{3}{*}{\textbf{Statistics}} & \multicolumn{10}{|c||}{\textbf{Integrator used with sampling filter}}     & \multicolumn{2}{|c|}{\multirow{2}{*}{\textbf{Traditional Filters}}}  \\ \cline{3-12} 
 &  & \multicolumn{2}{|c||}{\textbf{Verlet}}   & \multicolumn{2}{|c||}{\textbf{Two-stage}}& \multicolumn{2}{|c||}{\textbf{Three-stage}}    & \multicolumn{2}{|c||}{\textbf{Four-stage}}     & \multicolumn{2}{|c||}{\textbf{Integrator on Hilbert space}}     & \multicolumn{2}{|c|}{\textbf{  }}\\ \cline{3-14}
 &  & \textit{Fixed step} & \textit{Different step}  & \textit{Fixed step} & \textit{Different step}  & \textit{Fixed step} & \textit{Different step}  & \textit{Fixed step} & \textit{Different step}  & \textit{Fixed step} & \textit{Different step} &    \textbf{EnKF}&\textbf{MLEF}   \\ \hline \hline
   \multirow{6}{2.5cm}{\centering Linear observation operator}     &     Min& 0.20266 &  0.192911    & 0.22185 &  0.19728     &0.218158 &  0.193404    &0.209657 &   0.194968   & 1.176385&    1.225406 &0.153021   &   1.167385     \\ \cline{2-14} 
 &     Max& 0.334077&  0.275264    & 0.371612&  0.271278    &0.345005 &  0.27841     &0.350564 &   0.262573   & 1.740247&    1.749126 &0.477702   &   5.341185     \\ \cline{2-14} 
 &     Mean     & 0.266498&  0.2247& 0.269914&  0.225638    &0.263563 &  0.230312    &0.263878 &   0.227828   & 1.413573&    1.411903 &0.305293   &   3.318066     \\ \cline{2-14} 
 &     Std& 0.024486&  0.016978    & 0.026238&  0.013978    &0.02482  &  0.017086    &0.028621 &   0.016688   & 0.097418&    0.094974 &0.091245   &   1.413271     \\ \cline{2-14} 
 & Mean$+2*$std & 0.31547 &  0.258656    & 0.32239 &  0.253594    &0.313203 &  0.264484    &0.32112  &   0.261204   & 1.608409&    1.601851 &0.487783   &   6.144607     \\ \cline{2-14} 
 & Mean$-2*$std & 0.217526&  0.190744    & 0.217438&  0.197682    &0.213923 &  0.19614     &0.206636 &   0.194452   & 1.218737&    1.221955 &0.122803   &   0.491525     \\ \hline \hline
   \multirow{6}{2.5cm}{\centering Quadratic observation operator}  &     Min& 0.293207&  0.223915    & 0.229499&  0.227357    &0.257983 &  0.225245    &0.26389  &   0.216563   & 1.803022&    1.804532 &3.340558   &   4.669170     \\ \cline{2-14} 
 &     Max& 5.437528&  5.710717    & 3.525457&  5.148642    &3.286706 &  5.412255    &3.489366 &   5.042837   & 3.449921&    3.414033 &4.555472   &   6.005837     \\ \cline{2-14} 
 &     Mean     & 4.492176&  2.802898    & 0.65887 &  1.161938    &0.577684 &  1.132138    &0.645417 &   0.80452    & 2.487458&    2.328798 &3.927612   &   5.118004 \\ \cline{2-14} 
 &     Std& 0.836105&  2.326364    & 0.828443&  1.711295    &0.662738 &  1.642162    &0.8385   &   1.295648   & 0.432504&    0.401300 &0.323017   &   0.377321     \\ \cline{2-14} 
 & Mean$+2*$std & 6.164386&  7.455626    & 2.315756&  4.584528    &1.90316  &  4.416462    &2.322417 &   3.395816   & 3.352466&    3.131399 &4.573646   &   5.872646     \\ \cline{2-14} 
 & Mean$-2*$std & 2.819966& -1.84983     &-0.998016& -2.260652    &     -0.747792 & -2.152186    &     -1.031583 &  -1.786776   & 1.62245 &    1.526198 &3.281578   &   4.363361     \\ \hline \hline
   \multirow{6}{2.5cm}{\centering Cubic observation operator}&     Min& 3.005206&  0.453015    & 0.295747&  0.475533    &0.310439 &  0.526365    &0.293922 &   0.426434   & 1.673592&    1.249802 &5.747101   &   4.720723     \\ \cline{2-14} 
 &     Max& 5.195681&  5.083948    & 3.904205&  2.310545    &2.606434 &  2.510949    &1.32128  &   2.505575   & 2.659555&    2.595746 &     11.620567   &   6.672957     \\ \cline{2-14} 
 &     Mean     & 4.159783&  1.514528    & 1.300494&  1.006574    &0.606089 &  1.235655    &0.454872 &   1.071686   & 2.213129&    2.078567 &8.768142   &   5.691641     \\ \cline{2-14} 
 &     Std& 0.382443&  0.970227    & 1.153967&  0.482539    &0.482256 &  0.519058    &0.166534 &   0.481047   & 0.220747&    0.248998 &1.61792    &   0.665145 \\ \cline{2-14} 
 & Mean$+2*$std & 4.924669&  3.454982    & 3.608428&  1.971652    &1.570601 &  2.273771    &0.78794  &   2.03378    & 2.654623&    2.576563 &     12.003982   &   7.021931     \\ \cline{2-14} 
 & Mean$-2*$std & 3.394897& -0.425926    &-1.00744 &  0.041496    &     -0.358423 &  0.197539    &0.121804 &   0.109592   & 1.771635&    1.580570 &5.532302   &   4.361352     \\ \hline \hline
   \multirow{6}{2.5cm}{\centering Absolute value observation operator}   &     Min& 0.223854&  0.215914    & 0.230685&  0.215965    &0.221819 &  0.207496    &0.213172 &   0.21736    & 1.594316&    1.350040 &0.156432   &   2.814780     \\ \cline{2-14} 
 &     Max& 3.770117&  4.278896    & 4.221607&  3.440444    &3.783911 &  3.414783    &3.255274 &   4.605238   & 3.186142&    3.809884 &0.489043   &   5.822207\\ \cline{2-14} 
 &     Mean     & 0.390026&  0.693854    & 0.488504&  0.591975    &0.514439 &  0.569576    &0.401922 &   0.761355   & 2.240063&    2.184913 &0.235906   &   4.566194     \\ \cline{2-14} 
 &     Std& 0.472502&  0.938676    & 0.70202 &  0.74757     &0.719887 &  0.721405    &0.52001  &   0.993143   & 0.410658&    0.493300 &0.101655   &   0.794186     \\ \cline{2-14} 
 & Mean$+2*$std & 1.33503 &  2.571206    & 1.892544&  2.087115    &1.954213 &  2.012386    &1.441942 &   2.747641   & 3.061379&    3.171513 &0.439216   &   6.154566     \\ \cline{2-14} 
 & Mean$-2*$std &-0.554978& -1.183498    &-0.915536& -0.903165    &     -0.925335 & -0.873234    &     -0.638098 &  -1.224931   & 1.418747&    1.198312 &0.032596   &   2.977822     \\ \hline \hline
   \multirow{6}{2.5cm}{\centering Quadratic observation operator with threshold}     &     Min& 0.256795&  0.208163    & 0.25226 &  0.201484    &0.253754 &  0.207151    &0.229083 &   0.203665   & 1.348305&    1.326491 & 2.018643   &   4.691330     \\ \cline{2-14} 
 &     Max& 4.585849&  4.775978    & 0.512091&  3.665279    &0.402913 &  2.117683    &0.415055 &   0.283818   & 1.91278 &    3.690276 &3.296156   &   7.047109     \\ \cline{2-14} 
 &     Mean     & 3.406461&  1.920009    & 0.303039&  0.448207    &0.295141 &  0.290576    &0.303422 &   0.240194   & 1.579326&    1.853404 &2.801247   &   5.918242     \\ \cline{2-14} 
 &     Std& 1.067004&  1.927817    & 0.044306&  0.686636    &0.03201  &  0.235245    &0.037552 &   0.016292   & 0.119657&    0.456134 &0.311357   &   0.603508     \\ \cline{2-14} 
 & Mean$+2*$std & 5.540469&  5.775643    & 0.391651&  1.821479    &0.359161 &  0.761066    &0.378526 &   0.272778   & 1.81864 &    2.765672 &3.423961   &   7.125258     \\ \cline{2-14} 
 & Mean$-2*$std & 1.272453& -1.935625    & 0.214427& -0.925065    &0.231121 & -0.179914    &0.228318 &   0.20761    & 1.340012&    0.941136 &2.178533   &   4.711226\\ \hline \hline
   \multirow{6}{2.5cm}{\centering Exponential observation operator with $r=0.2$}     &     Min& 0.318723&  0.315953    & 0.313345&  0.296624    & 0.309291&  0.277073    &0.321489 &   0.291672   & 1.42193 &    1.408169 &2.028516   &   4.460268\\ \cline{2-14} 
 &     Max& 2.563976&  3.075637    & 0.688663&  0.43783     & 0.643646&  0.612475    &0.674257 &   0.439936   & 1.97735 &    2.081439 &3.911317   &   7.736204     \\ \cline{2-14} 
 &     Mean     & 0.433829&  0.453889    & 0.408104&  0.348357    & 0.405423&  0.349503    &0.408271 &   0.348964   & 1.610456&    1.661009 &3.153979   &   5.713367     \\ \cline{2-14} 
 &     Std& 0.267084&  0.320384    & 0.05718 &  0.028835    & 0.055325&  0.041517    &0.05946  &   0.02787    & 0.10526 &    0.139997 &0.551218   &   0.934198     \\ \cline{2-14} 
 & Mean$+2*$std & 0.967997&  1.094657    & 0.522464&  0.406027    & 0.516073&  0.432537    &0.527191 &   0.404704   & 1.820976&    1.941003 &4.256415   &   7.581764     \\ \cline{2-14} 
 & Mean$-2*$std &-0.100339& -0.186879    & 0.293744&  0.290687    & 0.294773&  0.266469    &0.289351 &   0.293224   & 1.399936&    1.381015 &2.051543   &   3.844971\\ \hline \hline
   \end{tabular} 
   } 
   
   \label{table:RMSE_Statistics}
   \end{sidewaystable}

   \begin{table}[!h]
   \scriptsize
   \centering
   \caption{RMS error statistics of experiments for assimilation time points $8\leq t\leq10$ (after filter stabilizes). The exponential observation operator with factor $r=0.5$ is used.}

   \begin{tabular}{|c|c|c|}   
   \hline
   \multirow{2}{*}{\textbf{Statistics}} & \multicolumn{2}{|c|}{\textbf{Integrator used with sampling filter}}\\ \cline{2-3}
    &  \textbf{Three-stage}; $h=0.01,\ m=60$  &  \textbf{Hilbert} ; $h=0.001,\ m=30$ \\ \hline
 Min&0.304178   &1.234498\\ \hline
 Max&2.671971   &2.350684\\ \hline
 Mean     &0.439776   &1.699096\\ \hline
 Std&0.274643   &0.250088\\ \hline 
   Mean$+2*$std &0.989062   &2.199272\\ \hline
   Mean$-2*$std &     -0.109510   &1.198920\\ \hline
   \end{tabular} 
   \label{table:RMSE_Statistics_Exponential_divide_by_2_Full}
   \end{table}

   \begin{table}[!h]
   \centering
   \scriptsize
   \caption{RMS error statistics of experiments for assimilation time points $8\leq t\leq10$ (after filter stabilizes). Fixed step size $h=0.01,\ m=10$.}

   \begin{tabular}{|l|c|c|c|c|c|c|c|c|}   
   \hline
   \multirow{2}{2.2cm}{\textbf{Observation Operator} }     & \multirow{2}{*}{\textbf{Statistics}} & \multicolumn{5}{|c|}{\textbf{Integrator used with sampling filter}}     & \multicolumn{2}{|c|}{\textbf{Traditional Filters}}   \\ \cline{3-9}
     &  &   \textbf{Verlet}   & \textbf{Two-stage}  & \textbf{Three-stage}& \textbf{Four-stage} & \textbf{Hilbert}    & \textbf{EnKF}   & \textbf{MLEF}    \\ \hline
   \multirow{2}{2.2cm}{Linear}     &     Mean     & 0.266498& 0.269914&0.263563 &0.263878 & 1.413573&0.305293   &   3.318066 \\ \cline{2-9}
     &     Std& 0.024486& 0.026238&0.02482  &0.028621 & 0.097418&0.091245   &   1.413271 \\ \hline
   \multirow{2}{2.2cm}{Quadratic}  &     Mean     & 4.492176& 0.65887 &0.577684 &0.645417 & 2.487458&3.927612   &   5.118004   \\ \cline{2-9}
     &     Std& 0.836105& 0.828443&0.662738 &0.8385   & 0.432504&0.323017   &   0.377321 \\ \hline
   \multirow{2}{2.2cm}{Cubic}&     Mean     & 4.159783& 1.300494&0.606089 &0.454872 & 2.213129&8.768142   &   5.691641 \\ \cline{2-9}
     &     Std& 0.382443& 1.153967&0.482256 &0.166534 & 0.220747&1.61792    &   0.665145  \\ \hline
   \multirow{2}{2.2cm}{Absolute value}   &     Mean     & 0.390026& 0.488504&0.514439 &0.401922 & 2.240063&0.235906   &   4.566194 \\ \cline{2-9}
     &     Std& 0.472502& 0.70202 &0.719887 &0.52001  & 0.410658&0.101655   &   0.794186 \\ \hline
   \multirow{2}{2.2cm}{Quadratic with threshold}     &     Mean     & 3.406461& 0.303039&0.295141 &0.303422 & 1.579326&2.801247   &   5.918242 \\ \cline{2-9}
     &     Std& 1.067004& 0.044306&0.03201  &0.037552 & 0.119657&0.311357   &   0.603508 \\ \hline
   \multirow{2}{2.2cm}{Exponential with $r=0.2$}     &     Mean     & 0.433829& 0.408104& 0.405423&0.408271 & 1.610456&3.153979   &   5.713367 \\ \cline{2-9}
     &     Std& 0.267084& 0.05718 & 0.055325&0.05946  & 0.10526 &0.551218   &   0.934198 \\ \hline
   \end{tabular} 
   \label{table:RMSE_Statistics_Fixed_Step}
   \end{table}

   \begin{table}[!h]
   \centering
   \scriptsize
   \caption{RMS error statistics of experiments for assimilation time points $8\leq t\leq10$ (after filter stabilizes). The exponential observation operator with factor $r=0.5$ is used.}

   \begin{tabular}{|c|c|c|}   
   \hline
   \multirow{2}{*}{\textbf{Statistics}} & \multicolumn{2}{|c|}{\textbf{Integrator used with sampling filter}}\\ \cline{2-3}
    &  \textbf{Three-stage}; $h=0.01,\ m=60$  &  \textbf{Hilbert} ; $h=0.001,\ m=30$ \\ \hline
 Mean     &0.439776   &1.699096\\ \hline
 Std&0.274643   &0.250088\\ \hline 
   \end{tabular} 
   \label{table:RMSE_Statistics_Exponential_divide_by_2_Short}
   \end{table}
   
\section{Conclusions and Future Work}\label{sec:conclusions}
%
This paper proposes a sampling filter for data assimilation where the analysis scheme is replaced by sampling directly from the posterior distribution. A Hybrid MCMC technique is employed to generate a representative analysis ensemble at each time. The sampling filter avoids the need to develop tangent linear or adjoint models of the model solution operator. The sampling filter can work with highly nonlinear observation operators and provides analysis ensembles that describe non-Gaussian posterior probability densities. The mean of the generated posterior ensemble provides the analysis (a minimum variance estimate of the state). The ensemble covariance offers an estimate of the analysis error covariance matrix and can be used to quantify the uncertainty associated with the analysis state.
The implementation does not require the construction of full covariance matrices, which makes the method attractive for large scale data assimilation problems with operational models and complex observation operators.

Numerical experiments are carried out with the Lorenz-96 model with several observation operators with different levels of non-linearity and smoothness. The sampling filter competes with EnKF for linear observations. For nonlinear observations the results are very promising, and the new filter outperforms both EnKF and MLEF. In addition, sampling filter continues to produce satisfactory results in cases where EnKF and MLEF fail.

Large scale ensemble filtering data assimilation problems are typically run on large parallel machines.
One important challenge is the failure of subsets of nodes, which terminates some of the ensemble member runs, and leads to fewer 
ensemble members being available at the next time. Over several cycles the number of ensemble members can decrease considerably.
The sampling strategy proposed herein can be used to replace dead ensemble members in any parallel implementation of the EnKF. 
In addition, the sampling filter can be used in combination with classical filters by building analysis ensembles that have members 
given by EnKF analysis (these members retain the history of the system) mixed with sampling members (which are consistent with the posterior probability density, but add new directions to explore and can therefore avoid filter divergence).

The computational performance of the sampling filter depends on tuning its parameters, especially the symplectic integration time step and the number of steps taken in the Markov chain between successive accepted ensemble members. Future work will focus on refining the strategies for parameter tuning  in the context of large operational models at high-resolution. We also plan to perform a side-by-side comparison between the proposed filter and the implicit sampling filter.

%


\section*{Acknowledgments}
This work was supported in part by awards NSF CCF--1218454 and AFOSR FA9550--12--1--0293--DEF, and by the Computational Science Laboratory at Virginia Tech.

%

%
\begin{appendices}
\section{Symplectic numerical integrators}\label{sec:numerical-integrators}

Here we present the five numerical integrators employed in this work. 
 We start with the standard position Verlet integrator in \ref{subsubsec:Position_Verlet}. The results of the standard Verlet are very sensitive to the choice of the time step. Three higher order integrators namely, two-stage (\ref{subsubsec:Two_Stage}), three-stage (\ref{subsubsec:Three_Stage}),
 and four-stage  (\ref{subsubsec:Four_Stage}) position splitting integrators, are taken from ~\citep{blanes2013numerical}. These higher-order integrators lead to filters that are more stable and efficient than Verlet. The last integrator tested (\ref{subsubsec:Hilbert_Integrator}) is from \cite{beskos2011hybrid} and is designed to work efficiently in infinite dimensional state spaces, and to avoid problems resulting from subtracting infinitely large numbers related to the total energy of the Hamiltonian system for infinite
 dimensional state spaces.
 
All the integrators are applied to a Hamiltonian system of the form \eqref{eqn:hamiltonian_vector_dynamics} \cite{sanz2014Markov,sanz1994numerical}. 
 
 \subsection{Position Verlet integrator}\label{subsubsec:Position_Verlet}
   
   One step of the position Verlet algorithm advances the solution of the Hamiltonian equations \eqref{eqn:hamiltonian_vector_dynamics} from time $t_k$ to time
   $t_{k+1} = t_k + h$ as follows~\cite{sanz2014Markov}:
   \begin{subequations}
   \label{eqn:Verlet}
   \begin{eqnarray}
\x_{k+1/2} &=&  \x_k + \frac{h}{2}\, \mathbf{M}^{-1}\, \p_k     \,,   \\
\p_{k+1}   &=&  \p_k - h \,\nabla_\x \mathcal{J}(\x_{k+1/2})  \,,   \\
\x_{k+1}   &=&  \x_{k+1/2} + \frac{h}{2}\, \mathbf{M}^{-1} \,\p_{k+1}. 
   \end{eqnarray}
   \end{subequations} 

   The optimal time step $h$ is   $h \propto {(1/{\nvar})}^{1/4}$ ~\cite{beskos2013optimal}.
   The experiments show that the step size should be small (close to zero) to make this integrator stable. It may still fail
   for high dimensionality and whenever complications are present in the target distributions.    
   The weakness of this simple integrator is illustrated in our numerical experiments with highly nonlinear observation operators.
   
 \subsection{Two-stage integrator}\label{subsubsec:Two_Stage}
   One step of the two-stage algorithm advances the solution of the Hamiltonian equations \eqref{eqn:hamiltonian_vector_dynamics} from time $t_k$ to time $t_{k+1} = t_k + h$
   as follows~\cite{blanes2013numerical}:
   \begin{subequations}
   \label{eqn:two_stage}
   \begin{eqnarray}
\x_1     &=& \x_k + (a_1 h) \mathbf{M}^{-1} \p_k    \,, \\ 
\p_1     &=& \p_k - (b_1 h) \nabla_\x \mathcal{J}(\x_1)   \,, \\ 
\x_2     &=& \x_1 + (a_2 h) \mathbf{M}^{-1} \p_1    \,, \\ 
\p_{k+1} &=& \p_1 - (b_1 h) \nabla_\x \mathcal{J}(\x_2)   \,, \\ 
\x_{k+1} &=& \x_2 + (a_2 h) \mathbf{M}^{-1} \p_{k+1}\,,  
   \end{eqnarray}
   \end{subequations} 
   where $a_1=0.21132$, $a_2=1-2a_1$, and $b_1=0.5$.
   The stability of this time integrator is achieved for time step that lies in the interval $\left(0 ,\, 2.6321480259\right)$ (units), that is, $h$ should be chosen  such that $0 < h < 2.6321480259$~\cite{blanes2013numerical}.

\subsection{Three-stage integrator}\label{subsubsec:Three_Stage}
   One step of the three-stage algorithm advances the solution of the Hamiltonian equations  \eqref{eqn:hamiltonian_vector_dynamics}
   from time $t_k$ to time $t_{k+1} = t_k + h$ by the set of equations~\cite{blanes2013numerical}:
   \begin{subequations}
   \label{eqn:three_stage}
   \begin{eqnarray}
\x_1     &=& \x_k + (a_1 h) \mathbf{M}^{-1} \p_k     \,, \\ 
\p_1     &=& \p_k - (b_1 h) \nabla_\x \mathcal{J}(\x_1)    \,, \\ 
\x_2     &=& \x_1 + (a_2 h) \mathbf{M}^{-1} \p_1     \,, \\ 
\p_2     &=& \p_1 - (b_2 h) \nabla_\x \mathcal{J}(\x_2)    \,, \\ 
\x_3     &=& \x_2 + (a_2 h) \mathbf{M}^{-1} \p_2     \,, \\ 
\p_{k+1} &=& \p_2 - (b_1 h) \nabla_\x \mathcal{J}(\x_3)    \,, \\ 
\x_{k+1} &=& \x_3 + (a_1 h) \mathbf{M}^{-1} \p_{k+1} \,,
   \end{eqnarray}
   \end{subequations}
   where: $a_1=0.11888010966548$, $a_2=0.5-a_1$, $b_1=0.29619504261126$, and $b_2=1 - 2 b_1$.
   
   The stability interval of the time step associated with this time integrator is of length $\approx 4.67$, that is, $h$ should be chosen  such that $0 < h < 4.67$~\cite{blanes2013numerical}.  
   
 \subsection{Four-stage integrator}\label{subsubsec:Four_Stage}
   One step of the four-stage algorithm advances the solution of the Hamiltonian equations  \eqref{eqn:hamiltonian_vector_dynamics} from time $t_k$ to time
   $t_{k+1} = t_k + h$ as follows~\cite{blanes2013numerical}:
   \begin{subequations}
   \label{eqn:four_stage}
   \begin{eqnarray}
\x_1     &=& \x_k + (a_1 h) \mathbf{M}^{-1} \p_k    \,, \\ 
\p_1     &=& \p_k - (b_1 h) \nabla_\x \mathcal{J}(\x_1)   \,, \\ 
\x_2     &=& \x_1 + (a_2 h) \mathbf{M}^{-1} \p_1    \,, \\ 
\p_2     &=& \p_1 - (b_2 h) \nabla_\x \mathcal{J}(\x_2)   \,, \\ 
\x_3     &=& \x_2 + (a_3 h) \mathbf{M}^{-1} \p_2    \,, \\ 
\p_3     &=& \p_2 - (b_2 h) \nabla_\x \mathcal{J}(\x_3)   \,, \\ 
\x_4     &=& \x_3 + (a_2 h) \mathbf{M}^{-1} \p_3    \,, \\ 
\p_{k+1} &=& \p_3 - (b_1 h) \nabla_\x \mathcal{J}(\x_4)   \,, \\ 
\x_{k+1} &=& \x_4 + (a_1 h) \mathbf{M}^{-1} \p_{k+1}\,,
   \end{eqnarray}
   \end{subequations}
   where: $a_1=0.071353913450279725904$, $a_2=0.268458791161230105820$, $a_3=1-2a_1-2a_2$, $b_1=0.1916678$, and $b_2=0.5-b_1$.
   
   This integrator has a stability interval of length $\approx 5.35$, that is, $h$ should be chosen  such that $0 < h < 5.35$~\cite{blanes2013numerical}. The time here has unspecified units.                    
                    Generally speaking, the high order integrators (\ref{eqn:two_stage}, \ref{eqn:three_stage}, \ref{eqn:four_stage}), provide more favorable and wider stability ranges for the time step.
                    For more on the stability intervals of the time step settings of these high-order integrators, see~\cite{blanes2013numerical}.

\subsection{General integrator defined on Hilbert space}\label{subsubsec:Hilbert_Integrator}
  One step of the Hilbert integrator advances the solution of the Hamiltonian equations  \eqref{eqn:hamiltonian_vector_dynamics} from time $t_k$ to time
   $t_{k+1} = t_k + h$ as follows~\cite{beskos2011hybrid}:
   \begin{subequations}
   \label{eqn:hilbert_integrator}
   \begin{eqnarray}
\p_1     &=& \p_k -\frac{h}{2} \mathbf{M}^{-1} \nabla_\x \mathcal{J}(\x_k) \,, \\ 
\x_{k+1} &=& \cos{(h)} \x_k + \sin{(h)} \p_1 \,, \\ 
\p_2     &=& -\sin{(h)} \x_k + \cos{(h)} \p_1\,, \\ 
\p_{k+1} &=& \p_2 - \frac{h}{2} \mathbf{M}^{-1} \nabla_\x \mathcal{J}(\x_{k+1}).
   \end{eqnarray}
   \end{subequations}
   As with the standard position Verlet integrator  the selection
   criterion of step size is not precisely defined,  however, it is designed to work with finite (non-zero) steps in infinite dimensional settings. Numerical results presented 
   in Section \ref{sec:Results} show that with careful tuning this integrator provides satisfactory results. 

\end{appendices}



\end{document}

%% file: logo.tex
\thispagestyle{empty}
\setcounter{page}{0}

\begin{Huge}
\begin{center}
Computer Science Technical Report CSTR-{4/2014} \\
\today
\end{center}
\end{Huge}
\vfil
\begin{huge}
\begin{center}
{\tt Ahmed Attia and Adrian Sandu}
\end{center}
\end{huge}

\vfil
\begin{huge}
\begin{it}
\begin{center}
``{\tt A Hybrid Monte Carlo Sampling Filter for Non-Gaussian Data Assimilation}''
\end{center}
\end{it}
\end{huge}
\vfil

\begin{large}
\begin{center}
Computational Science Laboratory \\
Computer Science Department \\
Virginia Polytechnic Institute and State University \\
Blacksburg, VA 24060 \\
Phone: (540)-231-2193 \\
Fax: (540)-231-6075 \\ 
Email: \url{sandu@cs.vt.edu} \\
Web: \url{http://csl.cs.vt.edu}
\end{center}
\end{large}

\vspace*{1cm}

\begin{tabular}{ccc}
\includegraphics[width=2.5in]{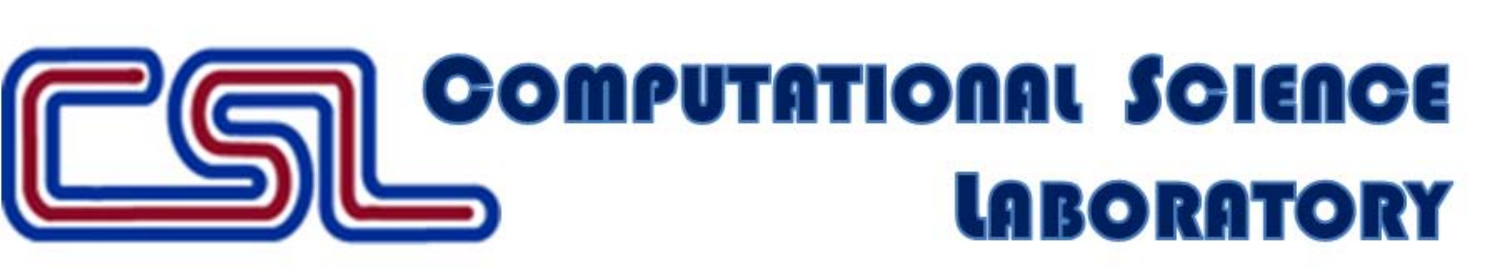}
&\hspace{2.5in}&
\includegraphics[width=2.5in]{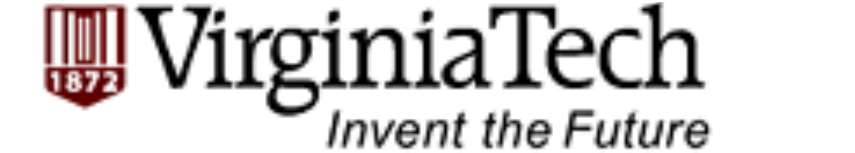} \\
{\bf\em Innovative Computational Solutions} &&\\
\end{tabular}

\newpage